\documentclass[11pt,english,oneside]{report}
\usepackage[a4paper,left=1.5in,right=1in,top=1in,bottom=1in]{geometry}
\usepackage{fancyhdr}
\newcommand{\changefont}{%
    \fontsize{9}{10}\selectfont
}
\usepackage[T1]{fontenc}
\usepackage[utf8]{inputenc}
\usepackage{graphicx}
\usepackage[space]{grffile}
\usepackage{babel}
\usepackage{float}
\usepackage{textcomp}
\usepackage{epsfig}
\usepackage{epstopdf}
\usepackage{url}
\usepackage{amsmath}
\usepackage{graphicx}
\usepackage{setspace}
\PassOptionsToPackage{normalem}{ulem}
\usepackage{ulem}
\makeatletter
\def\env@cases{%
  \let\@ifnextchar\new@ifnextchar
  \left\lbrace
  \def\arraystretch{0.8}%
  \array{l@{\quad}l@{}}
}
\providecommand{\tabularnewline}{\\}

\numberwithin{figure}{section}
\numberwithin{equation}{section}
\numberwithin{table}{section}

\usepackage{afterpage}

\renewcommand\[{\begin{equation}}
\renewcommand\]{\end{equation}}

\usepackage{xcolor}
\usepackage{hyperref}
\usepackage[all]{hypcap}
\hypersetup{
    colorlinks,
    linkcolor={red!50!black},
    citecolor={blue!50!black},
    urlcolor={blue!80!black}
}

\usepackage{colortbl}

\usepackage{tikz}
\usetikzlibrary{calc}
\usepackage{bm}
\usepackage{xcolor}
\usetikzlibrary{calc}
\usetikzlibrary{shapes.misc}
\tikzset{cross/.style={cross out, draw=black, minimum size=2cm,scale=0.3, inner sep=0pt, outer sep=0pt},
cross/.default={1pt}}
\usepackage[titletoc]{appendix}
\renewcommand{\arraystretch}{0.75}
\usepackage{listings}
\usepackage{tocloft}
\fancypagestyle{plain}{%
  \fancyhf{}%
  \fancyfoot[C]{\footnotesize\thepage}%
}
\def\l@figure{\@dottedtocline{1}{1.5em}{3em}}
\def\l@table{\@dottedtocline{1}{1.5em}{3em}}
\@ifundefined{showcaptionsetup}{}{%
 \PassOptionsToPackage{caption=false}{subfig}}
\usepackage{subfig}

\newcommand\Prefix[3]{\vphantom{#3}#1#2#3}

\makeatother
\usepackage{listings}

\usepackage{cleveref}
\begin{document}
\clearpage 

\thispagestyle{empty}

\begin{center}
ÉCOLE POLYTECHNIQUE DE MONTRÉAL
\par\end{center}

\begin{center}
ELECTRICAL ENGINEERING DEPARTMENT
\par\end{center}

\begin{center}
AUTOMATION SECTION
\par\end{center}

~

\begin{center}
~\\
~\\
\includegraphics[scale=0.6]{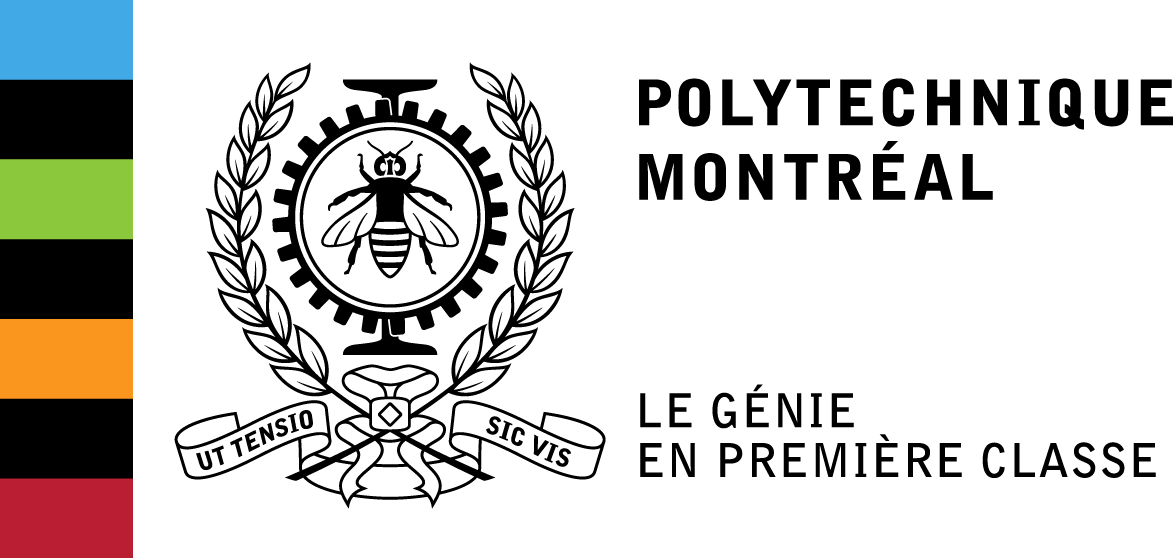}
\par\end{center}

\begin{center}
~\\
~~\\
\par\end{center}

\begin{center}
\textbf{\LARGE{}\uline{DESIGN OF A TRAJECTORY TRACKING CONTROLLER FOR A NANOQUADCOPTER}}
\par\end{center}{\LARGE \par}

\begin{center}
~~\\
~\\
~\\
~\\
{\small Luis, C., \& Le Ny, J. (August, 2016). \textit{Design of a Trajectory Tracking Controller for a Nanoquadcopter}. Technical report, Mobile Robotics and Autonomous Systems Laboratory, Polytechnique Montreal.}
\par\end{center}

~\\
~\\
~\\
~~~~~~~Author:~~~~~~~~~~~~~~~~~~~~~~~~~~~~~~~~~~~~~~~~~~~~~~~~~~~~~~~~~~~~~~~~~~~~~~~~~~~~~~~~~~~~~Supervisor:

\begin{flushleft}
Carlos Luis~~~~~~~~~~~~~~~~~~~~~~~~~~~~~~~~~~~~~~~~~~~~~~~~~~~~~~~~~~~~~~~~~~~~~~~~~~~~~~~Jérôme
Le Ny
\par\end{flushleft}

\begin{abstract}
\thispagestyle{plain}
\pagenumbering{roman}
\setcounter{page}{2}

The primary purpose of this study is to investigate the system modeling
of a nanoquadcopter as well as designing position and trajectory control algorithms,
with the ultimate goal of testing the system both in simulation and on a real
platform.

The open source nanoquadcopter platform named Crazyflie 2.0 was chosen
for the project. The first phase consisted in the development of a
mathematical model that describes the dynamics of the quadcopter. Secondly,
a simulation environment was created to design two different control architectures: cascaded PID position tracker and LQT trajectory tracker. Finally, the implementation
phase consisted in testing the controllers on the chosen platform and comparing their performance in trajectory tracking.

Our simulations agreed with the experimental results, and further refinement of the model is proposed as future work through closed-loop model identification techniques. The results show that the LQT controller performed better at tracking trajectories, with RMS errors in position up to four times smaller than those obtained with the PID. LQT control effort was greater, but eliminated the high control peaks that induced motor saturation in the PID controller. The LQT controller was also tested using an ultra-wide band two-way ranging system, and comparisons with the more precise VICON system indicate that the controller could track a trajectory in both cases despise the difference in noise levels between the two systems.
\end{abstract}

\newpage
\pagenumbering{roman}

\setcounter{page}{3}

\phantomsection
\addtocontents{toc}{\protect{\pdfbookmark[0]{\contentsname}{toc}}}
\tableofcontents{}
\phantomsection
\addcontentsline{toc}{chapter}{List of Figures}
\listoffigures
\phantomsection
\addcontentsline{toc}{chapter}{List of Tables}
\listoftables

\pagebreak{}

\pagenumbering{arabic}

\chapter{Introduction}

In the past decade quadcopters have been studied due to their
relative simple fabrication in comparison to other aerial vehicles,
which turns them into ideal platforms for modeling, simulation and
implementation of control algorithms. The fact that they are unmanned
vehicles naturally invites developers to explore tasks that require
a high degree of autonomy.
~\\

Past works such as \cite{key-17,key-30} have set the base for developing quadcopter platforms from their construction to the automation techniques necessary to control the highly non-linear dynamics that characterize these vehicles. 

~\\
The scope of the quadcopter technology has changed over the years. The cost
and sizes have been reduced, it is now a platform affordable for a
broad type of public, from researchers to hobbyists. But beyond the
economic revenue these vehicles generate, the manufacturers are
searching for more autonomy, longer flight time, high data processing
capabilities and adaptation to changing environments, hence the active research on quadcopters.

~\\
A fairly new type of quadcopters are the so called ``nanoquads''
that are of considerably low size and weight, making them an ideal
platform for indoor usage. The project proposed here considers the
study of a commercial platform of a nanoquadcopter called ``Crazyflie
2.0'' developed by Bitcraze company \cite{key-33}. Weighting only 27 grams and
having 9.2 cm of length and width, this nanoquad has rapidly become
one of the preferred platforms for quadcopter research.

~\\
For indoor control of quadcopters different localization techniques can be employed, for example the VICON motion capture system \cite{key-33} is one of the preferred systems for precise localization and it has been used widely in recent quadcopter studies \cite{key-11,key-26}. A recent low-cost technology based on ultra-wide band radio modules has proven effective for indoor localization in robotics systems and specially in quacopters \cite{key-35}. Its low costs are inviting developers to create their own
implementations and the system is getting more precise and robust.
In a few words, the system measures the distance
between two ultra-wide band modules, normally called anchor and tag,
by measuring the time of flight of an electromagnetic wave. Thus,
by the simple relationship between time, distance and velocity (in
this case, the speed of light), then the distance can be easily determined.
By having at least three anchors constantly calculating the distance
between them and a certain tag, a triangulation allows 
to calculate the position in space of the tag, knowing beforehand
the fixed position of each anchor with respect to a frame.

~\\
The UWB system can be implemented using a two-way ranging protocol or a one-way ranging protocol. In two-way ranging, the tag communicates with each anchor individually following a sequence to go through all the anchors and calculate each distance. On the other hand, in one-way ranging the tag constantly broadcasts messages that are received by every anchor and by precisely synchronising the clocks of the anchors then the distance between each of them and the tag are calculated. One-way ranging is particular useful for multi-robot localization applications as there exists no bottle-neck in the number of tags the system can support. In particular, for this project the two-way ranging system developed in \cite{key-36} was used to test the control loop behavior using different localization techniques. This system was developed using the decaWave DMW1000 ultra-wide band module \cite{key-37} which offers an accuracy of 10-20 centimeters in distance measurements.
\section{Main Objectives}

The main objectives of the research project were:
\begin{enumerate}
\item Develop the mathematical model that describes the dynamics of the
Crazyflie 2.0 quadcopter.
\item Create a simulation environment for testing position and trajectory tracking control algorithms.
\item Implement, test and compare different control architectures.
\item Evaluate the performance of a low-cost UWB-based localization system when integrated in the control loop.
\end{enumerate}

\section{Secondary Objectives}
A set of small milestones were defined to help achieve the main objectives of the project:
\begin{enumerate}
\item Investigate past works to identify the physical and aerodynamical
parameters of the Crazyflie 2.0.
\item Linearize the quadcopter's dynamics around hover state.
\item Study and identify the control architecture inside the Crazyflie's
firmware.
\item Design, simulate and implement an off-board position controller using
data from the VICON positioning system.
\item Conceive a second control system, from simulation to implementation, to track more demanding trajectories.
\item Compare the performance of both controllers with in-flight data.
\item Compare the performance of the LQT controller using both the VICON and the UWB systems.
\end{enumerate}

\chapter{Model of the Quadcopter}

In this section a mathematical model of the Crazyflie 2.0 is proposed.
This study was the basis on which the simulation environment was built and an important component in the design of controllers. Thus,
it was important to dedicate enough time to understand how the system
works and identify correctly some physical parameters that were relevant
for the simulation to be useful in the real case scenario.

\section{Coordinate Frames}

\begin{figure}[H]
\begin{centering}
\includegraphics[scale=0.6]{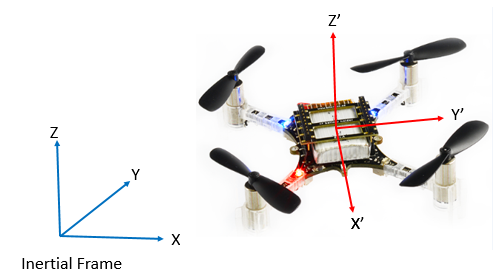}
\par\end{centering}
\caption{\label{fig:inertial}Body-fixed frame and Inertial frame.}

\end{figure}

Before any dynamic study of the quadcopter begins, it is necessary
to define the coordinate frames of the body of the quadcopter (non-inertial
frame) as well as the inertial frame, also called ``world frame'',
which in the case of this project refers to the coordinate frame set
by the external positioning system (VICON/UWB). Following the conventions
set by the ``Bitcraze'' company when designing their quadcopter, as seen in
\autoref{fig:inertial} the body-fixed frame is defined.

In the aeronautic systems, a popular axes convention is to define
a positive altitude downwards, the Y axis pointing towards the east and the X axis pointing towards the true north. These types of frames are called
NED frames (North, East, Down). It was decided to follow the convention
used in the Crazyflie 2.0 firmware, meaning a positive altitude upwards,
which defines an ENU frame (East, North, Up). Another remark is that
the origin of the body-fixed frame matches with the center of gravity
of the quadcopter.
~\\

\begin{figure}[H]
\begin{centering}
\includegraphics[scale=0.45]{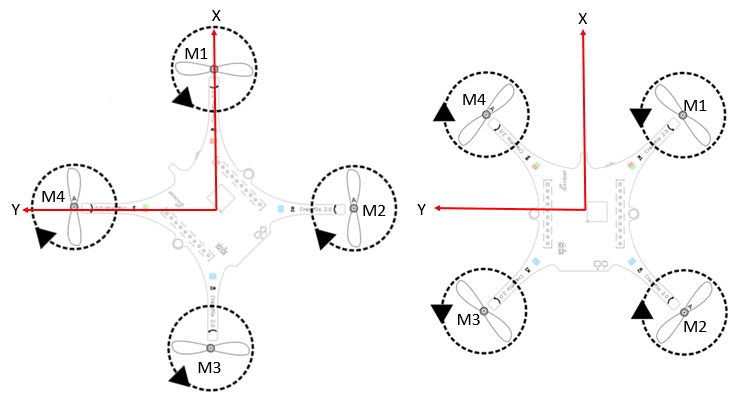}
\par\end{centering}
\caption{\label{fig:ix}``+'' configuration at the left and ``X'' configuration at the right.}

\end{figure}

Another important remark is knowing the flight configuration of the
quadcopter as there are two of them: configuration ``+'' or configuration
``X''. The difference between them is the orientation of the X-Y
frame in terms of the arms of the quadcopter, as shown in \autoref{fig:ix} taken from the manufacturer's website \cite{key-32} and modified accordingly.
In the modern conceptions of quadcopters the ``X'' configuration
is prefered over the ``+'' configuration, mainly because in ``X'' it is easier to add a camera functionality as
the quadcopter's arms will not be interfering with the images captured.
By default the Crazyflie 2.0 is in X mode, so for the rest of this
project and during the mathematical modeling it will be considered
that the quadcopter is in this configuration.

\section{\label{subsec:Dynamic-equations}Dynamic Equations}

The dynamic equations of the quadcopter proposed here take into account
certain physical properties that are not necessarily perfectly valid in the
real platform that is being used in this work, but they are good approximations
that simplify greatly the study and comprehension of this type of
vehicles. Here are the hypothesis:
\begin{enumerate}
\item The quadcopter is a rigid body that cannot be deformed, thus it is
possible to use the well-known dynamic equations of a rigid body.
\item The quadcopter is symmetrical in its geometry, mass and propulsion system.
\item The mass is constant (i.e its derivative is 0).
\end{enumerate}

\begin{figure}[H]
\begin{centering}
\includegraphics[scale=0.5]{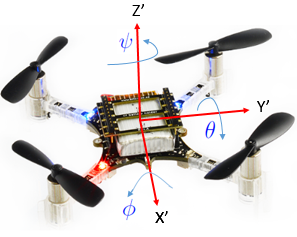}
\par\end{centering}
\caption{\label{fig:euler}Euler angles in the quadcopter's body.}
\end{figure}

The mechanical classic laws of motion are valid in inertial systems,
so to be able to translate these equations into the body frame it is
necessary to define a rigid transformation matrix from the inertial frame
to the body-fixed frame, in which only the rotational part is meaningful to the discussion and is given by three successive rotations: first a rotation of an angle $\psi$
around the $z$ axis, then a rotation of an angle $\theta$ around
the intermediate $y$ axis and finally a rotation of an angle $\phi$ around the intermediate
$x$ axis. Once these three rotations are calculated, the resulting
transformation matrix is defined as:
\begin{equation}
\resizebox{.8\hsize}{!}{$\boldsymbol{R}_{i}^{b}=\left[\begin{array}{ccc}
\cos\theta\cos\psi & \cos\theta\sin\psi & -\sin\theta\\
\sin\phi\sin\theta\cos\psi-\cos\phi\sin\psi & \sin\phi\sin\theta\sin\psi+\cos\phi\cos\psi & \sin\phi\cos\theta\\
\cos\phi\sin\theta\cos\psi+\sin\phi\sin\psi & \cos\phi\sin\theta\sin\psi-\sin\phi\cos\psi & \cos\phi\cos\theta
\end{array}\right]$}\label{eq:euler}
\end{equation}
where $\phi$, $\theta$ and $\psi$ represent the roll, pitch and
yaw angles of the quadcopter's body. \autoref{fig:euler} shows the direction of said angles in the Crazyflie
2.0 body-fixed frame defined previously.

The notation convention used during the mathematical analysis of the quadcopter's dynamics is exhibited in \autoref{tab:notation}, where the state variables are defined.
\renewcommand{\arraystretch}{1.05}
\setlength{\arrayrulewidth}{1pt}
\begin{table}[H]
\begin{centering}
\begin{tabular}{|c|c|c|}
\hline 
\rowcolor[gray]{0.7}  \textbf{Vector} & \textbf{State} & \textbf{Description}\tabularnewline
\hline 
 & {\footnotesize{}$x$} & {\footnotesize{}X position of CoG in the inertial frame}\tabularnewline
\cline{2-3} 
{\footnotesize{}$\boldsymbol{p}_{CG/o}$} & {\footnotesize{}$y$} & {\footnotesize{}Y position of CoG in the inertial frame}\tabularnewline
\cline{2-3} 
 & {\footnotesize{}$z$} & {\footnotesize{}Z position of CoG in the inertial frame}\tabularnewline
\hline 
 & {\footnotesize{}$\phi$} & {\footnotesize{}Roll angle }\tabularnewline
\cline{2-3} 
{\footnotesize{}$\boldsymbol{\Phi}$} & {\footnotesize{}$\theta$} & {\footnotesize{}Pitch angle }\tabularnewline
\cline{2-3} 
 & {\footnotesize{}$\psi$} & {\footnotesize{}Yaw angle }\tabularnewline
\hline 
 & {\footnotesize{}$u$} & {\footnotesize{}X linear velocity of CoG in the body-fixed frame w.r
to the inertial frame}\tabularnewline
\cline{2-3} 
{\footnotesize{}$\boldsymbol{V}_{CG/o}$} & {\footnotesize{}$v$} & {\footnotesize{}Y linear velocity of CoG in the body-fixed frame w.r
to the inertial frame}\tabularnewline
\cline{2-3} 
 & {\footnotesize{}$w$} & {\footnotesize{}Z linear velocity of CoG in the body-fixed frame w.r
to the inertial frame}\tabularnewline
\hline 
 & {\footnotesize{}$p$} & {\footnotesize{}Roll angular velocity in the body-fixed frame w.r
to the inertial frame}\tabularnewline
\cline{2-3} 
{\footnotesize{}$\boldsymbol{\omega}_{b/o}$} & {\footnotesize{}$q$} & {\footnotesize{}Pitch angular velocity in the body-fixed frame w.r
to the inertial frame}\tabularnewline
\cline{2-3} 
 & {\footnotesize{}$r$} & {\footnotesize{}Yaw angular velocity in the body-fixed frame w.r to
the inertial frame}\tabularnewline
\hline 
\end{tabular}
\par\end{centering}
w.r = with respect.
\caption{\label{tab:notation}Notation for vectors and states.}
\end{table}
\renewcommand{\arraystretch}{0.75}
Furthermore, a left superindex such as $\Prefix^{o}{\dot{\boldsymbol{V}}_{CG/o}}$ will indicate in what frame a derivative is taken, while a right superindex indicate a vector coordinates into the specified frame. If a right superindex is not specified as in the example, then the vector does not experience any rotations after the derivative is taken.
\subsection{Force Equations}

According to Newton's Second Law, the expression for the sum of forces
is:

\begin{equation}
\sum\boldsymbol{F}=m\Prefix^{o}{\dot{\boldsymbol{V}}_{CG/o}}\label{eq:2}
\end{equation}

\begin{figure}[H]
\begin{centering}
\includegraphics[scale=0.5]{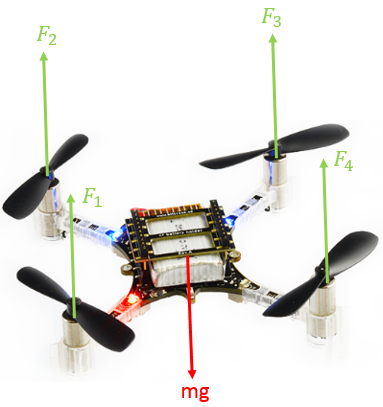}
\par\end{centering}
\caption{Force diagram in the body-fixed frame.\label{fig:Force-diagram-in}}
\end{figure}

The expression of this derivative of velocity can be determined using
the Coriolis equation, which gives the following dynamic expression
in the body-fixed frame:
\begin{equation}
\sum\boldsymbol{F}=m^{o}\dot{\boldsymbol{V}}_{CG/o}=m\left(^{b}\dot{\boldsymbol{V}}_{CG/o}+\boldsymbol{\omega}_{b/o}\times\boldsymbol{V}_{CG/o}\right)\label{eq:3}
\end{equation}
Each propeller of the quadcopter creates an aerodynamical force as shown in \autoref{fig:Force-diagram-in} that
acts upwards in the body-fixed frame.

In a situation where the quadcopter is parallel to the ground, meaning
its roll and pitch angles are zero, the aerodynamic forces created
by the propellers will search to counteract the effect of the weight
and then make the quadcopter move upwards, downwards or stay in a hover position. In
\autoref{fig:Force-diagram-in} the vector ``mg'' actually
represents the projection of the weight vector from the inertial frame
to the body-fixed frame. That being said, this qualitative analysis
of how the forces work in the quadcopter's body can be translated into
\eqref{eq:3} as:
\begin{equation}
\left[\begin{array}{c}
0\\
0\\
F_{z}
\end{array}\right]-\boldsymbol{R}_{o}^{b}\left[\begin{array}{c}
0\\
0\\
mg
\end{array}\right]=m\left({\left[\begin{array}{c}
\dot{u}\\
\dot{v}\\
\dot{w}
\end{array}\right]}+\left[\begin{array}{c}
p\\
q\\
r
\end{array}\right]\times\left[\begin{array}{c}
u\\
v\\
w
\end{array}\right]\right)\label{eq:4}
\end{equation}
From \eqref{eq:4} it is possible to isolate the vector $\Prefix^{b}{\dot{\boldsymbol{V}}_{CG/o}}$:
\begin{equation}
{\left[\begin{array}{c}
\dot{u}\\
\dot{v}\\
\dot{w}
\end{array}\right]}=\left[\begin{array}{c}
0\\
0\\
F_{z}/m
\end{array}\right]-\boldsymbol{R}_{o}^{b}\left[\begin{array}{c}
0\\
0\\
g
\end{array}\right]-\left[\begin{array}{c}
p\\
q\\
r
\end{array}\right]\times\left[\begin{array}{c}
u\\
v\\
w
\end{array}\right]\label{eq:5}
\end{equation}
This equation dictates how the velocity of the center of gravity of
the quadcopter evolves in its body-fixed frame. To determine another
set of state space variables it is necessary to project this vector
in the inertial frame to calculate the velocity in this coordinate
system. Note: the matrix $\boldsymbol{R}_{o}^{b}$ is a rotation
matrix, so it has the following property: $\left(\boldsymbol{R}_{o}^{b}\right)^{-1}=\left(\boldsymbol{R}_{o}^{b}\right){}^{T}=\boldsymbol{R}_{b}^{o}$. Applying it, the projection is calculated:
\begin{equation}
\Prefix^{o}{\dot{\boldsymbol{p}}_{CG/o}^{b}}=\left(\boldsymbol{R}_{o}^{b}\right)\Prefix^{o}{\dot{\boldsymbol{p}}_{CG/o}}\Longleftrightarrow\Prefix^{o}{\dot{\boldsymbol{p}}_{CG/o}}=\boldsymbol{R}_{b}^{o}\boldsymbol{V}_{CG/o}^{b}\Longleftrightarrow
{\left[\begin{array}{c}
\dot{x}\\
\dot{y}\\
\dot{z}
\end{array}\right]}=\boldsymbol{R}^{o}_{b}\left[\begin{array}{c}
u\\
v\\
w
\end{array}\right]\label{eq:6}
\end{equation}
By integrating \eqref{eq:6} it is possible to know the position
of the quadcopter in the inertial frame. 

~\\
Concerning the equations of forces and their state variables, it is
necessary to specify the form of the aerodynamical force generated
by the propellers. Following the diagram in \autoref{fig:Force-diagram-in},
the force generated by each propeller has the form:

\begin{equation}
\boldsymbol{F}_{i}^{b}=\left[\begin{array}{c}
0\\
0\\
T_{i}
\end{array}\right]\label{eq:7}
\end{equation}
where $T_{i}$ represents the upward thrust force in Newtons generated
by each propeller. It is widely known that the thrust generated by
a propeller is a function of the square of its angular speed:
\begin{equation}
T_{i}=C_{T}\omega_{i}^{2}\label{eq:8}
\end{equation}
$C_{T}$ is a thrust coefficient that will be specified in
\Cref{sec:Physical-parameters-estimation} and $\omega_{i}$ is the rotation speed of the i-th motor, in revolutions per minute. As \autoref{fig:Force-diagram-in}
suggests, each propeller generates a thrust force following
\eqref{eq:8} and all in the same direction, which leads to a sum of
all thrust forces:
\begin{equation}
\sum\boldsymbol{F}_{i}^{b}=\left[\begin{array}{c}
0\\
0\\
C_{T}\left(\omega_{1}^{2}+\omega_{2}^{2}+\omega_{3}^{2}+\omega_{4}^{2}\right)
\end{array}\right]
\end{equation}
\subsection{Momentum Equations}

\begin{figure}[H]
\begin{centering}
\includegraphics[scale=0.6]{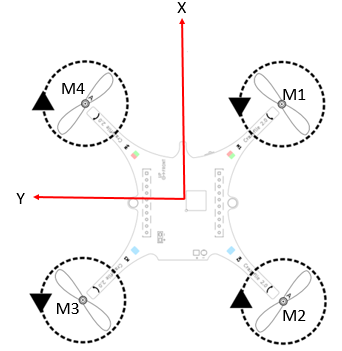}
\par\end{centering}
\caption{\label{fig:rotation direction}{{\small Rotation direction of each motor,
courtesy of Bitcraze ``Crazyflie 2.0 user guide''.}}}
\end{figure}

These equations dictate the rotational dynamics of the quadcopter. Following
the theorem of angular momentum:
\begin{equation}
\sum\boldsymbol{M}^{o}=\Prefix^{o}{\dot{\boldsymbol{h}}}
\end{equation}
where $\boldsymbol{h}$ denotes the angular momentum around the center
of gravity. Using Coriolis equation:
\begin{equation}
\sum\boldsymbol{M}^{o}=\Prefix^{o}{\dot{\boldsymbol{h}}}=\Prefix^{b}{\dot{\boldsymbol{h}}}+\boldsymbol{\omega}_{b/o}\times\boldsymbol{h}\label{eq:11}
\end{equation}
It is desirable to express \eqref{eq:11} in the body-fixed
frame as the momentum equations are more easily calculated, as explained in \cite{key-31}:
\begin{equation}
\sum\boldsymbol{M}^{b}=\boldsymbol{J}\Prefix^{b}{\dot{\boldsymbol{\omega}}_{b/o}}+\boldsymbol{\omega}_{b/o}\times\boldsymbol{J}\boldsymbol{\omega}_{b/o}\label{eq:12}
\end{equation}
here $\boldsymbol{J}$ denotes the inertia matrix of the quadcopter,
which in general can be expressed as:
\begin{equation}
\boldsymbol{J}=\left[\begin{array}{ccc}
I_{xx} & -I_{xy} & -I_{xz}\\
-I_{xy} & I_{yy} & -I_{yz}\\
-I_{xz} & -I_{yz} & I_{zz}
\end{array}\right]
\end{equation}
but from the hypothesis that the body of the quadcopter is symmetrical
around all its axes, the inertia matrix has all crossed terms equal
to zero, i.e.,
\begin{equation}
\boldsymbol{J}=\left[\begin{array}{ccc}
I_{xx} & 0 & 0\\
0 & I_{yy} & 0\\
0 & 0 & I_{zz}
\end{array}\right]
\end{equation}
From equation \eqref{eq:12} it is possible to isolate the vector $\Prefix^{b}{\dot{\boldsymbol{\omega}}_{b/o}}$:
\begin{equation}
{\left[\begin{array}{c}
\dot{p}\\
\dot{q}\\
\dot{r}
\end{array}\right]}=\left(\boldsymbol{J}\right)^{-1}\left(\left[\begin{array}{c}
M_{x}\\
M_{y}\\
M_{z}
\end{array}\right]-\left[\begin{array}{c}
p\\
q\\
r
\end{array}\right]\times\boldsymbol{J}\left[\begin{array}{c}
p\\
q\\
r
\end{array}\right]\right)
\end{equation}
The last state equations come from the relation between $\boldsymbol{\omega}_{b/o}$
and the Euler angles derivative $\dot{\boldsymbol{\Phi}}$
\begin{equation}
\left[\begin{array}{c}
p\\
q\\
r
\end{array}\right]=\left[\begin{array}{ccc}
1 & 0 & -\sin\theta\\
0 & \cos\phi & \sin\phi\cos\theta\\
0 & -\sin\phi & \cos\phi\cos\theta
\end{array}\right]\left[\begin{array}{c}
\dot{\phi}\\
\dot{\theta}\\
\dot{\psi}
\end{array}\right]
\end{equation}
with the inverse relation the state vector $\dot{\boldsymbol{\Phi}}$
is isolated:
\begin{equation}
\hspace{2.2cm}
\left[\begin{array}{c}
\dot{\phi}\\
\dot{\theta}\\
\dot{\psi}
\end{array}\right]=\left[\begin{array}{ccc}
1 & \sin\phi\tan\theta & \cos\phi\tan\theta\\
0 & \cos\phi & -\sin\phi\\
0 & \sin\phi/\cos\theta & \cos\phi/\cos\theta
\end{array}\right]\left[\begin{array}{c}
p\\
q\\
r
\end{array}\right]\,\textrm{for}\,\theta\neq\frac{\pi}{2}
\end{equation}
To calculare the total momentum generated in the quadcopter
system, it is imperative to know the rotation direction of each motor. As seen in \autoref{fig:rotation direction}, the manufacturer of the Crazyflie 2.0 provides this information \cite{key-32}.
The expression for the momentum is given as:
\begin{equation}
\boldsymbol{M}=\sum_{i=1}^{4}\boldsymbol{P}_{i}\times\boldsymbol{F}_{i}+\sum_{i=1}^{4}\boldsymbol{\tau}_{i}
\end{equation}
where $\boldsymbol{P}_{i}$ represents the position of each motor
in the body-fixed frame and $\boldsymbol{\tau}_{i}$ represents the
induced momentum in the quadcopter's body generated by the i-th motor. When a rotor turns in a given direction, conservation of angular momentum dictates that the quadcopter's body will have a tendency to counteract the generated angular momentum, being consistent with Newton's third law of action and reaction. This reaction momentum due to the spin of a rotor is the induced moment $\boldsymbol{\tau}_{i}$.
~\\

If $d$ denotes the distance from the center of gravity to the center
of each motor, the position of each motor is:
\begin{equation}
\resizebox{.8\hsize}{!}{$
\boldsymbol{P}_{1}=\left[\begin{array}{c}
d/\sqrt{2}\\
-d/\sqrt{2}\\
0
\end{array}\right]\text{  },\text{  }\boldsymbol{P}_{2}=\left[\begin{array}{c}
-d/\sqrt{2}\\
-d/\sqrt{2}\\
0
\end{array}\right]\text{  },\text{  }\boldsymbol{P}_{3}=\left[\begin{array}{c}
-d/\sqrt{2}\\
d/\sqrt{2}\\
0
\end{array}\right]\text{  },\text{  }\boldsymbol{P}_{4}=\left[\begin{array}{c}
d/\sqrt{2}\\
d/\sqrt{2}\\
0
\end{array}\right]$}
\end{equation}
Then the mometum generated by the thrust force of each motor can be
calculated:

\begin{center}
\resizebox{.65\hsize}{!}{
$\boldsymbol{P}_{1}^{b}\times\boldsymbol{F}_{1}^{b}=\left[\begin{array}{c}
\left(-C_{T}\omega_{1}^{2}\right)d/\sqrt{2}\\
\left(-C_{T}\omega_{1}^{2}\right)d/\sqrt{2}\\
0
\end{array}\right]$~$\boldsymbol{P}_{2}^{b}\times\boldsymbol{F}_{2}^{b}=\left[\begin{array}{c}
\left(-C_{T}\omega_{2}^{2}\right)d/\sqrt{2}\\
\left(C_{T}\omega_{2}^{2}\right)d/\sqrt{2}\\
0
\end{array}\right]$}
\par\end{center}

\begin{center}
\resizebox{.65\hsize}{!}{
$\boldsymbol{P}_{3}^{b}\times\boldsymbol{F}_{3}^{b}=\left[\begin{array}{c}
\left(C_{T}\omega_{3}^{2}\right)d/\sqrt{2}\\
\left(C_{T}\omega_{3}^{2}\right)d/\sqrt{2}\\
0
\end{array}\right]$~$\boldsymbol{P}_{4}^{b}\times\boldsymbol{F}_{4}^{b}=\left[\begin{array}{c}
\left(C_{T}\omega_{4}^{2}\right)d/\sqrt{2}\\
\left(-C_{T}\omega_{4}^{2}\right)d/\sqrt{2}\\
0
\end{array}\right]$}
\par\end{center}

~\\
The induced moments $\bm{\tau}_{i}$ act only in the Z axis and have an opposite magnitude
from the moment generated by each propeller, due to the conservation of angular momentum. In this particular case,
given the axis convention that is being used (z axis pointing upwards),
applying the right-hand rule indicate that a clockwise spinning rotor yields a negative momentum (one thumb points downward, in the opposite direction of the z axis), thus the induced momentum will be positive. Then, the sum of induced moments in the quadcopter's body is calculated:

\begin{equation}
\sum_{i=1}^{4}\boldsymbol{\tau}_{i}^{b}=\left[\begin{array}{c}
0\\
0\\
C_{D}\left(-\omega_{1}^{2}+\omega_{2}^{2}-\omega_{3}^{2}+\omega_{4}^{2}\right)
\end{array}\right]\label{eq:19}
\end{equation}
where $C_{D}$ denotes the aerodynamic drag coefficient that will
be specified in \Cref{sec:Physical-parameters-estimation}.
Finally, the total moment has the following form:

\begin{equation}
\boldsymbol{M}^{b}=\left[\begin{array}{c}
M_{x}\\
M_{y}\\
M_{z}
\end{array}\right]=\left[\begin{array}{c}
dC_{T}/\sqrt{2}\left(-\omega_{1}^{2}-\omega_{2}^{2}+\omega_{3}^{2}+\omega_{4}^{2}\right)\\
dC_{T}/\sqrt{2}\left(-\omega_{1}^{2}+\omega_{2}^{2}+\omega_{3}^{2}-\omega_{4}^{2}\right)\\
C_{D}\left(-\omega_{1}^{2}+\omega_{2}^{2}-\omega_{3}^{2}+\omega_{4}^{2}\right)
\end{array}\right]
\end{equation}
In the total momentum equation there are certain terms that include
angular accelerations that have been neglected as they tend to be
small compared to the other terms of the equation. Gyroscopic moments
have also been neglected using the argument that the inertia moment
of each motor tends to be small thus their contribution in the total
momentum is also small \cite{key-21,key-30}.

\section{\label{sec:Physical-parameters-estimation}Physical Parameters}

The precise measurement of certain physical parameters is the key
to create a simulation environment that correctly describes the behavior
of the quadcopter. In \cite{key-1} a study of said physical parameters
was undertaken for the Crazyflie 2.0. The aerodynamical coefficients
were studied in \cite{key-2} for the Crazyflie 1.0, but they are the same or at least close to those of the Crazyflie 2.0 given the fact that these coefficients only depend on the geometry of the propellers \cite{key-3}, which remained unchanged between the two models. Results of both works are summarized
in \autoref{tab:Physical-parameters-for}.
\renewcommand{\arraystretch}{1.1}
\begin{table}[H]
\resizebox{\textwidth}{!}{
\begin{centering}
\begin{tabular}{|c|c|c|}
\hline 
\rowcolor[gray]{0.7} \textbf{Parameter} & \textbf{Description} & \textbf{Value}\tabularnewline
\hline 
$m_{quad}$ & Mass of the quadcopter alone & $0.27\,[\textrm{Kg}]$\tabularnewline
\hline 
$m_{uwb}$ & Mass of the UWB module  & $0.04\,[\textrm{Kg}]$\tabularnewline
\hline 
$m_{vicon}$ & Mass of one VICON marker & $0.02\,[\textrm{Kg}]$\tabularnewline
\hline 
$m$ & Total mass & $0.33\,[\textrm{Kg}]$\tabularnewline
\hline 
$d$ & Arm length & $39.73\times10^{-3}\,[\textrm{m}]$\tabularnewline
\hline 
$r$ & Rotor radius & $23.1348\times10^{-3}\,[\textrm{m}]$\tabularnewline
\hline 
$I_{xx}$ & Principal Moment of Inertia around x axis & $1.395\times10^{-5}\,[\textrm{Kg}\times \textrm{m}^{2}]$\tabularnewline
\hline 
$I_{yy}$ & Principal Moment of Inertia around y axis & $1.436\times10^{-5}\,[\textrm{Kg}\times \textrm{m}^{2}]$\tabularnewline
\hline 
$I_{zz}$ & Principal Moment of Inertia around z axis & $2.173\times10^{-5}\,[\textrm{Kg}\times \textrm{m}^{2}]$\tabularnewline
\hline 
$k_{T}$ & Non-dimensional thrust coefficient & $0.2025$\tabularnewline
\hline 
$k_{D}$ & Non-dimensional torque coefficient & $0.11$\tabularnewline
\hline 
\end{tabular}
\par\end{centering}}
\caption{\label{tab:Physical-parameters-for}Physical parameters for the Crazyflie
2.0.}
\end{table}
\renewcommand{\arraystretch}{0.75}
In addition, as explained in \cite{key-3}, the thrust generated by
the propeller is often expressed as:
\begin{equation}
T=k_{T}\rho n^{2}D^{4}\label{eq:20}
\end{equation}
where $k_{T}$ is the non-dimensional thrust coefficient, $\rho$
is the density of air, $n$ is the propeller speed in revolutions
per second and $D$ is the diameter of the rotor. As it will be evident
later, it is convenient to express the propeller speed in RPM's. Knowing
that 1 revolution per second is the same as 60 revolutions per minute, \eqref{eq:20} becomes:
\begin{equation}
T=k_{T}\mbox{\ensuremath{\rho}}\left(\omega/60\right)^{2}D^{4}
\end{equation}
where $\omega$ is the angular speed of the propeller in RPM. Comparing
the above with \eqref{eq:8}, it is possible to determine the
thrust coefficient $C_{T}$ as:
\begin{equation}
C_{T}=k_{T}\rho\left(2r\right)^{4}/3600\label{eq:22}
\end{equation}
taking the value of air density constant $\rho=1.225\,[\textrm{Kg/m}^{3}]$
and all the other constants defined previously, finally this coefficient
is:
\begin{equation}
\boxed{C_{T}=3.1582\times10^{-10}\,[\textrm{N/rpm}^{2}]}\label{eq:23}
\end{equation}
Now for the torque coefficient, as specified in \cite{key-3}, the
torque created by the propellers is described by this equation:
\begin{equation}
Q=k_{D}\rho n^{2}D^{5}
\end{equation}
Operating the same variable change as in \eqref{eq:22}, then:
\begin{equation}
C_{D}=k_{D}\rho\left(2r\right)^{5}/3600
\end{equation}
\begin{equation}
\boxed{C_{D}=7.9379\times10^{-12}\,[\textrm{Nm/rpm}^{2}]}\label{eq:26}
\end{equation}
With the parameters specified in \autoref{tab:Physical-parameters-for}
and the constants calculated in \eqref{eq:23} and \eqref{eq:26},
all the basic physical parameters were determined. Here we use
the word ``basic'' as these parameters are the minimum necessary
to be able to simulate the behavior of a quadcopter and because in most
applications, this one included, are a good approximation of the real
physical system. 

\section{\label{sec:Linearization-and-State}Linearization and State Space Representation}

The state space representation of a system gives an idea of how the
system evolves in time by the following equations:
\begin{equation}
\begin{cases}
\dot{\boldsymbol{x}}\left(t\right)=\boldsymbol{A}\left(t\right)\boldsymbol{x}\left(t\right)+\boldsymbol{B}\left(t\right)\boldsymbol{u}\left(t\right)\\
\boldsymbol{y}\left(t\right)=\boldsymbol{C}\left(t\right)\boldsymbol{x}\left(t\right)+\boldsymbol{D}\left(t\right)\boldsymbol{u}\left(t\right)
\end{cases}
\end{equation}
In general, this equation describes the evolution of a linear time-varying
system, where $\boldsymbol{x}\left(t\right)$ is the vector of states,
$\boldsymbol{y}\left(t\right)$ is the output vector and $\boldsymbol{u}\left(t\right)$
is the input vector. For the state space representation of a quadcopter
it is conventional to consider the following linear time-invariant
realisation of the system, meaning that matrices A, B, C and D are
static and don't change over time:
\begin{equation}
\begin{cases}
\Delta\dot{\boldsymbol{x}}=\boldsymbol{A}\Delta\boldsymbol{x}+\boldsymbol{B}\Delta\boldsymbol{u}\\
\Delta\boldsymbol{y}=\boldsymbol{C}\Delta\boldsymbol{x}+\boldsymbol{D}\Delta\boldsymbol{u}
\end{cases}
\end{equation}
where the prefix $\Delta$ means that the vector is the result of
a linearisation process.

~\\
When linearizing a system the important question becomes at around which so-called "trim" trajectory we desire to linearize, or better yet, what is the trim trajectory
most well-suited given the needs of the system.
This trim trajectory has to be associated with an equilibrium point
in which the states of the system do not change over time, meaning
$\dot{\boldsymbol{x}}_{e}=0$. In the case of the quadcopter, the common
trim trajectory is the hover, in which the drone stays stationary
at a certain altitude. This fact can be translated as an equilibrium
state:

\begin{equation}
\boldsymbol{x}_{e}=\left[\begin{array}{cccccccccccc}
x_{e} & y_{e} & z_{e} & \psi_{e} & \theta_{e} & \phi_{e} & u_{e} & v_{e} & w_{e} & r_{e} & q_{e} & p_{e}\end{array}\right]^{T}
\end{equation}
At the equilibrium point, the quadcopter's linear position and yaw
angle are indifferent in terms of the linearization calculus, so they
can be considered arbitrary constants. For the roll and pitch angles, they need
to be zero in order for the quadcopter to keep the stationary position,
and so does any linear or angular velocity. Finally the equilibrium
vector is as simple as:
\begin{equation}
\boldsymbol{x}_{e}=\left[\begin{array}{cccccccccccc}
x_{e} & y_{e} & z_{e} & \psi_{e} & 0 & 0 & 0 & 0 & 0 & 0 & 0 & 0\end{array}\right]^{T}\label{eq:30}
\end{equation}
In order to keep the quadcopter flying in hover mode, knowing that
the body is levelled with the floor and there are no gravity components
other that in the z axis, the force generated by the propellers need
to compensate exactly for the weight of the quadcopter to stay stationary
in the air, this means:
\begin{equation}
C_{T}\left(\omega_{e1}^{2}+\omega_{e2}^{2}+\omega_{e3}^{2}+\omega_{e4}^{2}\right)=mg\label{eq:31}
\end{equation}
From the hypothesis that the quadcopter's body is perfectly symmetrical,
a quick deduction is that all motors need to rotate with the same
speed in order to maintain the body levelled and don't create any angular
momentum, this means that in equilibrium:
\begin{equation}
\omega_{e1}=\omega_{e2}=\omega_{e3}=\omega_{e4}=\omega_{e}\label{eq:32}
\end{equation}
Combining \eqref{eq:31} and eqref{eq:32} gives as result:
\begin{equation}
\boxed{\omega_{e}=\sqrt{\frac{mg}{4C_{T}}}=16073\,\textrm{[rpm]}}\label{eq:2.34}
\end{equation}
This constant specifies the required speed of each rotor in order
to maintain the hover position and so the input vector in equilibrium
is:
\begin{equation}
u_{e}=\left[\begin{array}{cccc}
\omega_{e} & \omega_{e} & \omega_{e} & \omega_{e}\end{array}\right]^{T}
\end{equation}
After applying a Taylor's first order expansion, and taking into account
the equilibrium state vector specified in \eqref{eq:30}, the linearized
equations are:
\begin{equation}
\begin{cases}
\Delta F_{x}=m\Delta\dot{u}-mg\Delta\theta\\
\Delta F_{y}=m\Delta\dot{v}+mg\Delta\phi\\
\Delta F_{z}=m\Delta\dot{w}\\
\Delta M_{x}=I_{xx}\Delta\dot{p}\\
\Delta M_{y}=I_{yy}\Delta\dot{q}\\
\Delta M_{z}=I_{zz}\Delta\dot{r}
\end{cases}\label{eq:35}
\end{equation}
The linearized forces and moments are:
\begin{equation}
\boldsymbol{\Delta F}^{b}=\left[\begin{array}{c}
\Delta F_{x}\\
\Delta F_{y}\\
\Delta F_{z}
\end{array}\right]=\left[\begin{array}{c}
0\\
0\\
2C_{T}\omega_{e}\left(\Delta\omega_{1}+\Delta\omega_{2}+\Delta\omega_{3}+\Delta\omega_{4}\right)
\end{array}\right]\label{eq:36}
\end{equation}
\begin{equation}
\boldsymbol{\Delta M}^{b}=\left[\begin{array}{c}
\Delta M_{x}\\
\Delta M_{y}\\
\Delta M_{z}
\end{array}\right]=\left[\begin{array}{c}
\sqrt{2}dC_{T}\omega_{e}\left(-\Delta\omega_{1}-\Delta\omega_{2}+\Delta\omega_{3}+\Delta\omega_{4}\right)\\
\sqrt{2}dC_{T}\omega_{e}\left(-\Delta\omega_{1}+\Delta\omega_{2}+\Delta\omega_{3}-\Delta\omega_{4}\right)\\
2C_{D}\omega_{e}\left(-\Delta\omega_{1}+\Delta\omega_{2}-\Delta\omega_{3}+\Delta\omega_{4}\right)
\end{array}\right]\label{eq:37}
\end{equation}
In hover position the body-fixed frame coincides with the inertial
frame, meaning:
\begin{equation}
\begin{cases}
\Delta\dot{x}=\Delta u\\
\Delta\dot{y}=\Delta v\\
\Delta\dot{z}=\Delta w
\end{cases}\,\,\,\,\,\,\,\,\,\,\,\,\,\,\,\,\,\begin{cases}
\Delta\dot{\phi}=\Delta p\\
\Delta\dot{\theta}=\Delta q\\
\Delta\dot{\psi}=\Delta r
\end{cases}\label{eq:38}
\end{equation}
Merging \eqref{eq:35} to \eqref{eq:38} allows to write the
state space representation of the linearized quadcopter:
\begin{equation}
\resizebox{1\hsize}{!}{$\left[\begin{array}{c}
\Delta\dot{x}\\
\Delta\dot{y}\\
\Delta\dot{z}\\
\Delta\dot{\psi}\\
\Delta\dot{\theta}\\
\Delta\dot{\phi}\\
\Delta\dot{u}\\
\Delta\dot{v}\\
\Delta\dot{w}\\
\Delta\dot{r}\\
\Delta\dot{q}\\
\Delta\dot{p}
\end{array}\right]=\left[\begin{array}{cccccccccccc}
0 & 0 & 0 & 0 & 0 & 0 & 1 & 0 & 0 & 0 & 0 & 0\\
0 & 0 & 0 & 0 & 0 & 0 & 0 & 1 & 0 & 0 & 0 & 0\\
0 & 0 & 0 & 0 & 0 & 0 & 0 & 0 & 1 & 0 & 0 & 0\\
0 & 0 & 0 & 0 & 0 & 0 & 0 & 0 & 0 & 1 & 0 & 0\\
0 & 0 & 0 & 0 & 0 & 0 & 0 & 0 & 0 & 0 & 1 & 0\\
0 & 0 & 0 & 0 & 0 & 0 & 0 & 0 & 0 & 0 & 0 & 1\\
0 & 0 & 0 & 0 & g & 0 & 0 & 0 & 0 & 0 & 0 & 0\\
0 & 0 & 0 & 0 & 0 & -g & 0 & 0 & 0 & 0 & 0 & 0\\
0 & 0 & 0 & 0 & 0 & 0 & 0 & 0 & 0 & 0 & 0 & 0\\
0 & 0 & 0 & 0 & 0 & 0 & 0 & 0 & 0 & 0 & 0 & 0\\
0 & 0 & 0 & 0 & 0 & 0 & 0 & 0 & 0 & 0 & 0 & 0\\
0 & 0 & 0 & 0 & 0 & 0 & 0 & 0 & 0 & 0 & 0 & 0
\end{array}\right]\left[\begin{array}{c}
\Delta x\\
\Delta y\\
\Delta z\\
\Delta\psi\\
\Delta\theta\\
\Delta\phi\\
\Delta u\\
\Delta v\\
\Delta w\\
\Delta r\\
\Delta q\\
\Delta p
\end{array}\right]+\omega_{e}\left[\begin{array}{cccc}
0 & 0 & 0 & 0\\
0 & 0 & 0 & 0\\
0 & 0 & 0 & 0\\
0 & 0 & 0 & 0\\
0 & 0 & 0 & 0\\
0 & 0 & 0 & 0\\
0 & 0 & 0 & 0\\
0 & 0 & 0 & 0\\
2C_{T}/m & 2C_{T}/m & 2C_{T}/m & 2C_{T}/m\\
-2C_{D}/I_{zz} & 2C_{D}/I_{zz} & -2C_{D}/I_{zz} & 2C_{D}/I_{zz}\\
-\sqrt{2}dC_{T}/I_{yy} & \sqrt{2}dC_{T}/I_{yy} & \sqrt{2}dC_{T}/I_{yy} & -\sqrt{2}dC_{T}/I_{yy}\\
-\sqrt{2}dC_{T}/I_{xx} & -\sqrt{2}dC_{T}/I_{xx} & \sqrt{2}dC_{T}/I_{xx} & \sqrt{2}dC_{T}/I_{xx}
\end{array}\right]$$\left[\begin{array}{c}
\Delta\omega_{1}\\
\Delta\omega_{2}\\
\Delta\omega_{3}\\
\Delta\omega_{4}
\end{array}\right]$}\nonumber
\end{equation}

\section{\label{sec:Mouvement-decoupling}Movement Decoupling}

In the state space realization of the quadcopter's linear model, the
four inputs of the system act directly in just four states of the system.
Rewriting together \eqref{eq:36} and \eqref{eq:37} show how each input of the system contributes to each force and momentum.
\begin{equation}
\left[\begin{array}{c}
F_{z}\\
M_{x}\\
M_{y}\\
M_{z}
\end{array}\right]=2\omega_{e}\left[\begin{array}{cccc}
C_{T} & C_{T} & C_{T} & C_{T}\\
-dC_{T}/\sqrt{2} & -dC_{T}/\sqrt{2} & dC_{T}/\sqrt{2} & dC_{T}/\sqrt{2}\\
-dC_{T}/\sqrt{2} & dC_{T}/\sqrt{2} & dC_{T}/\sqrt{2} & -dC_{T}/\sqrt{2}\\
-C_{D} & C_{D} & -C_{D} & C_{D}
\end{array}\right]\left[\begin{array}{c}
\Delta\omega_{1}\\
\Delta\omega_{2}\\
\Delta\omega_{3}\\
\Delta\omega_{4}
\end{array}\right]\label{eq:trans}
\end{equation}
From \eqref{eq:trans} a transformation matrix can be defined
between the forces acting on the quadcopter's body and the angular
speed from the motors:
\begin{equation}
\bm{\Gamma}=2\omega_{e}\left[\begin{array}{cccc}
C_{T} & C_{T} & C_{T} & C_{T}\\
-dC_{T}/\sqrt{2} & -dC_{T}/\sqrt{2} & dC_{T}/\sqrt{2} & dC_{T}/\sqrt{2}\\
-dC_{T}/\sqrt{2} & dC_{T}/\sqrt{2} & dC_{T}/\sqrt{2} & -dC_{T}/\sqrt{2}\\
-C_{D} & C_{D} & -C_{D} & C_{D}
\end{array}\right]
\end{equation}
If matrix $\bm{\Gamma}$ is invertible (i.e: $\bm{\Gamma}^{-1}$exists) that means that
all four lines are independent and thus the vertical and angular forces
of the quadcopter act independently from each other. Inverting the matrix:
\begin{equation}
\bm{\Gamma}^{-1}=\frac{1}{2\omega_{e}}\left[\begin{array}{cccc}
1/\left(4C_{T}\right) & -\sqrt{2}/\left(4dC_{T}\right) & -\sqrt{2}/\left(4dC_{T}\right) & -1/\left(4C_{D}\right)\\
1/\left(4C_{T}\right) & -\sqrt{2}/\left(4dC_{T}\right) & \sqrt{2}/\left(4dC_{T}\right) & 1/\left(4C_{D}\right)\\
1/\left(4C_{T}\right) & \sqrt{2}/\left(4dC_{T}\right) & \sqrt{2}/\left(4dC_{T}\right) & -1/\left(4C_{D}\right)\\
1/\left(4C_{T}\right) & \sqrt{2}/\left(4dC_{T}\right) & -\sqrt{2}/\left(4dC_{T}\right) & 1/\left(4C_{D}\right)
\end{array}\right]\label{eq:2.42}
\end{equation}
Now taking the result in \eqref{eq:2.42}, it is possible to
find the inverse relation of \eqref{eq:trans}:
\begin{equation}
\resizebox{0.85\hsize}{!}{$
\left[\begin{array}{c}
\Delta\omega_{1}\\
\Delta\omega_{2}\\
\Delta\omega_{3}\\
\Delta\omega_{4}
\end{array}\right]=\frac{1}{2\omega_{e}}\left[\begin{array}{cccc}
1/\left(4C_{T}\right) & -\sqrt{2}/\left(4dC_{T}\right) & -\sqrt{2}/\left(4dC_{T}\right) & -1/\left(4C_{D}\right)\\
1/\left(4C_{T}\right) & -\sqrt{2}/\left(4dC_{T}\right) & \sqrt{2}/\left(4dC_{T}\right) & 1/\left(4C_{D}\right)\\
1/\left(4C_{T}\right) & \sqrt{2}/\left(4dC_{T}\right) & \sqrt{2}/\left(4dC_{T}\right) & -1/\left(4C_{D}\right)\\
1/\left(4C_{T}\right) & \sqrt{2}/\left(4dC_{T}\right) & -\sqrt{2}/\left(4dC_{T}\right) & 1/\left(4C_{D}\right)
\end{array}\right]\left[\begin{array}{c}
F_{z}\\
M_{x}\\
M_{y}\\
M_{z}
\end{array}\right]\label{eq:2.43}$}
\end{equation}
Equation \eqref{eq:2.43} dictates how each motor contributes to each
of the forces acting on the quadcopter's body. This study confirms
that the vertical, lateral, longitudinal and directional (yaw) forces
act independently from each other in the mathematical model and thus the quadcopter's dynamics
are decoupled and can be studied as sub-systems.
~\\
\pagebreak

\begin{itemize}
\item \textbf{Vertical Subsystem}
\end{itemize}
This subsystem describes the dynamics of the upward movements of the
quadcopter, following this state space equation:
\begin{equation}
\left[\begin{array}{c}
\Delta\dot{w}\\
\Delta\dot{z}
\end{array}\right]=\left[\begin{array}{cc}
0 & 0\\
1 & 0
\end{array}\right]\left[\begin{array}{c}
\Delta w\\
\Delta z
\end{array}\right]+\left[\begin{array}{c}
1/m\\
0
\end{array}\right]\Delta F_{z}
\end{equation}
\begin{itemize}
\item \textbf{Directional Subsystem}
\end{itemize}
The yaw angle and its velocity dictate the dynamics of the quadcopter
direction in the XY plane, as suggests this following state space
equation:
\begin{equation}
\left[\begin{array}{c}
\Delta\dot{r}\\
\Delta\dot{\psi}
\end{array}\right]=\left[\begin{array}{cc}
0 & 0\\
1 & 0
\end{array}\right]\left[\begin{array}{c}
\Delta r\\
\Delta\psi
\end{array}\right]+\left[\begin{array}{c}
1/I_{zz}\\
0
\end{array}\right]\Delta M_{z}
\end{equation}
\begin{itemize}
\item \textbf{Lateral Subsystem}
\end{itemize}
The lateral dynamic governs the pitch movement of the quadcopter, as well
as its Y position in the inertial frame:
\begin{equation}
\left[\begin{array}{c}
\Delta\dot{p}\\
\Delta\dot{\phi}\\
\Delta\dot{v}\\
\Delta\dot{y}
\end{array}\right]=\left[\begin{array}{cccc}
0 & 0 & 0 & 0\\
1 & 0 & 0 & 0\\
0 & -g & 0 & 0\\
0 & 0 & 1 & 0
\end{array}\right]\left[\begin{array}{c}
\Delta p\\
\Delta\phi\\
\Delta v\\
\Delta y
\end{array}\right]+\left[\begin{array}{c}
1/I_{xx}\\
0\\
0\\
0
\end{array}\right]\Delta M_{x}
\end{equation}
\begin{itemize}
\item \textbf{Longitudinal Subsystem}
\end{itemize}
Similar to the lateral subsystem, it rules the movement around the
X axis of the body-fixed frame of the quadcopter, and its X position
and velocity in the inertial frame.
\begin{equation}
\left[\begin{array}{c}
\Delta\dot{q}\\
\Delta\dot{\theta}\\
\Delta\dot{u}\\
\Delta\dot{x}
\end{array}\right]=\left[\begin{array}{cccc}
0 & 0 & 0 & 0\\
1 & 0 & 0 & 0\\
0 & g & 0 & 0\\
0 & 0 & 1 & 0
\end{array}\right]\left[\begin{array}{c}
\Delta q\\
\Delta\theta\\
\Delta u\\
\Delta x
\end{array}\right]+\left[\begin{array}{c}
1/I_{yy}\\
0\\
0\\
0
\end{array}\right]\Delta M_{x}
\end{equation}

\section{\label{subsec:Motor-characterization}Motor Characterization}

The inputs of the above state space realization are the angular speed
of each motor in RPM's, but after exploring the Crazyflie's 2.0 firmware
it became evident that this is not the real input of the system. The
voltage sent to each DC motor is controlled using a PWM signal specified
as a 16 bit number, ranging from 0 to 65535, meaning that the actual
input of the system can be considered directly as this PWM signal
and not the actual voltage sent to the motors. In \cite{key-2} experimental
data was retrieved from the motors to identify the relationship between
the PWM signal sent to the motors and the RPM's generated. The experiments proved that the angular speed of the motors have a linear relationship with the PWM input of the system, following
the equation:
\begin{equation}
RPM= 0.2685\times PWM+4070.3 
\label{eq:3.1}
\end{equation}
The characterization of a DC motor usually derives in a first order
transfer function that specifies certain response time from the motors
as they do not react immediately to the commands sent, but it is a
good approximation to assume that this response time is fast enough
and that it will not cause much delays in the system, thus
\eqref{eq:3.1} serves as a good approximation for the motor characterization.

\chapter{Simulation}

In this section a simulation environment of the quadcopter's dynamics
is proposed with the intention of testing and designing control schemes.
Two different controllers are proposed: in the first phase of the
project a position PID tracker was considered and in a second phase
a trajectory tracker known as the Linear-Quadratic Tracker (LQT) was
studied.

\section{\label{sec:Cascaded-PID-Position}Cascaded PID Position Tracker}

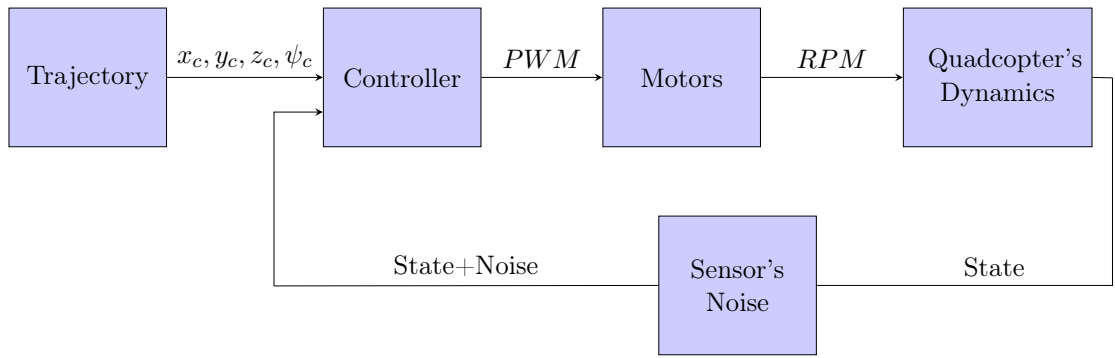
\begin{figure}[H]
\centering
\resizebox{1\textwidth}{!}{
\begin{tikzpicture}[>=stealth]
  \coordinate (orig)   at (0,0);
  \coordinate (LLD)    at (9,1);
  \coordinate (LLA)    at (-0.3,4);
  \coordinate (LLB)    at (4+0.2,4);
  \coordinate (LLC)    at (8+0.2,4);
  \coordinate (LLE)	   at (12+0.5,4);
  \coordinate (ED)     at (15+0.5,4+1);
  \coordinate (BoE)    at (3.5,2);

  \node[draw,fill=blue!20, minimum width=2cm, minimum height=2cm, anchor=south west, text width=2cm, align=center] (A) at (LLA) {Trajectory};
  \node[draw,fill=blue!20, minimum width=2cm, minimum height=2cm, anchor=south west, text width=2cm, align=center] (B) at (LLB) {Controller};
  \node[draw,fill=blue!20, minimum width=1.2cm, minimum height=2cm, anchor=south west, text width=2cm, align=center] (C) at (LLC) {Motors};
  \node[draw,fill=blue!20, minimum width=2cm, minimum height=2cm, anchor=south west, text width=2cm, align=center] (D) at (LLD) {Sensor's\\Noise};
  \node[draw,fill=blue!20, minimum width=2.7cm, minimum height=2cm, anchor=south west, text width=2cm, align=center] (E) at (LLE) {Quadcopter's\\Dynamics};
  
   \draw[->] (A.0) -- node[above] {$x_c,y_c,z_c,\psi_c$} (B.180);
   \draw[->] (B.0) -- node[above] {$PWM$} (C.180);
   \draw[->] (C.0) -- node[above] {$RPM$} (E.180);
   \draw[-] (E.0) -- (ED.0);
   \draw[-] (ED.0) |- node [above,pos = 0.7] {State}  (D.0);
   \draw[-] (D.180) -- node [above] {State+Noise} (BoE);
   \draw[->] (BoE.0) |- ($(B.180) + (0,-1/2)$);

\end{tikzpicture}}
\caption{\label{fig:Block-Diagram-of}Block Diagram of Simulation environment.}
\end{figure}

\autoref{fig:Block-Diagram-of} presents the simulation model created for this phase of the project.
Now an explanation of each simulation block will be given:
\begin{enumerate}
\item \textbf{Trajectory}: this block serves as the input of the overall
system, specifying a trajectory in the x,y,z and yaw positions. As
of now the commands sent by this block are mere constants or signal
inputs such as sinusoids, random signals, ramps, etc.
It serves as input commands to the controller.
\item \textbf{Controller:} takes the desired trajectory and the quadcopter's
states as inputs and computes the necessary 16-bit PWM signal to send to the motors.
\item \textbf{Motors:} implements the linear relation between the 16-bit
PWM signal sent to the motors and the actual angular speed in revolutions
per minutes generated by them, as specified in \Cref{subsec:Motor-characterization}.
\item \textbf{Quadcopter's dynamics:} this block implements the dynamic equations of \Cref{subsec:Dynamic-equations}. The vectorial form of the
equations, as they were developed, is the simplest, most common and
elegant way of constructing this block. The non-linear model was linearized
using MATLAB's command ``linmod'' to verify it was consistent with
the theoretical state space model found in the previous section.
\item \textbf{Sensor's noise:} allows to add additive white Gaussian noise
to specific states of the quadcopters, to simulate the sensors that
give the real states used by the control system. This block could be modified by the user to define more complex models, e.g., including sensor bias.
\end{enumerate}

\subsection{On-Board Control Architecture}

The first steps to control the quadcopter was understanding the already
implemented controllers that came with the stock firmware of the Crazyflie
2.0 (Firmware release 2016.02). As seen in \autoref{fig:On-board-control-architecture} the manufacturers specify the control
architecture used:

\begin{figure}[H]
\begin{centering}
\includegraphics[scale=0.6]{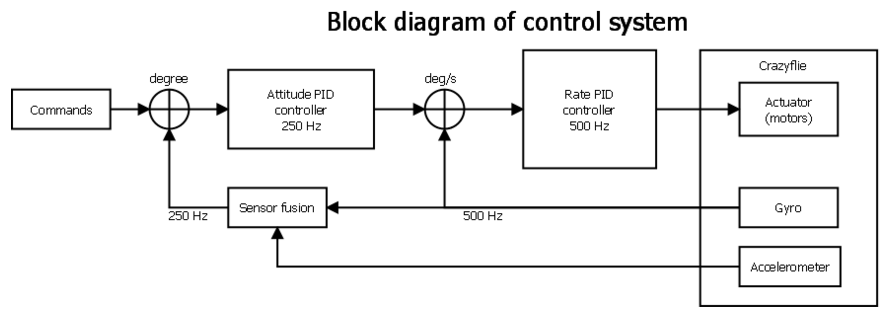}
\par\end{centering}
\caption{\label{fig:On-board-control-architecture}On-board control architecture,
image courtesy of Bitcraze.}

\end{figure}
A two cascaded PID control scheme was found in the Crazyflie's firmware
in order to control the pitch and roll angles. A cascaded control
structure can be analysed by decomposing the architecture in an inner
and outer control loops, in which the outer loop regulates the inner
loop, which in turn regulates the plant of the system. As a common
rule in cascaded structures, the inner loop needs to regulate at a
faster rate than the outer loop. It is ideal for the inner loop output
to reach an steady state value before the outer loop changes the setpoint
sent to the inner loop. Synchronization problems will occur between
the two controllers if the inner loop response is not as fast as it
should, or if the outer loop is faster than it should. In implementation
terms this is easily remediable by forcing the inner loop to be, as
in this case, twice as fast as the outer loop (Attitude controller
running at 250Hz and Rate controller running at 500Hz).

\subsubsection{Inner Loop: Rate Controller}

\begin{figure}[H]
\begin{centering}
\begin{tikzpicture}[>=stealth]
  \coordinate (orig)   at (0,0);
  \coordinate (LLA)    at (0,4);
  \coordinate (AroneA) at (-1.8,5.5);
  \coordinate (ArtwoA) at (-1.8,4.5);
  \coordinate (outA) at (5,5);

 \node[draw,fill=blue!20, minimum width=2cm, minimum height=2cm, anchor=south west, text width=2cm, align=center] (A) at (LLA) {Rate\\Controller};
   
   \draw[->] (AroneA)-- node[above]{$\color{black}\bm{p_c},\bm{q_c}, \bm{r_c}$} ($(A.180)+(0,0.5)$);
   \draw[->] (ArtwoA)-- node[above]{$p, q, r$} ($(A.180)+(0,-0.5)$);
   \draw[->] (A.0)-- node[above]{$\Delta_\phi, \Delta_\theta, \Delta_\psi$} (outA);

\end{tikzpicture}
\par\end{centering}
\caption{\label{fig:rate}Rate Controller diagram.}

\end{figure}
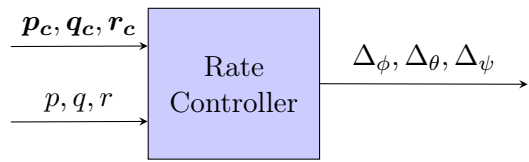

The inputs and outputs of this block are shown in \autoref{fig:rate}.

The goal of this controller is to calculate the input variation from the equilibrium point of the motors in order to create
the angular momentum required for the state variables $p$, $q$ and
$r$ to get the values $p_{c}$, $q_{c}$ and $r_{c}$ respectively.
For that, three independent controllers are used:
\begin{itemize}
\item \textbf{Roll Rate Proportional controller:} calculates $\Delta_{\phi}$
following this equation, the desired value $p_{c}$ is calculated
by the outer loop attitude controller:
\begin{equation}
\Delta_{\phi}\left(t\right)=K_{P,p}\left(p_{c}\left(t\right)-p\left(t\right)\right)
\end{equation}
\item \textbf{Pitch Rate Proportional controller:} very similar to the roll
rate controller, calculates $\Delta_{\theta}$ from the setpoint value
$q_{c}$:
\begin{equation}
\Delta_{\theta}\left(t\right)=K_{P,q}\left(q_{c}\left(t\right)-q\left(t\right)\right)
\end{equation}
\item \textbf{Yaw Rate Proportional-Integral controller:} calculates the
desired deviation from the base thrust, $\Delta_{\psi}$, from an
external setpoint $r_{c}$ that can be specified through a teleoperation
system or as it is going to be specified later, by an off-board controller.
The control law for this compensator is:
\begin{equation}
\Delta_{\theta}\left(t\right)=K_{P,r}\left(r_{c}\left(t\right)-r\left(t\right)\right)+K_{I,r}\int_{0}^{t}\left(r_{c}\left(\tau\right)-r\left(\tau\right)\right)d\tau
\end{equation}
\end{itemize}
\autoref{tab:Rate-Controller's-gains} contains the gains for each of these controllers, taken directly from the Crazyflie's firmware.
\renewcommand{\arraystretch}{1.1}
\begin{table}[H]
\begin{centering}
\begin{tabular}{|c|c|c|}
\hline 
Controller & $K_{P}$ & $K_{I}$\tabularnewline
\hline 
Roll Rate & $70$ & -\tabularnewline
\hline 
Pitch Rate & $70$ & -\tabularnewline
\hline 
Yaw Rate & $70$ & $16.7$\tabularnewline
\hline 
\end{tabular}
\par\end{centering}
\caption{\label{tab:Rate-Controller's-gains}Rate Controller's gains.}
\end{table}
\renewcommand{\arraystretch}{0.75}
\subsubsection{Outer Loop: Attitude Controller}

\begin{figure}[H]
\begin{centering}
\begin{tikzpicture}[>=stealth]
  \coordinate (orig)   at (0,0);
  \coordinate (LLA)    at (0,4);
  \coordinate (AroneA) at (-1.8,5.5);
  \coordinate (ArtwoA) at (-1.8,4.5);
  \coordinate (outA) at (5,5);

 \node[draw,fill=blue!20, minimum width=2cm, minimum height=2cm, anchor=south west, text width=2cm, align=center] (A) at (LLA) {Attitude\\Controller};
   
   \draw[->] (AroneA)-- node[above]{$\color{black}\bm{\phi_c},\bm{\theta_c}$} ($(A.180)+(0,0.5)$);
   \draw[->] (ArtwoA)-- node[above]{$\phi, \theta$} ($(A.180)+(0,-0.5)$);
   \draw[->] (A.0)-- node[above]{$p_c, q_c$} (outA);

\end{tikzpicture}
\par\end{centering}
\caption{\label{fig:rate_controller}Rate Controller diagram.}
\end{figure}
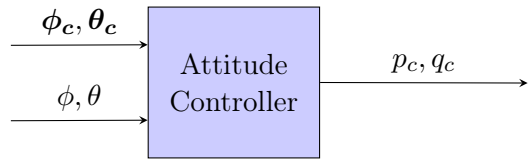

The inputs and outputs of the attitude controller are as show in \autoref{fig:rate_controller}. 

This controller act as a regulator of the rate controller, calculating
the appropriate setpoints for the angular velocities around the X
and Y axis, in order to stabilize the quadcopter at a certain desired
angular position. The attitude controller uses the pitch and roll
angles estimates from the sensor fusion algorithm, compares them
to the external commands $\phi_{c}$ and $\theta_{c}$ (coming from
teleoperation, off-board controller, etc) and feeds them to a controller
that calculates the desired angular velocities $p_{c}$ and $q_{c}$.
These controllers are as follows:
\begin{itemize}
\item \textbf{Roll Attitude Proportional-Integral controller:} computes
the desired roll rate in the body frame, $p_{c}$ ,using the control
law:
\begin{equation}
p_{c}\left(t\right)=K_{P,\phi}\left(\phi_{c}\left(t\right)-\phi\left(t\right)\right)+K_{I,\phi}\int_{0}^{t}\left(\phi_{c}\left(\tau\right)-\phi\left(\tau\right)\right)d\tau
\end{equation}
\item \textbf{Pitch Attitude Proportional-Integral controller:} works in
the same fashion as the roll controller, using the corresponding variables:
\begin{equation}
q_{c}\left(t\right)=K_{P,\theta}\left(\theta_{c}\left(t\right)-\theta\left(t\right)\right)+K_{I,\theta}\int_{0}^{t}\left(\theta_{c}\left(\tau\right)-\theta\left(\tau\right)\right)d\tau
\end{equation}
\end{itemize}
Once again, the gains for these controllers were already specified
in the Crazyflie's firmware release 2016.02, as seen in \autoref{tab:Attitude-Controller's-gains}. The same values were used during this project as they turned out to be well tuned according to the
simulations and the tests made with the real platform.
\renewcommand{\arraystretch}{1.1}
\begin{table}[H]
\begin{centering}
\begin{tabular}{|c|c|c|}
\hline 
Controller & $K_{P}$ & $K_{I}$\tabularnewline
\hline 
Roll Attitude & $3.5$ & $2$\tabularnewline
\hline 
Pitch Attitude & $3.5$ & $2$\tabularnewline
\hline 
\end{tabular}
\par\end{centering}
\caption{\label{tab:Attitude-Controller's-gains}Attitude Controller's gains.}
\end{table}
\renewcommand{\arraystretch}{0.75}
As it is evident from the gain values in \Cref{tab:Rate-Controller's-gains,tab:Attitude-Controller's-gains}, the roll and pitch gains for both controllers are the same, which is consistent with the initial
hypothesis that the quadcopter is a symmetrical body around all its
axes.

\subsubsection{Control Mixer}

The output of the rate controller is the total input
variation of the motors from the equilibrium state required to generate
a torque in the desired direction of movement. Afterwards this input
variation has to be distributed to the motors in the same fashion
as in \eqref{eq:37}, for it to move and rotate appropriately using the
PWM input it received. Given that the quadcopter is in X configuration,
the motor effort has to be distributed halfway in each motor for a
desired torque around the X or Y axis. All this analysis can be translated
in the following equations that are implemented on board of the Crazyflie
2.0, thus specifying a ``Control mixer'' block in the simulation
environment:
\begin{equation}
\begin{cases}
PWM_{motor_{1}}= & \Omega-\Delta_{\phi}/2-\Delta_{\theta}/2-\Delta_{\psi}\\
PWM_{motor_{2}}= & \Omega+\Delta_{\phi}/2-\Delta_{\theta}/2+\Delta_{\psi}\\
PWM_{motor_{3}}= & \Omega+\Delta_{\phi}/2+\Delta_{\theta}/2-\Delta_{\psi}\\
PWM_{motor_{4}}= & \Omega-\Delta_{\phi}/2-\Delta_{\theta}/2+\Delta_{\psi}
\end{cases}\label{eq:3.1-1}
\end{equation}
where $\Omega$ is the PWM base signal for maintaining a certain altitude,
a value that will be regulated from an altitude controller; $\Delta_{\phi}$,
$\Delta_{\theta}$ and $\Delta_{\psi}$ represent the outputs of the
Rate Controller and at the same time give an idea of a deviation needed
from the base thrust in order to obtain a certain torque in the X,
Y or Z axis. From \eqref{eq:3.1-1} it is more clearly how,
for example, a command $\Delta_{\phi}>0$ will be distributed 50/50
as a reduction of the PWM input supplied to motors 1 and 4, and an
increase of the power supplied to motors 2 and 3, thus creating the
appropriate angular momentum to obtain certain angle in the pitch
direction. 

\begin{figure}[H]
\centering
\resizebox{1\textwidth}{!}{
\begin{tikzpicture}[>=stealth]
  \coordinate (orig)   at (0,0);
  \coordinate (LLD)    at (9,1);
  \coordinate (LLA)    at (0,4);
  \coordinate (AroneA) at (-1.5,5.5);
  \coordinate (ArtwoA) at (-1.5,4.5);
  \coordinate (LLB)    at (5,4);
  \coordinate (AroneB) at (5-1.5,4.5);
  \coordinate (ArtwoB) at (5-1.5,5);
  \coordinate (LLC)    at (10,4);
  \coordinate (AroneC) at (10-1.5,4.5);
  \coordinate (outC) at (14,5);

 \node[draw,fill=blue!20, minimum width=2cm, minimum height=2cm, anchor=south west, text width=2cm, align=center] (A) at (LLA) {Attitude\\Controller};
 \node[draw,fill=blue!20, minimum width=2cm, minimum height=2cm, anchor=south west, text width=2cm, align=center] (B) at (LLB) {Rate\\Controller};
 \node[draw,fill=blue!20, minimum width=2cm, minimum height=2cm, anchor=south west, text width=2cm, align=center] (C) at (LLC) {Control\\Mixer};
  
   \draw[->] (AroneA)-- node[above]{$\color{red}\bm{\phi_c},\bm{\theta_c}$} ($(A.180)+(0,0.5)$);
   \draw[->] (ArtwoA)-- node[above]{$\phi, \theta$} ($(A.180)+(0,-0.5)$);
   \draw[->] ($(A.0)+(0,0.5)$)-- node[above]{$p_c, q_c$} ($(B.180)+(0,0.5)$);
   \draw[->] (AroneB)-- node[above]{$p, q, r$} ($(B.180)+(0,-0.5)$);
   \draw[->] (ArtwoB)-- node[above]{$\color{red}\bm{r_c}$} (B.180);
   \draw[->] ($(B.0)+(0,0.5)$)-- node[above]{$\Delta_\phi, \Delta_\theta, \Delta_\psi$} ($(C.180)+(0,0.5)$);
   \draw[->] (AroneC)-- node[above]{$\color{red}\bm{\Omega}$} ($(C.180)+(0,-0.5)$);
   \draw[->] (C.0)-- node[above]{$PWM$} (outC);

\end{tikzpicture}}
\caption{\label{fig:Onboard-control-architecture}Onboard control architecture
with Control mixer.}

\end{figure}
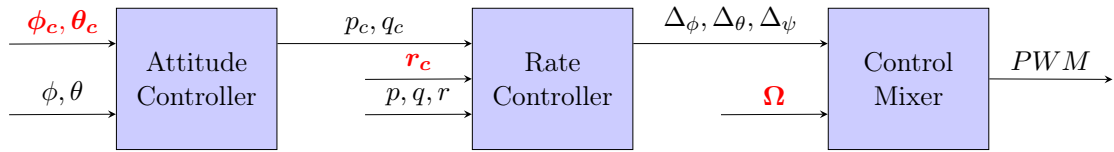

~\\
\autoref{fig:Onboard-control-architecture} shows a complete diagram of the on-board control scheme.

The red inputs in this diagram represent the actual inputs that can
be controlled from outside of the firmware, as it was suggested earlier,
either by a teleoperation system or by an automated position controller.
The other signals come from the sensor fusion algorithm or are intermediate
variables in the control process.

\subsection{\label{subsec:Off-Board-Position-controller}Off-Board Position Controller}

In hopes of controlling the quadcopter by sending waypoints or trajectories
in a tridimensional space, it became necessary to add a position controller
that in terms of the implementation will be running off-board unlike
the controller of the previous section. The job of this controller
can be divided into:
\begin{enumerate}
\item An altitude controller whose output is the thrust $\Omega$ required
to maintain a certain position in z.
\item An X-Y position controller whose outputs are the required roll
and pitch angles that will be regulated by the on-board controller.
\item A yaw position controller that sends the required angular velocity
to the on-board Yaw Rate Controller.
\end{enumerate}
As seen in the dynamic analysis of the quadcopter, there exists a theoretical
decoupling between the vertical, lateral, longitudinal and yaw movement,
meaning that each one of these controllers can be tuned independently
from each other, which simplifies the controller design task.

\subsubsection{Altitude Controller}

\begin{figure}[H]
\begin{centering}
\begin{tikzpicture}[>=stealth]
  \coordinate (orig)   at (0,0);
  \coordinate (LLA)    at (0,4);
  \coordinate (AroneA) at (-1.5,5.5);
  \coordinate (ArtwoA) at (-1.5,4.5);
  \coordinate (LLB)    at (5,5);
  \coordinate (Feed1)  at (5,4);
  \coordinate (Feed2)  at (5,3);
  \coordinate (out)    at (7,5);

 \node[draw,fill=blue!20, minimum width=2cm, minimum height=2cm, anchor=south west, text width=2cm, align=center] (A) at (LLA) {Altitude\\Controller};
 \node[draw, fill=blue!20, circle, minimum width=1cm, minimum height=1cm, anchor=center, align=center] (sum) at (LLB) {};
  
   \draw[->] (AroneA)-- node[above]{$\color{blue}\bm{z_c} $} ($(A.180)+(0,0.5)$);
   \draw[->] (ArtwoA)-- node[above]{$z$} ($(A.180)+(0,-0.5)$);
   \draw[->] (A.0)-- node[above]{$\Delta \Omega \hspace{1cm}+$} (sum.180);
   \draw[->] (Feed1)--node [left,pos=0.8]{$+$}(sum.270) ;
   \draw[-]  (Feed2)--node [above,pos=-0.5]{Feedforward $\Omega_e$} (Feed1);
   \draw[->] (sum)--node [above]{$\Omega$} (out);
    

\end{tikzpicture}
\par\end{centering}
\caption{\label{fig:Altitude-Controller}Altitude Controller.}
\end{figure}
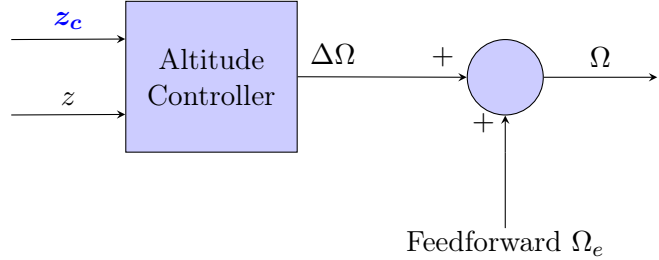

It is common practice to add
a feedforward term as in \cite{key-4} that compensates the weight
of the quadcopter, in order to avoid the use of large controller gains
that can lead to saturation problems. The structure in \autoref{fig:Altitude-Controller} was the one used for this controller.

The altitude controller is a simple PID compensator whose inputs are
the desired altitude that comes from the trajectory block, and the
altitude state that in terms of simulation comes from the quadcopter's
dynamics block. The equation that describes this PID controller is
the following:
\begin{equation}
\resizebox{.85\hsize}{!}{$
\Delta \Omega\left(t\right)=K_{P,z}\left(z_{c}\left(t\right)-z\left(t\right)\right)+K_{I,z}\int_{0}^{t}\left(z_{c}\left(\tau\right)-z\left(\tau\right)\right)d\tau+K_{D,z}\frac{d}{dt}\left(z_{c}\left(t\right)-z\left(t\right)\right)$}
\end{equation}
The output of this PID controller is the 16-bit thrust deviation from
the equilibrium point set at the hover state. That means that the
feedforward term $\Omega_{e}$ in \autoref{fig:Altitude-Controller}
is the necessary PWM 16-bit signal needed for the quad to maintain
its altitude. Using the calculations at the equilibrium, as seen in \eqref{eq:2.34} and \eqref{eq:3.1}, the feedforward term can be calculated as:
\begin{equation}
\boxed{\Omega_{e}=\frac{\omega_{e}-4070.3}{0.2685}=44705}
\end{equation}
\subsubsection{\label{subsec:X-Y-Position-Controller}X-Y Position Controller}

\begin{figure}[H]
\begin{centering}
\begin{tikzpicture}[>=stealth]
  \coordinate (orig)   at (0,0);
  \coordinate (LLD)    at (9,1);
  \coordinate (LLA)    at (0,4);
  \coordinate (AroneA) at (-1.5,5.5);
  \coordinate (ArthrA) at (-1.5,5);
  \coordinate (ArtwoA) at (-1.5,4.5);
  \coordinate (LLB)    at (5,4);
  \coordinate (AroneB) at (5-1.5,4.5);
  \coordinate (outB) at (9.5,5);

 \node[draw,fill=blue!20, minimum width=2cm, minimum height=2cm, anchor=south west, text width=2cm, align=center] (A) at (LLA) {Error to\\Body Frame};
 \node[draw,fill=blue!20, minimum width=2cm, minimum height=2cm, anchor=south west, text width=2cm, align=center] (B) at (LLB) {X-Y Position\\Controller};
  
   \draw[->] (AroneA)-- node[above]{$\color{blue}\bm{x_c},\bm{y_c}$} ($(A.180)+(0,0.5)$);
   \draw[->] (ArtwoA)-- node[above]{$\psi$} ($(A.180)+(0,-0.5)$);
   \draw[->] (ArthrA)-- node[above]{$x, y$} (A.180);
   
   \draw[->] ($(A.0)+(0,0.5)$)-- node[above]{$x_{e}^{b}, y_{e}^{b}$} ($(B.180)+(0,0.5)$);
   \draw[->] (AroneB)-- node[above]{$u, v$} ($(B.180)+(0,-0.5)$);
   \draw[->] (B.0)-- node[above]{$\color{red}\bm{\phi_c},\bm{\theta_c}$} (outB);

\end{tikzpicture}
\par\end{centering}
\caption{\label{fig:X-Y-Position-Controller}X-Y Position Controller.}
\end{figure}
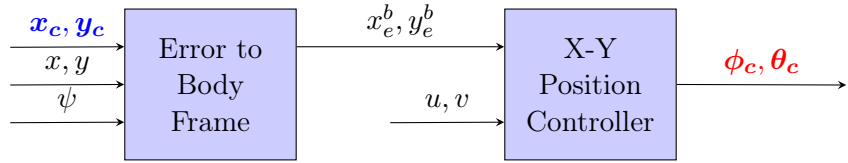

The objective of this controller is to regulate the on-board Attitude
Controller by calculating the necessary roll and pitch angles in order
to move the quadcopter between locations in the X-Y plane. The block diagram in \autoref{fig:X-Y-Position-Controller} shows the inputs and outputs of this controller.

In \cite{key-5} it was proven that a similar architecture as here
gives a good performance in position tracking. The first block calculates the error between
the desired and actual X-Y position and does the rotation operation
needed to project this error vector in the body frame. This operation
is given as:
\begin{equation}
\left[\begin{array}{c}
x_{e}\\
y_{e}
\end{array}\right]^{b}=\left[\begin{array}{cc}
\cos\left(\psi\right) & \sin\left(\psi\right)\\
-\sin\left(\psi\right) & \cos\left(\psi\right)
\end{array}\right]\left[\begin{array}{c}
x_{e}\\
y_{e}
\end{array}\right]^{o}
\end{equation}
doing the calculations, the error in the body frame is defined as:
\begin{equation}
\begin{cases}
x_{e}^{b}=x_{e}^{o}\cos\left(\psi\right)+y_{e}^{o}\sin\left(\psi\right)\\
y_{e}^{b}=-x_{e}^{o}\sin\left(\psi\right)+y_{e}^{o}\cos\left(\psi\right)
\end{cases}
\end{equation}
Then the error in the body-fixed frame becomes the setpoint to the
velocity in this same frame, the logic behind this apparently odd
choice of a setpoint is that the bigger the error is, the more rapidly
the quadcopter should move in order to arrive at the desired point
as quickly as possible. Otherwise, if the error is small, meaning
the drone is near the desired point, the setpoint for the velocity
should be also small. 

~\\
A more conventional method to do the position tracker is that
the error in the body-fixed frame dictates the setpoint of the position
instead of the velocity, but in practice it was found that in the
first method proposed is easier to tune the PID gains as there are
only 2 out of 3 gains that need to be adjusted (derivative gain is
not used), whereas for the traditional tracker all three gains have
to be used for a good performance. The only disadvantage that might
have the first method with respect to the second is that in the real system the velocity in the
body-fixed frame is not directly measured, thus it is necessary to
estimate this states through mathematical means.

~\\
The controllers that compute the desired attitude use the following
control laws:
\begin{itemize}
\item \textbf{X Position Proportional-Integral Controller:} 
\begin{equation}
\phi_{c}\left(t\right)=K_{P,x}\left(x_{e}^{b}\left(t\right)-u\left(t\right)\right)+K_{I,x}\int_{0}^{t}\left(x_{e}^{b}\left(\tau\right)-u\left(\tau\right)\right)d\tau
\end{equation}
\item \textbf{Y Position Proportional-Integral Controller:}
\begin{equation}
\theta_{c}\left(t\right)=K_{P,y}\left(y_{e}^{b}\left(t\right)-v\left(t\right)\right)+K_{I,y}\int_{0}^{t}\left(y_{e}^{b}\left(\tau\right)-v\left(\tau\right)\right)d\tau
\end{equation}
\end{itemize}

\subsubsection{Yaw Position Controller}

The inputs and outputs of this controller are reflected in \autoref{fig:Yaw-Position-Controller}:

\begin{figure}[H]
\begin{centering}
\begin{tikzpicture}[>=stealth]
  \coordinate (orig)   at (0,0);
  \coordinate (LLA)    at (0,4);
  \coordinate (AroneA) at (-1.5,5.5);
  \coordinate (AroneB) at (-1.5,4.5);
  \coordinate (outA) at (4,5);

 \node[draw,fill=blue!20, minimum width=2cm, minimum height=2cm, anchor=south west, text width=2cm, align=center] (A) at (LLA) {Yaw\\Position\\Controller};
  
   \draw[->] (AroneA)-- node[above]{$\color{blue}\bm{\psi_c}$} ($(A.180)+(0,0.5)$);
   \draw[->] (AroneB)-- node[above]{$\psi$} ($(A.180)+(0,-0.5)$);
   \draw[->] (A.0)-- node[above]{$\color{red}\bm{r_c}$} (outA);

\end{tikzpicture}
\par\end{centering}
\caption{\label{fig:Yaw-Position-Controller}Yaw Position Controller.}
\end{figure}
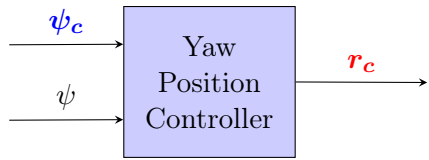
The controller computes the error between the desired yaw position
and the actual position and feeds it to a proportional controller
whose output is the desired yaw rate that is regulated by the on-board
rate controller. Thus the operation done by the controller is given
as:
\begin{equation}
r_{c}\left(t\right)=K_{P,\psi}\left(\psi_{c}\left(t\right)-\psi\left(t\right)\right)
\end{equation}

\subsubsection{Controllers Gains}

After validation both in simulation and in the real platform, the gains for each one of the off-board controllers are summarized in \autoref{tab:Gains-for-Off-Board}.
\renewcommand{\arraystretch}{1.1}
\begin{table}[H]
\begin{centering}
\begin{tabular}{|c|c|c|c|c|}
\hline 
Controller & $K_{P}$ & $K_{I}$ & $K_{D}$ & Output Limit\tabularnewline
\hline 
Altitude & $11000$ & $3500$ & $9000$ & $[-20000,15000]$\tabularnewline
\hline 
$X$ & $30$ & $2$ & - & $\pm30\,\textrm{[deg]}$ \tabularnewline
\hline 
$Y$ & $-30$ & $-2$ & - & $\pm30\,\textrm{[deg}]$ \tabularnewline
\hline 
Yaw & $3$ & - & - & $\pm200\,\textrm{[deg/s]}$ \tabularnewline
\hline 
\end{tabular}
\par\end{centering}
\caption{\label{tab:Gains-for-Off-Board}Gains for Off-Board Controllers.}
\renewcommand{\arraystretch}{0.75}
\end{table}
Note that the gains for the Y Position Controller are inverted, this
is because a positive roll angle makes the quadcopter move toward the
negative Y axis, so the solution is to invert the gains in order to
send the correct attitude commands to the on-board controller.

\subsection{Simulation Results}

The simulation environment was built in Simulink following the block
diagram in \autoref{fig:Block-Diagram-of}. The goal of said simulations
was to test the control system for tracking a desired position $[x_{c},y_{c},z_{c},\psi_{c}]$
and compare the behavior of the non-linear dynamics of the quadcopter
with the linear state model.
\begin{itemize}
\item\textbf{Linear trajectories} 
\end{itemize}

\begin{figure}[H]
\centering
\makebox[0pt]{
\includegraphics[width=1.2\textwidth]{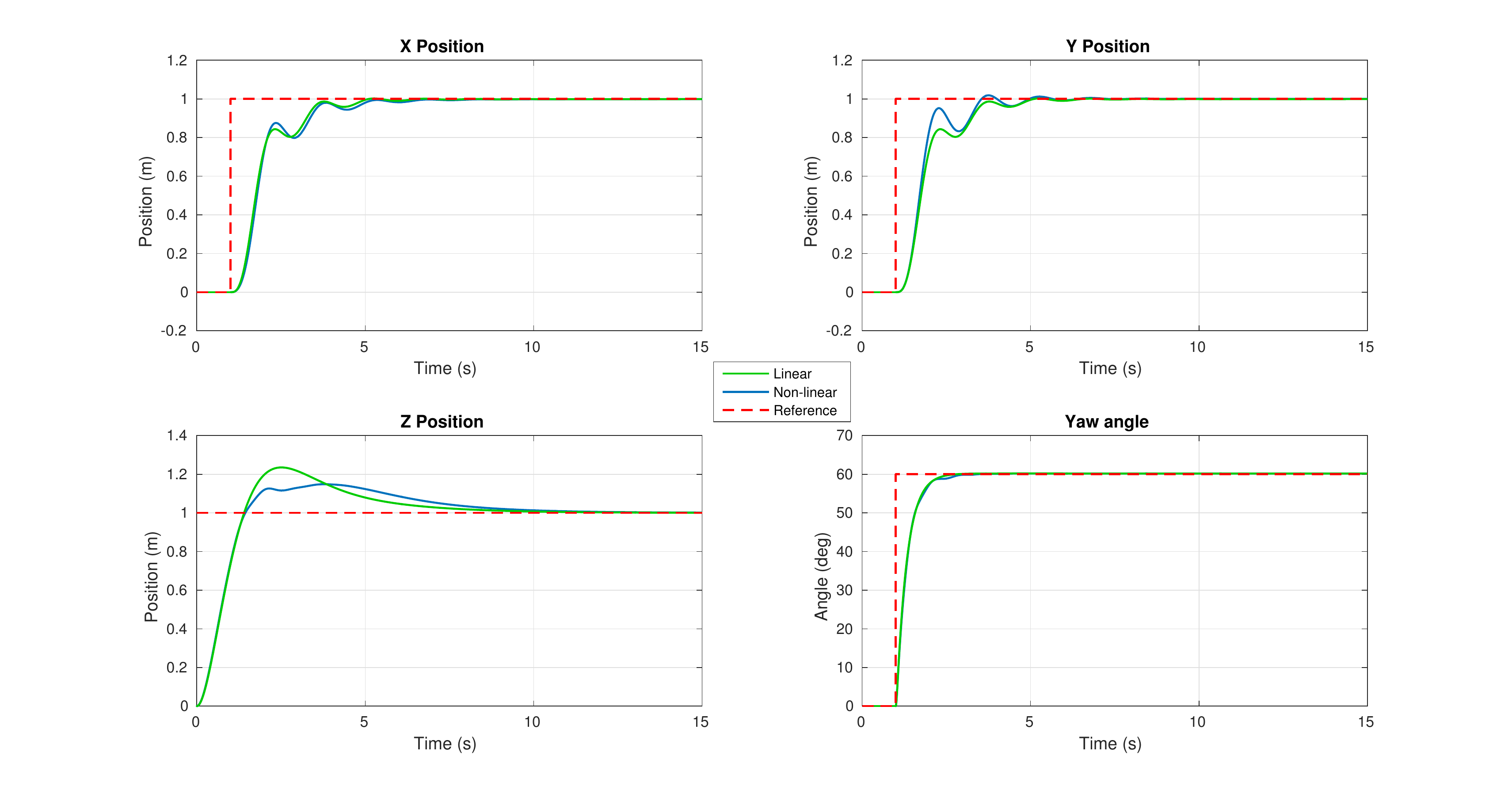}}
\caption{\label{fig:Simulation-results-Test}Simulation results for Test \#1.}
\end{figure}

This first simulation tests the response of the system when demanded
to follow step functions in all 4 trajectory inputs. The setpoints
for Test \#1 where: $x_{c}=1\textrm{m}$~~;~~$y_{c}=1\textrm{m}$~~;~~$z_{c}=1\textrm{m}$~~;~~$\psi_{c}=60\text{\textdegree}$.
\autoref{fig:Simulation-results-Test} shows the simulation of
a 15 seconds flight of the quadcopter given the desired trajectory.
For the X-Y position the time response is about 3 seconds, with almost
no overshoot, whereas for the Z position the response is slower at
roughly 8 seconds for a 2\% error margin and with a more pronounced overshoot.
With the experimental results, this values of PID gains gave the best
response after some trial and error in gain-tuning. As for the yaw response, it has a time response
of about 2 seconds with no overshoot.

\begin{figure}[H]
\centering
\makebox[0pt]{
\subfloat[\label{fig:Standard-View}Standard View.]{\includegraphics[width=0.55\textwidth]{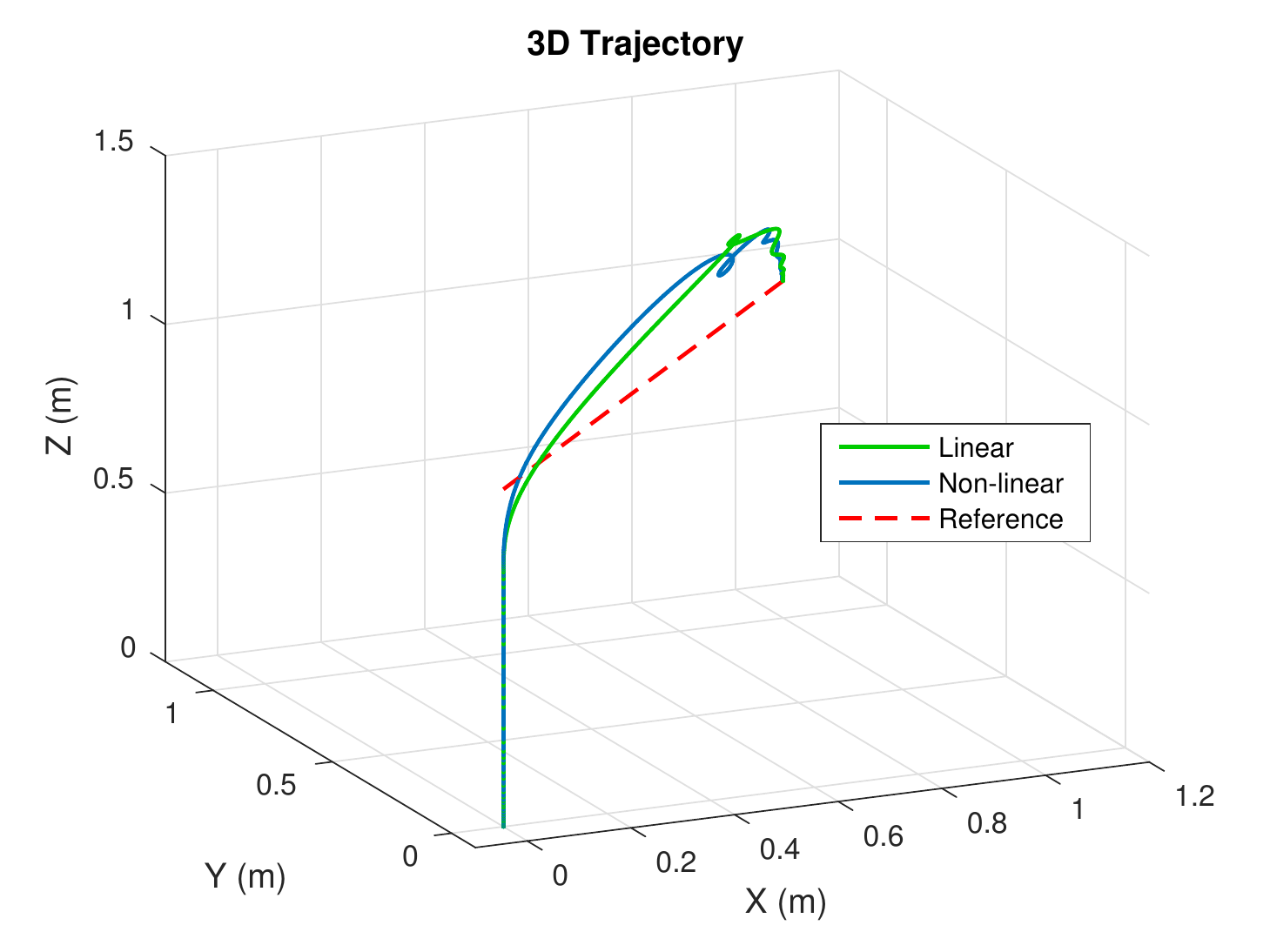}

}\subfloat[\label{fig:Top-View}Top View.]{\includegraphics[width=0.55\textwidth]{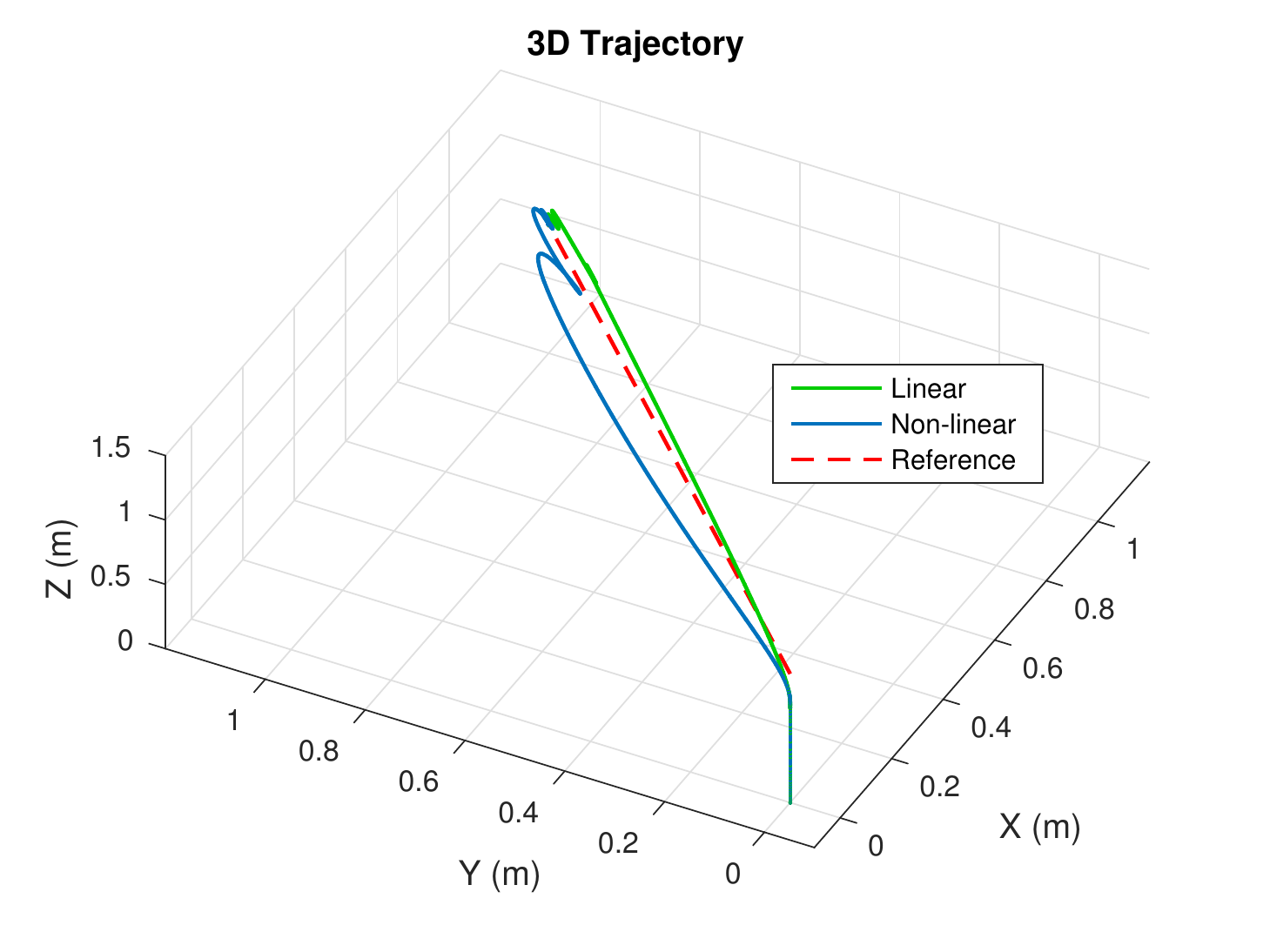}}}
\caption{3D Trajectory for Test \#1.}
\end{figure}

Comparing the behavior of the non-linear model with the linear state
space model, there are some notable differences. As shown in \Cref{fig:Standard-View,fig:Top-View}, the trajectory followed in the two cases is not exactly the same even though it is clear that
in both situations the desired final position was reached. The linear
system follows a perfect line from the starting point to the desired
point, meaning that all movements are decoupled. In the non-linear system there exist no such perfect
decoupling and as suggested by the trajectory followed, the motion
dynamics are intertwined and influence each other, meaning
for example that the movement in the quadcopter's X axis has some impact
in the Y axis and vice versa, even though they are small.

\begin{figure}[H]
\centering
\makebox[0pt]{
\subfloat[\label{fig:Standard-View-1}Time response.]{\includegraphics[width=0.55\textwidth]{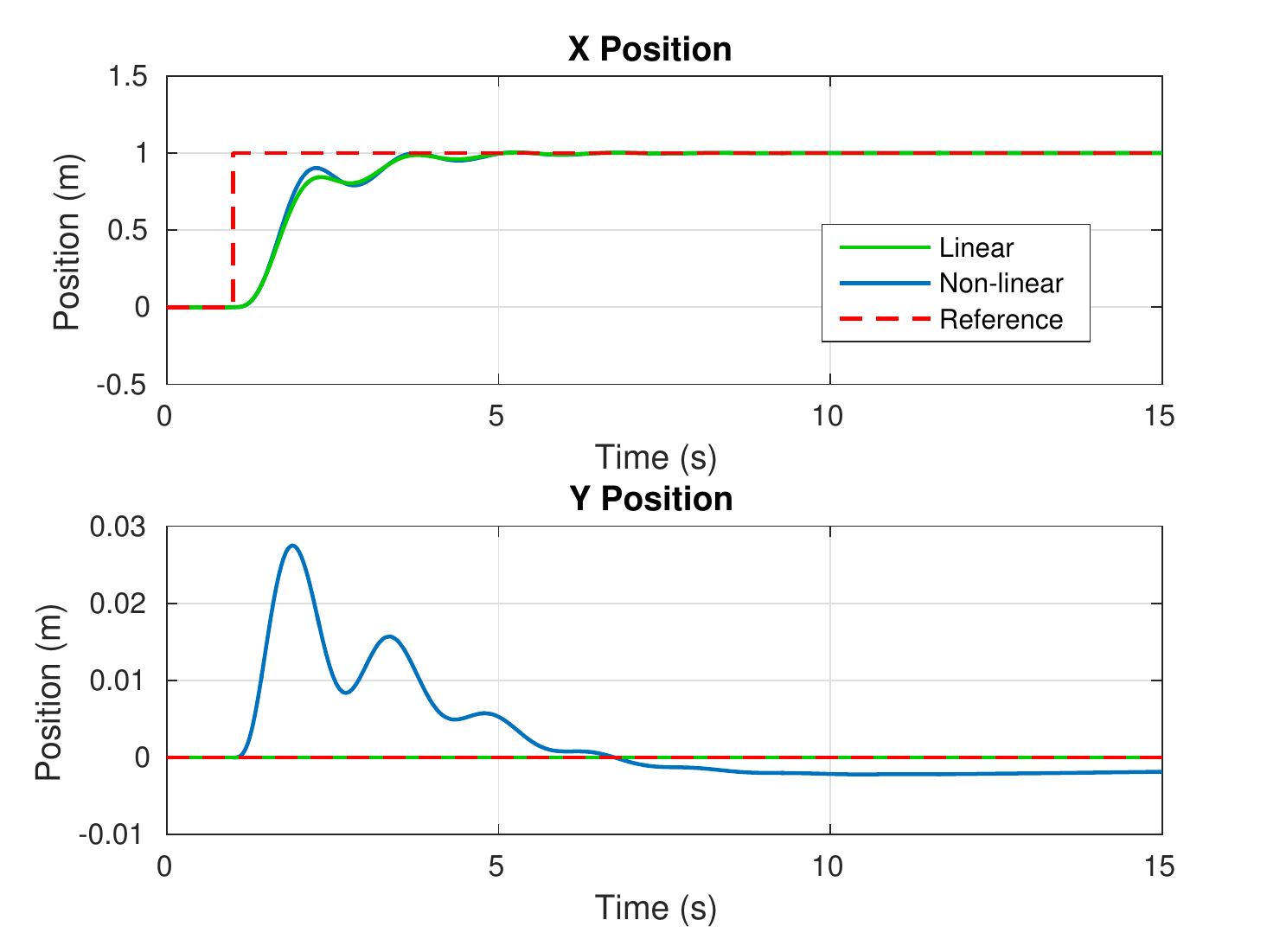}}
\subfloat[\label{fig:Top-View-1}Top View.]{\includegraphics[width=0.55\textwidth]{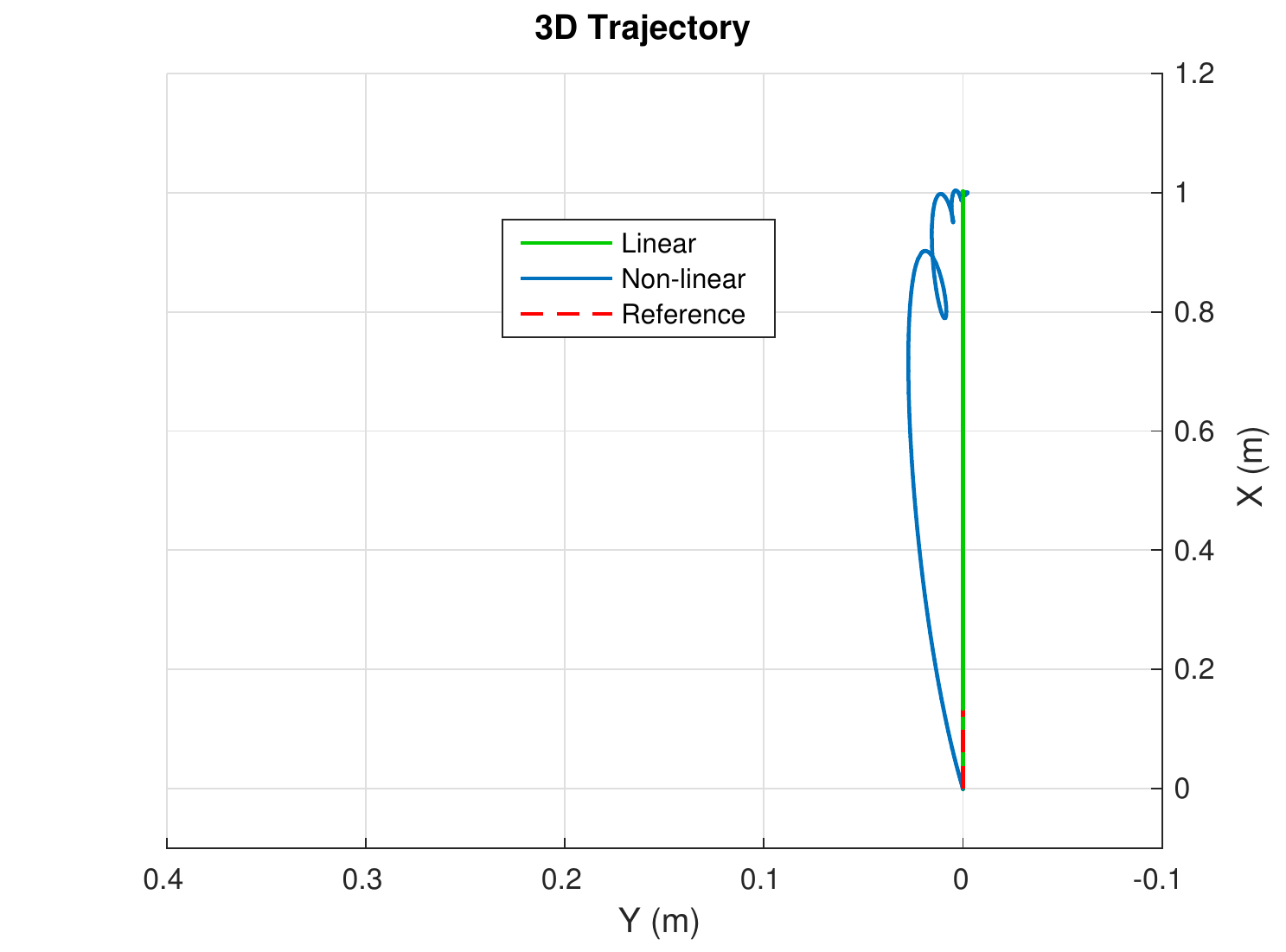}}}
\caption{Compound movement interference.}
\end{figure}

This coupled movement is better appreciated in \Cref{fig:Standard-View-1,fig:Top-View-1}, that show simulation results of a trajectory purely in the X axis, with a rotation of 60 degrees in the yaw angle,
in order to study the influence in the Y axis. Simulation results
confirm the theory of the interference between movements as this trajectory
generates a deviation of 3 centimeters in the Y axis that then returns
to zero with the controller action.

\begin{itemize}
\item\textbf{Circular trajectories} 
\end{itemize}

\begin{figure}[H]
\centering
\makebox[0pt]{
\includegraphics[width=1.2\textwidth]{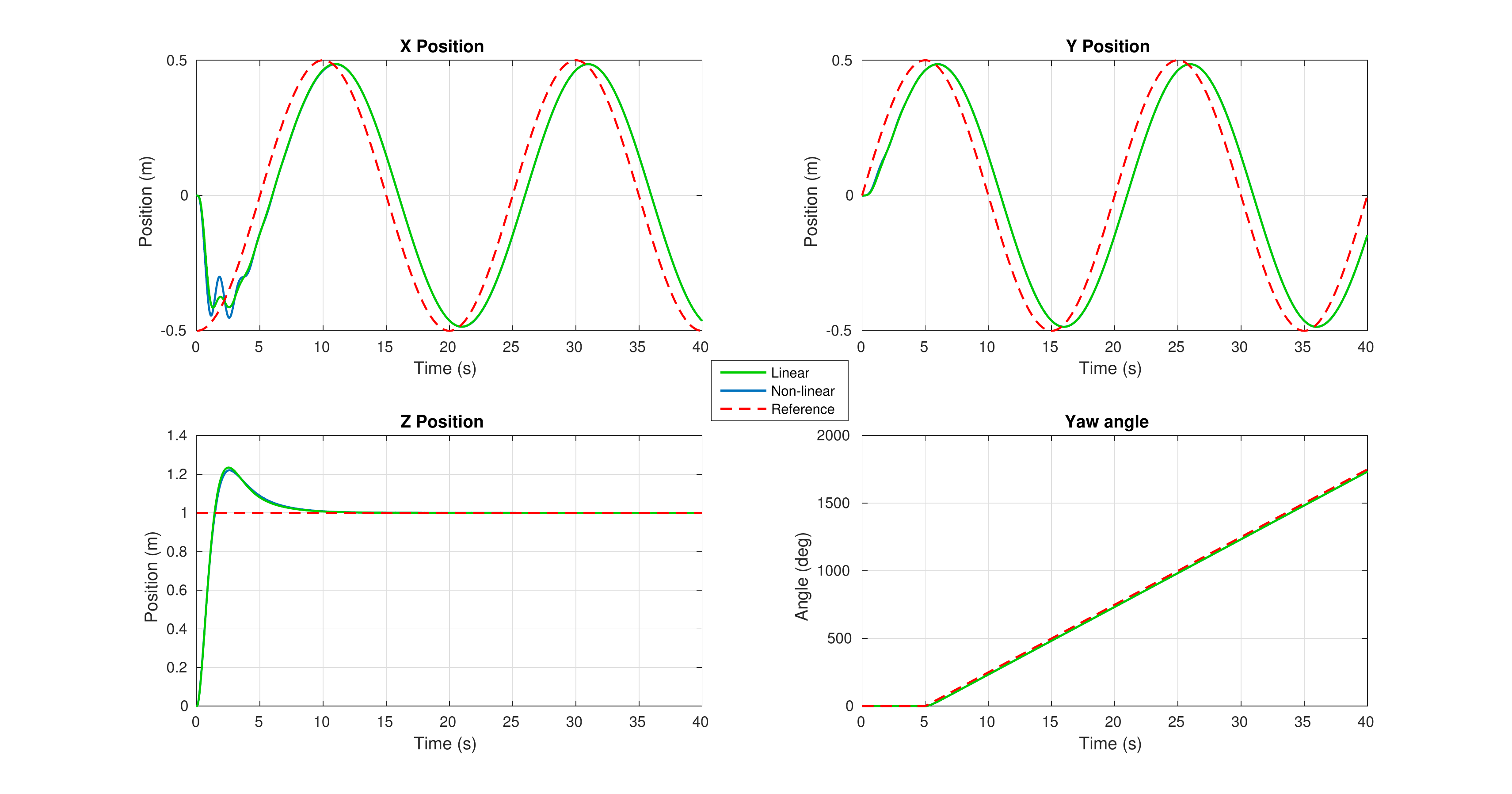}}
\caption{\label{fig:sim_circ}Simulation results for a circular trajectory.}
\end{figure}

The simulated system was also tested to track positions that change
over time, as a circular trajectory for example. This trajectory is
defined as:
\begin{equation}
\begin{cases}
x_{c}\left(t\right)=0.5\sin\left(2\pi0.05t-\pi/2\right)\\
y_{c}\left(t\right)=0.5\sin\left(2\pi0.05t\right)\\
z_{c}\left(t\right)=1\\
\psi\left(t\right)=50t
\end{cases}\label{eq:3.14}
\end{equation}
This trajectory specifies a circle of frequency 0.05Hz and radius
of 0.5 meters, at a constant altitude of 1 meter and a constant angular
velocity in the yaw angle of 50 degrees per second. \autoref{fig:sim_circ} displays the simulation results with these commands.

\begin{figure}[H]
\centering
\makebox[0pt]{
\includegraphics[width=1.1\textwidth]{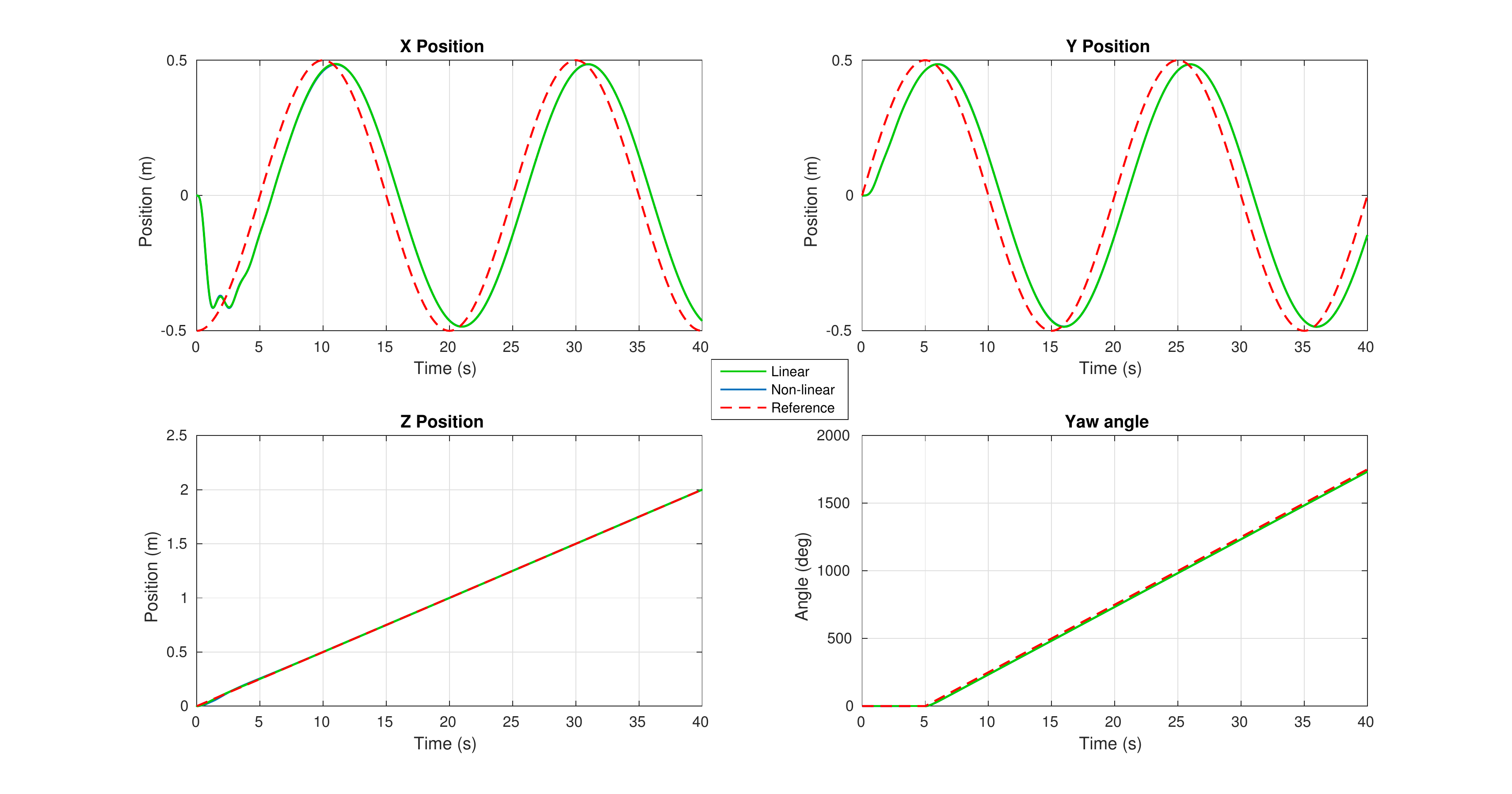}}
\caption{\label{fig:Helicoidal-Trajectory-Time}Helical Trajectory Time response.}
\end{figure}

\begin{figure}[H]
\begin{centering}
\includegraphics[scale=0.6]{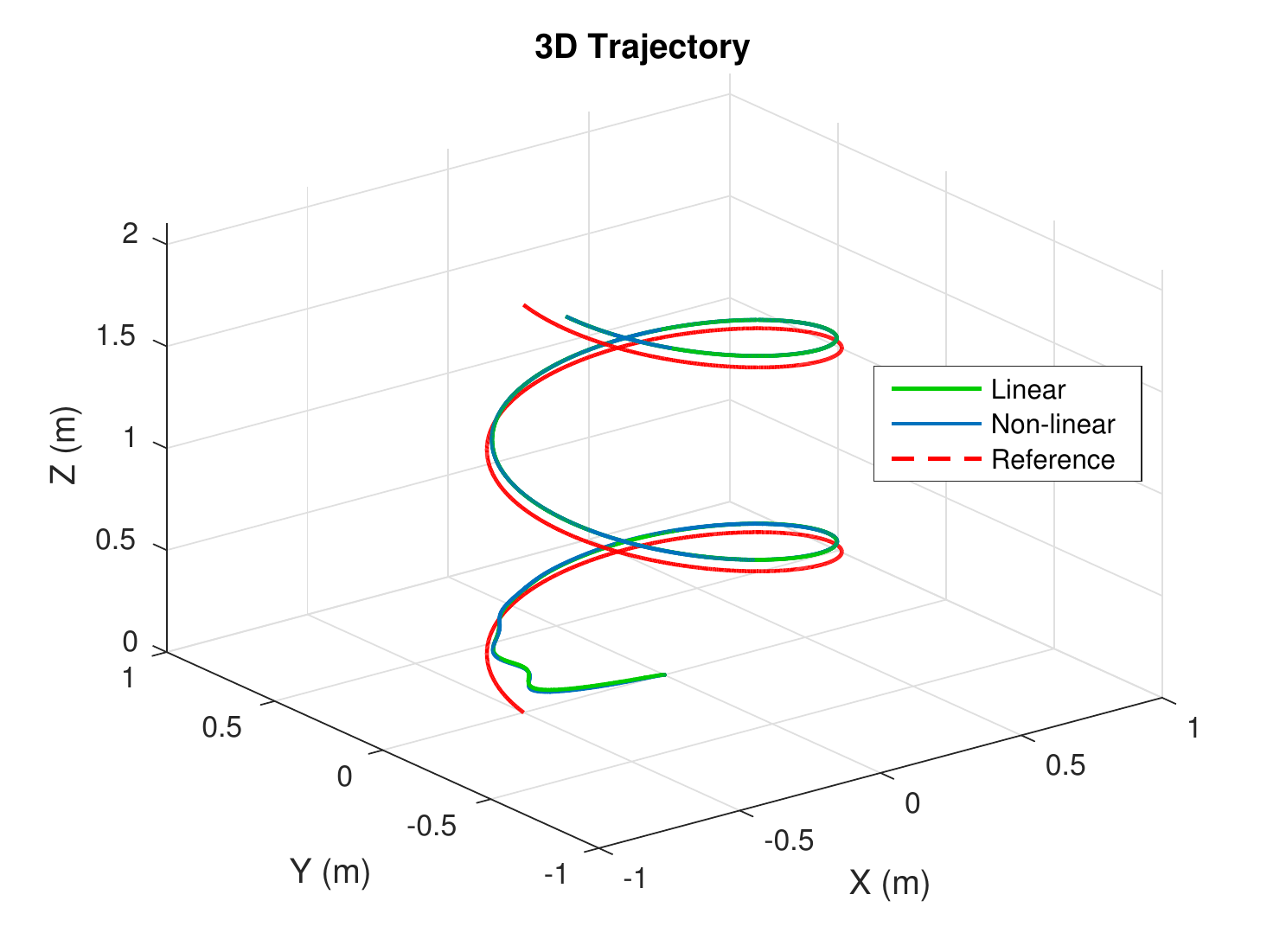}
\par\end{centering}
\caption{\label{fig:3D-Helicoidal-Trajectory}3D Helical Trajectory.}

\end{figure}

As there is no trajectory tracker in the control system, the path
taken to follow the trajectory specified in \eqref{eq:3.14}
never accomplishes the task of minimizing the error between the quadcopter's
trajectory and the desired trajectory. The position tracker by itself
cannot follow a time-varying trajectory when the change rate of said
trajectory is too fast. For example if the frequency of the circular
trajectory is augmented, the path following will be less precise.
\Cref{fig:Standard-View-circ,fig:Top-View-circ} show the 3D circular trajectory described by the quadcopter in this simulation:

\begin{figure}[H]
\centering
\makebox[0pt]{
\subfloat[\label{fig:Standard-View-circ}Standard view.]{\includegraphics[width=0.50\textwidth]{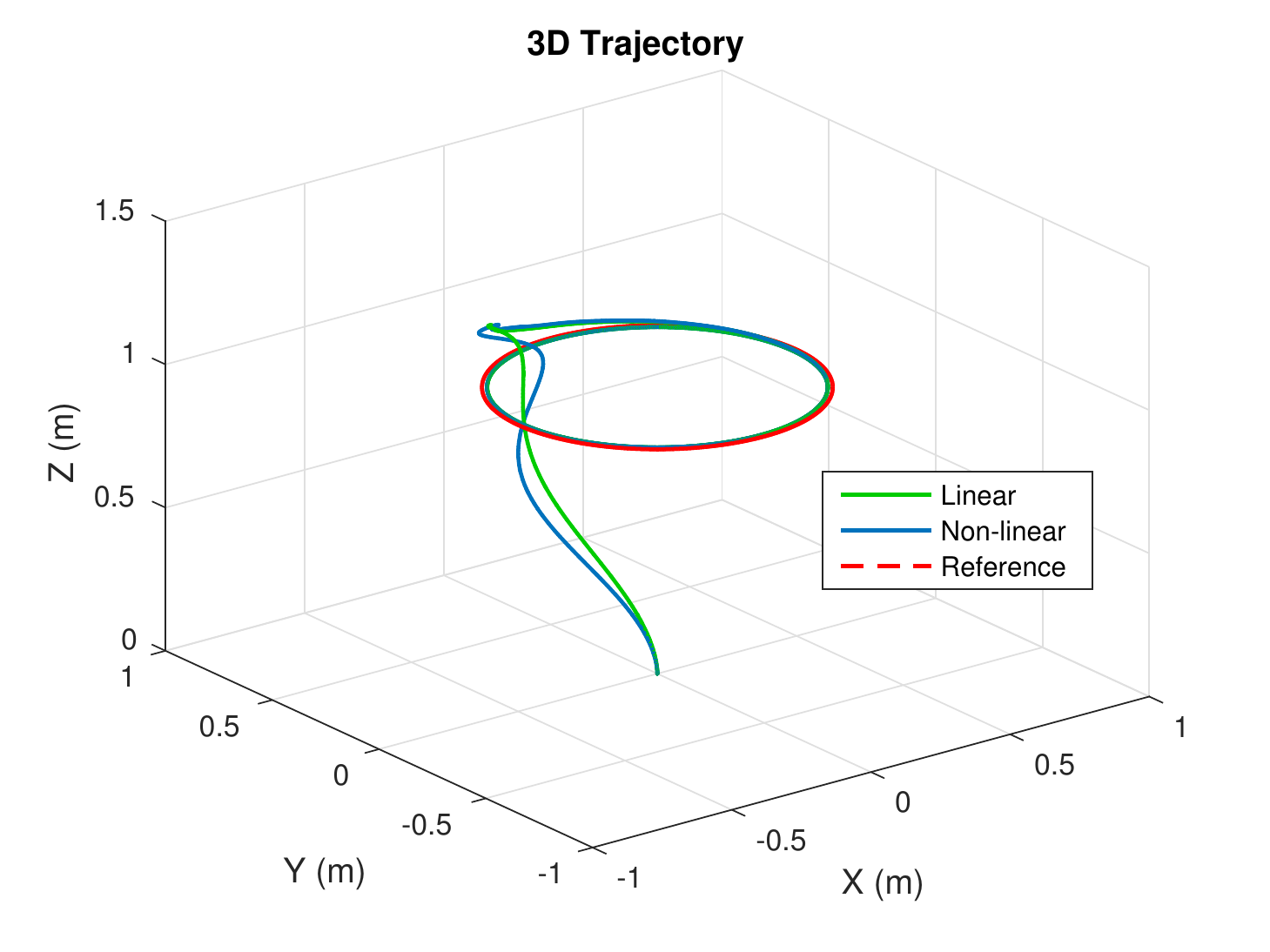}
}\subfloat[\label{fig:Top-View-circ}Top View.]{\includegraphics[width=0.50\textwidth]{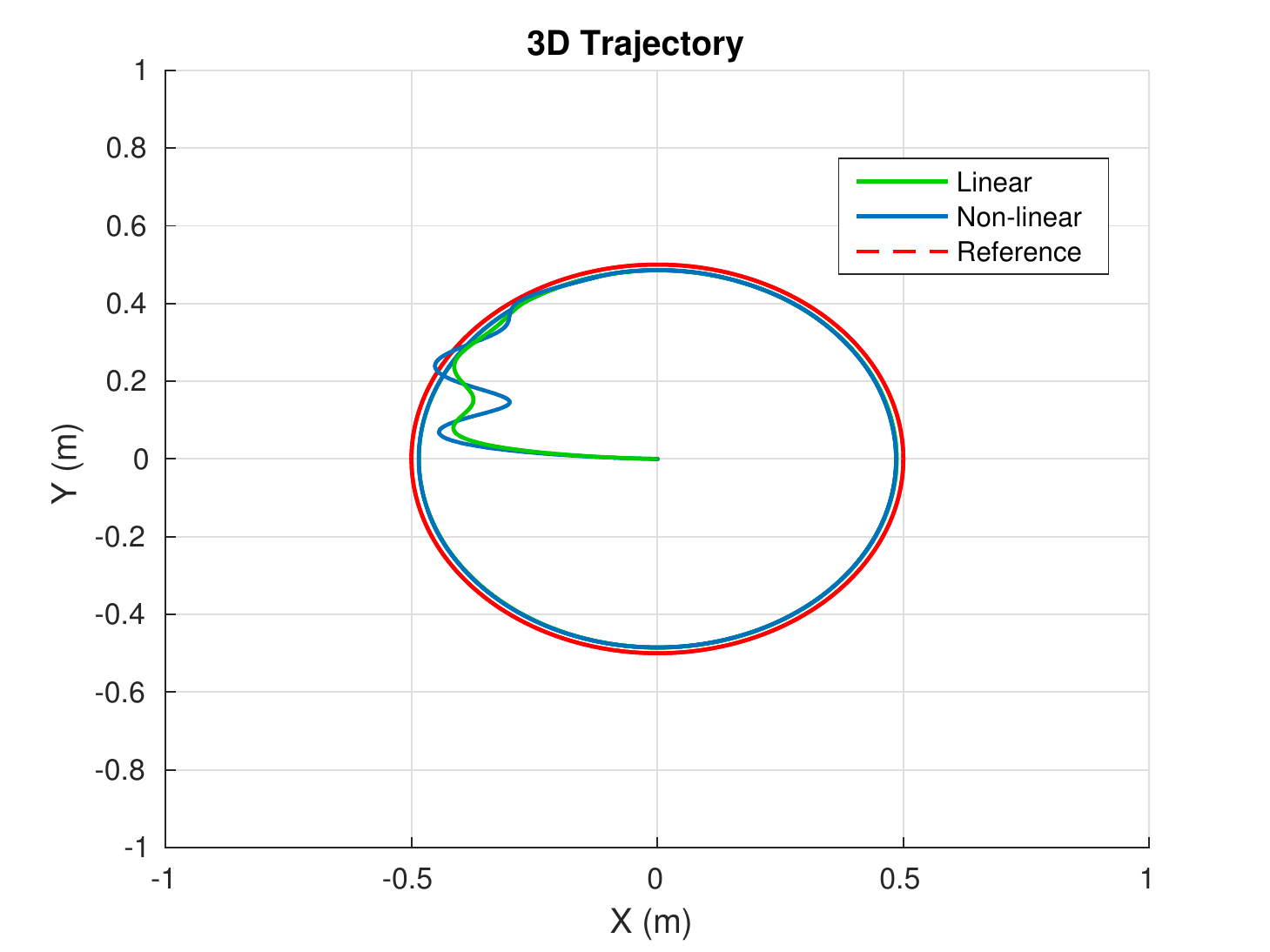}}}
\caption{3D Circular Trajectory.}
\end{figure}
A helical trajectory can be easily generated by making the altitude
command a ramp function, as specified in this next equation:
\begin{equation}
\begin{cases}
x_{c}\left(t\right)=0.5\sin\left(2\pi0.05t-\pi/2\right)\\
y_{c}\left(t\right)=0.5\sin\left(2\pi0.05t\right)\\
z_{c}\left(t\right)=0.05t\\
\psi\left(t\right)=50t
\end{cases}\label{eq:3.14-1}
\end{equation}
in which the altitude command refers to a linear velocity of 5 centimeters
per second in the vertical axis. Simulation results are presented
in \Cref{fig:Helicoidal-Trajectory-Time,fig:3D-Helicoidal-Trajectory}. The newly added time-varying command in altitude worked as expected, following the ramp.

The deficiencies of this control architecture to follow more complicated trajectories justify the need to conceive a higher performance controller for the task at hand.

\section{\label{sec:Linear-Quadratic-Tracker-(LQT)}Linear-Quadratic Tracker
(LQT)}

As the previous simulation results suggested, a more refined controller
is required in order to truly track trajectories that change over
time, which can be seen as a problem of being in the appropriate place,
at the appropriate time. There exists a wide variety of controllers suited to precisely track trajectories, e.g., Model-Predictive
controllers have proven to be quite robust in the case of quadcopters
\cite{key-26}. But this type of non-linear approach requires
heavy calculations and often require powerful processors in order
to be effective. On the other hand, the linear-quadratic tracker has
proven to be a versatile controller method for trajectory tracking
with quadcopters in previous works as \cite{key-11} and \cite{key-15},
with the advantages of linear controllers and their rapid prototyping
and implementation.

\subsection{The Optimization Problem Setup}

The LQT algorithm is formulated as an optimization problem to reduce
a cost function in terms of the plant's states, inputs and certain
weight functions that must be specified by the controller designer.
It is indeed a problem very much alike the well-known LQR, but in
this case the states considered for the resolution of the Algebraic
Ricatti Equation (ARE) are time-varying, which leads to gains that
also vary depending on the trajectory to follow. These characteristics
make the LQT controller more appropriate than the LQR when trying
to accomplish more precise trajectory following, which is exactly
the feature that the previous PID architecture lacked.

~\\
As the name suggests, the LQT algorithm is part of the family of linear
controllers, thus it uses linear state space models (or linearized
models, as in this case). The
design of this controller was done directly in the discrete-time
domain as it makes easier its implementation on the real platform. The linear state space realization obtained in \Cref{sec:Linearization-and-State}
was first discretized using a time step $T_{s}=0.01s$ that corresponds
to a frequency of 100 Hz. This time step was chosen as it corresponds
to the frequency the controller was going to be working on when
implemented in the real system. Considering the state vector:
\begin{equation}
\boldsymbol{\Delta x}=\left[\begin{array}{cccccccccccc}
\Delta x & \Delta y & \Delta z & \Delta\psi & \Delta\theta & \Delta\phi & \Delta u & \Delta v & \Delta w & \Delta r & \Delta q & \Delta p\end{array}\right]^{T}\label{eq:xaug}
\end{equation}
the state space realization obtained in subsection \ref{sec:Linearization-and-State}
went through a discretization process in MATLAB, using the zero-order
hold method and sample time $T_{s}$. The discrete-time system obeys
the following dynamic equation:
\begin{equation}
\begin{cases}
\Delta\dot{\boldsymbol{x}}[k+1]=\boldsymbol{A}_{d}\Delta\boldsymbol{x}[k]+\boldsymbol{B}_{d}\Delta\boldsymbol{u}[k]\\
\Delta\boldsymbol{y}[k]=\boldsymbol{C}_{d}\Delta\boldsymbol{x}[k]+\boldsymbol{D}_{d}\Delta\boldsymbol{u}[k]
\end{cases}\label{eq:augmentedsystem-1}
\end{equation}
where the matrices $\boldsymbol{A}_{d}$, $\boldsymbol{B}_{d}$, $\boldsymbol{C}_{d}$,
$\boldsymbol{D}_{d}$ are the result of the discretization. 

~\\
Following the procedure to set up the discrete-time linear quadratic
tracking system specified in \cite{key-23}, considering the state
space system described by \eqref{eq:augmentedsystem-1}, the performance
index to be minimized $J_{d}$ is defined as:{\small{}
\begin{align}
J_{d} & =\frac{1}{2}\left[\boldsymbol{C}_{d}\Delta\boldsymbol{x}\left[k_{f}\right]-\boldsymbol{z}\left[k_{f}\right]\right]^{T}\boldsymbol{F}\left[\boldsymbol{C}_{d}\Delta\boldsymbol{x}\left[k_{f}\right]-\boldsymbol{z}\left[k_{f}\right]\right]\\
 & +\frac{1}{2}\sum_{k=k_{0}}^{k_{f}-1}\left\{ \left[\boldsymbol{C}_{d}\Delta\boldsymbol{x}\left[k_{f}\right]-\boldsymbol{z}\left[k_{f}\right]\right]^{T}\boldsymbol{Q}\left[\boldsymbol{C}_{d}\Delta\boldsymbol{x}\left[k_{f}\right]-\boldsymbol{z}\left[k_{f}\right]\right]+\Delta\boldsymbol{u}^{T}\left[k\right]\boldsymbol{R}\Delta\boldsymbol{u}\left[k\right]\right\} \nonumber 
\end{align}
}where $\boldsymbol{F}$ and $\boldsymbol{Q}$ are state weight matrices,
$\boldsymbol{R}$ is the control weight matrix and $\boldsymbol{z}\left[k\right]$
is a 12x1 vector that specifies the time-varying trajectory for each
state. The final time step, $k_{f}$, is fixed and the final state
value $\Delta\boldsymbol{x}\left[k_{f}\right]$ is not fixed nor specified,
thus it is called in the literature ``free'' state. Weight matrices
have well-defined characteristics as to obtain a stable close-loop
system using the gains given by the optimization algorithm, mainly
that $\boldsymbol{F}$ and $\boldsymbol{Q}$ are both positive semidefinite
symmetric $n\times n$ matrices and $\boldsymbol{R}$ is a $p\times p$
positive definite symmetric matrix (in this case $n=12$ and $p=4$).

~\\
For simplicity in the next algorithm equations, the following matrices
are defined:
\begin{equation}
\boldsymbol{E}=\boldsymbol{B}_{d}\boldsymbol{R}^{-1}\boldsymbol{B}_{d}^{T}\,\,;\,\,\boldsymbol{V}=\boldsymbol{C}_{d}^{T}\boldsymbol{Q}\boldsymbol{C}_{d}\,\,;\,\,\boldsymbol{W}=\boldsymbol{C}_{d}^{T}\boldsymbol{Q}
\end{equation}
Now, using results from optimal control theory in \cite{key-23}
it is possible to establish a matrix Riccati Difference Equation (RDE)
as follows:
\begin{equation}
\boldsymbol{P}\left[k\right]=\boldsymbol{A}_{d}^{T}\left[\boldsymbol{P}\left[k+1\right]+\boldsymbol{E}\right]^{-1}\boldsymbol{A}_{d}+\boldsymbol{V}\label{eq:3.26}
\end{equation}
this equation must be solved backwards in time using the final condition
\begin{equation}
\boldsymbol{P}\left[k_{f}\right]=\boldsymbol{C}_{d}^{T}\bm{F}\bm{C}_d
\end{equation}
Also the algorithm requires to solve the following vector difference
equation:
\begin{equation}
\boldsymbol{g}\left[k\right]=\boldsymbol{A}_{d}\left[\boldsymbol{I}_{12\times12}-\left[\boldsymbol{P}^{-1}\left[k+1\right]+\boldsymbol{E}\right]^{-1}\boldsymbol{E}\right]\boldsymbol{g}\left[k+1\right]+\boldsymbol{W}\boldsymbol{z}\left[k\right]\label{eq:3.2.8}
\end{equation}
The vector $\boldsymbol{g}\left[k\right]$ depends on the desired
trajectory and thus contains all information about it. This equation
must also be solved backwards in time, with the final condition:
\begin{equation}
\boldsymbol{g}\left[k_{f}\right]=\boldsymbol{C}_{d}^{T}\boldsymbol{F}\boldsymbol{z}\left[k_{f}\right]
\end{equation}
After resolving \eqref{eq:3.26} and \eqref{eq:3.2.8}, the
optimal control law can be computed by:
\begin{equation}
\Delta\boldsymbol{u}\left[k\right]=-\boldsymbol{L}\left[k\right]\boldsymbol{x}+\boldsymbol{L}_{g}\left[k\right]\boldsymbol{g}\left[k+1\right]
\end{equation}
where the gain $\boldsymbol{L}$ corresponds to a state feedback gain
given by the expression:
\begin{equation}
\boldsymbol{L}\left[k\right]=\left[\boldsymbol{R}+\boldsymbol{B}_{d}^{T}\boldsymbol{P}\left[k+1\right]\boldsymbol{B}_{d}\right]^{-1}\boldsymbol{B}^{T}\boldsymbol{P}\left[k+1\right]\boldsymbol{A}_{d}
\end{equation}
and the gain $\boldsymbol{L}_{g}$ is a trajectory feedforward
gain that multiplies the vector $\boldsymbol{g}\left[k\right]$ which contains the trajectory information. This gain can be calculated from
the following equation:
\begin{equation}
\boldsymbol{L}_{g}\left[k\right]=\left[\boldsymbol{R}+\boldsymbol{B}_{d}^{T}\boldsymbol{P}\left[k+1\right]\boldsymbol{B}\right]^{-1}\boldsymbol{B}^{T}
\end{equation}
While developing the algorithm, it was noted that matrices $\bm{P}\left[k\right]$, $\bm{L}\left[k\right]$ and $\bm{L}_g\left[k\right]$ only varied at the end to enforce the terminal condition. It was preferred to remove the final "free state" enforcement of the algorithm as it lead to undesired behavior at the end of the trajectory. Hence, the aforementioned matrices were considered time-invariant by taking their constant values before the final state enforcement. Therefore, the only factor in the optimal control law
that changes over time is the feedforward vector $\boldsymbol{g}\left[k\right]$.
The versatility of this control method is that all gains can be computed offline, meaning before the trajectory is executed.

~\\
Being a model-based algorithm, the LQT controller performance will
be closely related to the accuracy of the linear model of the quadcopter.
The previous study with the PID architecture showed that the coupling
between the movements, as well as unmodeled phenomena such as blade
flapping, body and motor force asymmetry, all contribute as model
perturbations to the system. In the light of these real-life unfavorable
conditions, during the conception of the LQT controller the addition
of integral action was necessary to ensure disturbance rejection
and thus a zero steady-state error in the 3D position of the drone.

~\\
Normally the procedure dictate that an augmentation of the system
should be executed to include the state of the integral of the error,
but in the LQT algorithm these states can further complicate the
task of designing the trajectory $\boldsymbol{z}\left[k\right]$ ,
as it will also require to specify a trajectory for the position integral
error. The first simulation trials with this method were not satisfactory,
specifying a trivial zero trajectory for the integral of the error
states did not yield good results. Instead of searching for a model
that might describe the evolution over time of the error, which is
by itself a difficult task considering that the position trajectory
can take almost any form, a much more convenient solution to resolve
the disturbance rejection problem is to simply add the integral action
directly into the control vector $\Delta\boldsymbol{u}\left[k\right]$,
as proposed in a similar LQT formulation \cite{key-27}. 

\begin{figure}[H]
\begin{centering}
\begin{tikzpicture}[>=stealth]
  \coordinate (orig)   at (0,0);
  \coordinate (LLA)    at (0,4);
  \coordinate (AroneA) at (3,4.9);
  \coordinate (ArtwoA) at (5.8,4.9);
  \coordinate (LLB)    at (0.5,2.45);
  \coordinate (LLF)    at (-2,2.45);
  \coordinate (LLC)    at (2,2);
  \coordinate (LLD)    at (2,0.5);
  \coordinate (LLE)    at (-2,4.9);
  \coordinate (Mult2DW)at (4,0.7);
  \coordinate (Mult2UP)at (4,1.2);
  \coordinate (Mult1)  at (4,2.65);
  \coordinate (Eq)  at (-4,4.9);
  \coordinate (out1)    at (7,5);

 \node[draw,fill=blue!20, minimum width=3cm, minimum height=1.8cm, anchor=south west, text width=2cm, align=center] (A) at (LLA) {quadcopter's\\Dynamics};
 \node[draw, fill=blue!20, circle, minimum width=1cm, minimum height=1cm, anchor=center, align=center] (sum) at (LLB) {};
 \node[draw, fill=blue!20, circle, minimum width=1cm, minimum height=1cm, anchor=center, align=center] (sum3) at (LLF) {};
 \node[draw, fill=blue!20, circle, minimum width=1cm, minimum height=1cm, anchor=center, align=center] (sum2) at (LLE) {};
 \node[draw,fill=red!20, minimum width=3cm, minimum height=3cm,scale=0.3, anchor=south west, text width=1cm, align=center] (B) at (LLC) {};
 \draw (2+0.45,2+0.45) node[cross] {};
 \node[draw,fill=red!20, minimum width=3cm, minimum height=3cm,scale=0.3, anchor=south west, text width=1cm, align=center] (C) at (LLD) {};
 \draw (2+0.45,0.5+0.45) node[cross] {};
 \node[draw,fill=red!20, minimum width=3cm, minimum height=3cm,scale=0.3, anchor=south west, text width=1cm, align=center] (G) at ($(LLD)+(-6.5,0)$) {};
 \draw (2+0.45-6.5,0.5+0.45) node[cross] {};
  \node[draw,fill=red!20, minimum width=3cm, minimum height=3cm,scale=0.3, anchor=south west, text width=1cm, align=center] (I) at ($(LLD)+(-6.5,1.5)$) {};
 \draw (2+0.45-6.5,0.5+0.45+1.5) node[cross] {};
  
  \draw[-] (A.0)-- node[above, pos=0.8]{$\bm{x}$} (ArtwoA);
  \draw[->] (ArtwoA.0) |-  ($(B.0)+(0,-0.2)$);
  \draw[->] (Mult1.0) -- node[right,pos=0]{$\bm{L}[k]$}  ($(B.0)+(0,0.2)$);
  \draw[->] (Mult2DW.0) -- node[right,pos=0]{$\bm{g}[k+1]$}  ($(C.0)+(0,-0.25)$);
  \draw[->] (Mult2UP.0) -- node[right,pos=0]{$\bm{L}_g[k]$}  ($(C.0)+(0,0.25)$);
  \draw[->] (B.180) --  (sum.0);
  \draw[->] (C.180) -|  (sum.270);
  \draw[->] (sum2.0) -- node [above, pos = 0.5]{$\bm{u}$} (A.180);
  \draw[->] (sum.180) -- node [above, pos = 0.3]{$\Delta \bm{u}$} (sum3.0);
  \draw[->] (sum3.90) -- node [right, pos = 0.9]{$\Delta \bm{u}'$} (sum2.270);
  \draw[->] (G.0) -|  (sum3.270);
  \draw[->] (I.0) --  (sum3.180);
  \draw[->] (Eq.180) -- node [left, pos = 0]{$\bm{u}_e$} (sum2.180);
  \draw[red,very thick,dashed] (-1,-0.5) rectangle (6,3.5);
 \node[text width=6.5cm] at (2.5,-0.3) {LQT Algorithm calculated offline};
  \draw[->] ($(G)+(-1.5,-0.25)$) -- node [left,pos=0.1] {$\sum{(\bm{e}_{ang}\Delta k_{ang})}$} ($(G.180)+(0,-0.25)$);
  \draw[->] ($(G)+(-1.5,0.25)$) -- node [left,pos=0.1] {$\bm{K}_i^{ang}$} ($(G.180)+(0,0.25)$);
   \draw[->] ($(I)+(-1.5,-0.25)$) -- node [left,pos=0.1] {$\sum{(\bm{e}_{pos}\Delta k_{pos})}$} ($(I.180)+(0,-0.25)$);
  \draw[->] ($(I)+(-1.5,0.25)$) -- node [left,pos=0.1] {$\bm{K}_i^{pos}$} ($(I.180)+(0,0.25)$);
  \draw[olive,very thick,dashed] (-8.5,-0.5) rectangle (-3,3.5);
   \node[text width=3cm] at (-6,-0.3) {Integral Action};
   
  \node[text width=1cm] at ($(sum3)+(1,0.3)$) {$+$};
  \node[text width=1cm] at ($(sum3)+(-0.3,0.3)$) {$+$};
  \node[text width=1cm] at ($(sum3)+(0.6,-0.7)$) {$+$};
  
  \node[text width=1cm] at ($(sum2)+(-0.3,0.3)$) {$+$};
  \node[text width=1cm] at ($(sum2)+(0,-0.7)$) {$+$};

   \node[text width=1cm] at ($(sum)+(1,0.3)$) {$-$};
   \node[text width=1cm] at ($(sum)+(0.6,-0.7)$) {$+$};
\end{tikzpicture}
\par\end{centering}
\caption{\label{fig:lqt}LQT Closed-Loop System.}
\end{figure}
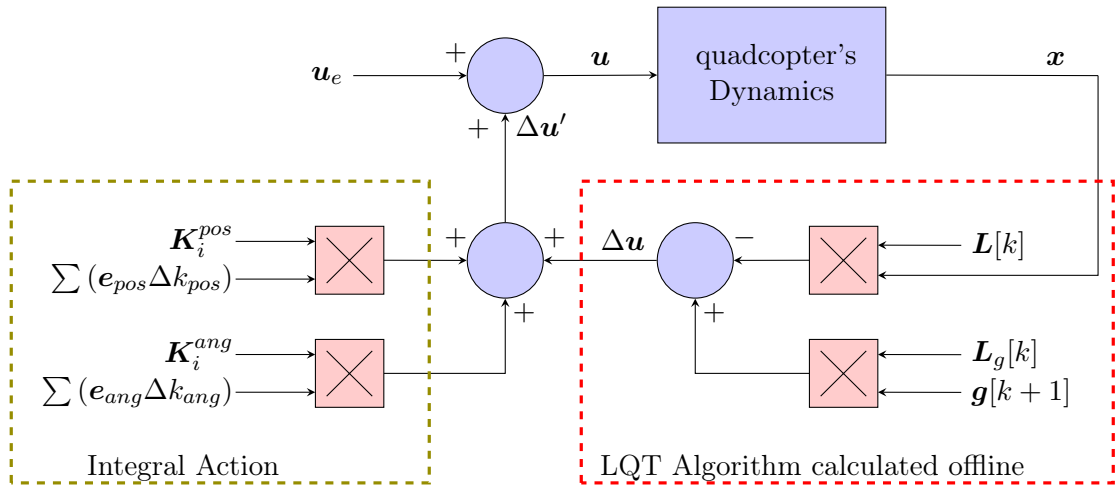

~\\
In addition to the integral action in the position, it was found in
practice that using integral action in the angular position improved
the overall performance of the system, by regulating the drone's Euler
angles to zero thus keeping it stable with an increased level of robustness.
Finally the command vector adopted the following form:
\begin{equation}
\resizebox{0.85\hsize}{!}{$
\Delta\boldsymbol{u}'\left[k\right]=-\boldsymbol{L}\left[k\right]\boldsymbol{x}+\boldsymbol{L}_{g}\left[k\right]\boldsymbol{g}\left[k+1\right]+\boldsymbol{K}_{i}^{ang}\sum\limits_{k=k_{0}}^{k_{f}-1}\left(\boldsymbol{e}_{ang}\left[k\right]\Delta k_{ang}\right)+\boldsymbol{K}_{i}^{pos}\sum\limits_{k=k_{0}}^{k_{f}-1}\left(\boldsymbol{e}_{pos}\left[k\right]\Delta k_{pos}\right)$}
\end{equation}
where $\boldsymbol{e}_{ang}\left[k\right]$ is the angular error vector
in regulation mode (error with respect to zero) defined as:
\begin{equation}
\boldsymbol{e}_{ang}\left[k\right]=\left[\begin{array}{ccc}
-\psi\left[k\right] & -\theta\left[k\right] & -\phi\left[k\right]\end{array}\right]^{T}
\end{equation}
and similarly $\boldsymbol{e}_{pos}\left[k\right]$ is the position
error vector with respect to the desired trajectory $\boldsymbol{z}\left[k\right]$:
\begin{equation}
\boldsymbol{e}_{pos}\left[k\right]=\left[\begin{array}{ccc}
z_{1}\left[k\right]-x\left[k\right] & z_{2}\left[k\right]-y\left[k\right] & z_{3}\left[k\right]-z\left[k\right]\end{array}\right]^{T}
\end{equation}
The coefficients $\Delta k_{ang}$ and $\Delta k_{pos}$ are the time steps corresponding to each integral gain. The block diagram in \autoref{fig:lqt} represents the closed-loop control system proposed.

With a better understanding of how to setup the LQT problem, the next
logical step was to test the algorithm in the quadcopter model and
study the feasibility for practical implementation in the Crazyflie
2.0 platform.

~\\
Before testing the actual control algorithm, it became necessary to
address the problem of the observation of the states, mainly to answer
the question of how to reconstruct all 12 states of the dynamic model
given the data coming from different sensors used in the implementation.
The Inertial Measurement Unit inside the Crazyflie 2.0 gives good
estimates of the Euler angles and the body-fixed frame angular velocities,
by a sensor fusion algorithm that merges data coming from the accelerometer
and the gyroscope. A localization system such as the VICON estimates
the position of an object with respect to a certain fixed inertial
frame, but a priori these type of systems do not directly measure nor estimate
the linear velocities. Thus the need of some algorithm that can reconstruct
the missing states from a model of how they evolve over time as well
from the sensors' data.

\subsection{\label{subsec:Kalman-Filter-for}Kalman Filter for Linear Velocity
Estimation}

With the objective of simulating as close as possible real-life
scenarios, white Gaussian noise was added to the position values coming
from the quadcopter non-linear dynamics, as to reproduce the uncertainties
given by any position-fixed system such as VICON or an ultra-wide
band system. In addition, the linear velocity is not directly given
by any of these systems, thus the need for a state observer capable
of filtering the noise coming from the position estimation while correctly
computing estimations for the linear velocities in the inertial frame
defined by the positioning system.

~\\
The well-known Kalman Filter is ideal for these type of tasks. Using
as reference the example found in \cite{key-24}, the Kalman Filter
problem was stated. First a vector containing the estimated variables
must be defined:
\begin{equation}
\hat{\boldsymbol{x}}\left[k\right]=\left[\begin{array}{cccccc}
\hat{x} & \hat{y} & \hat{z} & \hat{\dot{x}} & \hat{\dot{y}} & \hat{\dot{z}}\end{array}\right]^{T}\label{eq:ede}
\end{equation}
As the estimation process only addresses the state variables for the
position and linear velocities, this reduced state vector is ideal
to set-up the problem. The next step is to describe the state space
system based on the dynamics between the position and the velocity
of a given rigid body. The following equation defines the dynamics of the state space in discrete-time:
\begin{equation}
\begin{cases}
\hat{\boldsymbol{x}}[k+1]=\boldsymbol{A}_{hat}\hat{\boldsymbol{x}}[k]+\boldsymbol{G}\boldsymbol{w}\left[k\right]\\
\boldsymbol{y}_{hat}[k]=\boldsymbol{C}_{hat}\hat{\boldsymbol{x}}[k]+\boldsymbol{v}\left[k\right]
\end{cases}\label{eq:augmentedsystem-1-1}
\end{equation}
where $\boldsymbol{w}\left[k\right]$ is the process noise vector,
that multiplies a certain matrix $\boldsymbol{G}$ that models how
this noise interacts within the state evolution. Then $\boldsymbol{v}\left[k\right]$
is a random variable with certain variance that simulates
the measurements noise of the position-fixed system and $\boldsymbol{y}_{hat}\left[k\right]$
is the noise-infected output of the reduced position-velocity system.
Matrix $\boldsymbol{A}_{hat}$ is the state transition matrix, defined
as:
\renewcommand{\arraystretch}{0.75}
\[
\boldsymbol{A}_{hat}=\left[\begin{array}{cccccc}
1 & 0 & 0 & T_{s} & 0 & 0\\
0 & 1 & 0 & 0 & T_{s} & 0\\
0 & 0 & 1 & 0 & 0 & T_{s}\\
0 & 0 & 0 & 1 & 0 & 0\\
0 & 0 & 0 & 0 & 1 & 0\\
0 & 0 & 0 & 0 & 0 & 1
\end{array}\right]
\]
then matrix $\boldsymbol{G}$ was modeled:
\renewcommand{\arraystretch}{0.75}
\[
\boldsymbol{G}=\left[\begin{array}{ccc}
T_{s}/2 & 0 & 0\\
0 & T_{s}/2 & 0\\
0 & 0 & T_{s}/2\\
1 & 0 & 0\\
0 & 1 & 0\\
0 & 0 & 1
\end{array}\right]
\]
Thus, by examining the three lower rows of matrices $\boldsymbol{A}_{hat}$
and $\boldsymbol{G}$, the evolution of the linear velocities is defined
as:
\begin{equation}
\begin{cases}
\dot{\hat{x}}\left[k+1\right]=\dot{\hat{x}}\left[k\right]+w_{1}\left[k\right]\\
\dot{\hat{y}}\left[k+1\right]=\dot{\hat{y}}\left[k\right]+w_{2}\left[k\right]\\
\dot{\hat{z}}\left[k+1\right]=\dot{\hat{z}}\left[k\right]+w_{3}\left[k\right]
\end{cases}
\end{equation}
which describes a constant velocity model with and added random variable
that accounts for the process noise. As possible improvement, acceleration estimations in the world frame could be added to the model, taking for example the accelerometer measurements and the appropriate rotation estimate. Then, the position evolution
was modeled starting by the principle that the position in reality
is a continuous variable and the velocity is defined as the derivative
of the position:
\begin{equation}
\frac{d}{dt}\hat{x}=\dot{\hat{x}}
\end{equation}
this relationship can be approximated in the discrete domain as:
\begin{equation}
\frac{\hat{x}\left[n+1\right]-\hat{x}\left[n\right]}{T_{s}}=\frac{\dot{\hat{x}}\left[n+1\right]+\dot{\hat{x}}\left[n\right]}{2}\label{eq:3.20}
\end{equation}
Now, the left hand side of \eqref{eq:3.20} is the approximation
of the derivative in discrete time, while the right hand side describes
the approximation of a constant velocity between two time steps. Basically
the model proposed works with the idea of a constant velocity and
the accelerations that will introduce changes to the velocity are
modeled as random variables.

~\\
Matrix $\boldsymbol{C}_{hat}$ selects which states are truly measured
by the position-fixed system, in this particular case that corresponds
to the position states, then:
\[
\boldsymbol{C}_{hat}=\left[\begin{array}{cccccc}
1 & 0 & 0 & 0 & 0 & 0\\
0 & 1 & 0 & 0 & 0 & 0\\
0 & 0 & 1 & 0 & 0 & 0
\end{array}\right]
\]
Now that the system is clearly defined, a noise characterization needs
to be done in order to get the best possible estimator. In practice
that means to estimate the measurement noise covariance from experimentally-retrieved
data. This estimation gives a good approximation for the noise covariance
matrix $\boldsymbol{R}_{kal}$ that is assumed to be diagonal from
the principle that there exists no correlation between the noises
of the different positions. Then the noise covariance matrix takes
the following form:
\begin{equation}
\boldsymbol{R}_{kal}=diag\left[\begin{array}{ccc}
\sigma_{x}^{2} & \sigma_{y}^{2} & \sigma_{z}^{2}\end{array}\right]
\end{equation}
As for the process noise covariance $\boldsymbol{Q}_{kal}$, it is
analytically difficult to define or estimate its weights. In practice, the
values are often found by experimentation rather than modeling process
noise as a consequence of unmodeled phenomena. As the dynamical model
for each one of the axes is the same, and theoretically decoupled,
the matrix $\boldsymbol{Q}_{kal}$ takes also the form of a diagonal,
\begin{equation}
\boldsymbol{Q}_{kal}=diag\left[\begin{array}{ccc}
\sigma_{w_{1}}^{2} & \sigma_{w_{2}}^{2} & \sigma_{w_{3}}^{2}\end{array}\right]
\end{equation}
There exists an evident coupling between the position and the velocity
process noises, but this over complicates the task of a rather simple
method of state observation, thus the diagonal form was preferred
over a full scale 3x3 matrix. Also note that one advantage of using
the matrix $\boldsymbol{G}$ is that it reduces the tuning parameters of the Kalman filter algorithm, by considering a smaller dimension process covariance matrix. Normally, without a matrix $\boldsymbol{G}$
in the model, the process covariance matrix is $n\times n$, but if
added, a new matrix $\bar{\boldsymbol{Q}}_{kal}$ defined as:
\begin{equation}
\bar{\boldsymbol{Q}}_{kal}=\boldsymbol{G}\boldsymbol{Q}_{kal}\boldsymbol{G}^{T}
\end{equation}
is considered for the resolution of the Algebraic Riccati Equation
associated with the Kalman filter. Finally, the dimension of matrix
$\bar{\boldsymbol{Q}}_{kal}$ is $\left(n-r\right)\times\left(n-r\right)$,
where $r$ is the number of columns of matrix $\boldsymbol{G}$ and
therefore, the dimension of the vector $\boldsymbol{w}\left[k\right]$.

~\\
In order to complete the Kalman filter design, the values for
the noise matrices must be given. Two scenarios were taken in account:
one in which the data was taken from the VICON system and another
in which the position estimation came from an ultra-wide band system. 
\begin{enumerate}
\item \textbf{VICON system:} a simple test leaving the drone steady on the
floor allowed the retrieval of data during 1 minute to calculate through
MATLAB the variance of the position in each axis . The average results
were:
\begin{equation}
\sigma_{x}^{2}=\sigma_{y}^{2}=\sigma_{z}^{2}=5\times10^{-9}[\textrm{m}^2]\label{eq:noise_vicon}
\end{equation}
Having fixed the values of the noise covariance matrix $\boldsymbol{R}_{kal}$,
the process covariance matrix was hand-tuned through simulation and
experimentation. These values were:
\begin{equation}
\sigma_{w_{1}}^{2}=\sigma_{w_{2}}^{2}=\sigma_{w_{3}}^{2}=8\times10^{-8}\label{eq:process_vicon}
\end{equation}
The fact that the process covariance matrix has low values indicates
that the state space model is good at predicting future values for
the estimated state vector.
\item \textbf{Ultra-wide Band system: }the system was used to estimate the
X and Y position of the quadcopter, while the altitude was tracked
with the VICON. The same procedure as before was applied, giving the
following noise variance values:
\begin{equation}
\sigma_{x}^{2}=\sigma_{y}^{2}=5\times10^{-5}[\textrm{m}^2]\,\,;\,\,\sigma_{z}^{2}=5\times10^{-9}[\textrm{m}^2]\label{eq:noise_uwb}
\end{equation}
these values suggested that the standard deviation of the UWB is around
100 times greater that the VICON's. Then for the process covariance
matrix, the following values were used:
\begin{equation}
\sigma_{w_{1}}^{2}=\sigma_{w_{2}}^{2}=3\times10^{-5}\textrm{}\,\,;\,\,\sigma_{z}^{2}=8\times10^{-8}\textrm{}\label{eq:process_uwb}
\end{equation}
\end{enumerate}

The validation of the filter was done using real data from a dummy test using the Crazyflie 2.0 and making a simulated flight by hand, just grabbing the drone and moving it around the test area. Comparisons were made between the raw data of the position estimations with the filter output, and for the velocity a discrete time derivative of the incoming data estimations was taken to contrast it with the output of the filter.
~\\

\begin{figure}[H]
\centering
\makebox[0pt]{
\includegraphics[width=1.2\textwidth]{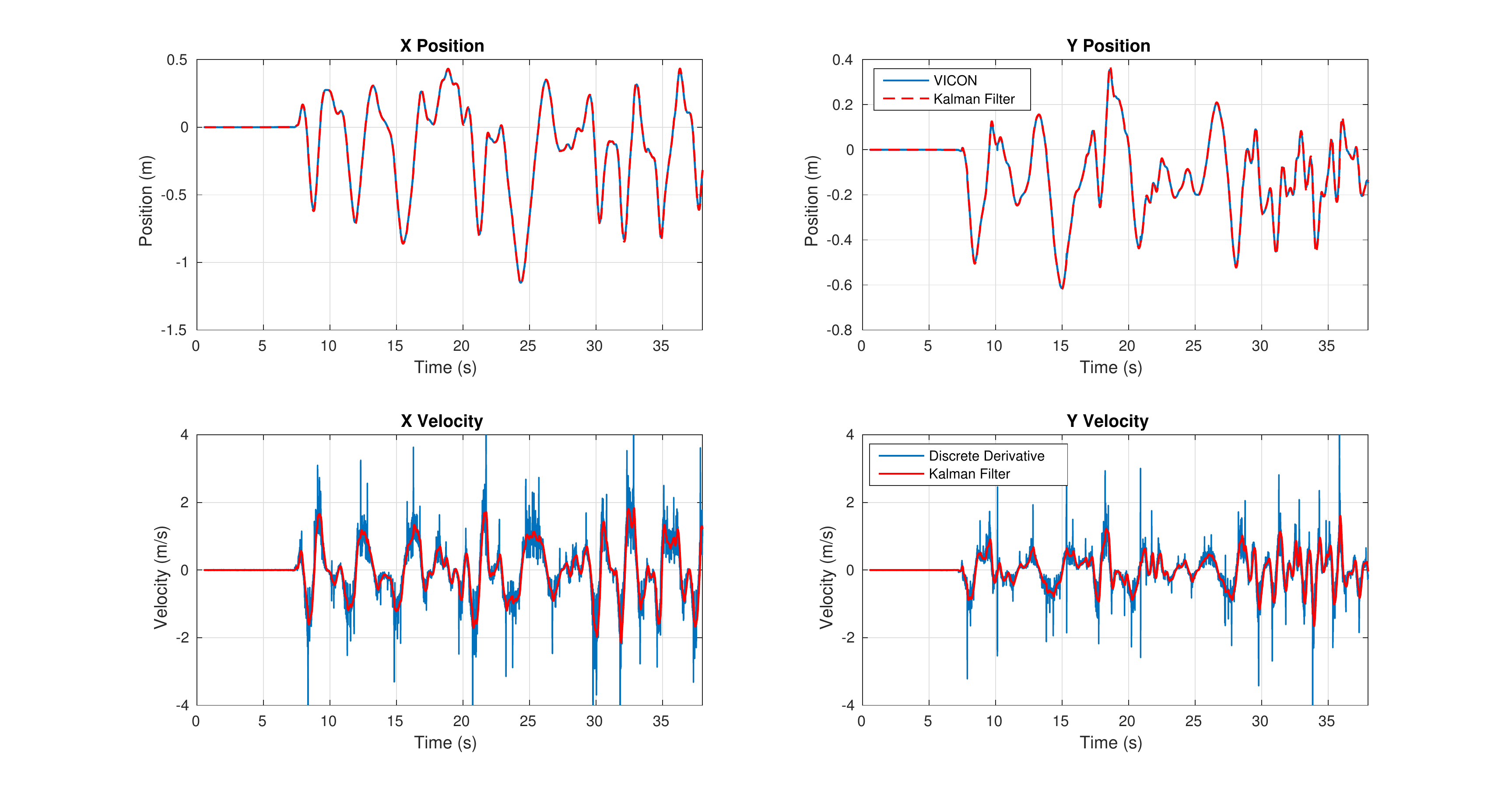}}
\caption{\label{fig:Kalman_validation}Experimental validation of the Kalman Filter using the VICON system raw data of X-Y positions.}
\end{figure}

First, while using the VICON positioning system and the values of variance in \eqref{eq:noise_vicon} and \eqref{eq:process_vicon}, the experimental results obtained are shown in \autoref{fig:Kalman_validation}. For the position estimation, the output of the filter superposes with the raw data as the VICON system does already a lot of filtering and the raw data has low levels of noise, hence the Kalman Filter algorithm output for the position is virtually the same as the raw data. As for the velocity estimations, the Kalman filter reduces the noise levels while being fast enough to follow the true dynamics.

\begin{figure}[H]
\centering
\makebox[0pt]{
\includegraphics[width=1.2\textwidth]{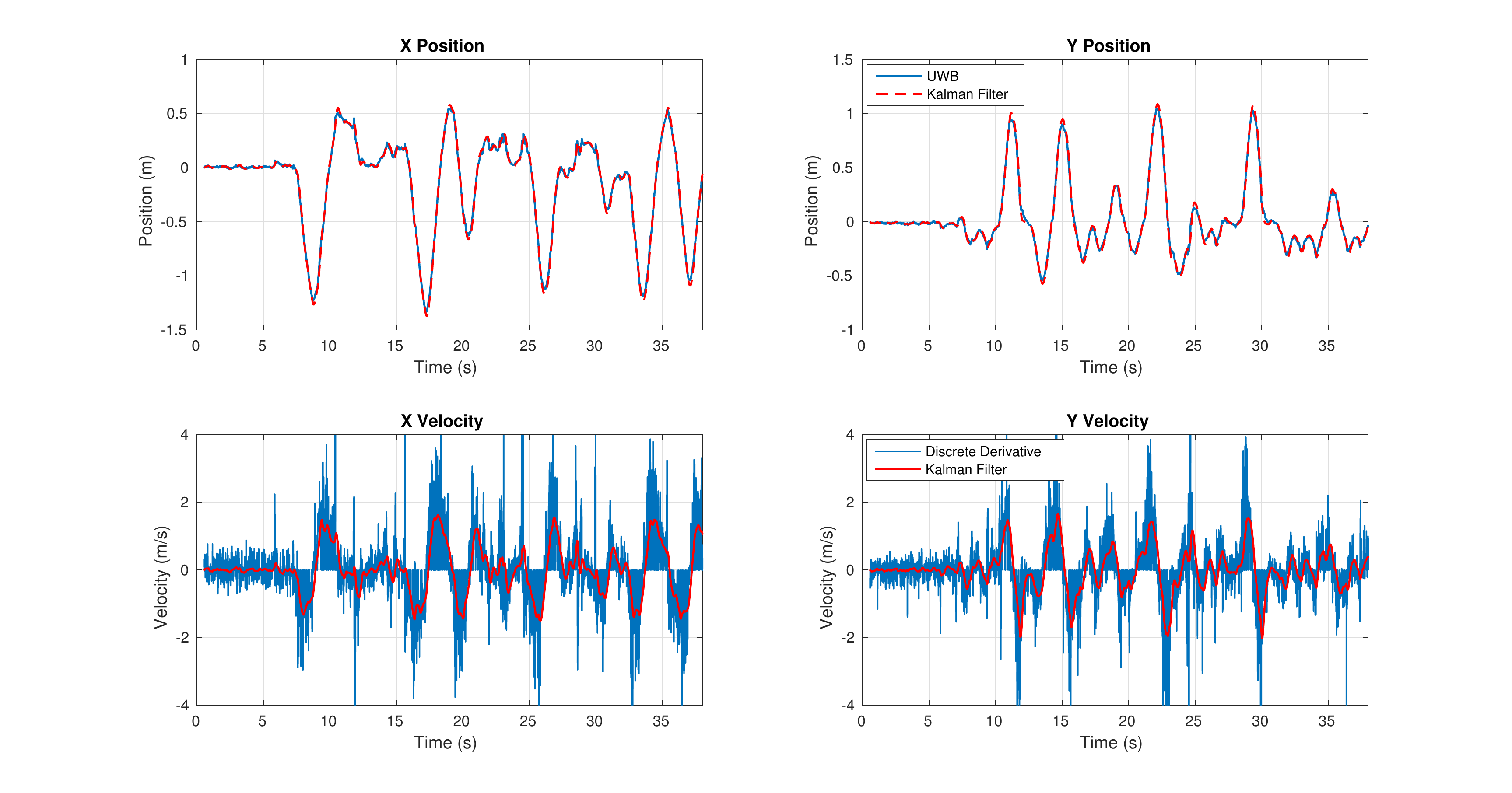}}
\caption{\label{fig:Kalman_uwb}Experimental validation of the Kalman Filter using the UWB system raw data of X-Y positions.}
\end{figure}

Then, using the UWB system, the filter was adjusted with the appropriate variance values for $\bm{R}_{kal}$ and $\bm{Q}_{kal}$, found in \eqref{eq:noise_uwb} and \eqref{eq:process_uwb}. The experimental results are displayed in the time plots of \autoref{fig:Kalman_uwb}.
~\\

The position raw data is lightly filtered in order to keep low levels of lag in the estimations, while the velocity estimations of the Kalman Filter are more heavily filtered and at the same time fast enough to keep a good convergence speed.

\begin{figure}[H]
\centering
\makebox[0pt]{
\includegraphics[scale=0.6]{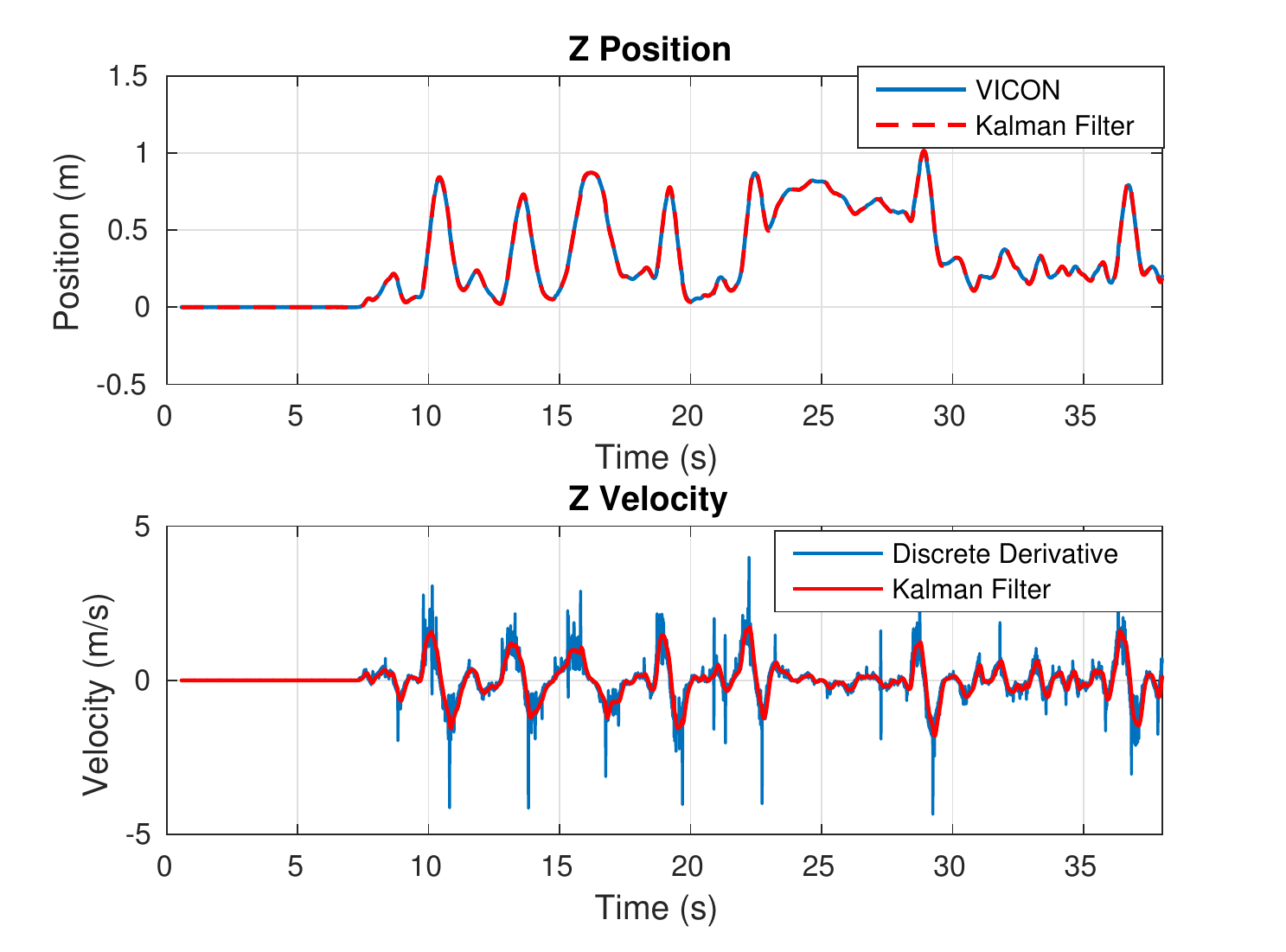}}
\caption{\label{fig:kalman_altitude}Experimental validation of the Kalman Filter altitude estimation from VICON raw data of Z position.}
\end{figure}

The last step of the filter validation was testing the altitude estimations, which in both cases came from the VICON system. \autoref{fig:kalman_altitude} displays the experimental results for this test.

Similar as the results seen in \autoref{fig:Kalman_validation}, the altitude estimation results confirm the quality of the filter developed and therefore validate the conception proposed.

\subsection{Weight Matrices and Integral Action }

A simulation environment in MATLAB was created to test the LQT algorithm
with the non-linear dynamics of the quadcopter, the added noise to
the position states and the Kalman Filter. The LQT algorithm is calculated
before the simulation runs, using the discrete time linear model of
the quadcopter and the following weight matrices:
\begin{equation}
\begin{cases}
\boldsymbol{Q}=diag\left(2000,2000,4000,4000,4000,4000,20,20,10,10,10,10\right)\\
\boldsymbol{F}=\boldsymbol{Q}\\
\boldsymbol{R}=0.00003\times\boldsymbol{I}_{4\times4}
\end{cases}\label{eq:khk}
\end{equation}
It is common practice to choose state weight matrices such as $\boldsymbol{Q}$
and $\boldsymbol{F}$ to be diagonal, that is a way of imposing individual
weights to each one of the states considering that all of them are
decoupled. In the case of the quadcopter it is known the existence
of some coupling between the movements, although it is minor. 
~\\

Another remark is that being a linear algorithm, the performance will hold as long as the quadcopter stays in the vicinity of the hover point, meaning by default that the three Euler angles will always be regulated at zero degrees. This implies that with the LQT algorithm, as proposed in this project, specifying a trayectory for the yaw angle will not be possible as the linearization will not longer be a valid approximation.  

~\\
Continuing with the setup of the simulation environment for the LQT controller, the angular integral action gain, the matrix $\boldsymbol{K}_{i}^{ang}$,
takes the following format:
\begin{equation}
\boldsymbol{K}_{i}^{ang}=\left[\begin{array}{ccc}
-k_{i}^{ang} & -k_{i}^{ang} & -k_{i}^{ang}\\
k_{i}^{ang} & k_{i}^{ang} & -k_{i}^{ang}\\
-k_{i}^{ang} & k_{i}^{ang} & k_{i}^{ang}\\
k_{i}^{ang} & -k_{i}^{ang} & k_{i}^{ang}
\end{array}\right]
\end{equation}
where $k_{i}^{ang}$ was a tuned parameter to compensate modeling
errors and other unmodeled phenomena. The tuning procedure is further
explained in \Cref{sec:LQT-Controller-Implementation}. The
value used both in simulation and in the experimental phase was $k_{i}^{ang}=8660$.

~\\
Similarly, the position integral action gain $\boldsymbol{K}_{i}^{pos}$
takes on the form:

\begin{equation}
\boldsymbol{K}_{i}^{pos}=\left[\begin{array}{ccc}
k_{i}^{pos} & -k_{i}^{pos} & -k_{i}^{pos}\\
-k_{i}^{pos} & -k_{i}^{pos} & -k_{i}^{pos}\\
-k_{i}^{pos} & k_{i}^{pos} & -k_{i}^{pos}\\
k_{i}^{pos} & k_{i}^{pos} & -k_{i}^{pos}
\end{array}\right]
\end{equation}
The value chosen for $k_{i}^{pos}$ was 5000, after thorough testing
both in simulation and practice. The approach to find the correct
weight matrices in \eqref{eq:khk} was a simple trial-and-error
method, taking in account some general knowledge about the dynamics
of the quadcopter, but most importantly through experimentation with
the actual platform. 

~\\
Basically the simulation served as a first good approximation to obtain
decent performance in practice and then as a tool to know which parameters
to tune further to improve the control system in the real-life scenario.

\subsection{\label{subsec:Trajectory-generation}Trajectory Generation}

A trajectory for the 12-state vector must be specified before running
the LQT algorithm. For the generation of the trajectory, a small angle
approach was taken meaning that the trajectory for angular velocities
and angular positions was considered as zero throughout the whole
trajectory. As for the trajectory of the states $[u,v,w]$, the position trajectory was
simply fed to a Kalman Filter similar as the one previously designed to obtain a velocity profile for each axis (in this case the filter acts only as a velocity estimator). Note
that in the small angle approximation, the vector $[u,v,w]$ coincides
with the vector $[\dot{x},\dot{y},\dot{z}]$, hence a projection from the inertial frame to the body-fixed frame was not necessary.

~\\
The generation of feasible trajectories for a quadcopter is a complex
and open research subject that has been treated before in works such
as \cite{key-25}. In the case of this project these types of studies
were not considered and are suggested as future work.

\begin{figure}[H]
\centering
\makebox[0pt]{
\subfloat[MATLAB GUI waypoint selection]{\includegraphics[width=0.55\textwidth]{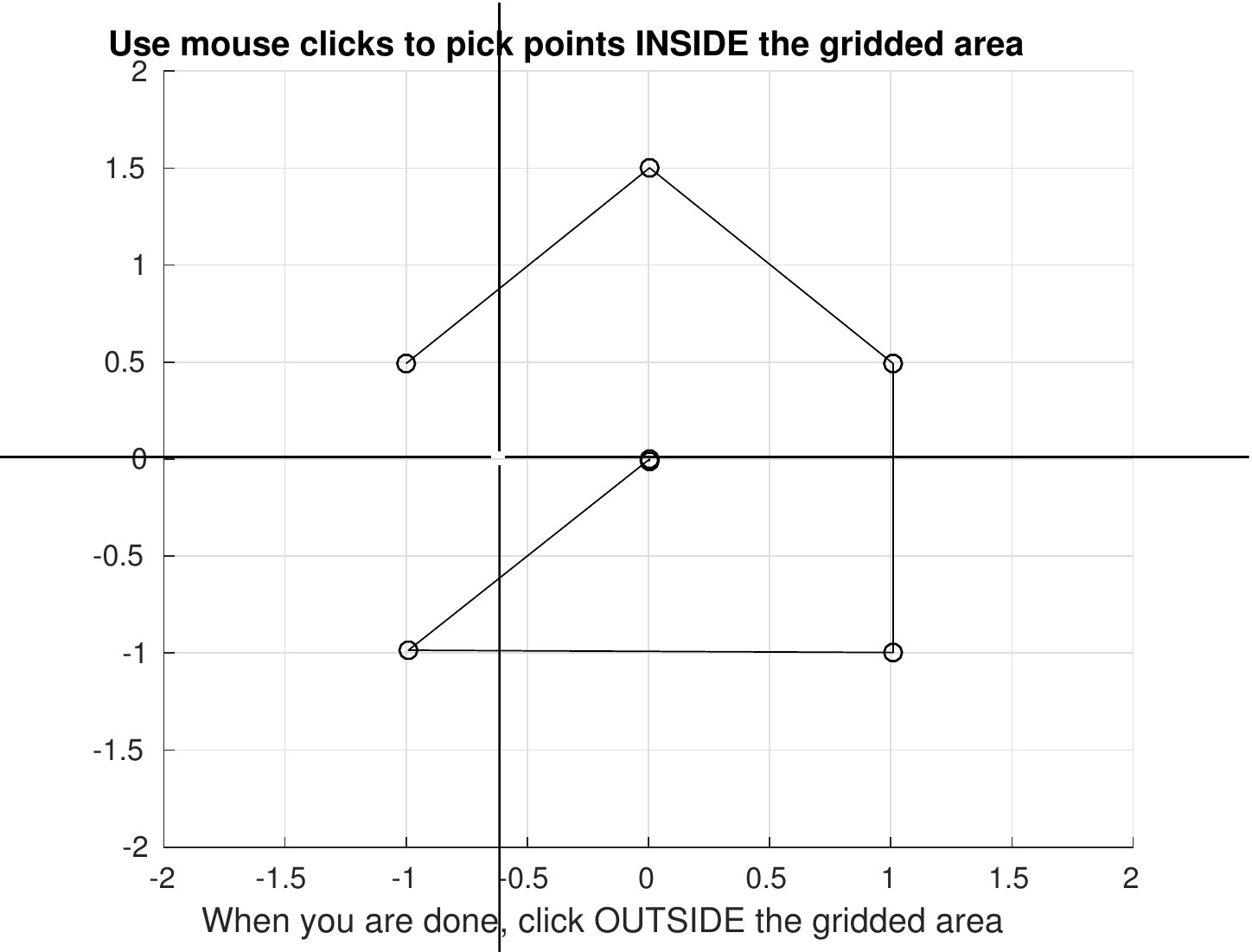}
}\subfloat[Generated Trajectory]{\includegraphics[width=0.55\textwidth]{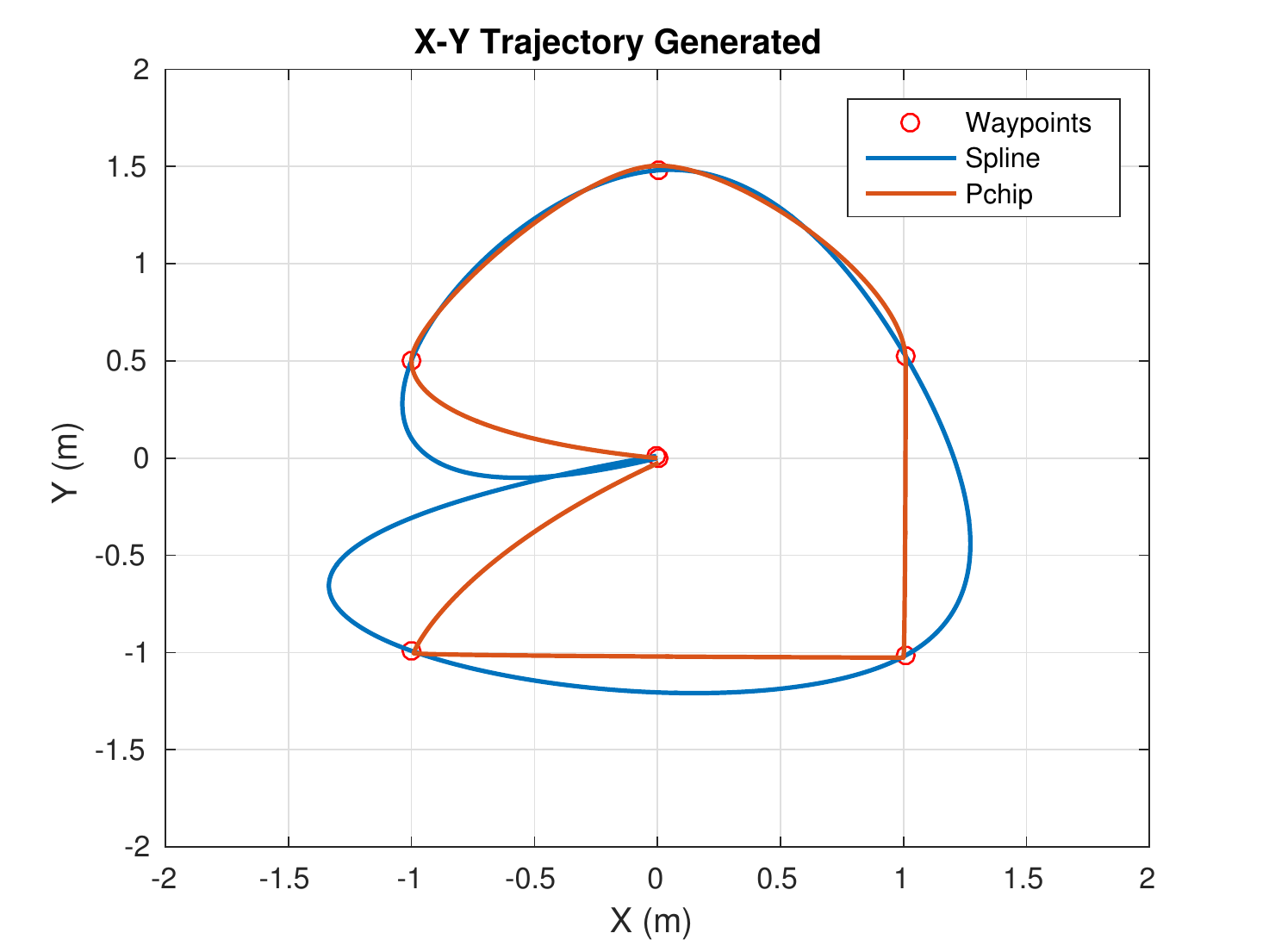}}}
\caption{\label{fig:Gui}Trajectory generation GUI.}
\end{figure}

~\\
The trajectories were generated via MATLAB through a GUI interface
(modified version of function get\_curve.m) that allowed the user
to specify a number of waypoints in the X-Y plane and then a cubic
interpolation calculated a trajectory between each one of the waypoints.
After this first step was done, a similar window opened to specify
the trajectory in the altitude of the drone. The interpolation was
done with respect to a time vector with a fixed time step of $T_{s}=0.01s$
and a range from 0 to a fixed final time in which the trajectory was
to be executed. Then the interpolation function created a position
vector of the same length as the time vector, specifying a discrete
time trajectory $z[k]$ between the waypoints chosen through the interface.
The interpolation method gives two variants to the trajectory generated:
\begin{enumerate}
\item ``spline'' does the classic piecewise cubic interpolation between
the waypoints.
\item ``pchip'' does a shape-preserving cubic interpolation.
\end{enumerate}
\autoref{fig:Gui} shows a series of points chosen by the user and the
corresponding interpolation and trajectory generated by the interface.
The user must then choose between the ``spline'' or ``pchip''
trajectories generated.
~\\

This method establishes an easy and automated way of generating trajectories,
but the feasibility of said trajectories depends mostly
on the time allowed to execute them.

\subsection{Simulation Results}

\begin{figure}[H]
\centering
\makebox[0pt]{
\includegraphics[width=1.2\textwidth]{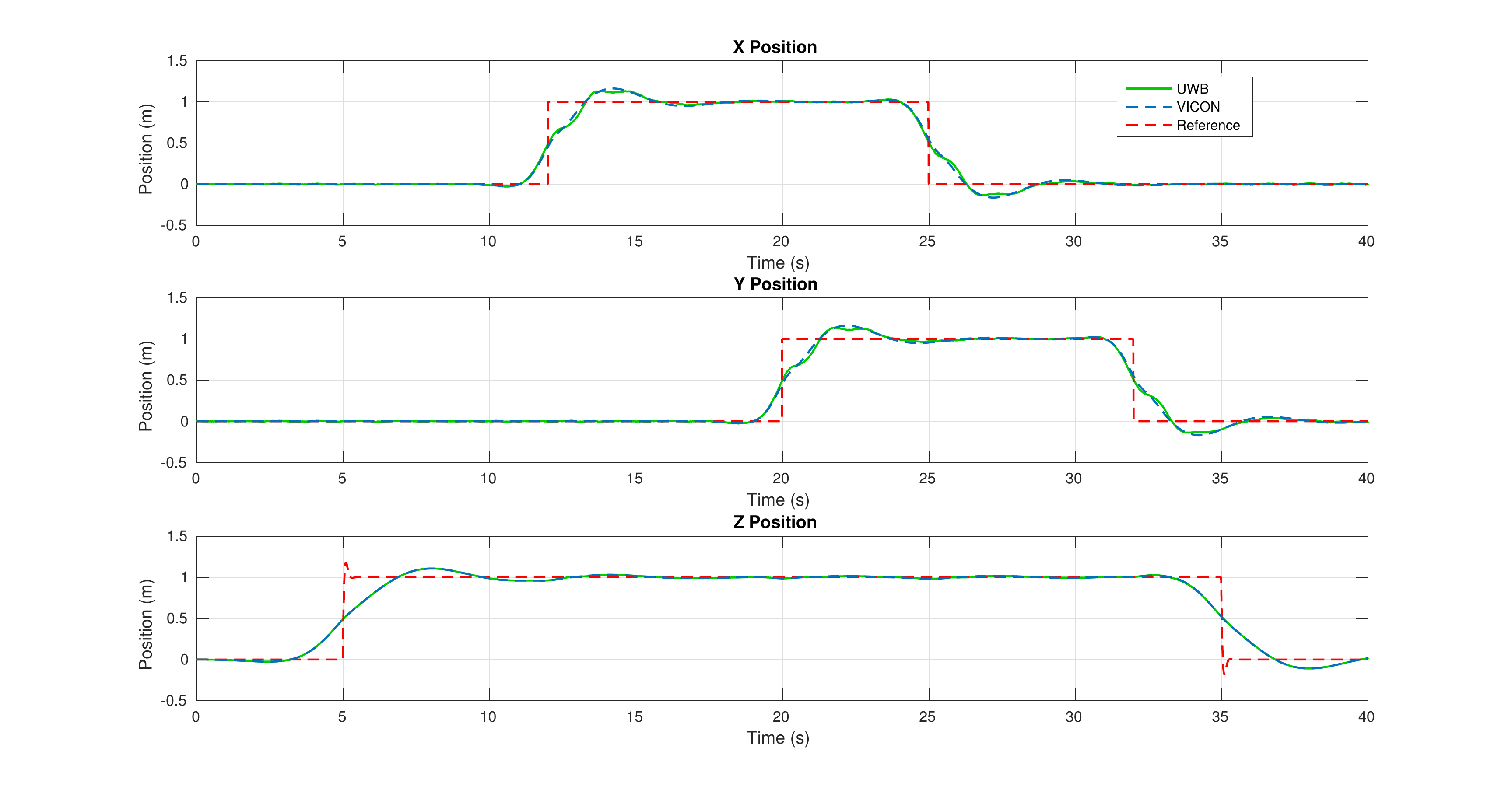}}
\caption{\label{fig:lqt_sim}Simulation for steps in $x$, $y$ and $z$ positions.}

\end{figure}

The control system was thoroughly tested in simulation before the
implementation phase, even though the final tuning of the controller
was done using and getting to know the experimental platform. First
the step response was computed for the $x$, $y$ and $z$ positions, using
simulated noise for both the VICON and UWB system. Simulation results are appreciated in \autoref{fig:lqt_sim}.

This first simulation of the LQT algorithm shows an interesting feature
for a step response. Instead of obtaining a classical response that
starts when the step is commanded, the quadcopter actually starts moving
beforehand as to minimize the overall error of the trajectory. This
feature is only possible because the trajectory was known by the controller
and the algorithm calculated the appropriate feedforward gain $g\left[k\right]$
to ensure the ``anticipatory'' feature seen in simulation.
Another important remark after playing around with the simulation
is that the overshoot can be reduced easily by decreasing the integral
gains $\boldsymbol{K}_{i}^{ang}$ and/or $\boldsymbol{K}_{i}^{pos}$,
but due to later difficulties in the implementation they remained
with the values specified before (this subject is further discussed
in \Cref{subsec:Simulation-vs-Experimental}). As to the UWB
vs VICON performance, the time plots suggests that both have almost the
same exact response in position. 

\begin{figure}[H]
\centering
\makebox[0pt]{
\includegraphics[width=0.8\paperwidth]{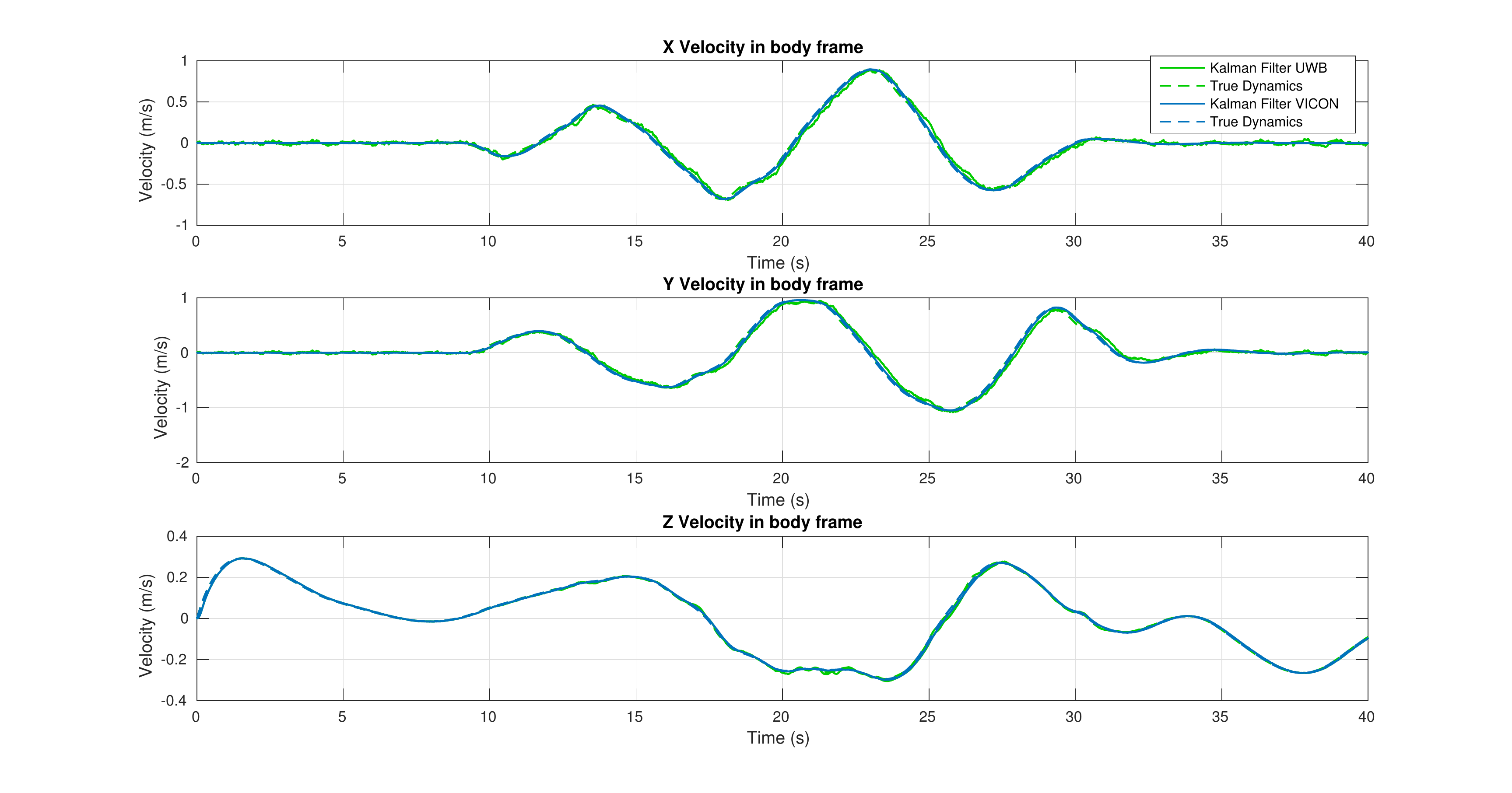}}
\caption{\label{fig:Kalman-Filter-simulation}Kalman Filter simulation for
a complex trajectory.}
\end{figure}

~\\
The Kalman Filter performance in simulation to estimate the linear velocities is presented in \autoref{fig:Kalman-Filter-simulation,}. The filters main job in the control system is to calculate reliable
state estimations for the linear velocities in the inertial frame,
then the appropriate rotation projects this estimations in the body
frame thus obtaining estimates for the states $[u,v,w]$. Simulation results
suggests that the levels of noise in the estimations for the UWB X-Y
velocities is greater, but this was expected knowing already that
the noise has at least two order of magnitude greater standard deviation
than the VICON system. 

\begin{figure}
\centering
\makebox[0pt]{
\includegraphics[width=1.2\textwidth]{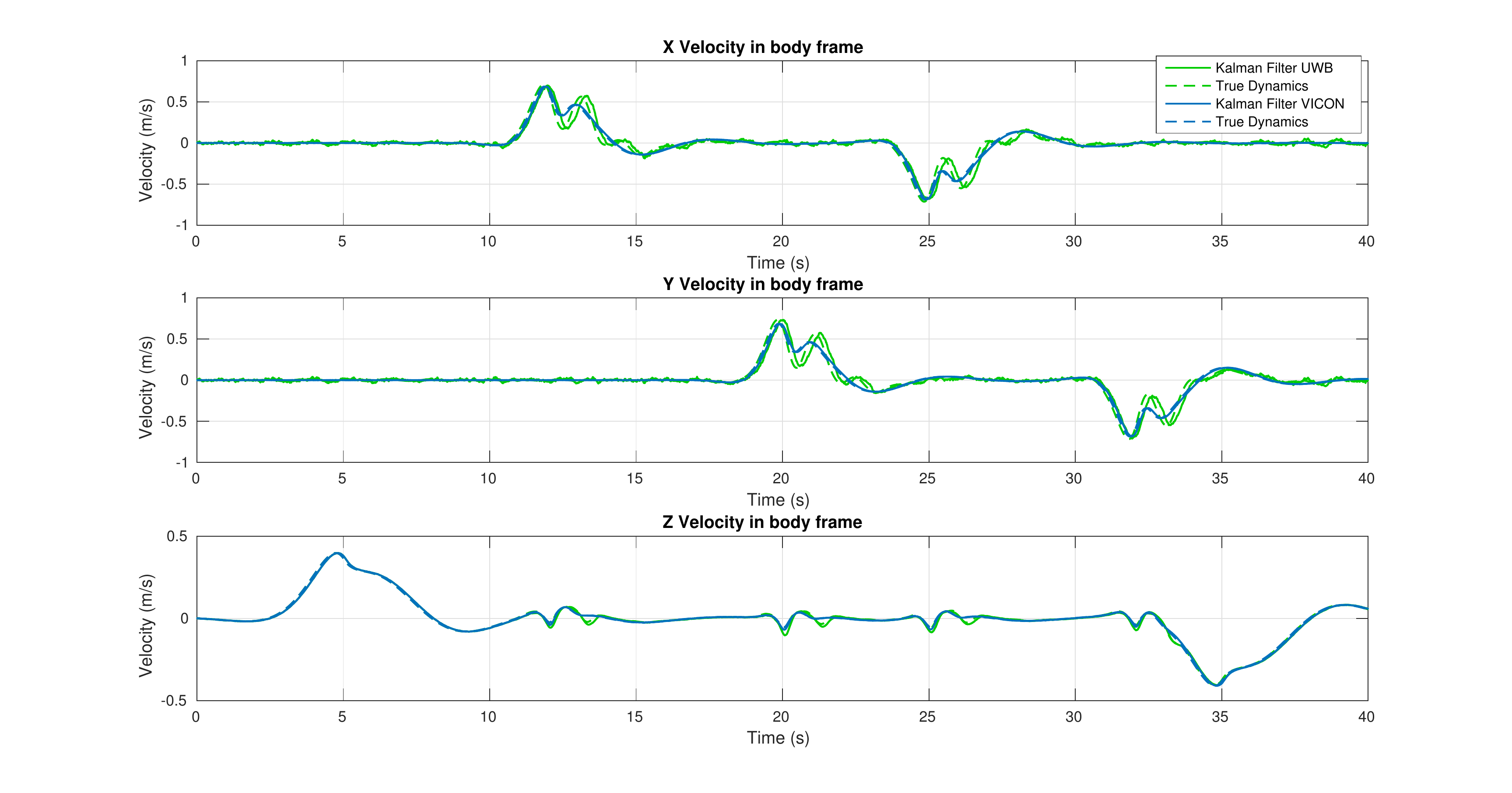}}
\caption{\label{fig:Kalman-Filter-simulation,}Kalman Filter simulation, with
VICON and UWB simulated noise.}
\end{figure}

Nonetheless, the Kalman Filter in both cases
manages a good compromise between the filtering and the convergence
speed. As seen in the plots of the X and Y velocities, the UWB Kalman Filter introduced some delay ({\raise.17ex\hbox{$\scriptstyle\sim$}}200ms) in the estimations with respect to the true dynamics of the quadcopter. This was the cost for a more aggressive filtering of the noise. In practice
the proposed gains provided a satisfactory performance despite the lag introduced, ultimately it was found that if the filtering was less aggressive then too much noise entered the system and the overall performance of the tracker degraded.

\begin{figure}[H]
\centering
\makebox[0pt]{
\includegraphics[width=1.2\textwidth]{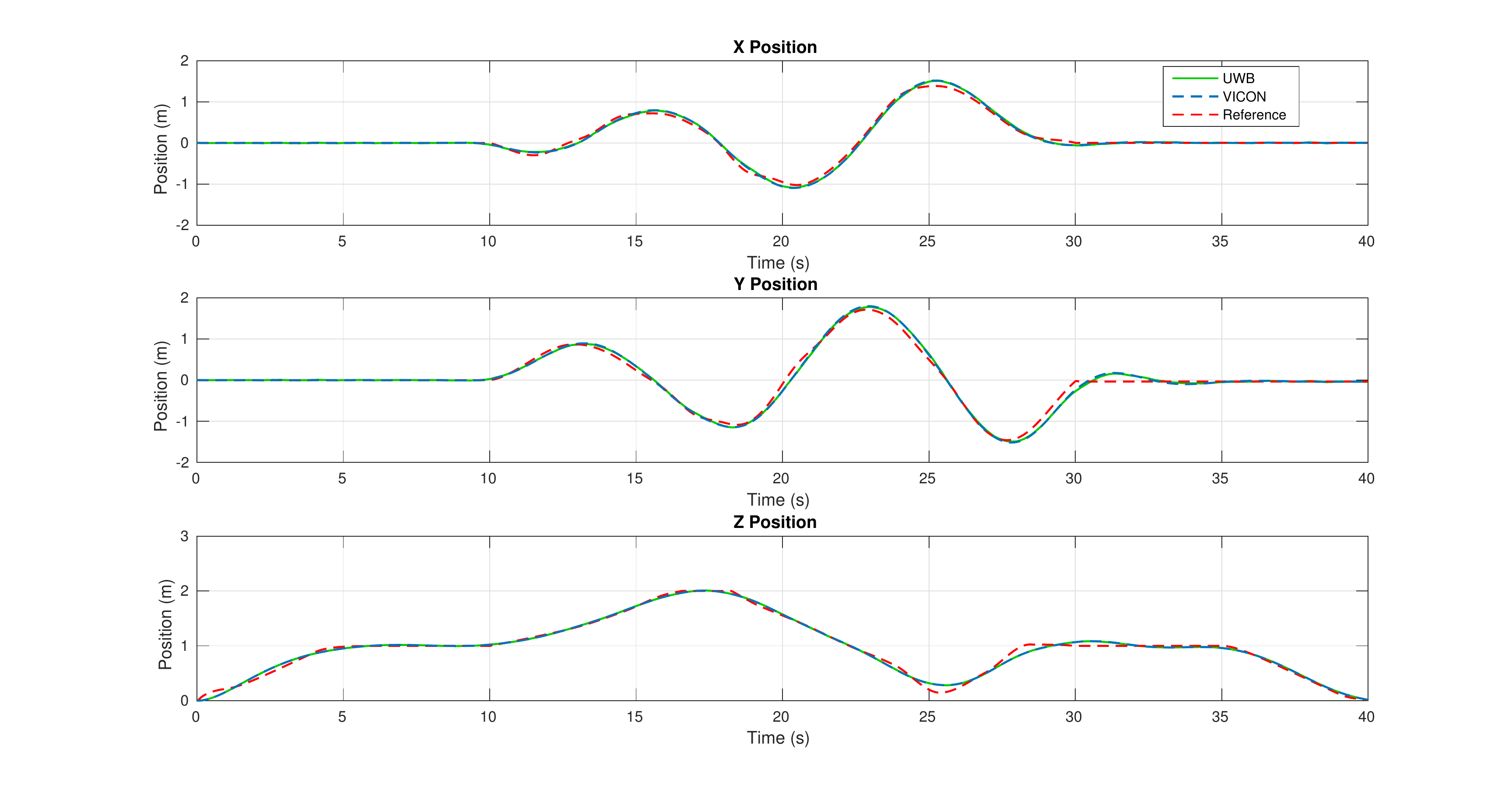}}
\caption{\label{fig:complex}Tracking for complex trajectories.}
\end{figure}

~\\
To truly justify the need of using a more refined trajectory tracker,
the simulated system was subject to follow a more complex trajectory
to test the tracking capabilities of the LQT algorithm and also to
evaluate the Kalman filter performance under more complex situations. \autoref{fig:complex} shows simulation results for a trajectory created by the user interface presented in \Cref{subsec:Trajectory-generation}.

\begin{figure}[H]
\centering
\makebox[0pt]{
\subfloat[\label{fig:Standard-View-ver-1-1-1-1}Standard view]{\includegraphics[width=0.55\textwidth]{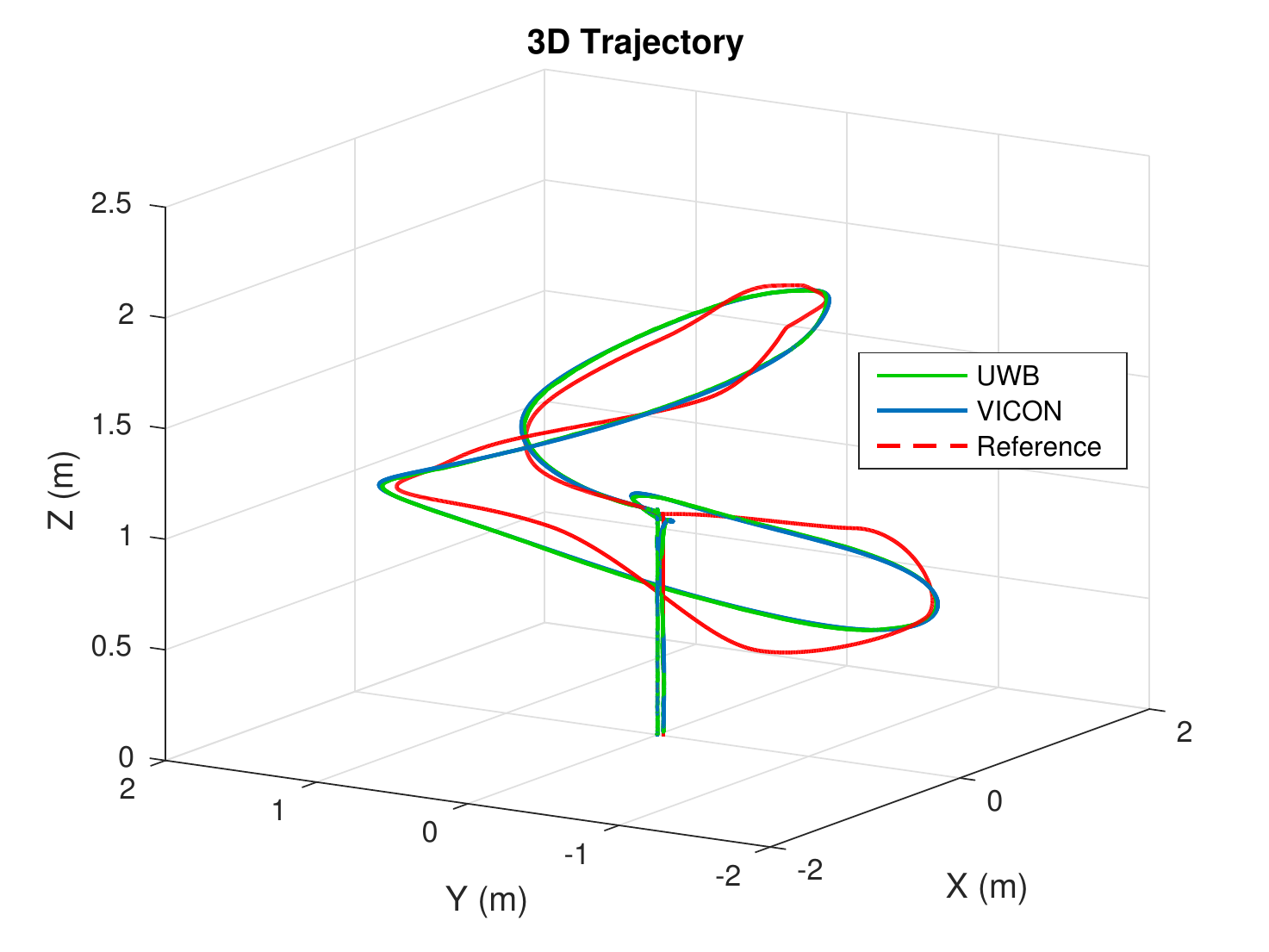}}
\subfloat[\label{fig:Top-View-ver-1-1-1-1}XY Plane]{\includegraphics[width=0.55\textwidth]{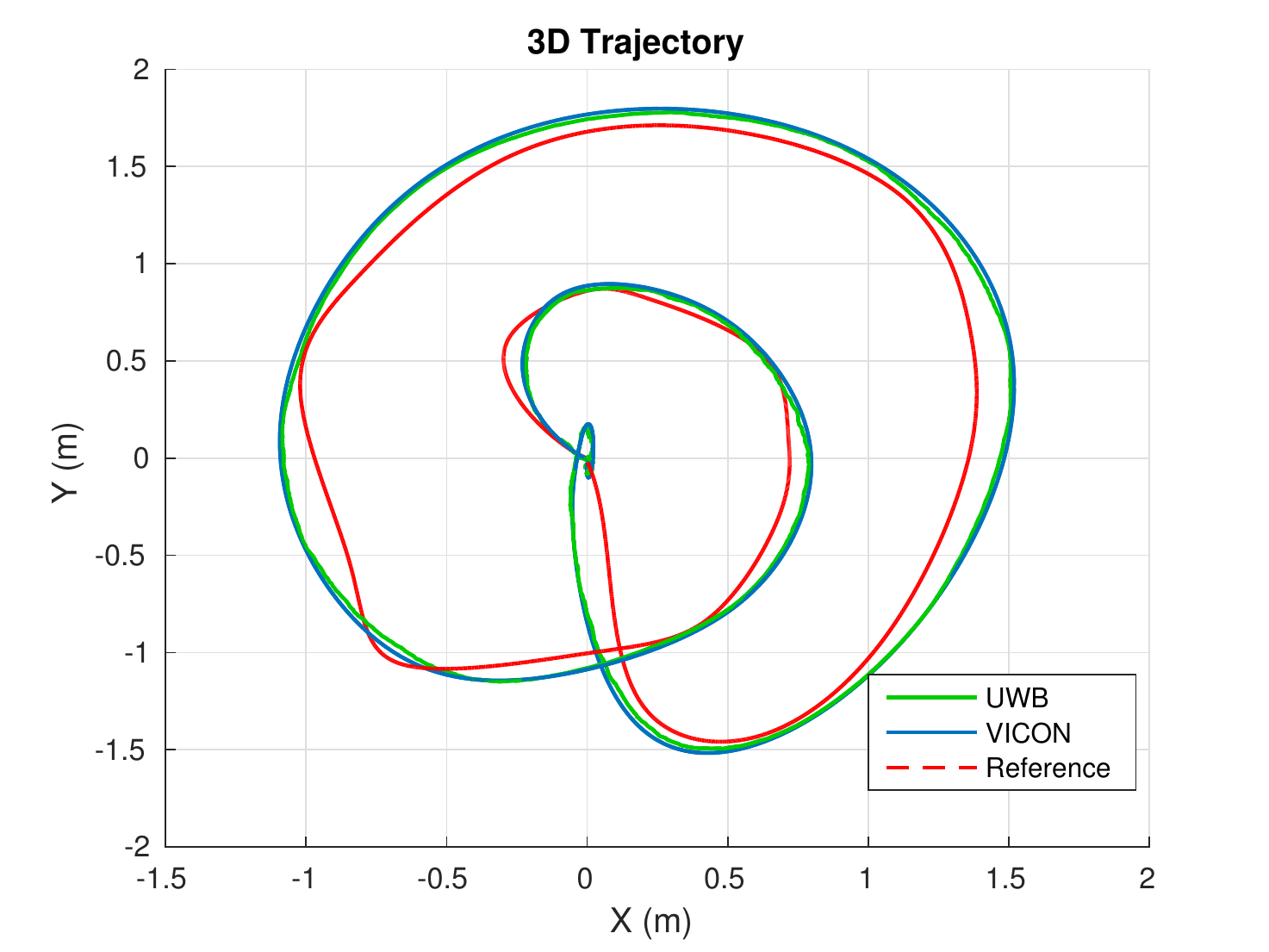}}}
\caption{\label{fig:3D-complex}3D Diagram for a complex trajectory.}
\end{figure}

~\\
The LQT controller was capable of tracking the desired trajectory,
with a low error even in closed curves as suggests the 3D diagrams
in \autoref{fig:3D-complex}.

As for the Kalman filter estimations, \autoref{fig:Kalman-Filter-simulation}
shows that the filter performed as expected when asked to track more
complex velocity profiles as the one generated by this trajectory.
The simulation environment for the LQT control system proved to be
a useful tool for developing and understanding the mechanics of the
newly adopted control technique for trajectory tracking. Going hand
by hand with the implementation phase, the simulation proposed in
this section served as a helpful guidance while fine tuning the controller in practice,
and at the same time serving as reference in terms of the performance to aim for in the control system.

\chapter{\label{sec:Implementation}Hardware Implementation and Experimental Results}

The implementation process was divided in two major phases: first
a familiarization with the Crazyflie 2.0 platform and implementation
of a PID position controller and secondly the implementation of a linear
quadratic algorithm.

\section{PID Controller}

After a thorough analysis of the original Crazyflie 2.0 firmware it
became evident that some minor changes were needed to be made because
the body-fixed frame defined in the embedded system did not match
the one adopted during the simulation phase, therefore, for the sake of
consistency with the body-fixed frame defined in \autoref{fig:rotation direction},
certain lines of the original firmware code were changed (see \hyperlink{appA}{Appendix A: Firmware Modifications}). The first step towards the implementation was to test the on-board
sensors such as the inertial measurement unit and all its components.
Even though the sensors' data analysis was not part of the project,
it was important to at least follow the idea of how the firmware captured
said data, ran it, for example, through the sensor fusion algorithm
and estimated the states of the quadcopter that were fed to the on-board
controllers. After this familiarization phase, the research moved
towards the off-board controller implementation using ROS.

\subsection{\label{sec:ROS}ROS Controller Node}

\begin{figure}[H]
\begin{centering}
\includegraphics[scale=0.7]{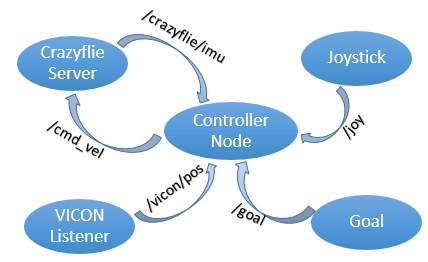}
\par\end{centering}
\caption{\label{fig:ROS-nodes-and}ROS nodes and topics.}

\end{figure}

Starting from the Open source ROS nodes presented in \cite{key-6},
the controller node was modified to implement the equations proposed
in \Cref{subsec:Off-Board-Position-controller}. \autoref{fig:ROS-nodes-and}
shows the ROS nodes and topics concerning the controller implemented.

A detailed explanation of the internal organization of each one of
this nodes and interactions is beyond the scope of this report, but
a qualitative analysis is adequate for an overall understanding of
the system:
\begin{itemize}
\item \textbf{Crazyflie Server:} this node defines the core interaction
between ROS and the Crazyflie 2.0, through a radio communication.
Its primary function in the control process is to serve as a data
bridge between the Crazyflie and the off-board controller. On the
one hand, it publishes sensors' readings coming from the quadcopter,
such as the IMU, that will serve as state estimations for the control
process. On the other hand, the node receives commands coming from
the controller node and sends them back to the Crazyflie. More specifically,
the data sent in the ``/cmd\_vel'' topic contains the outputs of
the off-board controller, that means, messages of the form $[\begin{array}{cccc}
\phi_{c} & \theta_{c} & r_{c} & \Omega\end{array}]$. This messages will then be the inputs of the on-board control system,
as shown previously in \autoref{fig:Onboard-control-architecture}.
\item \textbf{VICON Listener:} manages the communication with the VICON
positioning system. It publishes data concerning the inertial frame
coordinates $[x,y,z]$ of a reflective sphere as seen in \autoref{fig:Crazyflie-2.0-with}, sitting on top of the Crazyflie 2.0.

\begin{figure}[H]
\centering
\includegraphics[scale=0.1]{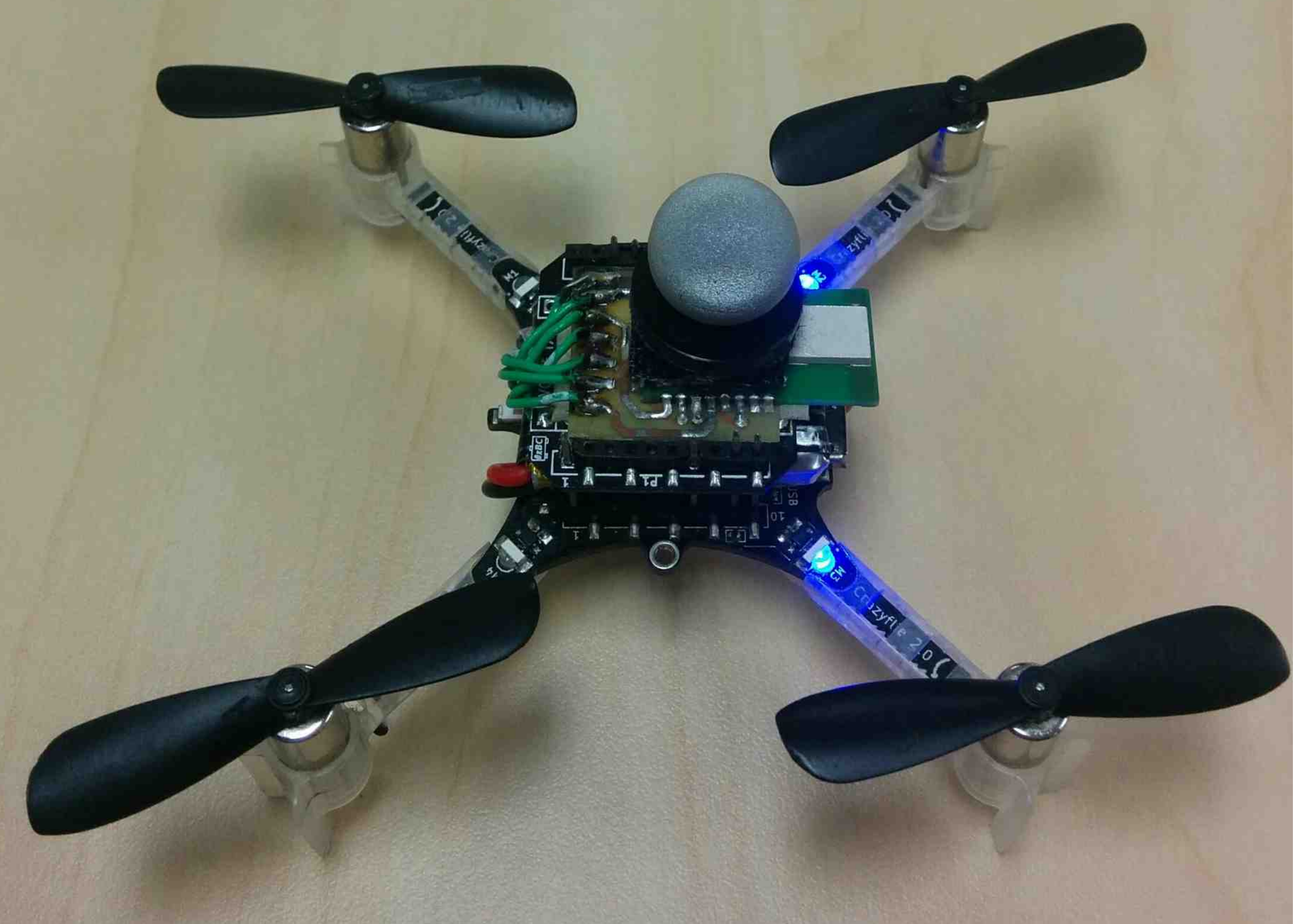}
\caption{\label{fig:Crazyflie-2.0-with}Crazyflie 2.0 with Vicon Sphere and UWB module.}

\end{figure}
Ideally, the VICON system should be used with three or more of these
spheres, in order for the position estimations to be more resilient
against other reflections in the laboratory or even for the Euler angles estimations. However, up to this point, all the tests were
done using the configuration shown in \autoref{fig:Crazyflie-2.0-with}.
and it did not present any major drawbacks during
the tests.
\item \textbf{Joystick:} this nodes serves basically as an external emergency
stop button. It is always a nice idea to keep the security measures
``software free'' in case something goes wrong during a flight,
thus the preference of using a hardware stop button instead of one,
probably more elegant, implemented in software such as MATLAB.
\item \textbf{Goal:} this is a user interactive topic created in MATLAB
that lets the user choose between a number of predefined trajectories
for the quadcopter to follow. Once the trajectory is selected, through
a certain ID number that can be changed in real-time, the MATLAB node
publishes the desired waypoint $\left[x_c,y_c,z_c,\psi_c\right]$. An additional
feature allows the user to choose whether to send position or velocity
commands to the yaw angle of the drone.
\item \textbf{Controller:} this is naturally the core of the real-time control
system. The node takes data from the IMU of the Crazyflie and from
the VICON system in order to have an estimate of the states of the
quadcopter in the control algorithm presented in \Cref{subsec:Off-Board-Position-controller}
. However this data only gives estimations of 9 states: the three linear
positions in the inertial frame, the three Euler angles and the three angular velocities coming from the gyroscope. The linear velocities in the body frame $[u,v,w]$ are not directly
measured by any of the sensors and they need to be estimated somehow
as they are used in the X-Y Position Controller. Usually, in this
type of scenarios, a state observer (Luenberger's,
Kalman Filter, etc) is required to reconstruct the missing states.
However, given the fact that the position estimations of the VICON
system are truly precise (error<1mm), the following
equations to estimate the velocities in the body-fixed frame using a pseudo-derivative in discrete time of the X-Y positions proved to be good enough for the control system to behave adequately:
\begin{equation}
V_{x}[k]=\left(\frac{x[k]-x[k-1]}{\Delta t}\right)
\end{equation}
\begin{equation}
V_{y}[k]=\left(\frac{y[k]-y[k-1]}{\Delta t}\right)
\end{equation}
where $V_{x}[k]$ and $V_{y}[k]$ are the linear velocities estimations
in the inertial frame. The term $\Delta t$ is the time step taken
between iterations in the control algorithm, given that the controller
node runs at 100 Hz, then $\Delta t=0.01s$. Normally to obtain the velocities
in the body-fixed frame a multiplication by the Euler matrix $\bm{R}^b_o$ should be executed, but taking a small angle approximation for the pitch and roll angles, then the calculation is simplified to just one rotation of the yaw angle around the Z axis, as shown in the following equation:
\begin{equation}
\left[\begin{array}{c}
u[k]\\
v[k]
\end{array}\right]=\left[\begin{array}{cc}
\cos\left(\psi[k]\right) & \sin\left(\psi[k]\right)\\
-\sin\left(\psi[k]\right) & \cos\left(\psi[k]\right)
\end{array}\right]\left[\begin{array}{c}
V_{x}[k]\\
V_{y}[k]
\end{array}\right]
\end{equation}
Finally the linear velocities in the body-fixed frame can be expressed
in terms of the measured states:
\begin{equation}
\begin{cases}
u[k]=\cos\left(\psi[k]\right)\left(\frac{x[k]-x[k-1]}{0.01}\right)+\sin\left(\psi[k]\right)\left(\frac{y[k]-y[k-1]}{0.01}\right)\\
v[k]=-\sin\left(\psi[k]\right)\left(\frac{x[k]-x[k-1]}{0.01}\right)+\cos\left(\psi[k]\right)\left(\frac{y[k]-y[k-1]}{0.01}\right)
\end{cases}
\end{equation}
This approximation proved to be good enough for the control architecture
proposed to work correctly.
\end{itemize}

\subsection{Experimental Results}

The flight data was retrieved using the MATLAB node mentioned earlier,
allowing for a more analytical interpretation of the results. The
following section presents a number of these flights with the appropriate
analysis.

\begin{itemize}
\item\textbf{Linear trajectories} 
\end{itemize}

\begin{figure}
\centering
\makebox[0pt]{
\includegraphics[width=1.2\textwidth]{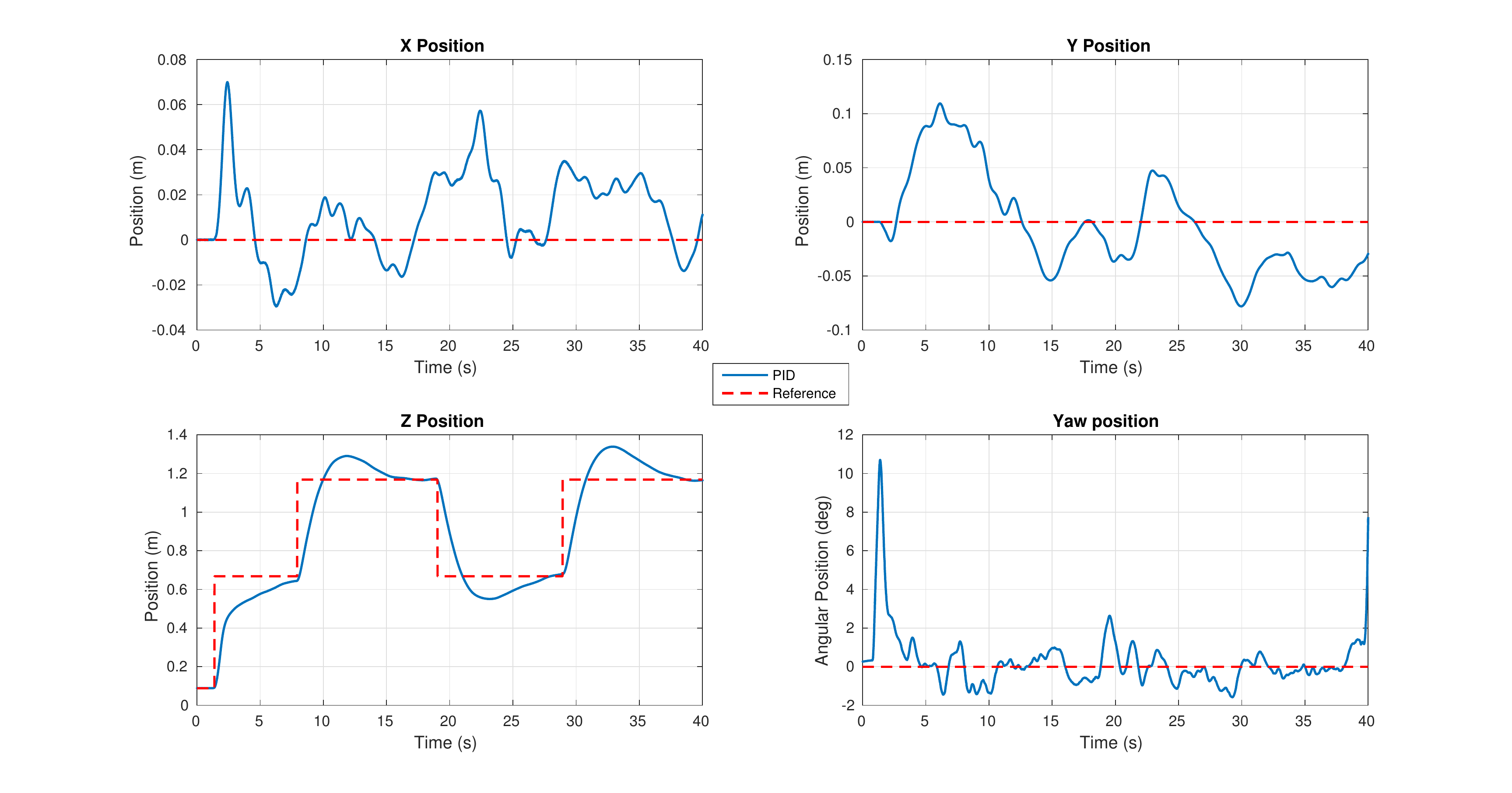}}
\caption{\label{fig:vertical}Steps in vertical command $z_{c}$.}
\end{figure}

For this type of trajectories the tests consisted basically in a take-off,
a linear movement in one or more directions, and a landing. For the
first test, steps of two different amplitudes were sent in the vertical
position of the drone, trying to maintain its X-Y position. The time plots of \autoref{fig:vertical} present the experimental data retrieved.
~\\

\begin{figure}[H]
\centering
\makebox[0pt]{
\subfloat[\label{fig:Standard-View-ver}Standard view]{\includegraphics[width=0.55\textwidth]{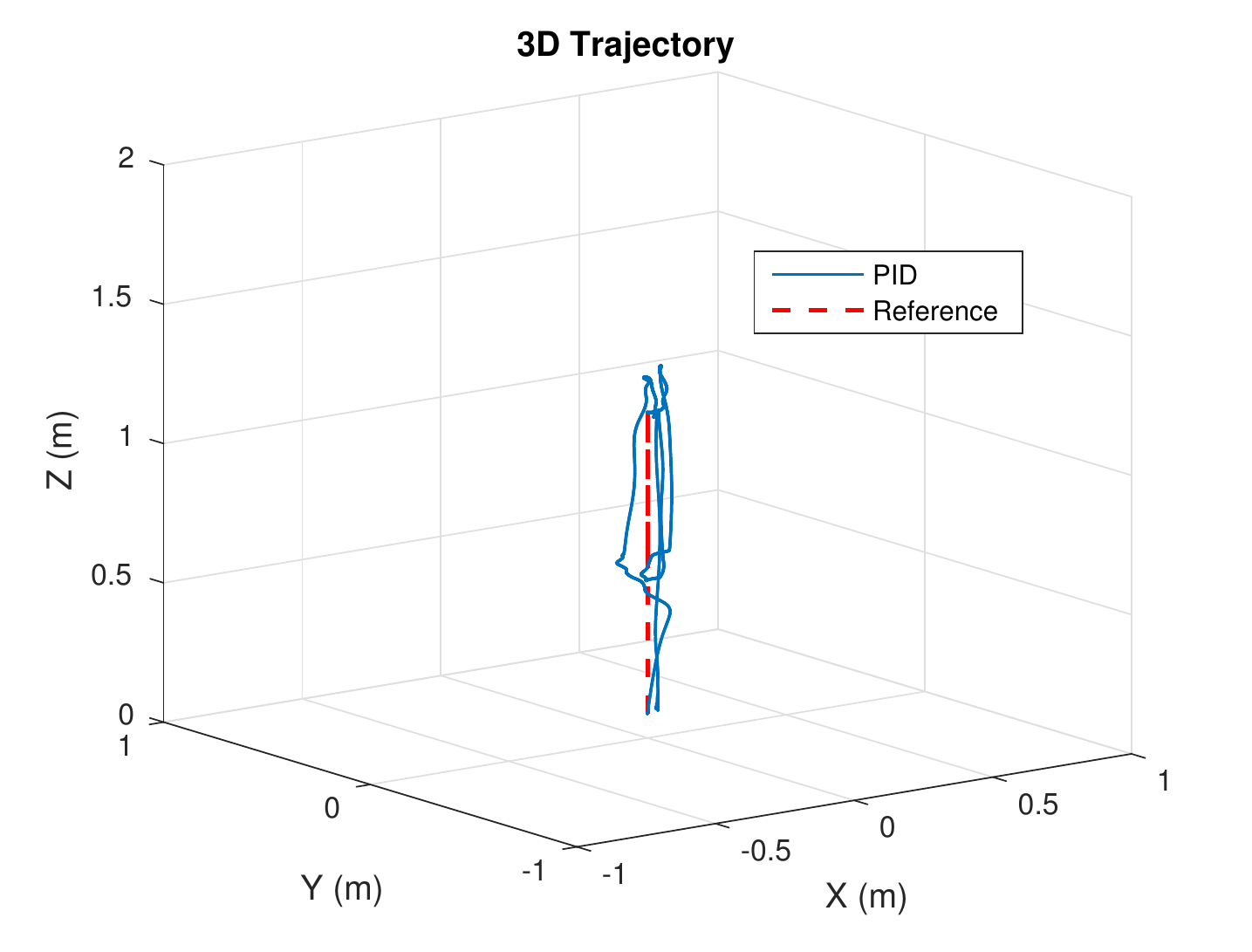}}
\subfloat[\label{fig:Top-View-ver}ZX Plane]{\includegraphics[width=0.55\textwidth]{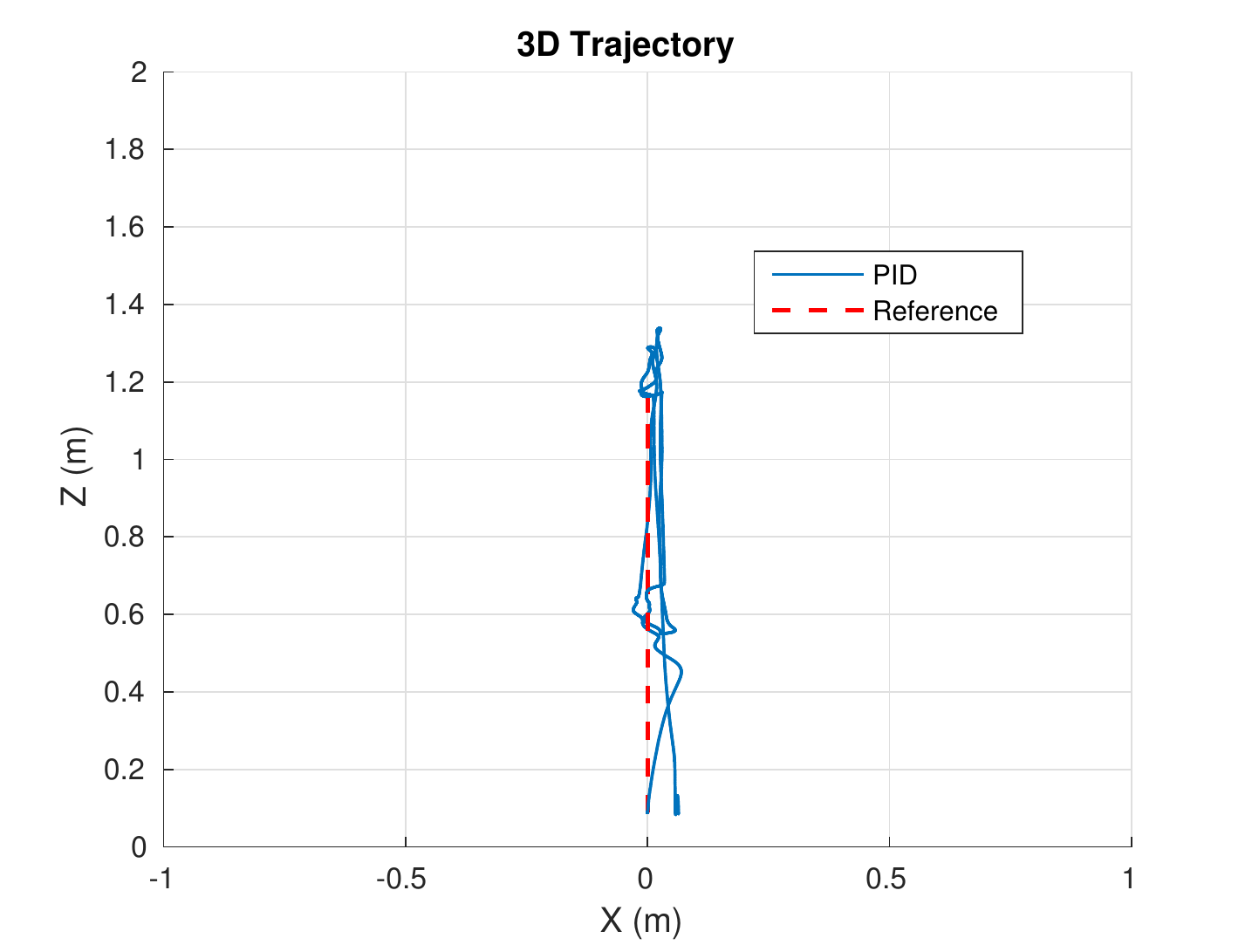}}}
\caption{\label{fig:3D-Vertical-Trajectory}3D Vertical Trajectory.}
\end{figure}

The two commands were sent to maintain an altitude of either 0.68 meters
or 1.18 meter, with all other commands set to zero. The X position stayed
within a 6cm margin of error while the Y position had a more prominent
error, roughly a 10 cm margin from the initial take-off position.
The fact that the drone used for these tests is not entirely symmetrical
(see \autoref{fig:Crazyflie-2.0-with}) means that the position holding
in one of the axes is better than in the other, as it was in fact the
case. \autoref{fig:3D-Vertical-Trajectory} shows different 3D
perspectives of the trajectory followed by the Crazyflie during the
test.

\begin{figure}[H]
\centering
\makebox[0pt]{
\includegraphics[width=1.2\textwidth]{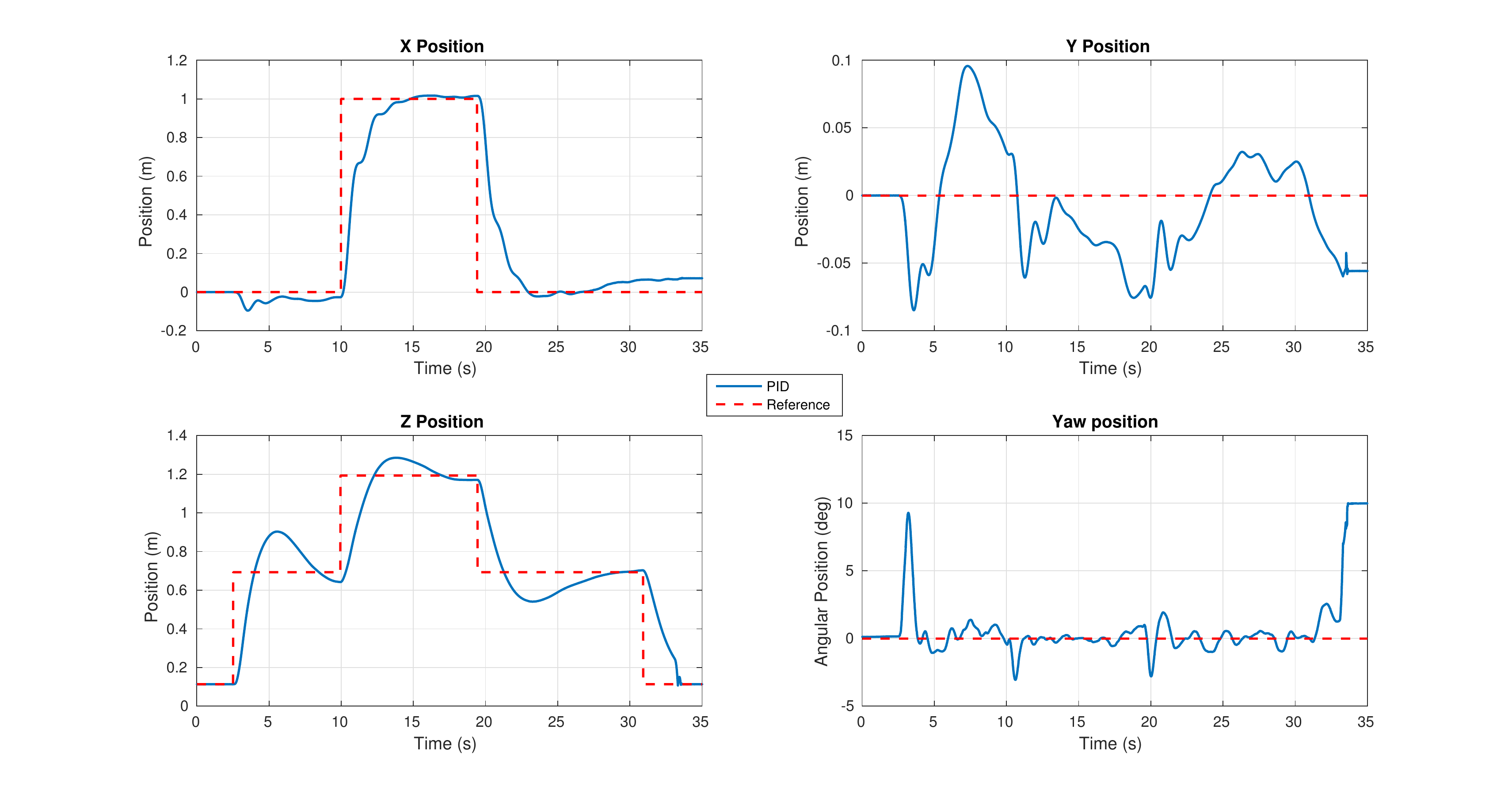}}
\caption{\label{fig:stepsz}Steps in $z_{c}$ and $x_{c}$.}
\end{figure}

The next test consisted in sending commands both in the X position
and in the altitude, in order to study the quadcopter's behavior when
sending mixed movements. Experimental data is plotted in \autoref{fig:stepsz}, showing the in-flight response of the quadcopter.

\begin{figure}[H]
\centering
\makebox[0pt]{
\subfloat[\label{fig:Standard-View-diag}Standard view.]{\includegraphics[width=0.55\textwidth]{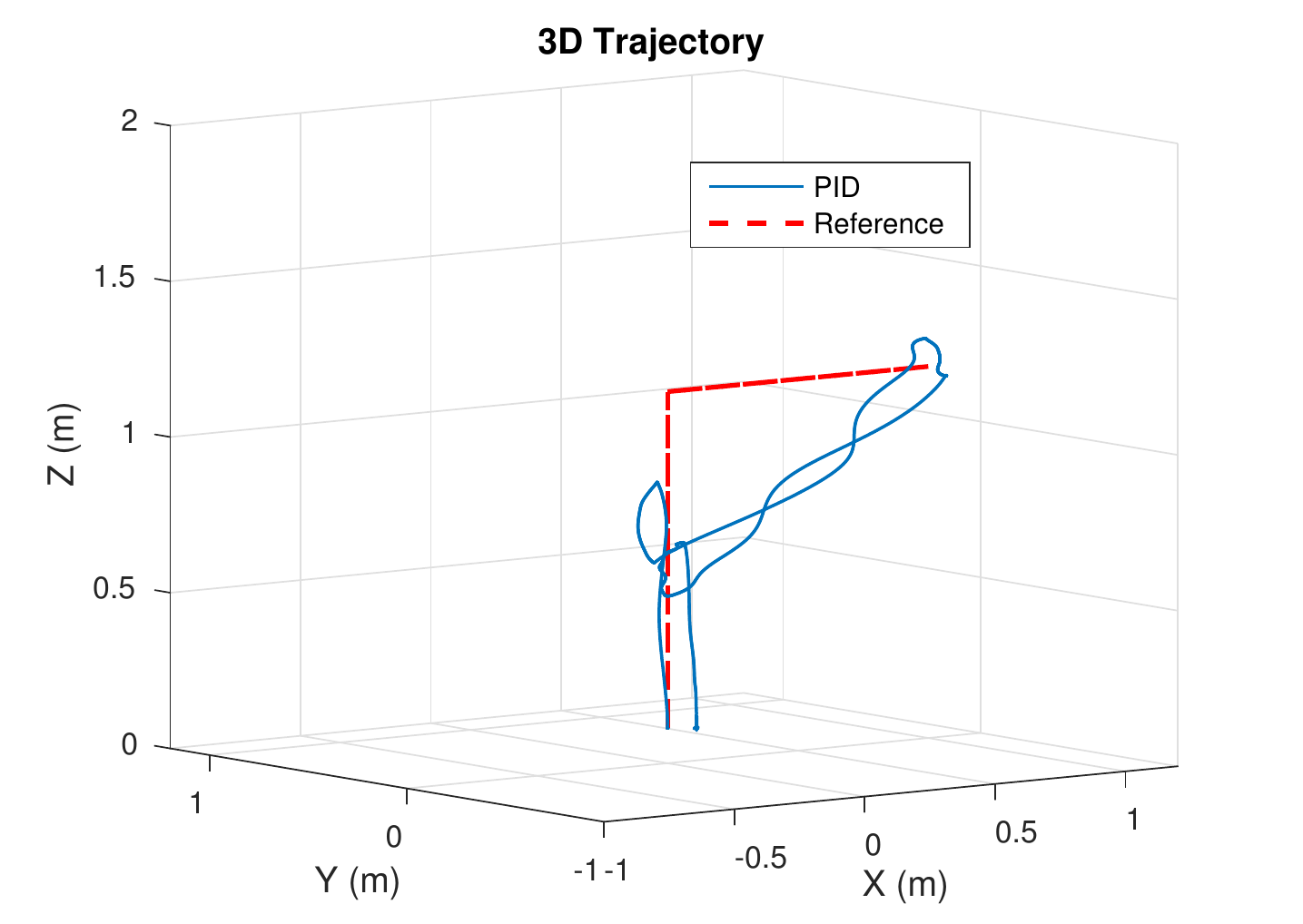}}
\subfloat[\label{fig:Top-View-diag}XY Plane.]{\includegraphics[width=0.55\textwidth]{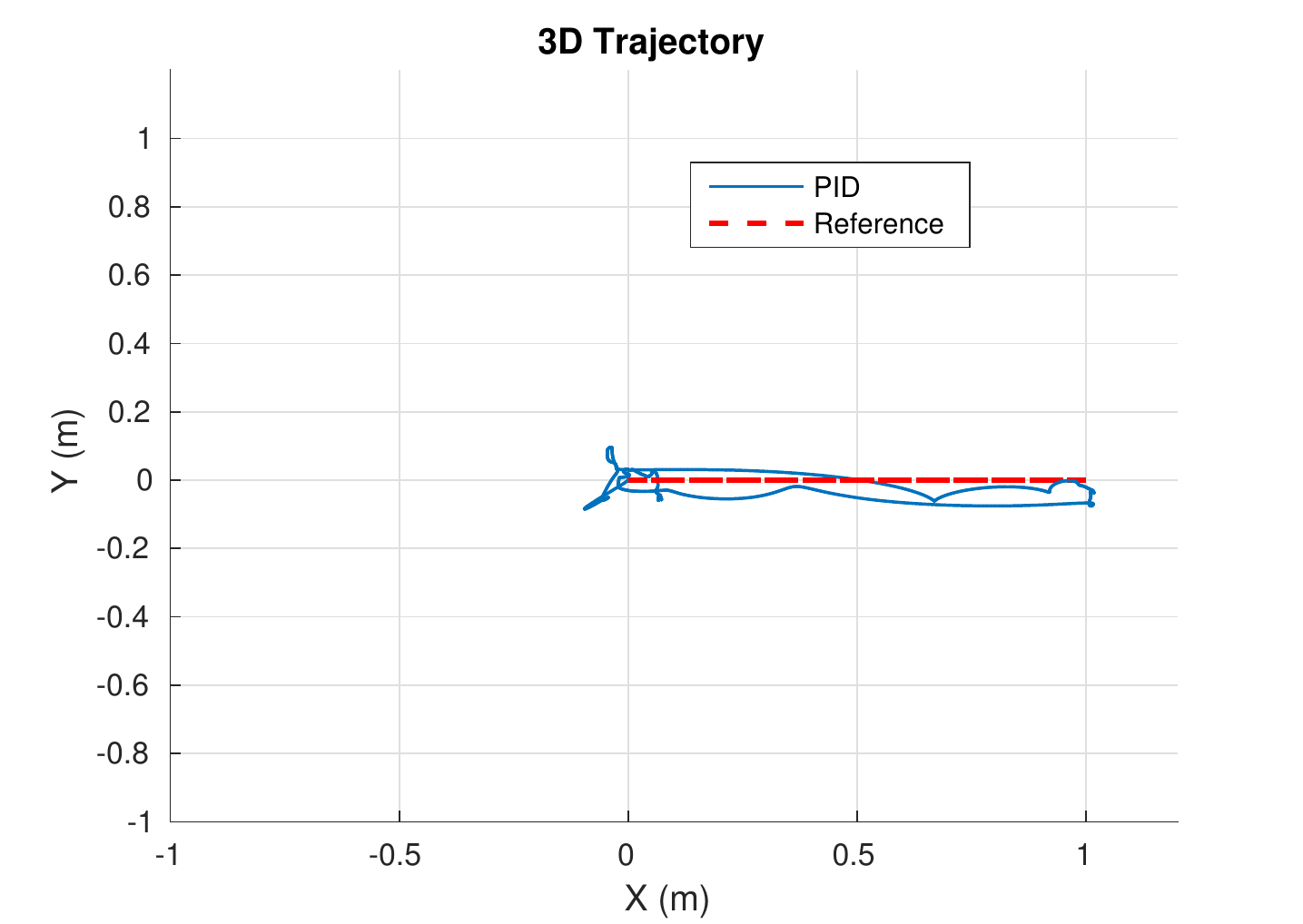}}}
\caption{\label{fig:3D-Diagonal-Trajectory}3D Diagonal Trajectory.}
\end{figure}

Similar to the simulations, the response time for a unity step in
the X position was around 4 seconds, whereas for the altitude is around
7 seconds. The Y position remained in a 10 cm error margin and the
yaw angle stayed well within a 3 degree margin after the initial take
off (which is represented in the peak at 3 seconds). \autoref{fig:3D-Diagonal-Trajectory} shows two different 3D perspectives of the experimental flight.

\begin{figure}[H]
\centering
\makebox[0pt]{
\includegraphics[width=1.2\textwidth]{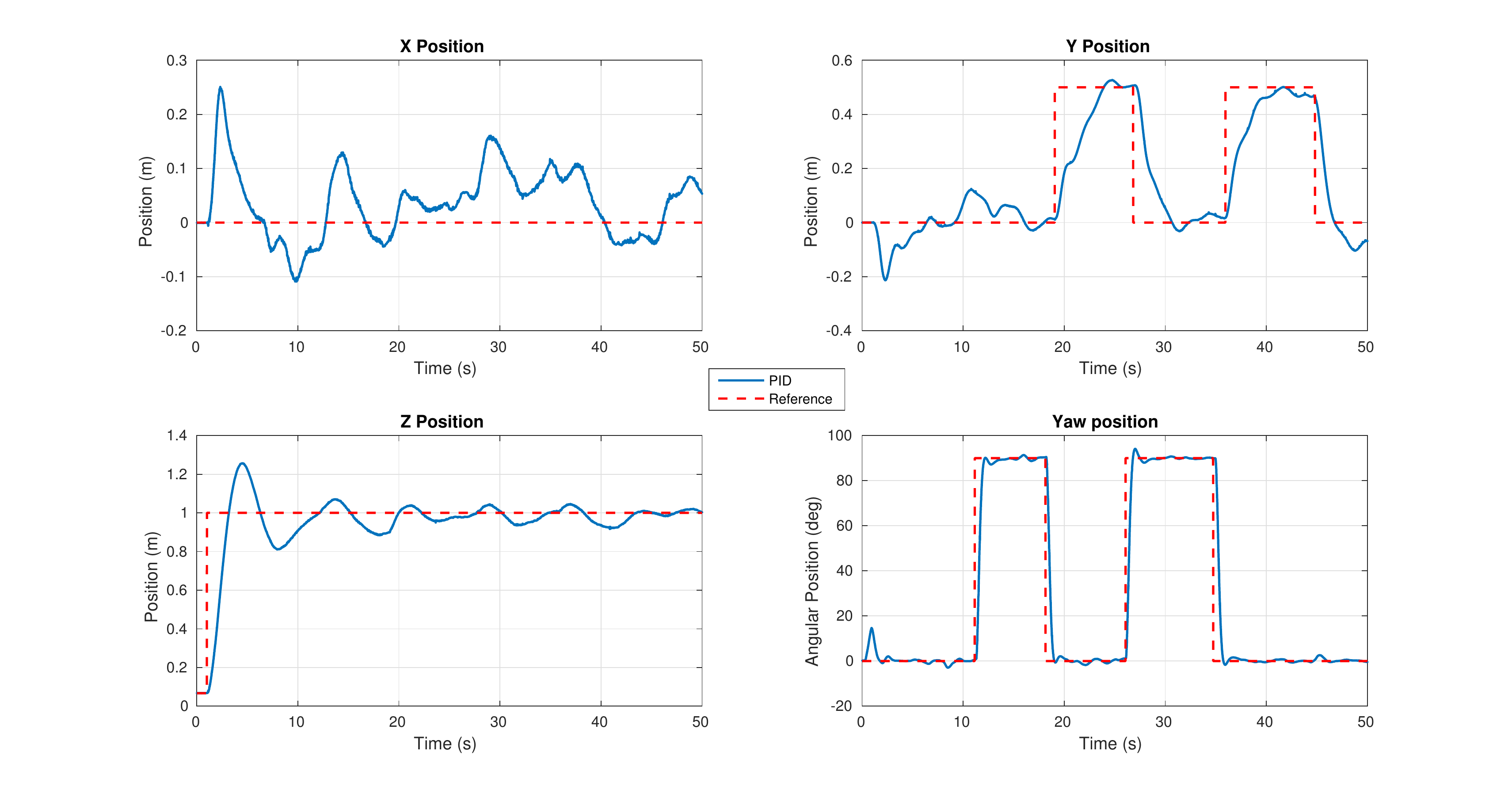}}
\caption{\label{fig:Steps-in-y}Steps in $y_{c}$ and $\psi_{c}$.}
\end{figure}

It remained only to test step commands in the Y axis and in the yaw
position, so this last test in the linear trajectories consisted in
sending both steps simultaneously and study the response of the system
knowing the movement interference analysed during the simulation phase
of the project. Test results are showcased in \autoref{fig:Steps-in-y}.

\begin{figure}[H]
\centering
\makebox[0pt]{
\subfloat[\label{fig:Standard-View-yaw}Standard view.]{\includegraphics[width=0.55\textwidth]{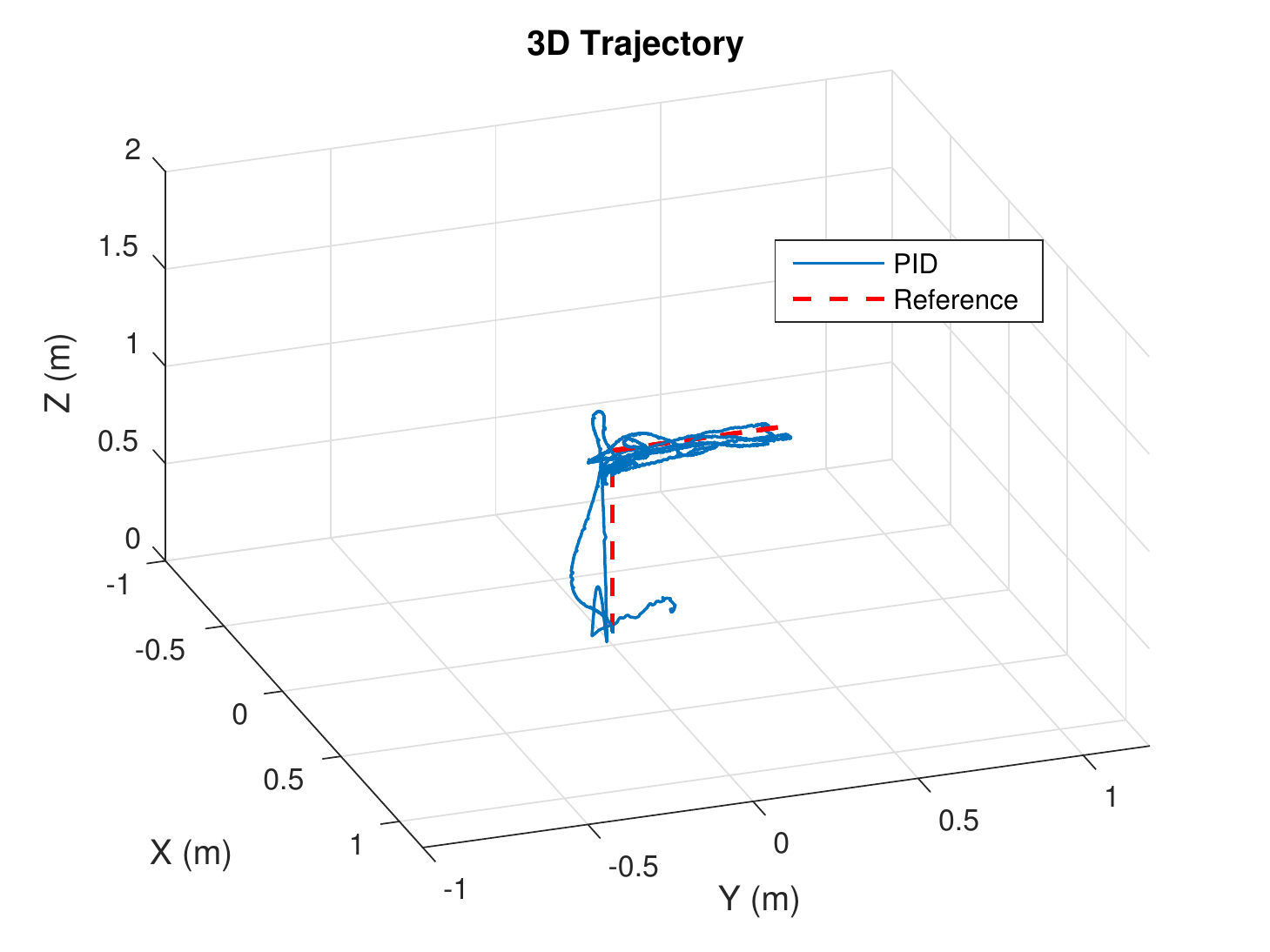}}
\subfloat[\label{fig:Top-View-yaw}XY Plane.]{\includegraphics[width=0.55\textwidth]{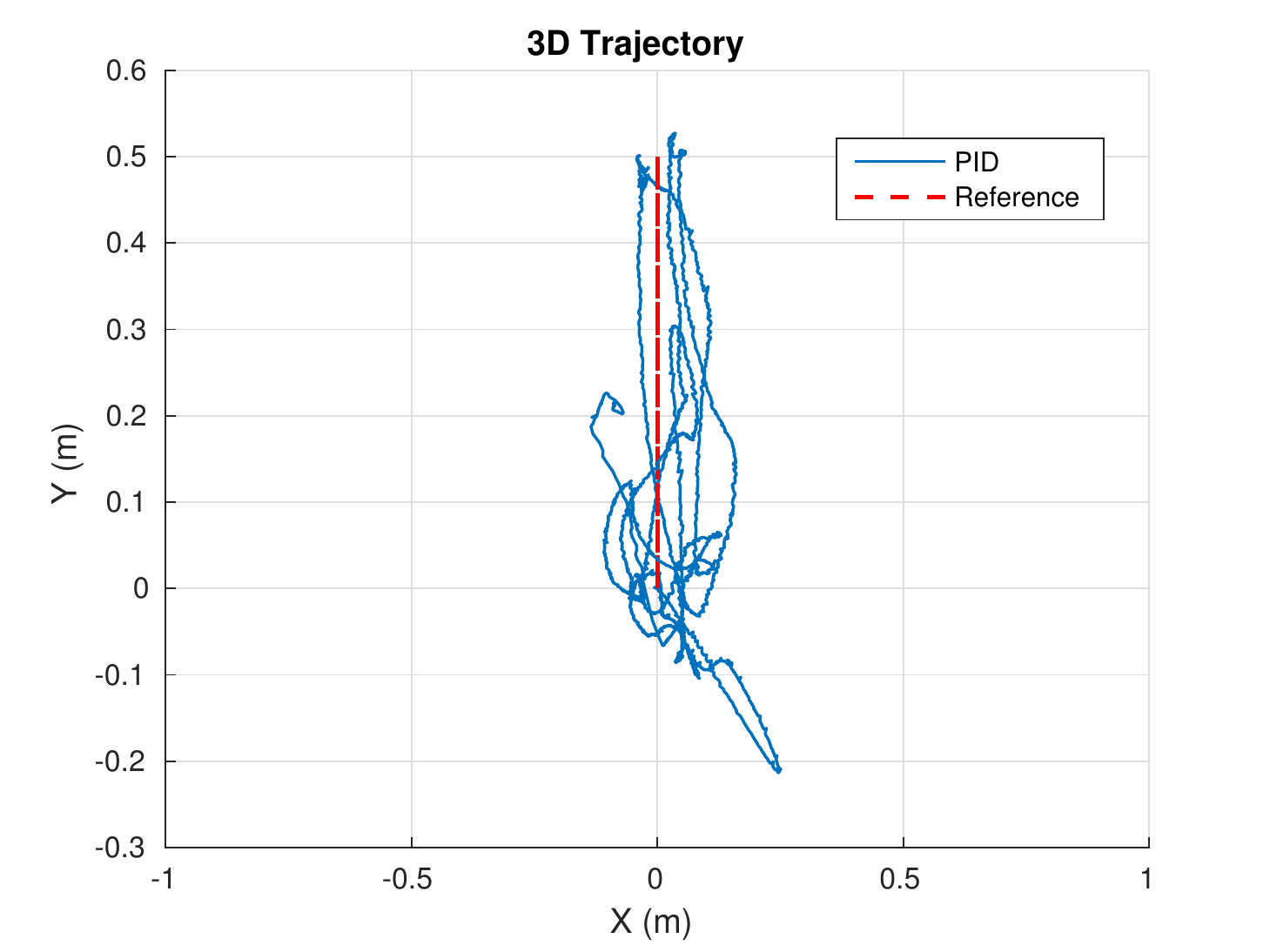}}}
\caption{\label{fig:3D-yaw}3D Trajectory with Y-Yaw compound movement.}
\end{figure}

While maintaining an altitude of 1 meter and a position in X of zero,
the commands sent were $y_{c}=0.5\textrm{m}$ and $\psi=90\text{\textdegree}$.
The response time for the Y position was, as in the case of the X
axis, around 4 seconds. On the other hand, the time response for a
command in yaw is much faster, about 1 second. Studying the impact these two movements had in the
other positions, for the altitude it is noted how it remained roughly
within the same level during the whole trial, meaning that those commands
did not have a major impact and the altitude controller compensated
any small interference those movements could have generated. As for
the X position, aside from the initial deviation due to take-off, the position
stayed around the 10 cm error margin, but in this case the effect
caused by the movements was much more notable, as seen in the 3D
perspectives of \autoref{fig:3D-yaw}.
~\\

The error in the X position around the initial position is more important
than in the other trajectories that were tested. Even as a qualitative
analysis from the experimental observations, the simultaneous movement
of the Y position and the yaw rotation generated light deviations in the X axis while moving from point A to point B of the trajectory.

\begin{itemize}
\item\textbf{Circular trajectories}
\end{itemize}

\begin{figure}[H]
\centering
\makebox[0pt]{
\includegraphics[width=1.2\textwidth]{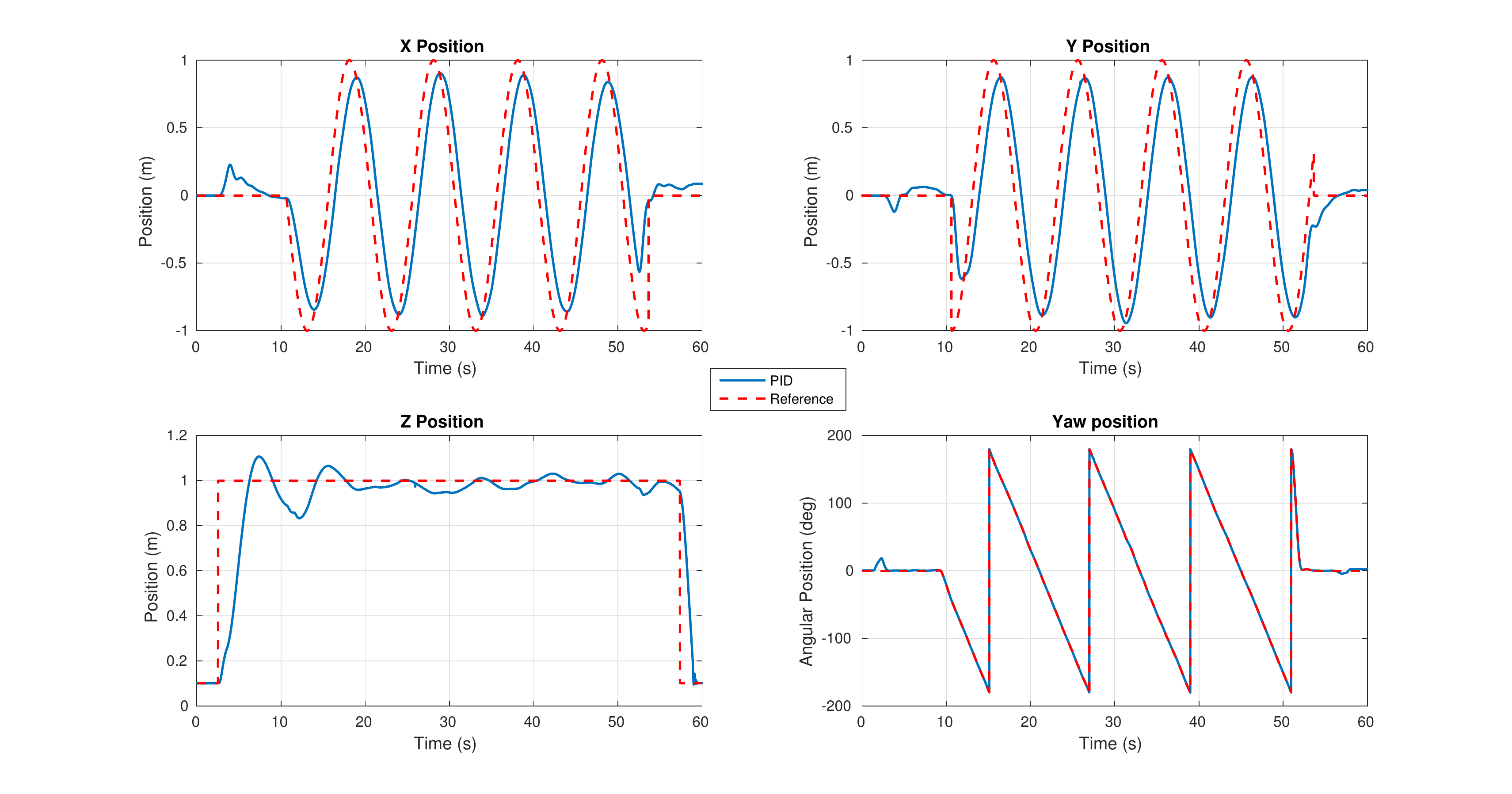}}
\caption{\label{fig:circle}Time Response for circular trajectory.}
\end{figure}

The next step for testing the position controller was to generate
more complicated trajectories and see if the system was capable of
following the desired path. To describe a circle, the following function
in discrete time was implemented:
\begin{equation}
\begin{cases}
x_{c}[k]=x_{0}+\sin\left(2\pi0.1k\right)\\
y_{c}[k]=y_{0}+\sin\left(2\pi0.1k+\pi/2\right)\\
z_{c}[k]=0.9\\
\psi[k]=-30k
\end{cases}\label{eq:3.14-2}
\end{equation}
where the values $x_{0}$ and $y_{0}$ represent the initial position
of the drone in the X-Y plane. The experimental data retrieved while performing the circular trajectory is presented in \autoref{fig:circle}.

~\\
In the X-Y position it is observed that the amplitude of the sine
waves did not reach the value of 1m, meaning that the position controller
failed to minimize the error as a consequence of not being fast enough for more complicated trajectories. A similar problem occurred with
the simulated model and actually the explanation can be reused. Given
the fact that the controller is a position tracker, it will try to
regulate at each time step of the process the actual position with
the desired position. In this case, the rate and amplitude of the
trajectory were too high for the position tracker developed to actually
make the quadcopter follow the path required. 

\begin{figure}[H]
\centering
\makebox[0pt]{
\subfloat[\label{fig:Standard-View-circle}Standard view.]{\includegraphics[width=0.55\textwidth]{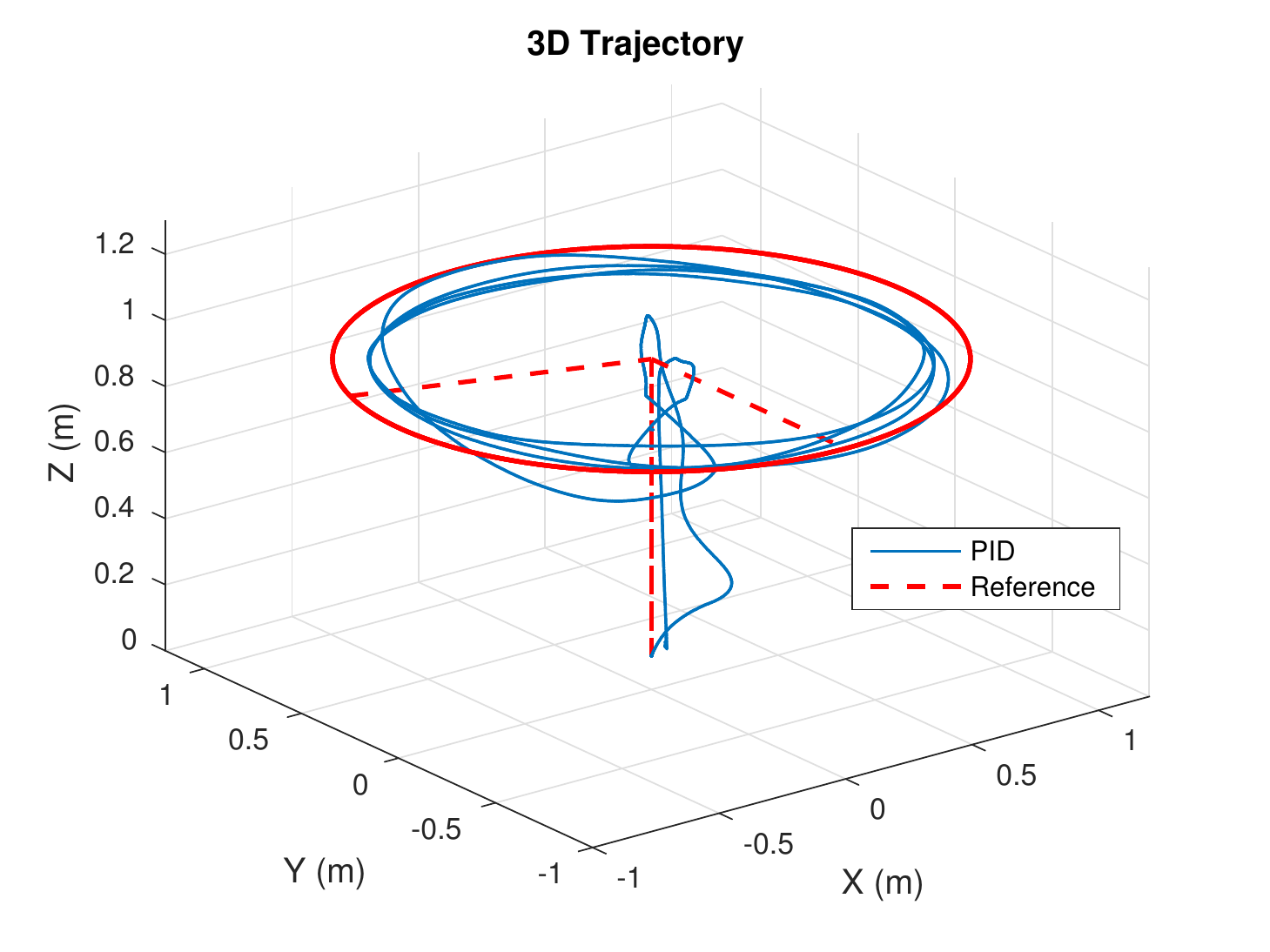}}
\subfloat[\label{fig:Top-View-circle}Top view.]{\includegraphics[width=0.55\textwidth]{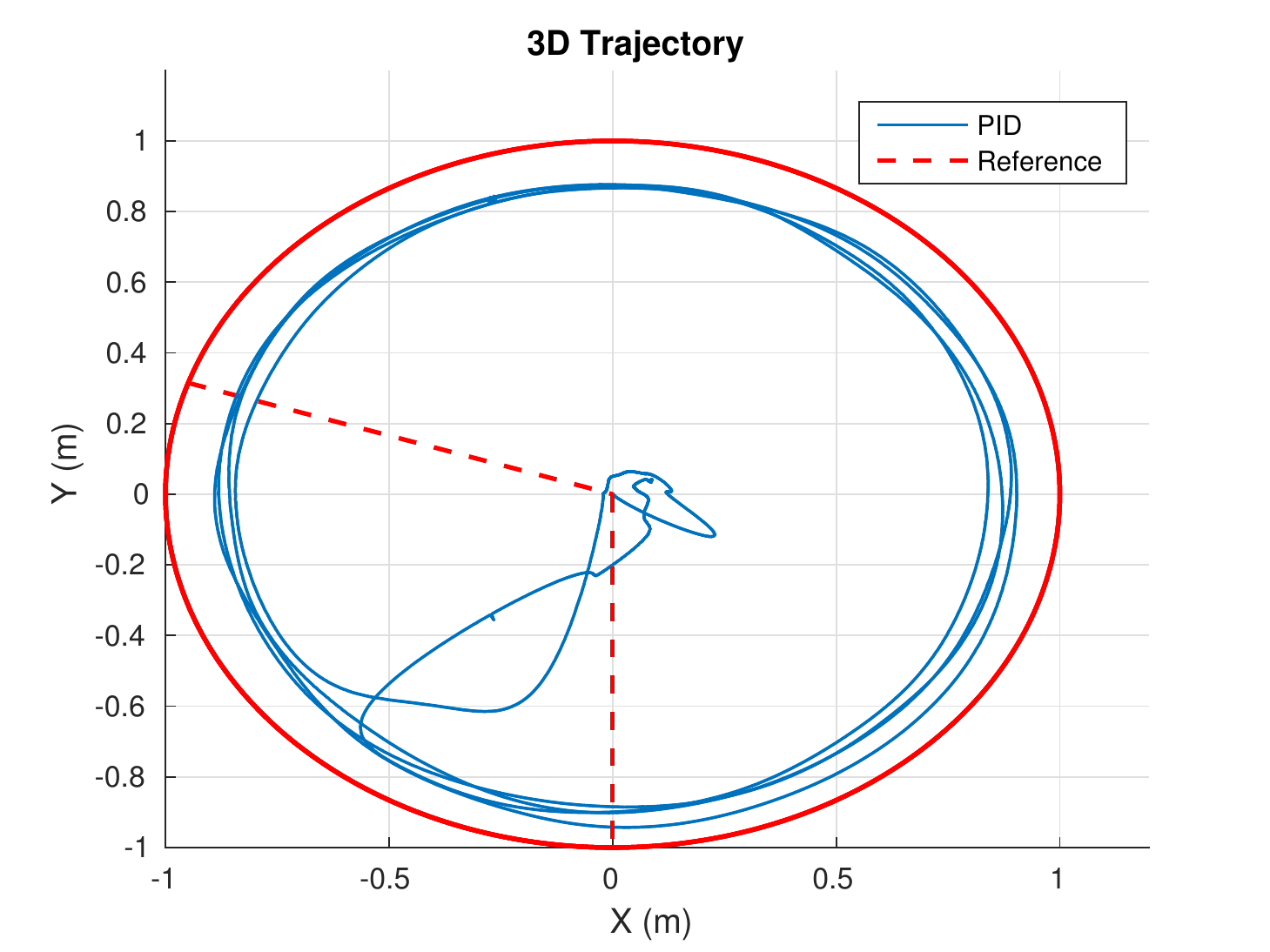}}}
\caption{\label{fig:3D-circle}3D Circular trajectory.}
\end{figure}

The 3D trajectories exhibited in \autoref{fig:3D-circle}
show more clearly the deficiencies in the path followed by the quadcopter.

\begin{figure}[H]
\centering
\makebox[0pt]{
\includegraphics[width=1.2\textwidth]{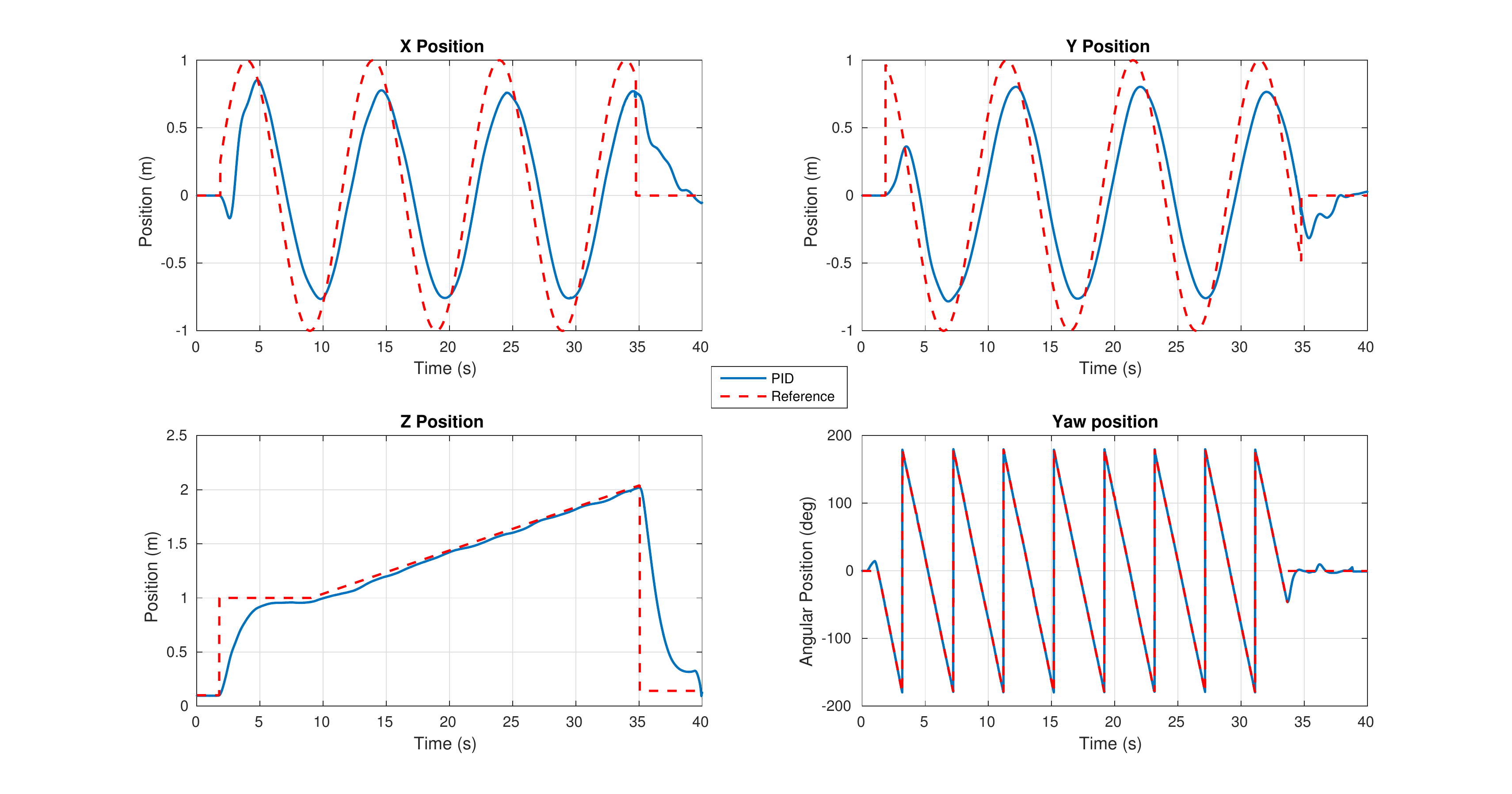}}
\caption{\label{fig:helical}Time Response for helix trajectory.}
\end{figure}

The last trajectory tested was the helix as seen during the simulation
phase, the discrete time function implemented in the controller was
defined as:

\begin{equation}
\begin{cases}
x_{c}[k]=x_{0}+\sin\left(2\pi0.1k\right)\\
y_{c}[k]=y_{0}+\sin\left(2\pi0.1k+\pi/2\right)\\
z_{c}[k]=0.9+0.04k\\
\psi[k]=-90k
\end{cases}\label{eq:3.14-2-1}
\end{equation}
meaning that in the X-Y plane it described the same circle as before,
but this time augmenting the altitude at a ratio of 4 centimeters
per second. Experimental data in \autoref{fig:helical} shows the performance of the quadcopter while following the helical trajectory.

\begin{figure}
\centering
\makebox[0pt]{
\subfloat[\label{fig:Standard-View-spiral}Standard view.]{\includegraphics[width=0.55\textwidth]{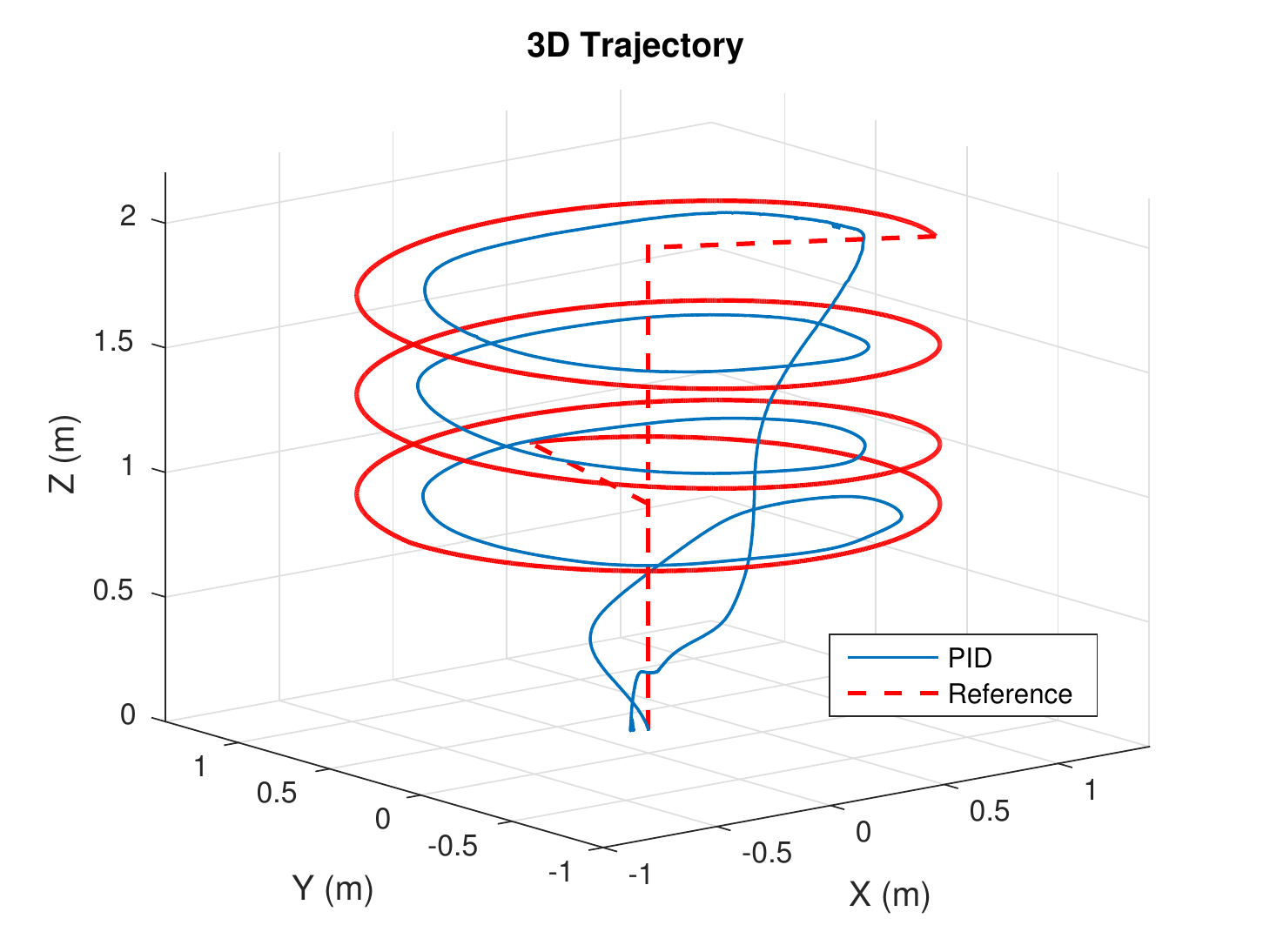}}
\subfloat[\label{fig:Top-View-circle-1}Top view.]{\includegraphics[width=0.55\textwidth]{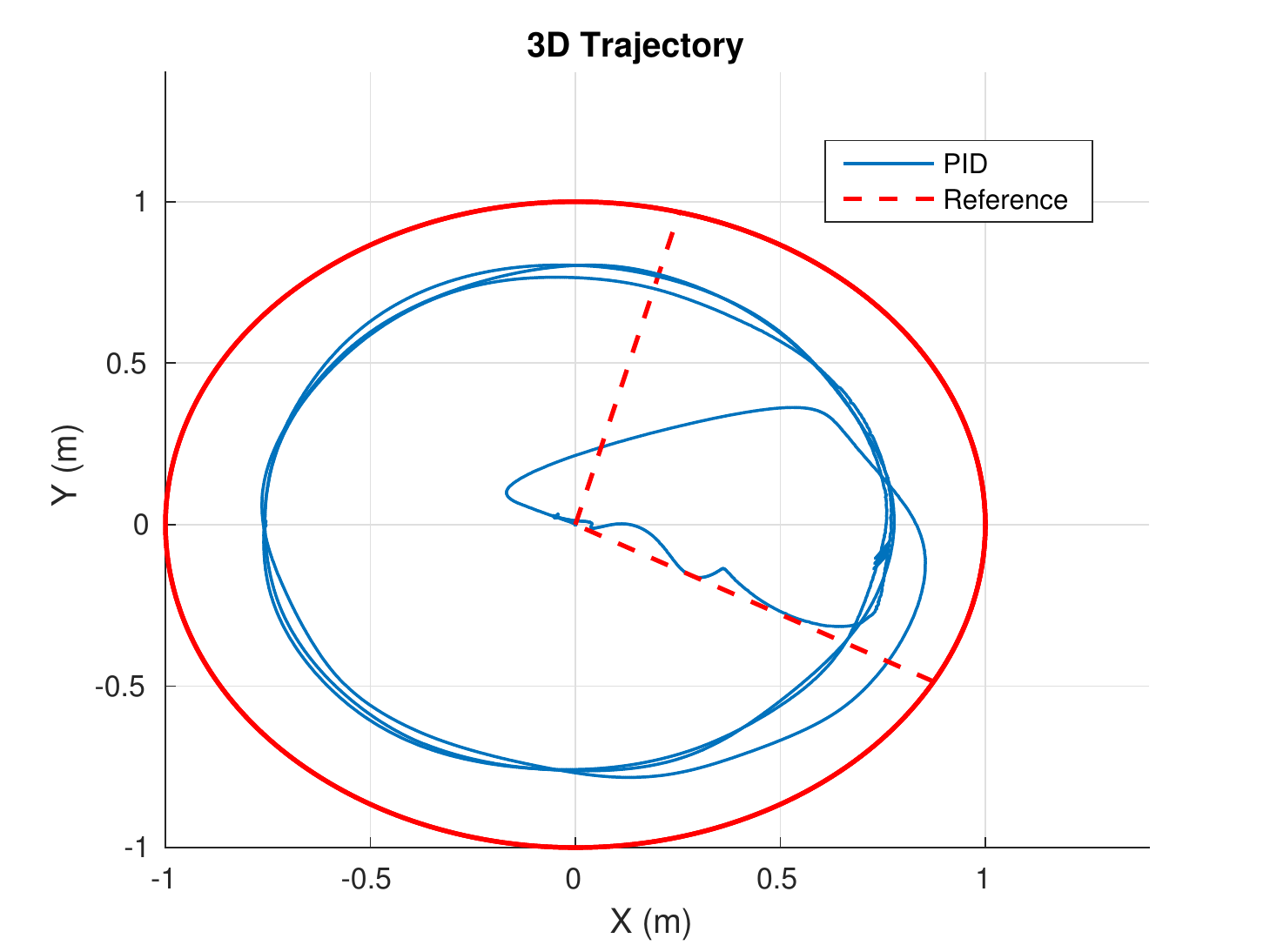}}}
\caption{\label{fig:3D-spiral}~~3D Helix Trajectory.}
\end{figure}

The increase of the yaw angle velocity did not have a major impact
in the performance of the system, the Crazyflie could turn at 90 degrees
per second without any complications. The results in the Z position
confirms that the altitude controller is good enough to follow
a time-varying function such as a straight line, even though the amplitude
variation of this trajectory was small. For the X-Y position the same
phenomenon occurred as in the case of the circular trajectory. The 3D perspectives in \autoref{fig:3D-spiral} displays the helicoidal trajectory followed.
~\\

After take-off, the quadcopter began the regulation of the desired trajectory which presents the same deficiencies as the previous test. The landing was successful as the data shows that the landing site was just a few centimeters away from the take-off point.

\section{\label{sec:LQT-Controller-Implementation}LQT Controller Implementation}

Having completed the first phase of the implementation process, a
broader knowledge of the embedded system and its limitations was gained
which was vital in the task of implementing the more delicate LQT
system. The word ``delicate'' in this context refers to the fact
that now the design included some low level control that was not done
in the previous phase. With the PID controller, the two cascaded architecture
embedded in the drone were already designed by the manufacturer, thus
the design was reduced just to the off-board position controller.
Instead, this new implementation required the careful thought of how
to implement every portion of the controller.

~\\
The design process began with the naive idea of implementing the whole
12 state feedback off-board using the MATLAB/ROS interface to send
directly the motor commands through the radio link communicating the
computer with the Crazyflie 2.0. This approach became rapidly discarded
after some failed trials, mainly due to the latency of the radio link
(around 2ms according to the manufacturer's specifications) and to
the refresh rate used in the MATLAB interface (100 Hz). The low level
stabilization, meaning the angle and angular velocities, is done in
almost every quadcopter inside the embedded system and not through
any external software, and there are two main reasons for it:
\begin{enumerate}
\item The control must be done at a high frequency rate to compensate the
quick angular dynamics of the quadcopter. These rates are seldom obtainable
through wireless communication protocols.
\item The low level control is not robust to the sort of delays introduced
by wireless links. If done within the embedded system, this latency
is easily eliminated.
\end{enumerate}
In order to resolve this issue, the elements of the time-invariant
state feedback gain $\bm{L}$ that corresponded to the angles and angular
velocities states were implemented in the embedded system while the
elements corresponding to the position and linear velocities of $\bm{L}$
and also the feedforward gains $\bm{g}\left[k\right]$ and $\bm{L}_{g}$ were
implemented in MATLAB (see \Cref{subsec:MATLAB-Interface-details}).

\begin{figure}[H]
\centering
\includegraphics[scale=0.1]{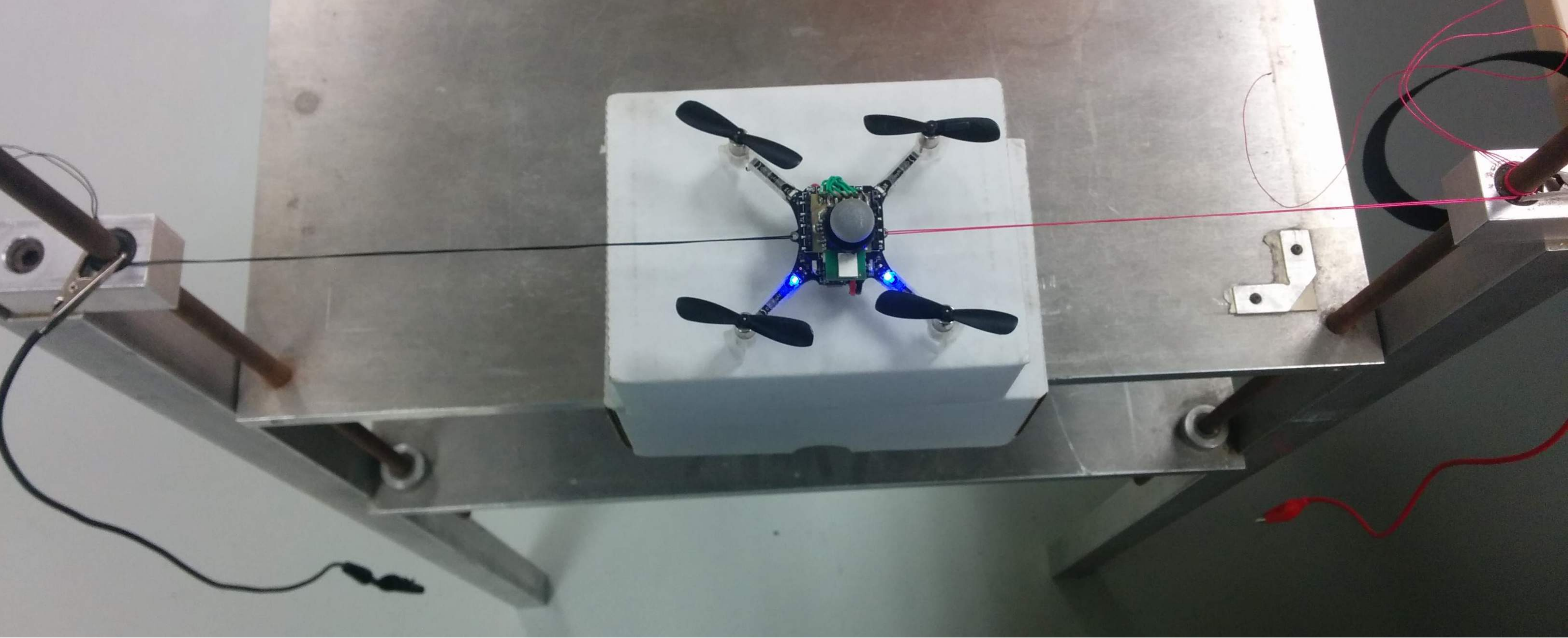}
\caption{\label{fig:rig}Pitch angle calibration rig.}
\end{figure}

~\\
After including the state feedback control in the Crazyflie's firmware, the first tests of the on-board attitude stabilizer were conducted using a calibration rig as seen in \autoref{fig:rig}. Although it could only be used to test the feedback of the pitch angle, the roll angle theoretically has similar dynamics and therefore the same tuned gains were used for both pitch and roll.

Even though there exists some non-negligible tension force induced
by the cables that attached the quadcopter's body to the posts, this
is a good first test trial to avoid potential crashes. After tuning
the gains of the roll and pitch angles the tests suggested that integral
action was needed to compensate for different sources of perturbation,
such as the asymmetry of the body and asymmetry of the motor's power.
While these effects might also be compensated by the position integral
action, the overall performance of the control system was improved
when adding the integral action to the angular positions. 

\begin{figure}[H]
\centering
\resizebox{0.8\hsize}{!}{
\begin{tikzpicture}[>=stealth]
  \coordinate (orig)   at (0,0);
  \coordinate (LLA)    at (0,4);
  \coordinate (AroneA) at (3,4.9);
  \coordinate (ArtwoA) at (4.8,4.9);
  \coordinate (LLB)    at (0,2.45);
  \coordinate (LLC)    at (2,2);
  \coordinate (LLD)    at (2,0.5);
  \coordinate (LLE)    at (-2,4.9);
  \coordinate (LLG)    at (-2,2.45);
  \coordinate (LLF)    at (2,7);
  \coordinate (Mult2DW)at (4,0.7);
  \coordinate (Mult2UP)at (4,1.2);
  \coordinate (Mult3)  at (4,7.65);
  \coordinate (Mult1)  at (4,2.65);
  \coordinate (Mult1DW)at (4,2.25);
  \coordinate (Eq)     at (-2,1);
  \coordinate (out1)   at (7,5);

 \node[draw,fill=blue!20, minimum width=3cm, minimum height=1.8cm, anchor=south west, text width=2cm, align=center] (A) at (LLA) {Crazyflie2.0};
 \node[draw, fill=blue!20, circle, minimum width=1cm, minimum height=1cm, anchor=center, align=center] (sum) at (LLB) {};
 \node[draw, fill=blue!20, circle, minimum width=1cm, minimum height=1cm, anchor=center, align=center] (sum2) at (LLE) {};
 \node[draw, fill=blue!20, circle, minimum width=1cm, minimum height=1cm, anchor=center, align=center] (sum3) at (LLG) {};
 \node[draw,fill=red!20, minimum width=3cm, minimum height=3cm,scale=0.3, anchor=south west, text width=1cm, align=center] (B) at (LLC) {};
 \draw (2+0.45,2+0.45) node[cross] {};
 \node[draw,fill=red!20, minimum width=3cm, minimum height=3cm,scale=0.3, anchor=south west, text width=1cm, align=center] (C) at (LLD) {};
 \draw (2+0.45,0.5+0.45) node[cross] {};
 \node[draw,fill=red!20, minimum width=3cm, minimum height=3cm,scale=0.3, anchor=south west, text width=1cm, align=center] (F) at (LLF) {};
 \draw (2+0.45,7+0.45) node[cross] {};
 \node[draw,fill=red!20, minimum width=3cm, minimum height=3cm,scale=0.3, anchor=south west, text width=1cm, align=center] (G) at ($(LLD)+(-6.5,3.95)$) {};
 \draw (2+0.45-6.5,0.5+0.45+4) node[cross] {};
  \node[draw,fill=red!20, minimum width=3cm, minimum height=3cm,scale=0.3, anchor=south west, text width=1cm, align=center] (I) at ($(LLD)+(-6.5,1.5)$) {};
 \draw (2+0.45-6.5,0.5+0.45+1.5) node[cross] {};
  
  \draw[-] (A.0)-- node[above, pos=0.5]{$\hat{\bm{x}}_{ang}[k]$} (ArtwoA);
  \draw[->] (ArtwoA.0) |-  ($(F.0)+(0,-0.2)$);
  \draw[->] (Mult3.0) -- node[right,pos=0]{$\bm{L}_{ang}[k]$}  ($(F.0)+(0,0.2)$);
  \draw[->] (Mult1.0) -- node[right,pos=0]{$\bm{L}_{pos}[k]$}  ($(B.0)+(0,0.2)$);
  \draw[->] (Mult1DW.0) -- node[right,pos=0]{$\hat{\bm{x}}_{pos}[k]$}  ($(B.0)-(0,0.2)$);
  \draw[->] (Mult2DW.0) -- node[right,pos=0]{$\bm{g}[k+1]$}  ($(C.0)+(0,-0.25)$);
  \draw[->] (Mult2UP.0) -- node[right,pos=0]{$\bm{L}_g[k]$}  ($(C.0)+(0,0.25)$);
  \draw[->] (B.180) --  (sum.0);
  \draw[->] (C.180) -| (sum.270);
  \draw[->] (F.180) -| node [left, pos = 0.9]{$\Delta \bm{u}_{ang}$} (sum2.90);
  \draw[->] (sum2.0) -- node [above, pos = 0.5]{$\bm{u}$} (A.180);
  \draw[->] (sum.180) -- node [above]{$\Delta \bm{u}$} (sum3.0);
  \draw[->] (sum3.90) -- node [left, pos = 0.9]{$\bm{u}_{pos}$} (sum2.270);
  \draw[->] (Eq.180) -- node [below, pos = 0]{$\bm{u}_e$} (sum3.270);
  \draw[->] ($(G)+(-1.5,-0.25)$) -- node [left,pos=0.1] {$\sum{(\bm{e}_{ang}\Delta k_{ang})}$} ($(G.180)+(0,-0.25)$);
  \draw[->] ($(G)+(-1.5,0.25)$) -- node [left,pos=0.1] {$\bm{K}_i^{ang}$} ($(G.180)+(0,0.25)$);
   \draw[->] ($(I)+(-1.5,-0.25)$) -- node [left,pos=0.1] {$\sum{(\bm{e}_{pos}\Delta k_{pos})}$} ($(I.180)+(0,-0.25)$);
  \draw[->] ($(I)+(-1.5,0.25)$) -- node [left,pos=0.1] {$\bm{K}_i^{pos}$} ($(I.180)+(0,0.25)$);
  \draw[->] (G.0) -- (sum2.180);
   \draw[->] (I.0) -- (sum3.180);
  \draw[red,very thick,dashed] (-8.5,-0.5) rectangle (6,3.5);
  \node[text width=6.5cm] at (-2,-0.3) {Off-Board @ 100Hz};
  \draw[olive,very thick,dashed] (-8.5,3.6) rectangle (6,8.6);
  \node[text width=6.5cm] at (-2,8.4) {On-Board @ 500Hz};
  \node[text width=1cm] at (-2,1.8) {$+$};
  \node[text width=1cm] at (-1,2.2) {$+$};
  \node[text width=1cm] at (-1.4,4.2) {$+$};
  \node[text width=1cm] at (-1.4,5.5) {$-$};
  \node[text width=1cm] at (1,2.2) {$-$};
  \node[text width=1cm] at (0,1.8) {$+$};
  \node[text width=1cm] at ($(sum3)+(-0.3,0.3)$) {$+$};
  
  \node[text width=1cm] at ($(sum2)+(-0.3,0.3)$) {$+$};

\end{tikzpicture}}
\caption{\label{fig:Implementation-diagram}Implementation diagram.}

\end{figure}
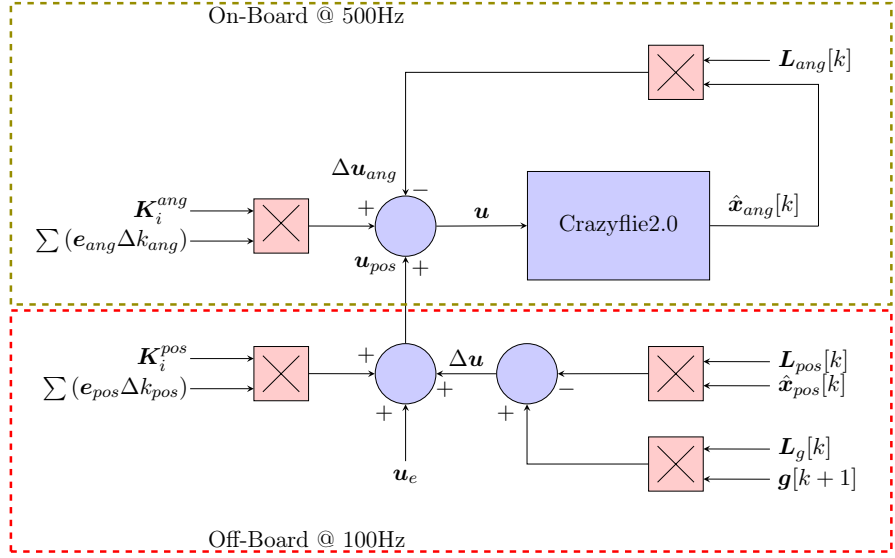

~\\
The implemented architecture is exposed in the block diagram of \autoref{fig:Implementation-diagram}. On top it shows the angular stabilization loop that runs in the embedded system at a faster rate of 500Hz, while below is the rest of the LQT architecture running off-board at a slower rate of 100 Hz.

The sensor fusion algorithm calculation for the three Euler angles
and the angular velocities coming from the gyroscope compose the
state vector $\hat{\boldsymbol{x}}_{ang}\left[k\right]$ in the following
manner:
\begin{equation}
\hat{\boldsymbol{x}}_{ang}=\left[\begin{array}{cccccc}
\hat{\psi} & \hat{\theta} & \hat{\phi} & \hat{r} & \hat{q} & \hat{p}\end{array}\right]^{T}
\end{equation}
The gain $\boldsymbol{L}_{ang}$ is a 4x6 matrix that multiplies accordingly
the state vector $\hat{\boldsymbol{x}}_{ang}.$ The system input $\Delta\boldsymbol{u}_{ang}$
is a deviation from the equilibrium point, $\boldsymbol{u}_{e}$,
to maintain an angle equal to zero in all three Euler angles. The
on-board controller runs at 500Hz, thus the time step $\Delta k_{ang}$
in \autoref{fig:Implementation-diagram} is equal to 0.002s.

~\\
The off-board section of the controller computes the state feedback
for the position and linear velocities states captured by the VICON or UWB
systems and the Kalman filter developed in \Cref{subsec:Kalman-Filter-for},
therefore composing the following state vector:

\begin{equation}
\hat{\boldsymbol{x}}_{pos}=\left[\begin{array}{cccccc}
\hat{x} & \hat{y} & \hat{z} & \hat{u} & \hat{v} & \hat{w}\end{array}\right]^{T}
\end{equation}
Matrix gain $\boldsymbol{L}_{pos}$ is also a 4x6 matrix that multiplies the state vector $\hat{\boldsymbol{x}}_{pos}$. For
the position integral action, the time step $\Delta k_{pos}$ is equal
to 0.01s

\subsection{ROS Controller Node Modifications}

For the LQT implementation, the ROS controller node presented in \Cref{sec:ROS} that previously implemented the PID controller was modified to only receive the incoming commands from the MATLAB interface and send them to the Crazyflie 2.0 through the radio link. The Crazyflie server node was then slightly modified to send the motor PWM signals instead of the desired Euler angles as it did with the previous control system. 

\subsection{\label{subsec:MATLAB-Interface-details}MATLAB Interface Details}

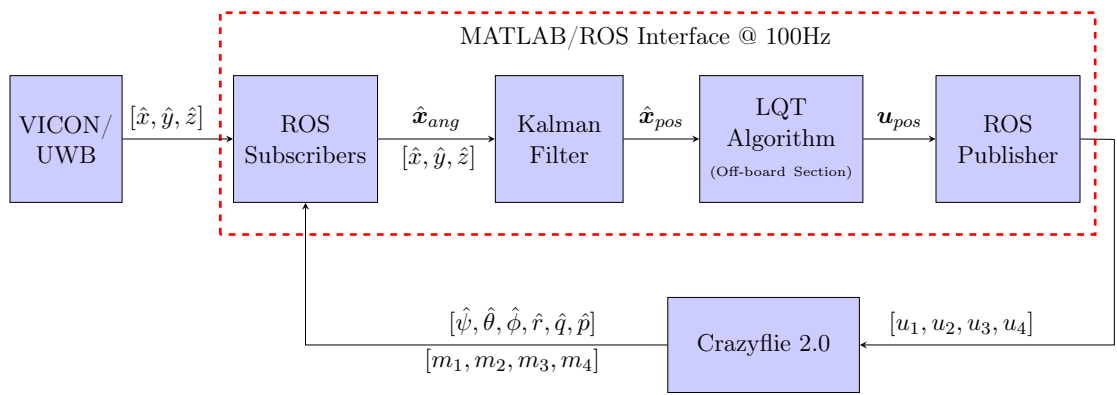
\begin{figure}[H]
\centering
\makebox[0pt]{
\resizebox{1\hsize}{!}{
\begin{tikzpicture}[>=stealth]
  \coordinate (orig)   at (0,0);
  \coordinate (LLD)    at (6.5,1);
  \coordinate (LLA)    at (-0.3,4);
  \coordinate (LLB)    at (3.8,4);
  \coordinate (LLC)    at (7,4);
  \coordinate (LLE)	   at (10+0.7,4);
  \coordinate (ED)     at (13+0.5,4+1);
  \coordinate (BoE)    at (3.5,2);
  \coordinate (LLF)    at (-3.8,4);

  \node[draw,fill=blue!20, minimum width=2cm, minimum height=2cm, anchor=south west, text width=2cm, align=center] (A) at (LLA) {ROS\\Subscribers};
  \node[draw,fill=blue!20, minimum width=2cm, minimum height=2cm, anchor=south west, text width=1.5cm, align=center] (B) at (LLB) {Kalman\\Filter};
  \node[draw,fill=blue!20, minimum width=1.4cm, minimum height=2cm, anchor=south west, text width=2.3cm, align=center] (C) at (LLC) {LQT\\Algorithm\\\tiny(Off-board Section)};
  \node[draw,fill=blue!20, minimum width=3cm, minimum height=1.5cm, anchor=south west, text width=2.5cm, align=center] (D) at (LLD) {Crazyflie 2.0};
  \node[draw,fill=blue!20, minimum width=2cm, minimum height=2cm, anchor=south west, text width=2cm, align=center] (E) at (LLE) {ROS\\Publisher};
  \node[draw,fill=blue!20, minimum width=1.5cm, minimum height=2cm, anchor=south west, text width=1.5cm, align=center] (F) at (LLF) {VICON/\\UWB};
  
   \draw[->] (A.0) -- node[above] {$\hat{\bm{x}}_{ang}$} (B.180);
   \node[text width=3cm] at ($(A)+(3,-0.3)$) {$[\hat{x},\hat{y},\hat{z}]$};
   \draw[->] (B.0) -- node[above] {$\hat{\bm{x}}_{pos}$} (C.180);
   \draw[->] (C.0) -- node[above] {$\bm{u}_{pos}$} (E.180);
   \draw[-] (E.0) -- (ED.0);
   \draw[->] (ED.0) |- node [above,pos = 0.8] {$[u_1,u_2,u_3,u_4]$}  (D.0);
   \draw[->] (D.180) -| node [above, pos = 0.2] {$[\hat{\psi},\hat{\theta},\hat{\phi},\hat{r},\hat{q},\hat{p}]$} (A.270);
   \draw[red,very thick,dashed] (-0.5,3.5) rectangle (13.2,7);
   \draw[->] (F.0)-- node[above, pos=0.4] {$[\hat{x},\hat{y},\hat{z}]$}(A.180);
   \node[text width=6.5cm] at (6.5,6.6) {MATLAB/ROS Interface @ 100Hz};
   \node[text width=3cm] at (4.2,1.5) {$[m_1,m_2,m_3,m_4]$};

\end{tikzpicture}}}
\caption{\label{fig:mat_interface}MATLAB Interface diagram to implement the LQT controller.}

\end{figure}

The MATLAB interface was the heart of the LQT algorithm implementation. The ROS and MATLAB communication was done using the Robotics System Toolbox\textsuperscript{\tiny TM}, in particular the Simulink blocks that implement ROS publishers and subscribers.
All of the off-board section of the controller was executed within
this interface, as well as the state vector reconstruction and flight
data retrieval for further analysis. \autoref{fig:mat_interface} exhibits the structure of the interface developed.

Now a breakdown of each block:
\begin{itemize}
\item \textbf{VICON/UWB:} block that gives the position estimations of the
drone in a previously defined inertial frame.
\item \textbf{ROS Subscribers:} these subscribers serve as bridge between
MATLAB, the positioning system and the Crazyflie angle data. Four subscribers are included: one for the VICON position estimations, one for the UWB system, another one for the IMU data of the Crazyflie 2.0 and the last one to retrieve the PWM commands sent to the motors.
\item \textbf{State observer:} takes data from the position, Euler angles
and angular velocities of the quadcopter to calculate an estimation
of the state vector $\hat{\boldsymbol{x}}_{pos}$.
This is done through the Kalman Filter developed in \Cref{subsec:Kalman-Filter-for}.
\item \textbf{LQT algorithm:} implements the off-board section of the LQT algorithm as suggested in \autoref{fig:Implementation-diagram}.
\item \textbf{ROS Publisher:} takes the input vector $\boldsymbol{u}_{pos}$
and decomposes in its four components, in order to send it to the
Crazyflie via radio link in the format $\left[u_{1},u_{2},u_{3},u_{4}\right]$.
\item \textbf{Crazyflie 2.0:} the drone takes the output message of the
ROS publisher and adds it to the on-board control section of the LQT
algorithm, as suggested once again in \autoref{fig:Implementation-diagram}.
The platform continuously outputs the IMU readings as well as the
total 16-bit PWM signal sent to the motors.
\end{itemize}

\subsection{\label{subsec:Experimental-Results}Experimental Results}

\begin{figure}[H]
\centering
\makebox[0pt]{
\includegraphics[width=1.2\textwidth]{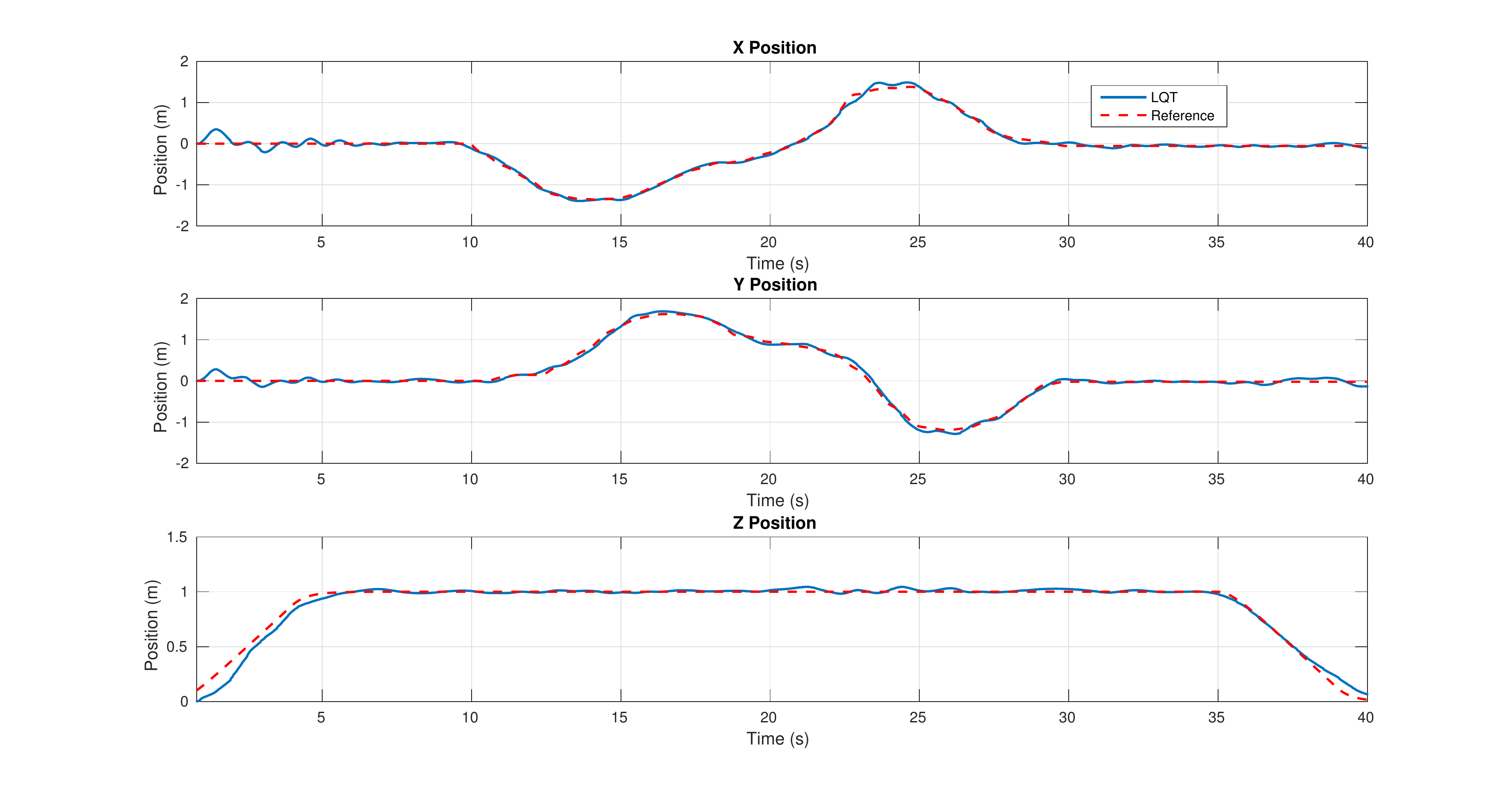}}
\caption{\label{fig:traj1}Position plots for Trajectory\#1.}
\end{figure}

\begin{figure}[H]
\centering
\makebox[0pt]{
\subfloat[\label{fig:Standard-View-ver-1-1-1}Standard view.]{\includegraphics[width=0.55\textwidth]{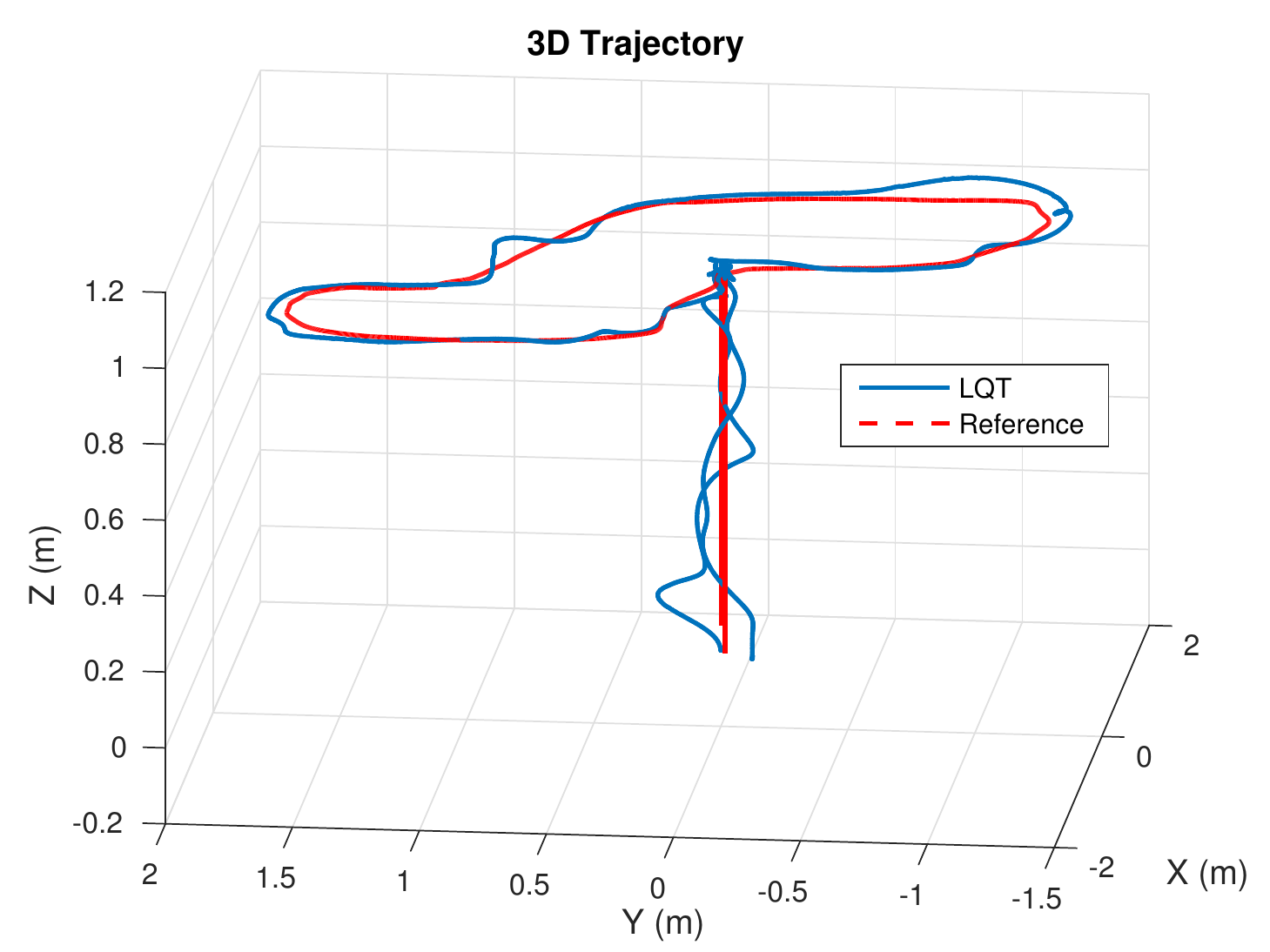}}
\subfloat[\label{fig:Top-View-ver-1-1-1}XY Plane.]{\includegraphics[width=0.55\textwidth]{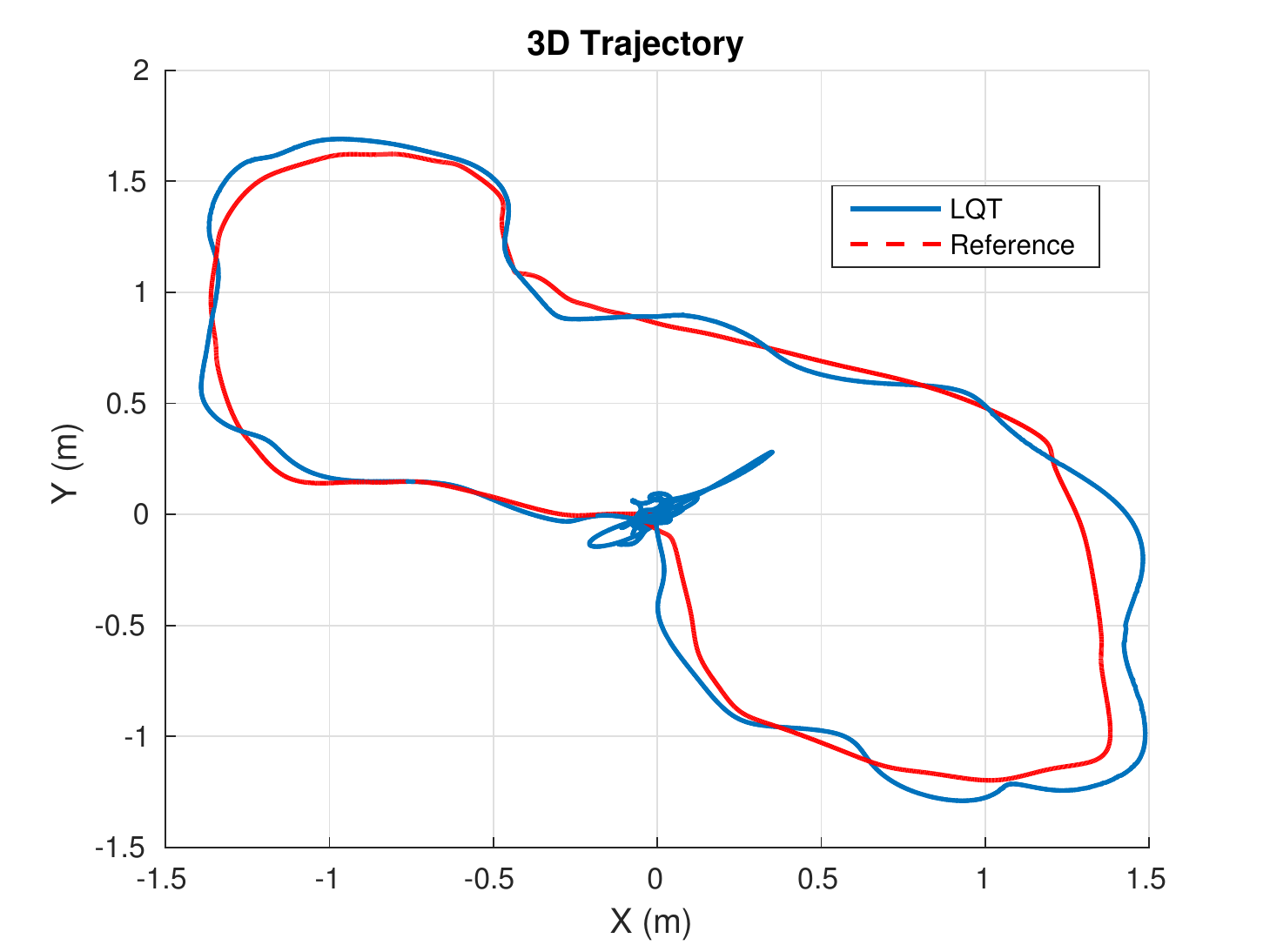}}}
\caption{\label{fig:3D-Vertical-Trajectory-1-1-1}3D Trajectory\#1.}
\end{figure}

A series of trajectories created using the GUI developed in \Cref{subsec:Trajectory-generation} were tested with the LQT algorithm.
The RMS error between the desired trajectory and the actual trajectory
followed by the drone was used as a measure of the controller's performance
as in past works in UAV control such as \cite{key-25,key-28} . Another
performance index was defined, from the basis that a 10 cm error margin
from the desired position is ideal, then it was relevant to calculate
for each one of the spatial coordinates of the drone the percentage
in which each coordinate stayed within this margin of the
desired position. These performance indices will be further known
as $\xi_{x}$, $\xi_{y}$ and $\xi_{z}$.
\begin{itemize}
\item \textbf{Trajectory \#1:} a simple trajectory with fixed altitude of
1 meter was commanded for the quadcopter to follow. The time plots in \autoref{fig:traj1} present the test results, while \autoref{fig:3D-Vertical-Trajectory-1-1-1} show different 3D perspectives of the flight.
\end{itemize}

\begin{figure}[H]
\centering
\makebox[0pt]{
\includegraphics[width=1.2\textwidth]{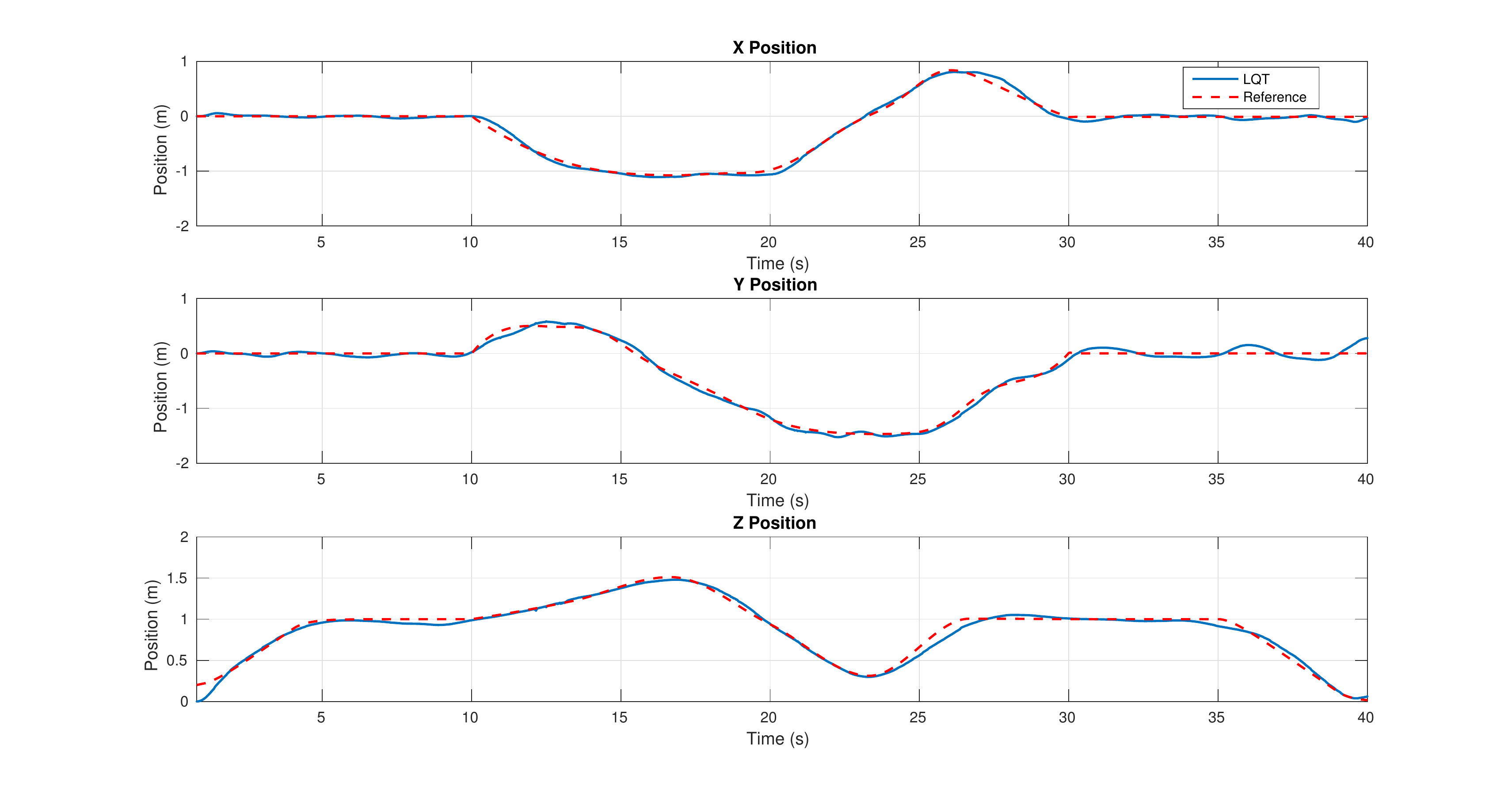}}
\caption{\label{fig:traj2}Position plots for Trajectory\#2.}
\end{figure}

\begin{figure}[H]
\centering
\makebox[0pt]{
\subfloat[\label{fig:Standard-View-ver-1-1}Standard view.]{\includegraphics[width=0.55\textwidth]{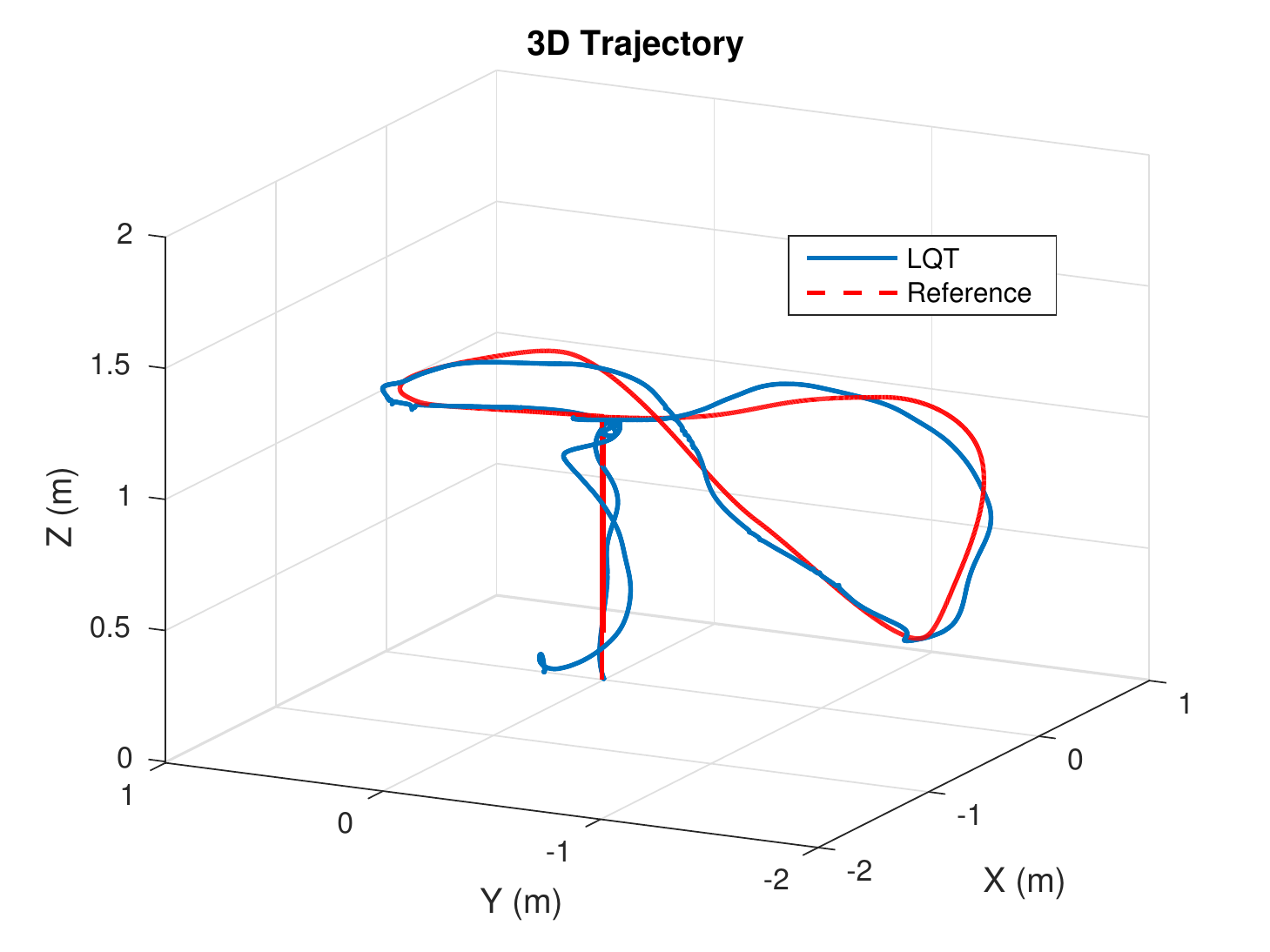}}
\subfloat[\label{fig:Top-View-ver-1-1}XY Plane.]{\includegraphics[width=0.55\textwidth]{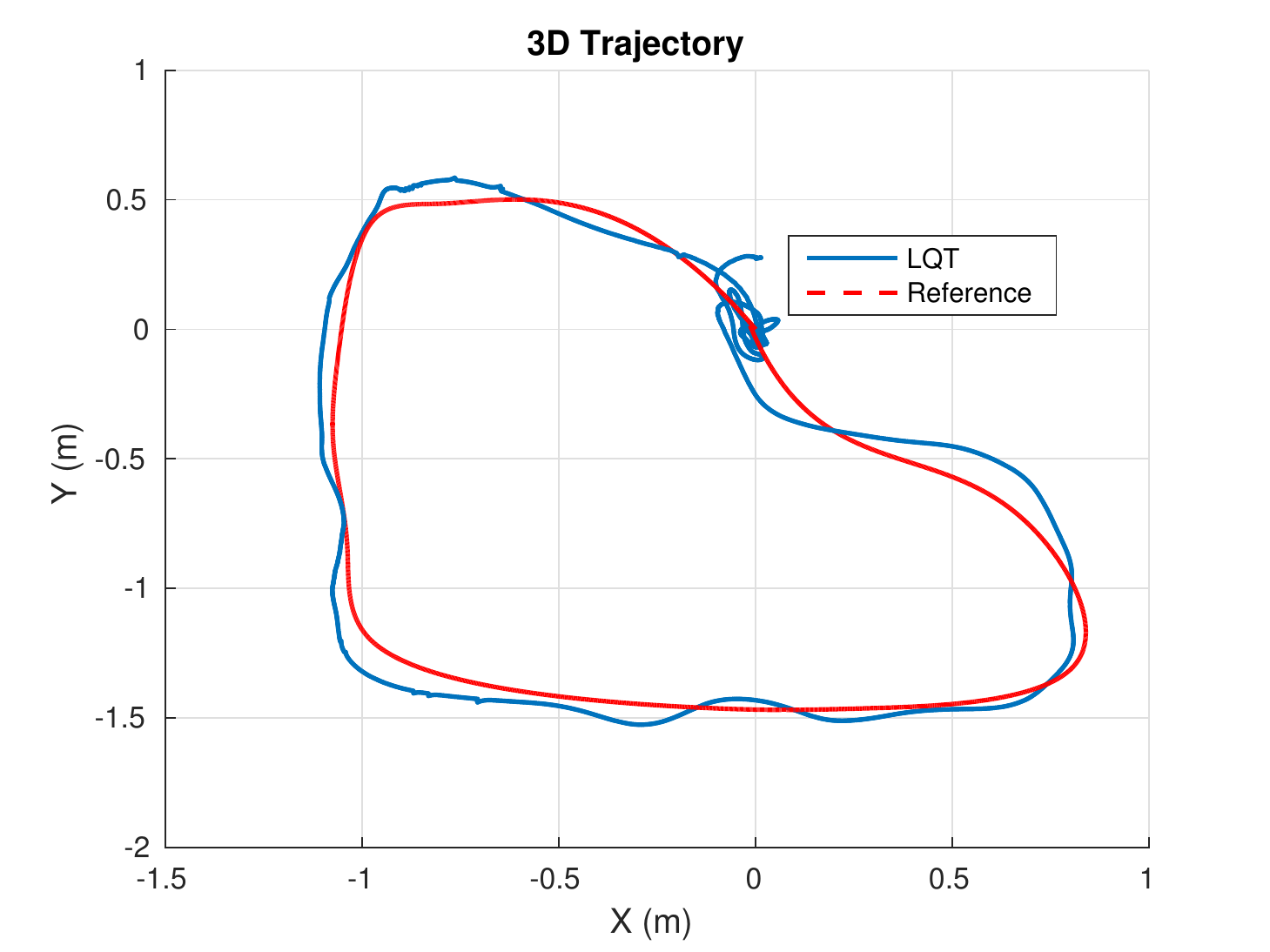}}}
\caption{\label{fig:3D-Vertical-Trajectory-1-1}3D Trajectory\#2.}
\end{figure}

For this simple trajectory the RMS errors were low, 6.2cm for the
X position, 6.3cm for the Y and 3.9cm for the altitude Z. The corresponding
performance indices were $\xi_{x}=92.65\%$, $\xi_{y}=90.72\%$ and
$\xi_{z}=95.87\%$. The tracking of these type of slow trajectories
had an outstanding level of performance.
\begin{itemize}
\item \textbf{Trajectory \#2:} for this second trajectory a varying altitude was also commanded to asses the behavior of the quadcopter while moving simultaneously in all three spatial coordinates. The flight data is displayed in \Cref{fig:traj2,fig:3D-Vertical-Trajectory-1-1}.
\end{itemize}

\begin{figure}[H]
\centering
\makebox[0pt]{
\includegraphics[width=1.2\textwidth]{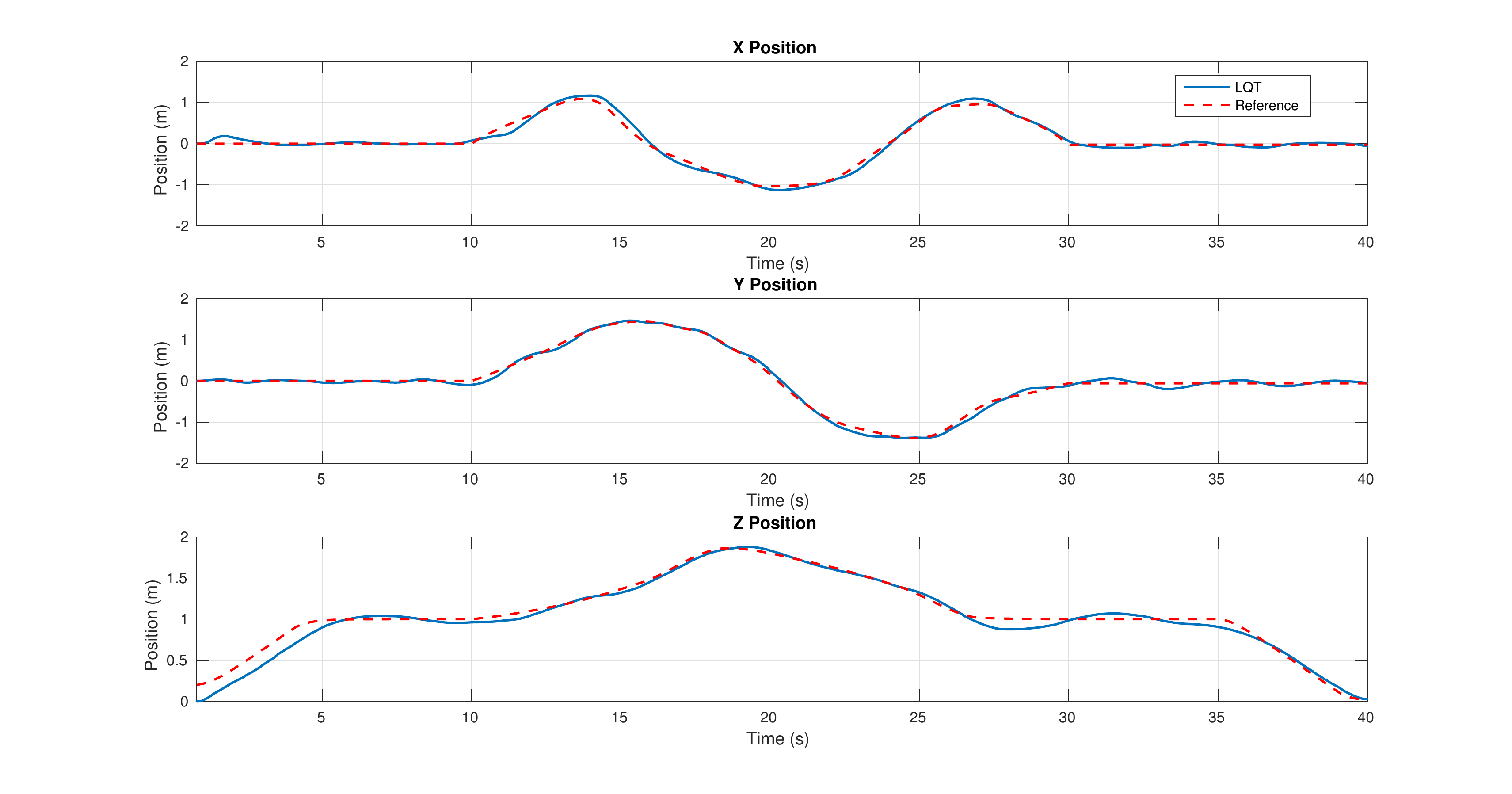}}
\caption{\label{fig:traj3}Position plots for Trajectory\#3.}
\end{figure}

\begin{figure}[H]
\centering
\makebox[0pt]{
\subfloat[\label{fig:Standard-View-ver-1}Standard view.]{\includegraphics[width=0.55\textwidth]{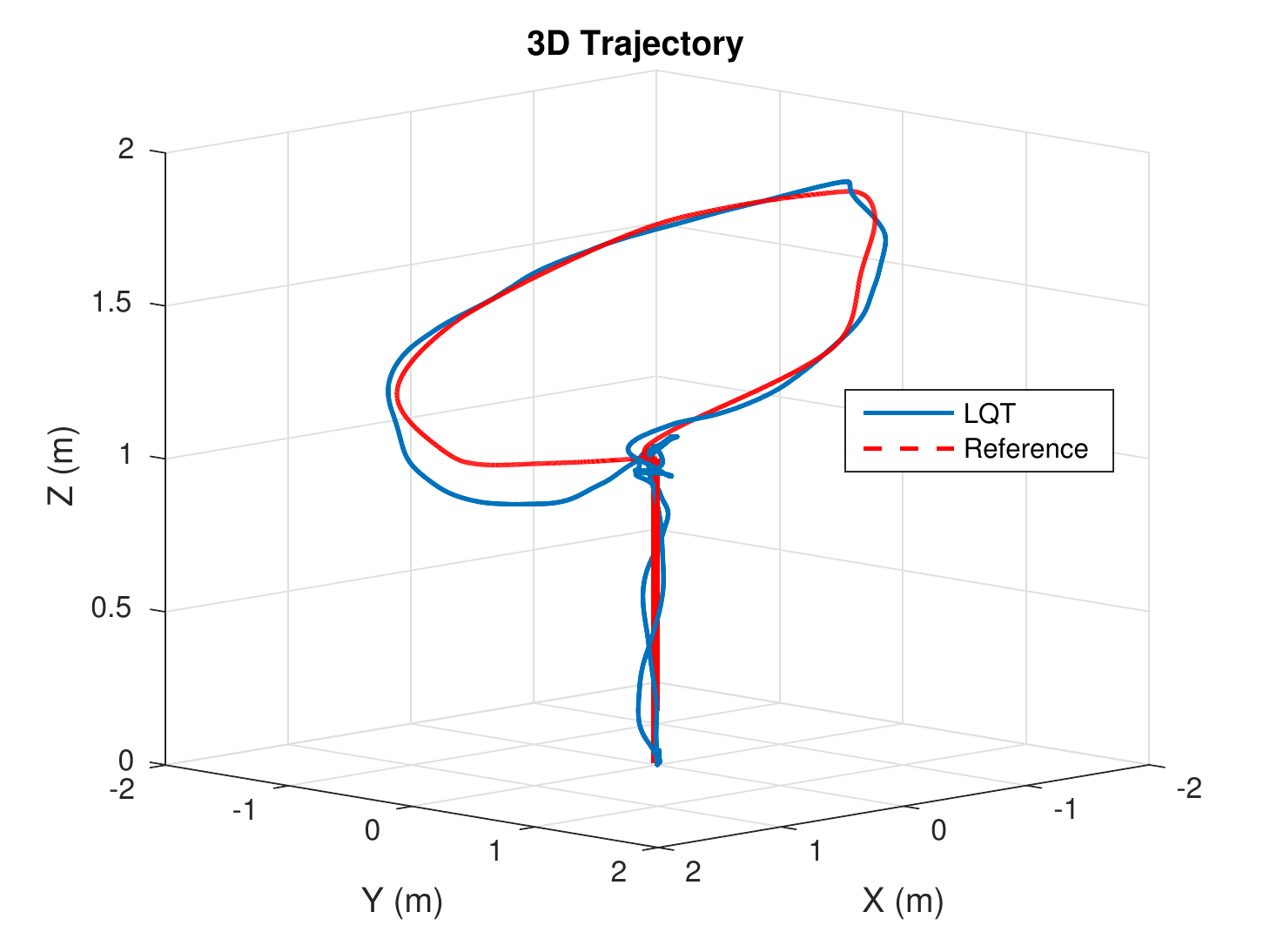}}
\subfloat[\label{fig:Top-View-ver-1}XY Plane.]{\includegraphics[width=0.55\textwidth]{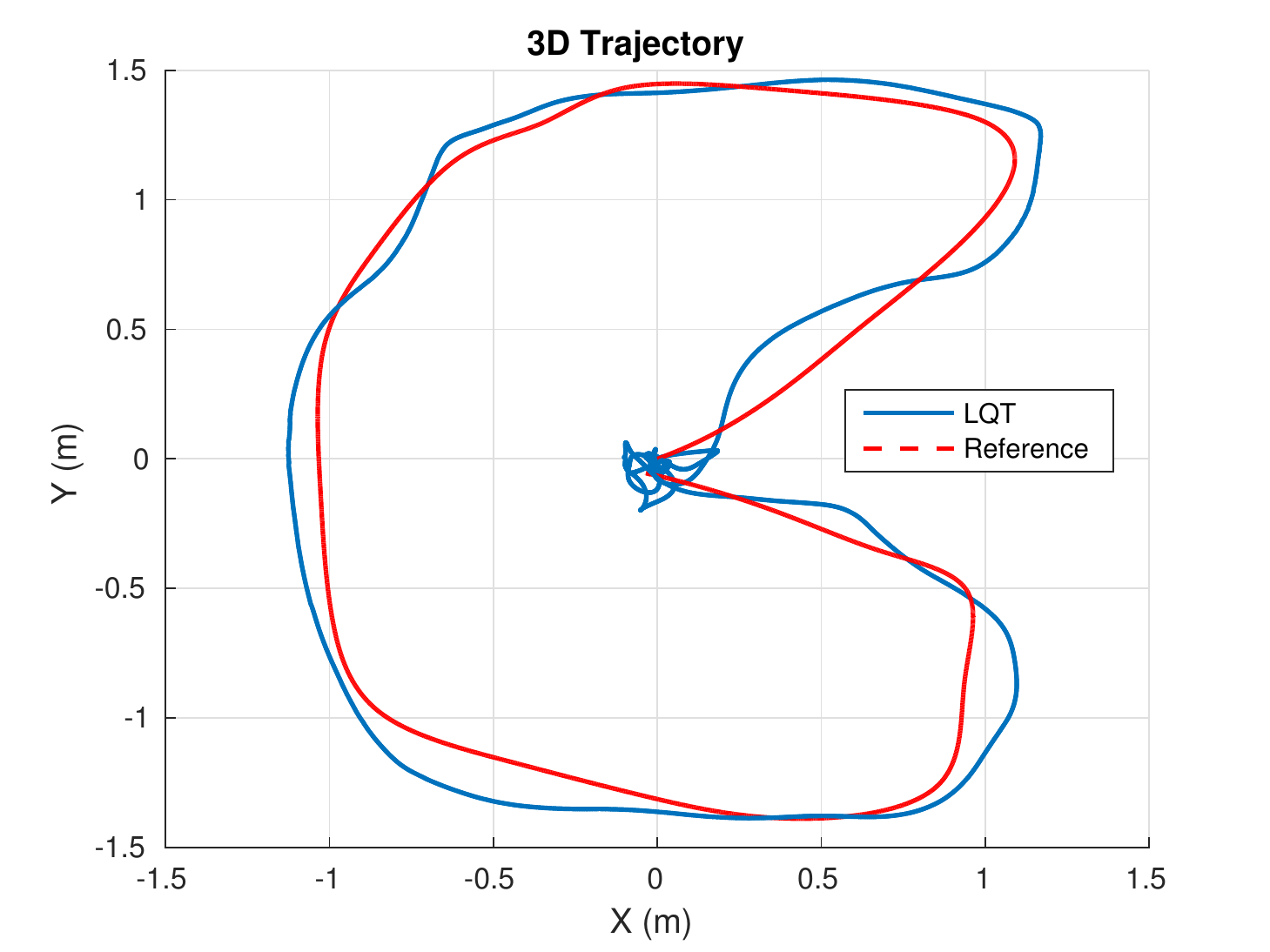}}}
\caption{\label{fig:3D-Vertical-Trajectory-1}3D Trajectory\#3.}
\end{figure}

Despise adding more difficulty to the trajectory the performance was
satisfactory, with RMS errors of 4.72cm, 7.81cm and 4.75cm respectively
for the x, y and z positions. The performance indices remained in
the high end, with $\xi_{x}=93.89\%$, $\xi_{y}=87.30\%$ and $\xi_{z}=94.63\%$.
\begin{itemize}
\item \textbf{Trajectory \#3:} a more complex trajectory is showcased in \Cref{fig:traj3,fig:3D-Vertical-Trajectory-1}
\end{itemize}

\begin{figure}[H]
\centering
\makebox[0pt]{
\includegraphics[width=1.2\textwidth]{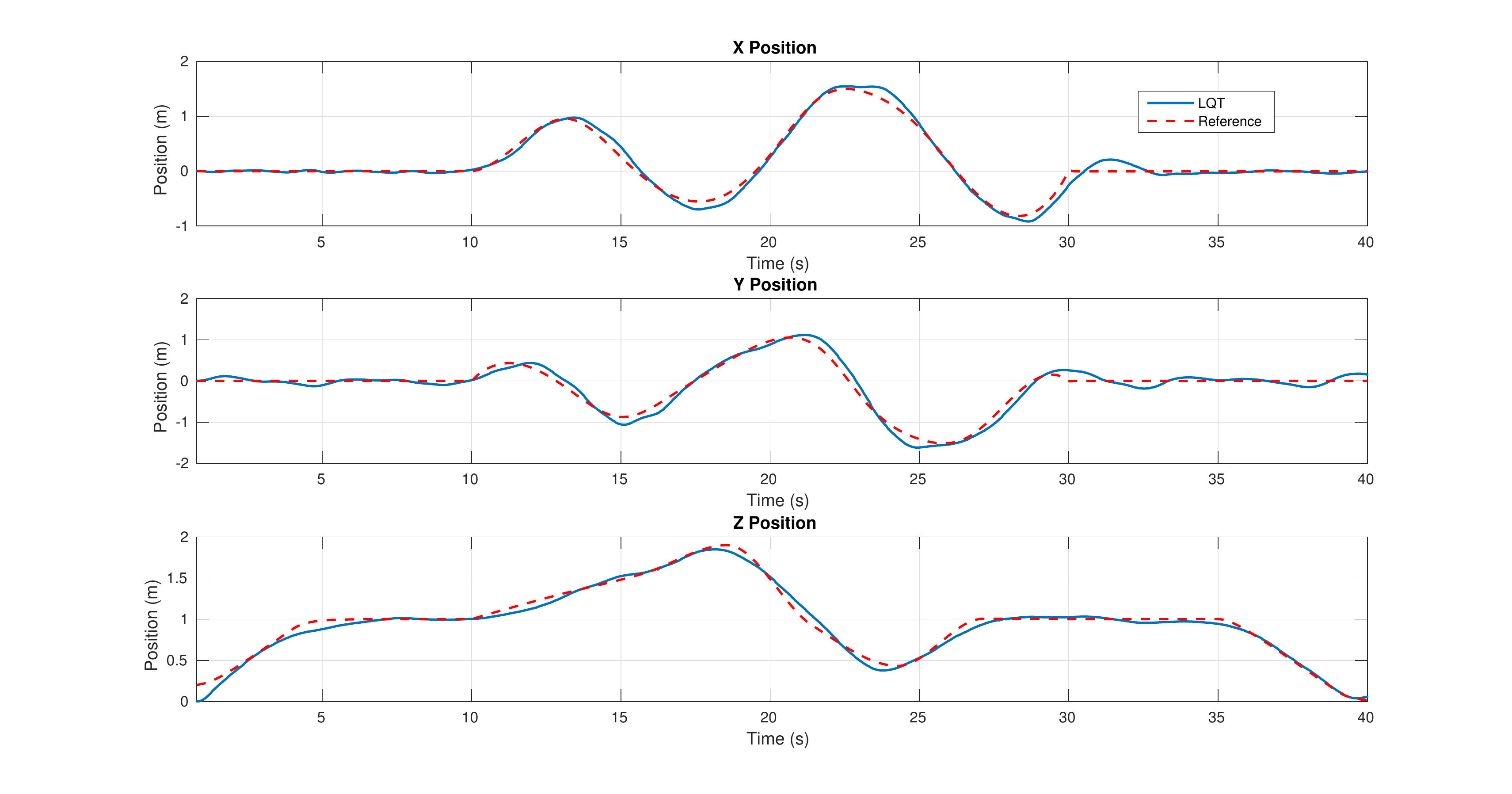}}
\caption{\label{fig:traj4}Position plots for Trajectory\#4.}
\end{figure}

\begin{figure}[H]
\centering
\makebox[0pt]{
\subfloat[\label{fig:Standard-View-ver-1-2}Standard view.]{\includegraphics[width=0.55\textwidth]{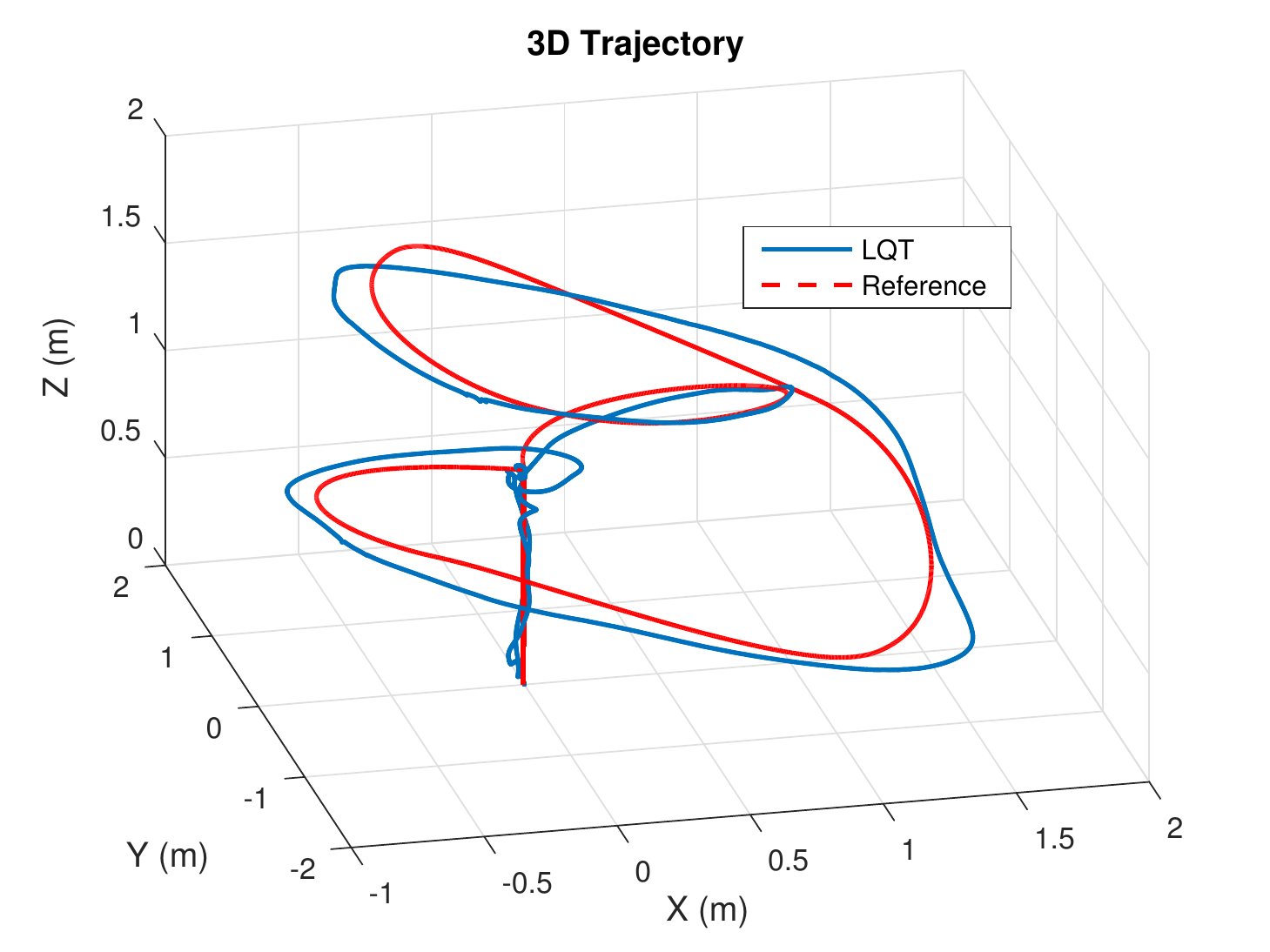}}
\subfloat[\label{fig:Top-View-ver-1-2}XY Plane.]{\includegraphics[width=0.55\textwidth]{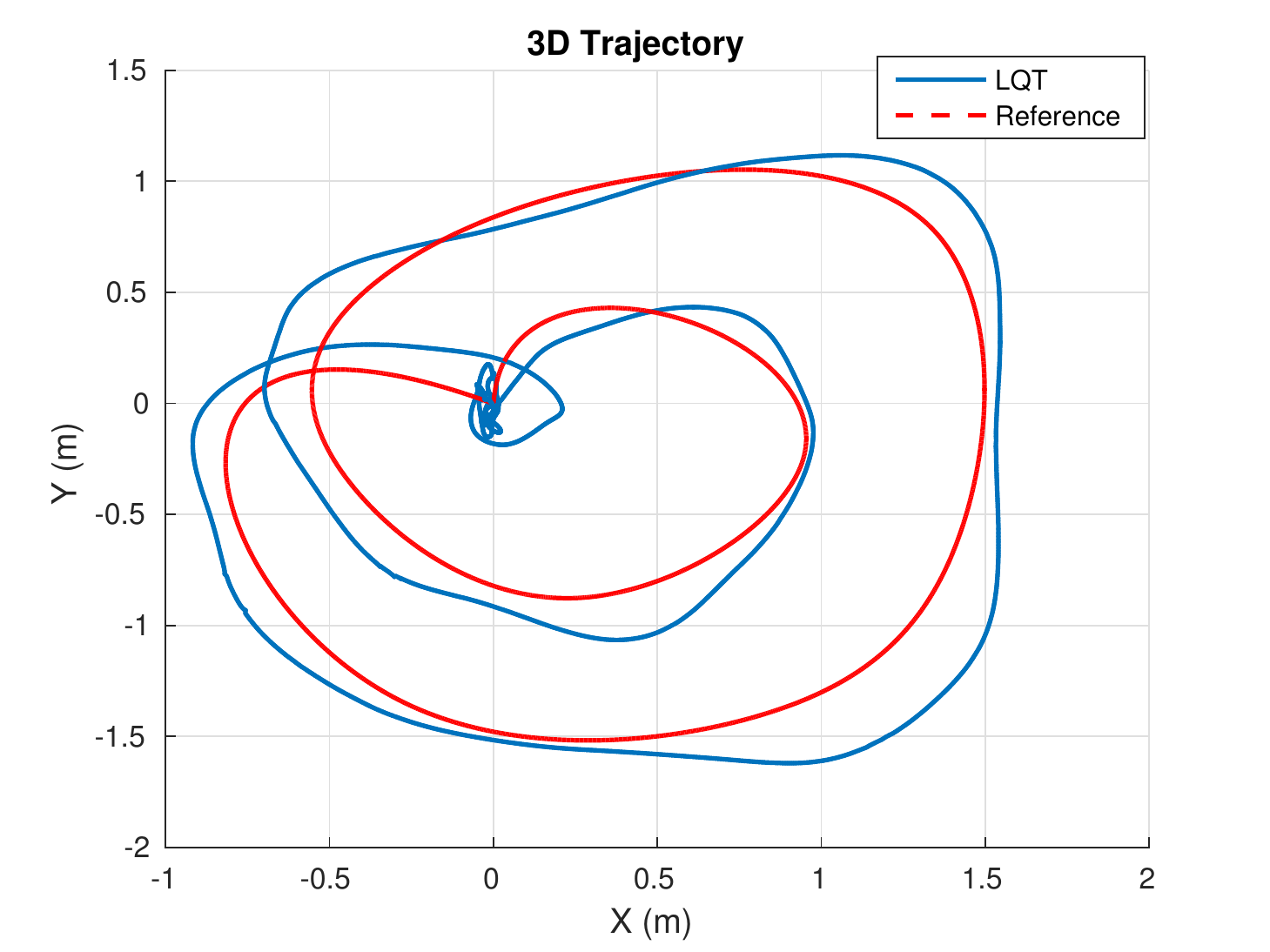}}}
\caption{\label{fig:3D-Vertical-Trajectory-1-2}3D Trajectory\#4.}
\end{figure}

Even though the controller kept a good trajectory tracking, it is
evident how the performance starts degrading when adding more complexity
to the trajectories. In this case the RMS erros were of 7.51cm, 5.97cm
and 7.34cm for the X, Y and Z positions. The performance indices were
lower than in the two previous trajectories, with $\xi_{x}=83.47\%$,
$\xi_{y}=88.49\%$ and $\xi_{z}=85.12\%$.
\begin{itemize}
\item \textbf{Trajectory \#4:} a complex spiral trajectory in 3D was commanded. The experimental results can be appreciated in \Cref{fig:traj4,fig:3D-Vertical-Trajectory-1-2}.
\end{itemize}

The RMS error incurred when following this trajectory was of 7.93cm
in X, 11.79cm in Y and 5.44cm in Z. The performance indices were $\xi_{x}=79.55\%$,
$\xi_{y}=58.19\%$ and $\xi_{z}=93.12\%$. Even more clear than with
Trajectory\#3, this spiral trajectory shows a performance degradation
when the trajectories demand faster movements, closer curves or harder
brakes. For example, at the 30 second mark in the X position a sudden
brake was required but the controller was not as fast which caused
an overshoot. As seen in the simulation phase these type of overshoots
are caused mainly by the integral action, but otherwise, if lowered,
in practice the position regulation would worsen. 

\section{Controller Comparisons}

In this section the step response of the control system is used as
a measure to compare the performance of the two controllers synthesized
in this project, as well as comparing the simulated model
and the experimental data to determine how close the simulation predicted
the actual test results. Then, to compare trajectory tracking capabilities,
the two controllers were tested on identical trials to follow sinusoidal
waves.

\subsection{Simulation vs Experimental - PID}

Comparing the step response of the simulation model developed in subsection
\ref{sec:Cascaded-PID-Position} and the one obtained during a real
flight of the drone was the chosen method to verify the accuracy of
the mathematical model of the quadcopter. Three different flights
were executed, sending individual steps in the direction of X, Y and
Z.
~\\

\begin{figure}[H]
\centering
\makebox[0pt]{
\includegraphics[width=1.2\textwidth]{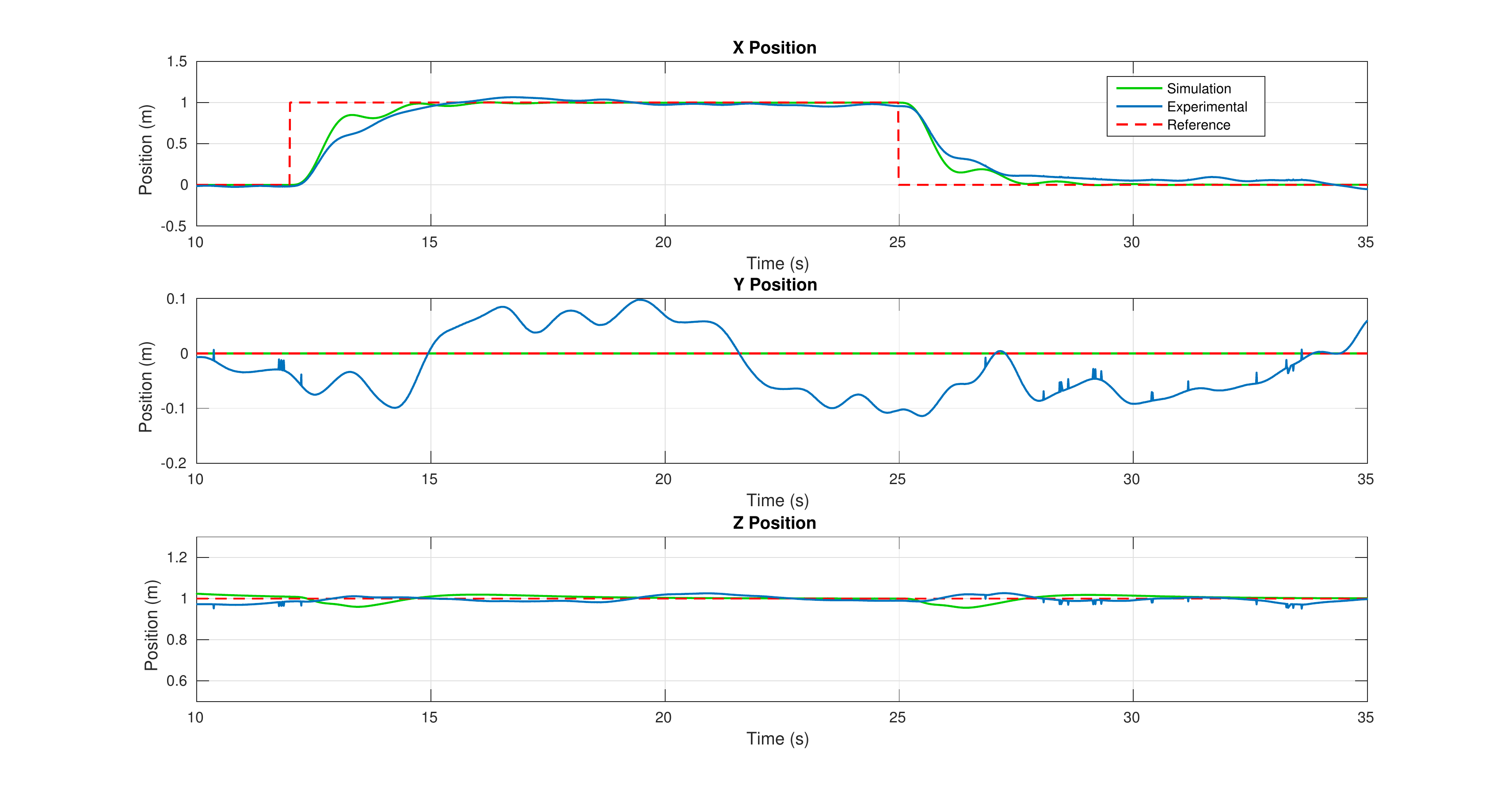}}
\caption{\label{fig:pidx}Trajectory using the PID controller to follow a Step in the X position.}
\end{figure}

\begin{figure}[H]
\centering
\makebox[0pt]{
\includegraphics[width=1.2\textwidth]{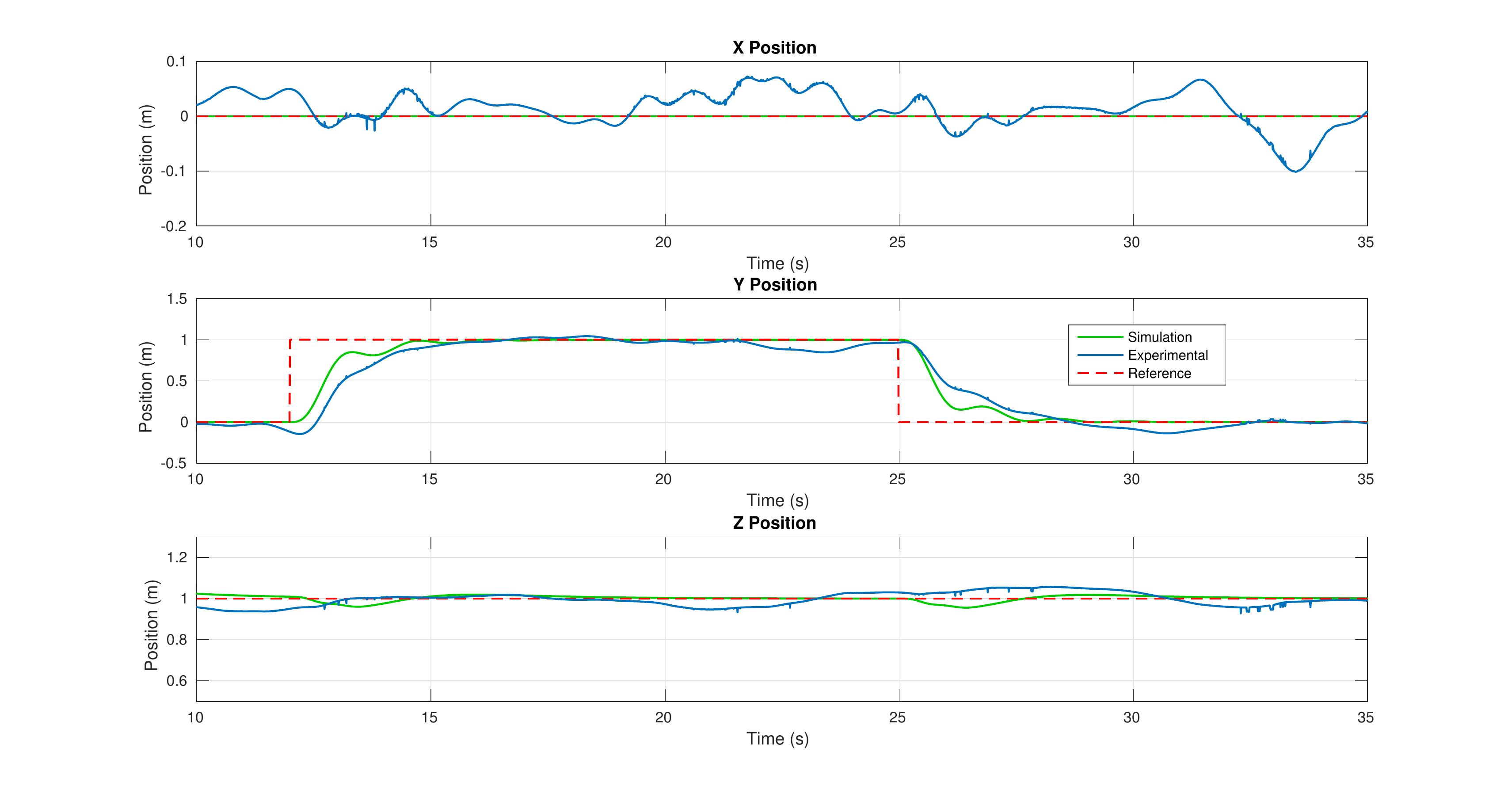}}
\caption{\label{fig:pidy}Trajectory using the PID controller to follow a Step in the Y position.}
\end{figure}

\begin{itemize}
\item \textbf{Step command in the X direction}
\end{itemize}
A step command of 1 meter was commanded in the X coordinate, while maintaining a 1 meter altitude and a zero position in the Y coordinate. The experimental results are presented in \autoref{fig:pidx}.
The simulation and experimental step responses had almost identical
response time of around 3 seconds with almost to no overshoot. The
simulation response shows less damping in the response. The clear
difference is in the Y response, the simulation predicted a perfect
zero which is clearly impossible in practice, nonetheless the drone
maintained a 10 centimeters error margin within the initial Y position.
The movement had little impact in the altitude.
\begin{itemize}
\item \textbf{Step command in the Y direction}
\end{itemize}
The test was conducted in a similar fashion as the previous one, but this time sending the appropriate command to the Y coordinate. Simulation and experimental results were plotted together as show in \autoref{fig:pidy}.

Similar to the X step response, the dynamics for the Y direction matches
up to some extent those predicted by the simulation.

\begin{figure}[H]
\centering
\makebox[0pt]{
\includegraphics[width=1.2\textwidth]{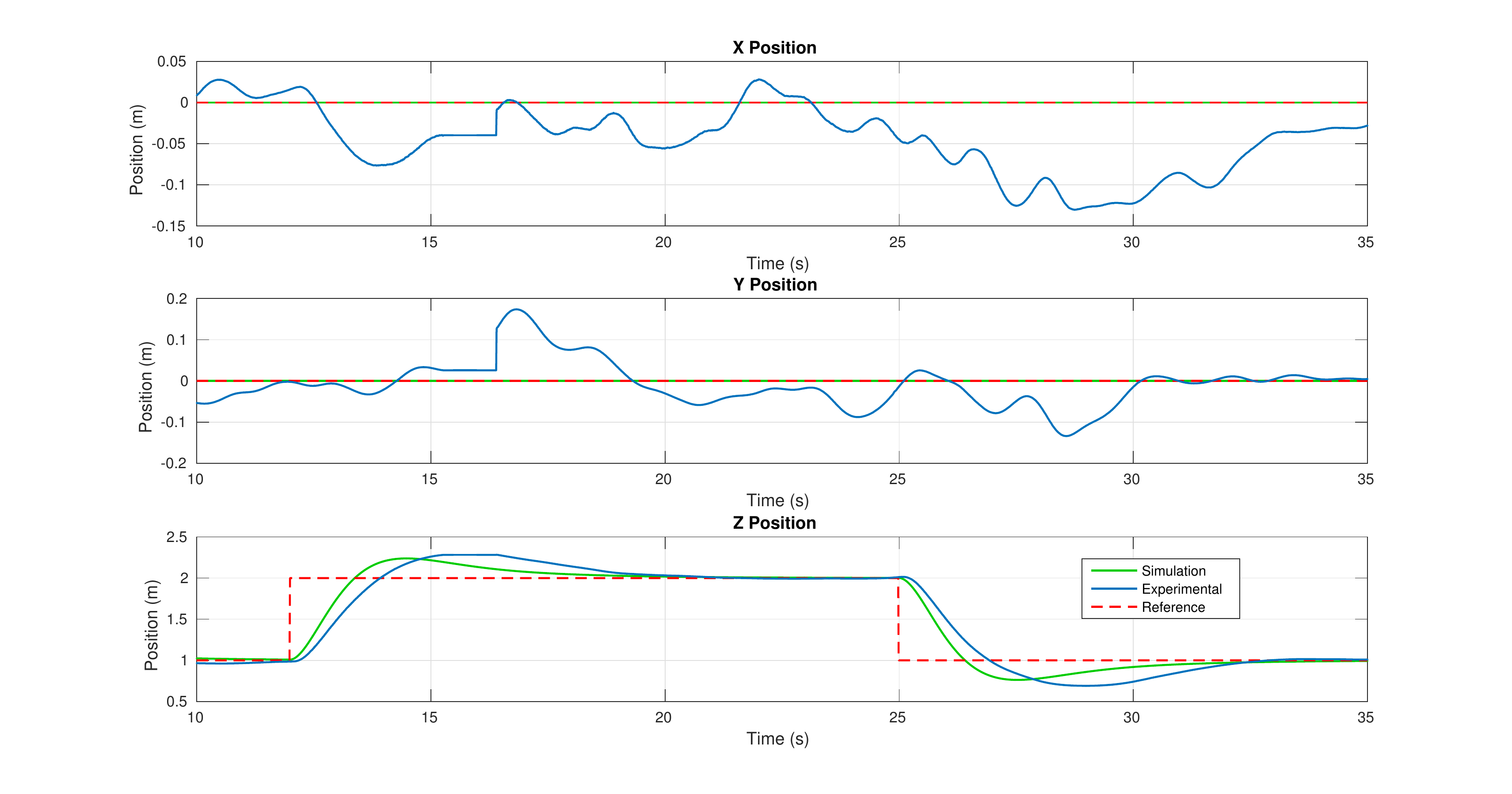}}
\caption{\label{fig:pidz}Trajectory using the PID controller to follow a Step in the Z position.}
\end{figure}

\begin{itemize}
\item \textbf{Step command in the Z direction}
\end{itemize}
Finally a step command of 1 meter in altitude was tested, as seen in \autoref{fig:pidz}.

The altitude dynamics were slower in practice than in simulation and with a the experimental data exposes a more pronounced overshoot.

\subsection{\label{subsec:Simulation-vs-Experimental}Simulation vs Experimental
- LQT}

\begin{figure}[H]
\centering
\makebox[0pt]{
\includegraphics[width=1.2\textwidth]{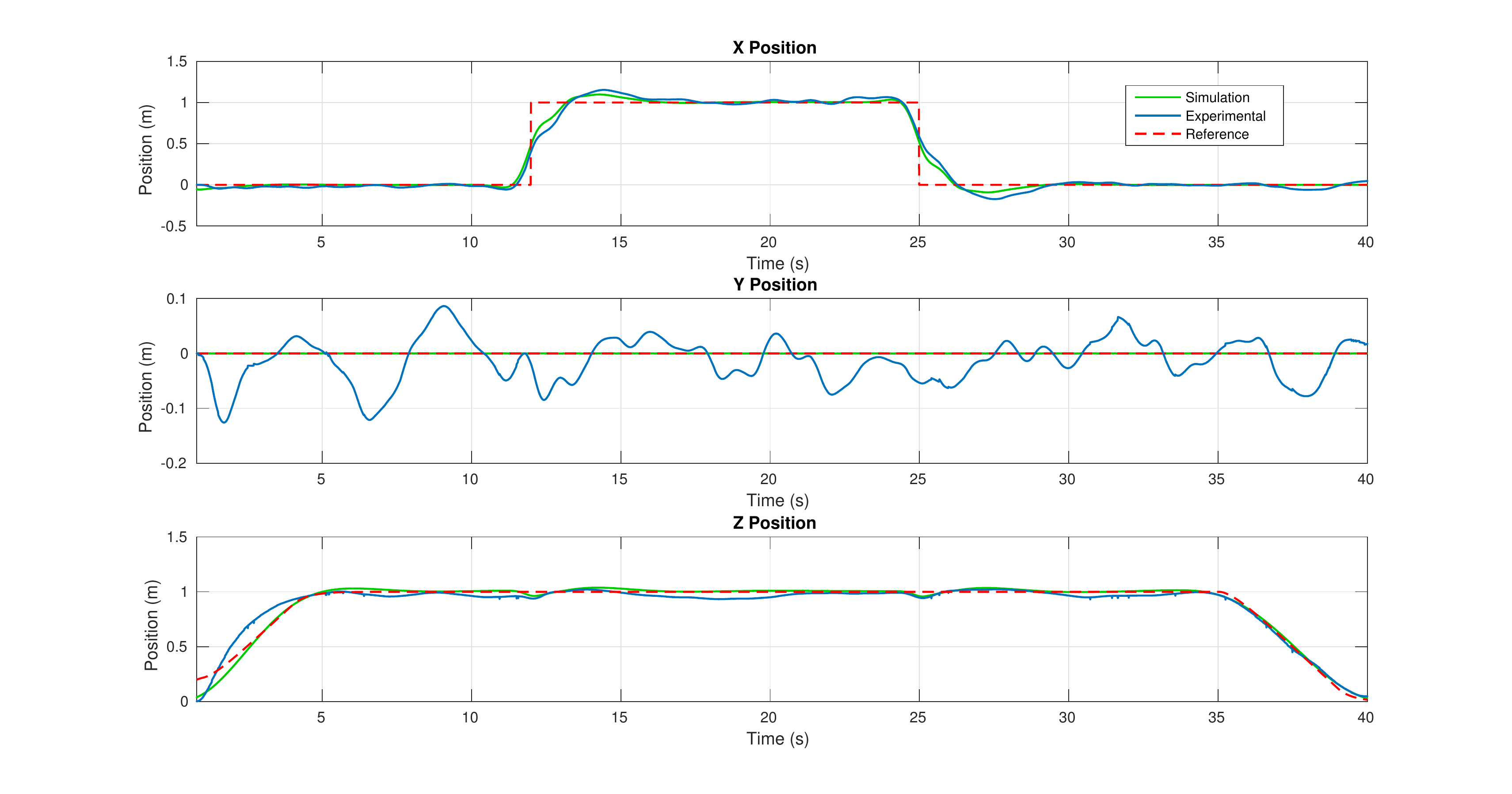}}
\caption{\label{fig:lqtx}Trajectory using the LQT controller to follow a Step in the X position.}
\end{figure}

Similarly, with the simulation model developed in \Cref{sec:Linear-Quadratic-Tracker-(LQT)},
the following plots show the step response comparisons between
the simulation and the experimental data.
\begin{itemize}
\item \textbf{Step command in the X direction}
\end{itemize}
\autoref{fig:lqtx} exhibits a comparative plot between the simulation and the experimental data.

\begin{figure}[H]
\centering
\makebox[0pt]{
\includegraphics[width=1.2\textwidth]{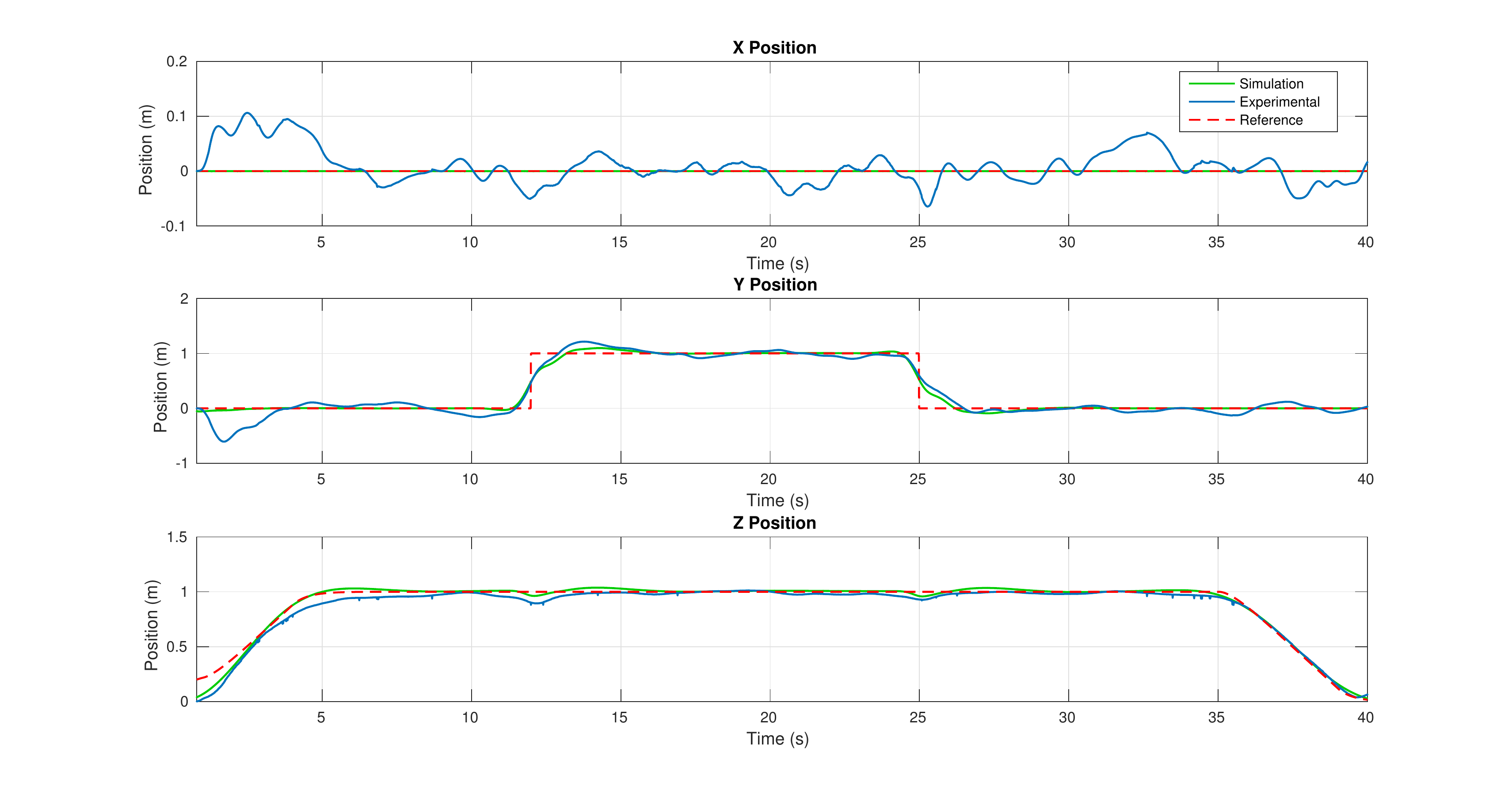}}
\caption{\label{fig:lqty}Trajectory using the LQT controller to follow a Step in the Y position.}
\end{figure}

The simulation scenario in the X and Z positions matches the results obtained
during the experiment. The main difference is the perturbation
around the zero position of the Y coordinate, the simulation predicted an almost perfect hold of this position while in reality the quadcopter oscillated in a 10 cm error margin. 
\begin{itemize}
\item \textbf{Step command in the Y direction}
\end{itemize}
Doing a similar test, but sending a command to the Y coordinate gave the results seen in \autoref{fig:lqty}.

\begin{figure}[H]
\centering
\makebox[0pt]{
\includegraphics[width=1.2\textwidth]{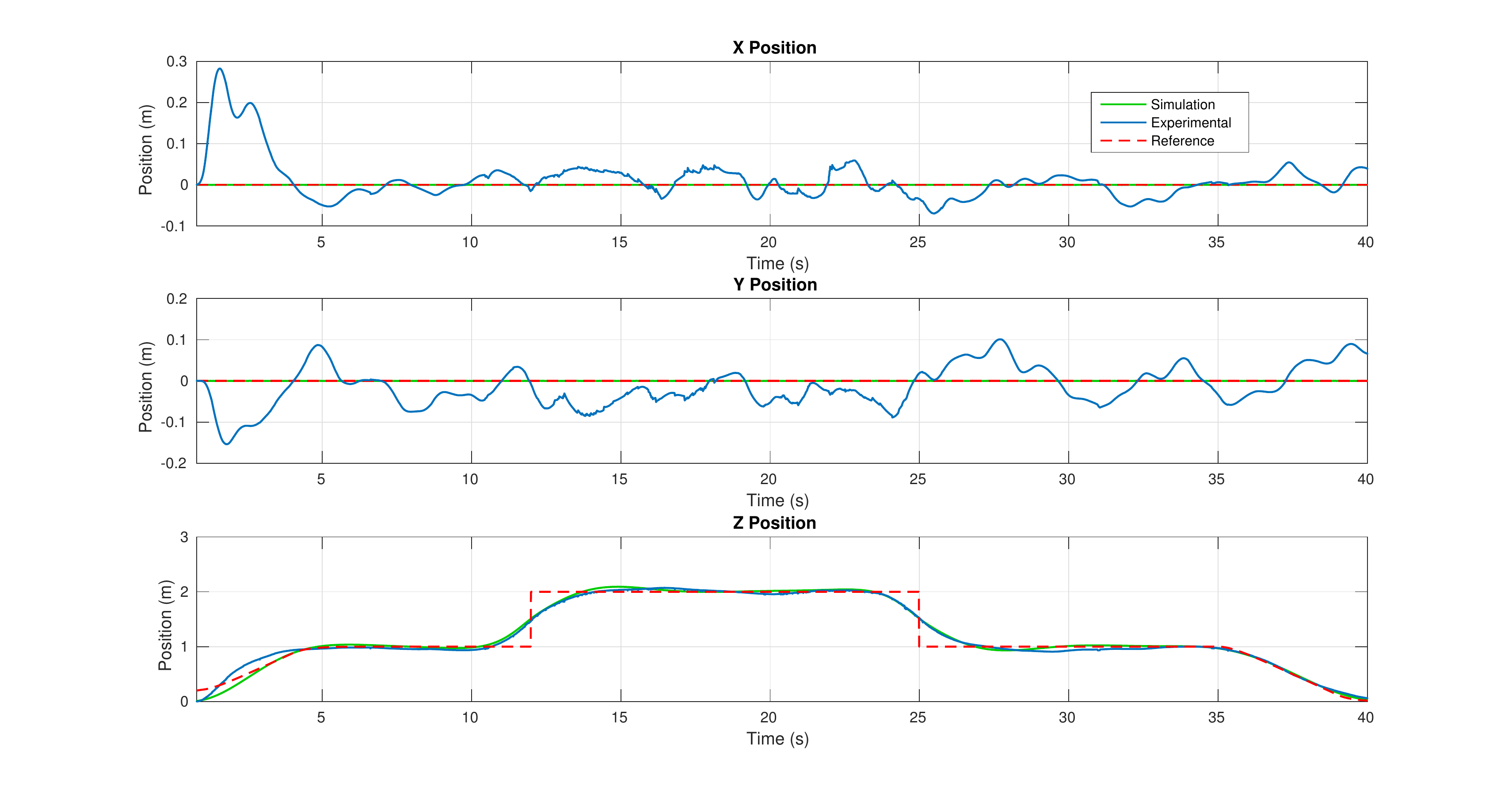}}
\caption{\label{fig:lqtz}Trajectory using the LQT controller to follow a Step in the Z position.}
\end{figure}

The dynamics were similar to those of the X step, the simulation proved
once again to be accurate in predicting the test results.
\begin{itemize}
\item \textbf{Step command in the Z direction}
\end{itemize}
The behavior in \autoref{fig:lqtz} corresponds with a 1 meter command in the Z coordinate.

The altitude step simulation matched almost perfectly the one obtained
in practice, as the comparison demonstrate.

~\\
After reviewing this data a few conclusions can be drawn. On the first
hand, the PID simulation is precise up to some extent at predicting
the response of the system when applying step commands, although it
has some considerable differences on the altitude estimation. On the
other hand the LQT simulation proved to be more precise, giving
accurate information of how the system will behave in the real scenario.
Both of these simulations serve the common purpose of validating the
mathematical model of the quadcopter.

~\\
In the LQT simulations there is an important remark to be made before
jumping to an early conclusion about the accuracy of the mathematical
model. In the X-Y step responses there is an overshoot introduced
mainly by the high integral gains $\boldsymbol{K}_{i}^{ang}$ and
$\boldsymbol{K}_{i}^{pos}$, if those gains were to be lowered at
least in simulation the overshoot would be reduced giving more pleasant
results, however various experimental results suggest that lowering
these gains will worsen the overall performance of the controller
in practice. The conclusion
is that the high integral gain is necessary to compensate all the
model deficiencies and future efforts should be made in refining
the model in order to lower the integral gains without loosing performance. 

\subsection{Controller Performance: PID vs LQT}

\begin{figure}[H]
\centering
\makebox[0pt]{
\includegraphics[width=1.2\textwidth]{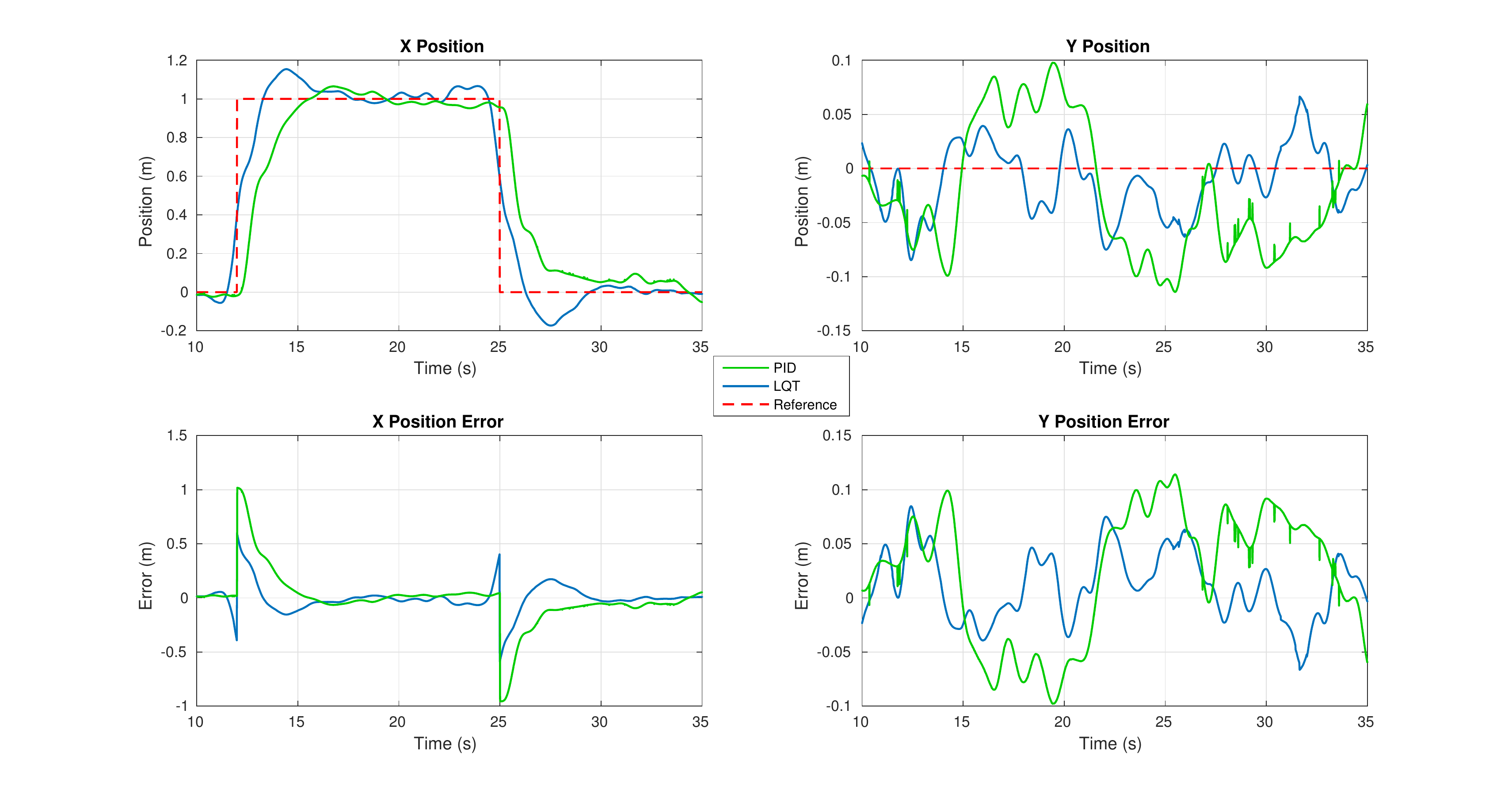}}
\caption{\label{fig:compx}X-Y Position and error comparison when following a unit step in the
X position.}
\end{figure}

One of the main goals of the project was to objectively compare the
performance of the two controllers synthesized, in terms of tracking
accuracy and command effort. First the step responses were compared
and then a sinusoidal trajectory showed the tracking capabilities of
each controller. For each controller, the results presented correspond to the flight with the best performance after performing a series of trials under the same conditions (gains, weights, etc.). The same performance indices as in \Cref{subsec:Experimental-Results} were used. To quantify the control
effort, the two-norm is used as in \cite{key-29}, then the control
effort is defined as:
\begin{equation}
U_{i}=\sum_{k=0}^{k=k_{f}}u_{i}^{2}\left[k\right]
\end{equation}
where $u_{i}\left[k\right]$ is the PWM signal ranging from 0-65536
sent to the i-th motor, and $U_{i}$ is the associated control effort.
Also, the ratio percentage $\%\left(\frac{U_{LQT}-U_{PID}}{U_{PID}}\right)$
is used to quantify the increase or decrease percentage in control
effort of the LQT with respect to the PID.
~\\
\begin{itemize}
\item \textbf{Step command in the X direction}
\end{itemize}
\autoref{fig:compx} exhibits the experimental results for both controllers when commanded to follow a 1 meter step in the X coordinate.

The error terms for the LQT controller are lower, at the expenditure
of a higher overshoot than the PID. By moving before the step command,
the LQT controller manages to reduce an otherwise big error of 1 meter.
\autoref{tab:tabcompx} summarizes the performance for both controllers.

\renewcommand{\arraystretch}{1.5}

\begin{table}[H]
\begin{centering}
\begin{tabular}{|c|c|c|c|c|}
\cline{2-5} 
\multicolumn{1}{c|}{} & $\textrm{RMS}\left(e_{x}\right)\,\textrm{[cm]}$ & $\textrm{RMS}\left(e_{y}\right)\,\textrm{[cm]}$ & $\%\xi_{x}$ & $\%\xi_{y}$\tabularnewline
\hline 
LQT & $12.58$ & $3.37$ & $76.53$ & $100$\tabularnewline
\hline 
PID & $24.63$ & $6.41$ & $74.98$ & $95.08$\tabularnewline
\hline 
\end{tabular}
\par\end{centering}
\caption{\label{tab:tabcompx}Error comparison when following a unit step in X position.}

\end{table}
The RMS error values for the X-Y position with the LQT controller
were almost half of those obtained with the PID. The performance indices
were slightly better with the LQT controller.
~\\

\begin{figure}[H]
\centering
\makebox[0pt]{
\includegraphics[width=1.2\textwidth]{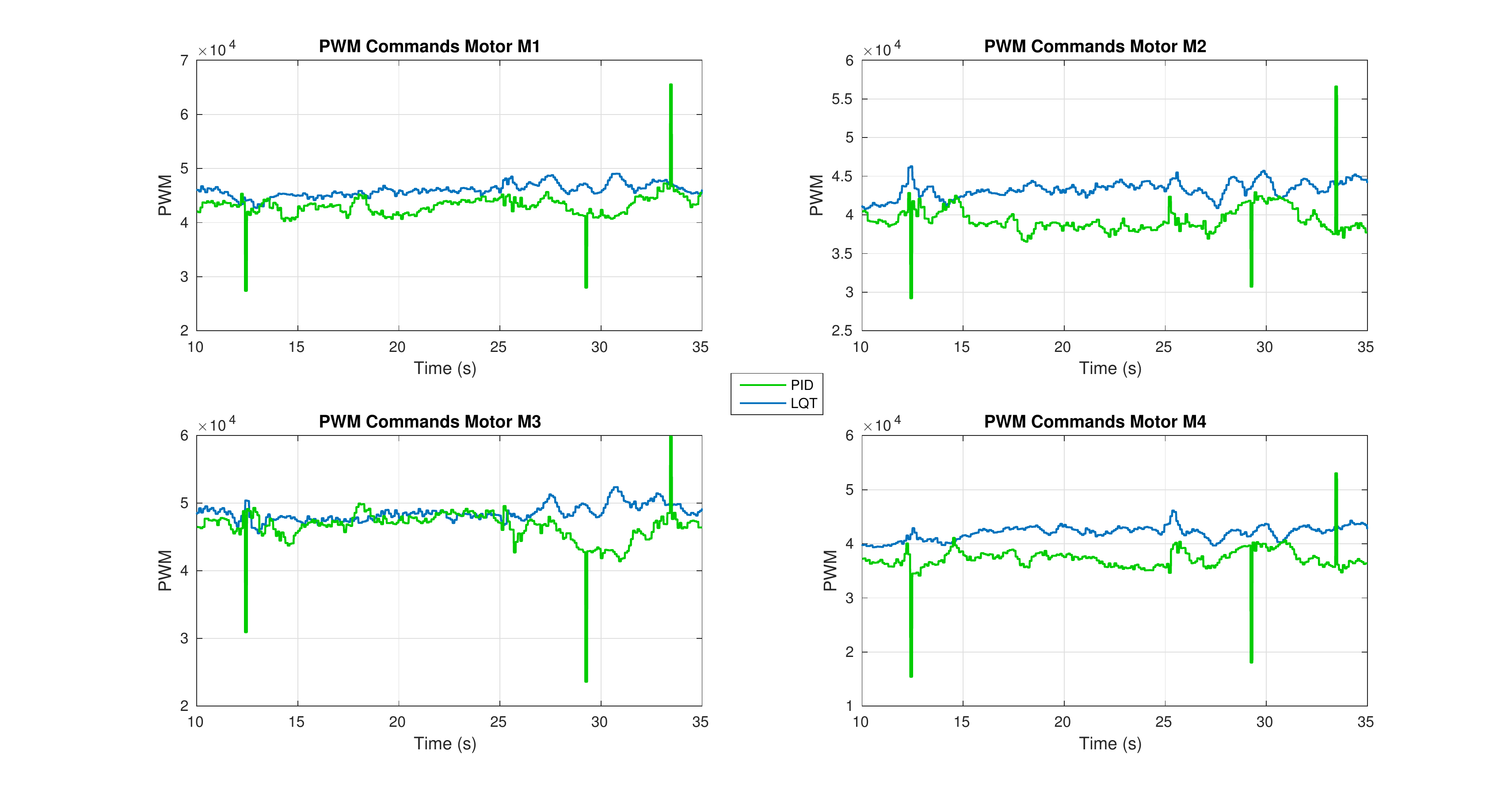}}
\caption{\label{fig:motor_x}Motor commands comparison when following a unit step in the X position.}
\end{figure}

The motor commands data for both trials were registered and compared as shown in \autoref{fig:motor_x}.

The control effort is overall greater with the LQT controller, but
the PID has control effort discontinuities when the step command goes
into action. Around the 32 second mark the motor M1 reached saturation
using the PID controller. Furthermore, \autoref{tab:tabcompxmotor} quantifies the motor's command effort in both cases.

\begin{figure}[H]
\centering
\makebox[0pt]{
\includegraphics[width=1.2\textwidth]{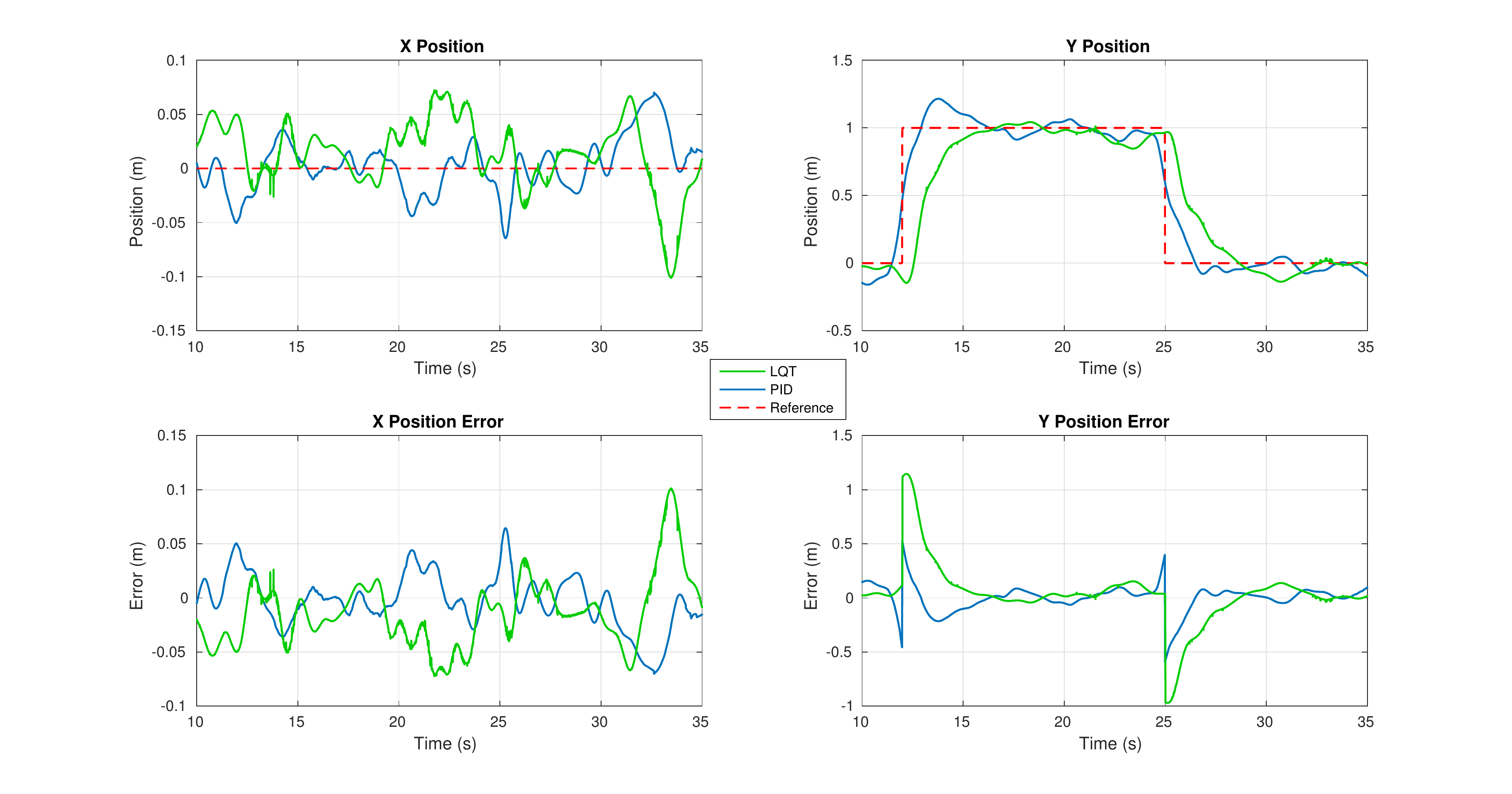}}
\caption{\label{fig:compy}X-Y Position and error comparison when following a unit step in the
Y position.}
\end{figure}

\begin{table}[H]
\begin{centering}
\begin{tabular}{|c|c|c|c|c|}
\cline{2-5} 
\multicolumn{1}{c|}{} & $U_{1}\textrm{[\ensuremath{\times10^{12}}]}$ & $U_{2}[\ensuremath{\times10^{12}}]$ & $U_{3}[\ensuremath{\times10^{12}}]$ & $U_{4}[\ensuremath{\times10^{12}}]$\tabularnewline
\hline 
LQT & $5.26$ & $4.66$ & $5.86$ & $4.40$\tabularnewline
\hline 
PID & $4.66$ & $3.87$ & $5.42$ & $3.46$\tabularnewline
\hline 
$\%\left(\frac{U_{LQT}-U_{PID}}{U_{PID}}\right)$ & $12.88\%$ & $20.41\%$ & $8.12\%$ & $27.17\%$\tabularnewline
\hline 
\end{tabular}
\par\end{centering}
\caption{\label{tab:tabcompxmotor}Motor effort comparison when following a unit step in the
X position.}
\end{table}
The results show a clear tendency of control effort increase while
using the LQT controller with respect to the PID controller.
\begin{itemize}
\item \textbf{Step command in the Y direction}
\end{itemize}
When commanded to follow a 1 meter step in the Y coordinate, the quadcopter behaved as suggested in \autoref{fig:compy}.

Similar as the last test, the anticipatory feature of the LQT controller
allows to reduce the big error the system incurs when sending step
commands. The graphical evidence is supported by the error comparison contained in \autoref{tab:tabcompy}.

\begin{table}[H]
\begin{centering}
\begin{tabular}{|c|c|c|c|c|}
\cline{2-5} 
\multicolumn{1}{c|}{} & $\textrm{RMS}\left(e_{x}\right)\,\textrm{[cm]}$ & $\textrm{RMS}\left(e_{y}\right)\,\textrm{[cm]}$ & $\%\xi_{x}$ & $\%\xi_{y}$\tabularnewline
\hline 
LQT & $2.53$ & $12.25$ & $100$ & $78.34$\tabularnewline
\hline 
PID & $3.50$ & $28.39$ & $99.68$ & $65.69$\tabularnewline
\hline 
\end{tabular}
\par\end{centering}
\caption{\label{tab:tabcompy}Error comparison when following a unit step in Y position.}
\end{table}
Once again, the RMS error in the direction the step was send was cut
by more than half with the LQT controller. Performance indices show
that the LQT was superior in terms of reducing the tracking error.
~\\

\begin{figure}[H]
\centering
\makebox[0pt]{
\includegraphics[width=1.2\textwidth]{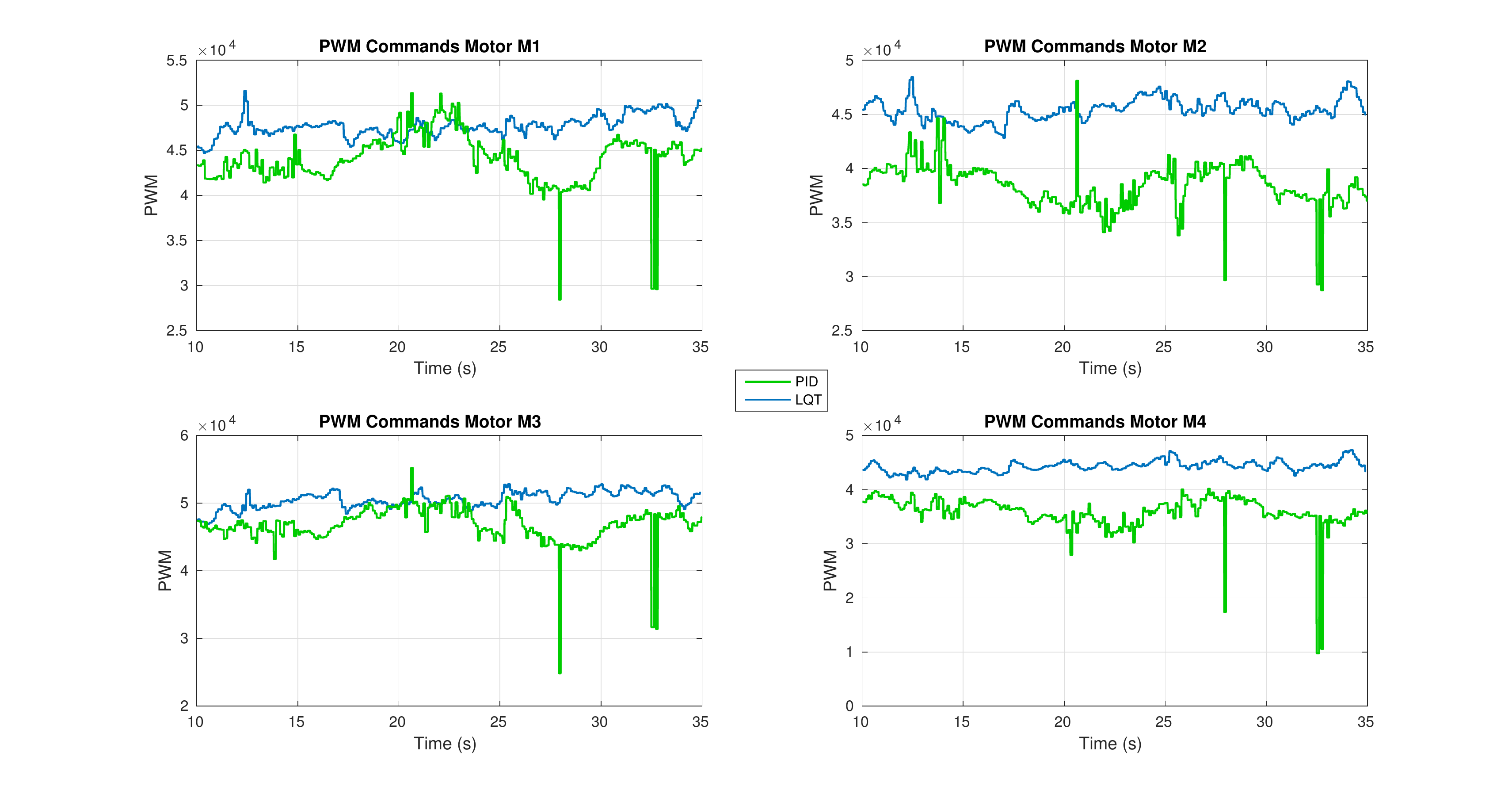}}
\caption{\label{fig:motorcompy}Motor commands comparison when following a unit step in the Y position.}
\end{figure}

As for the motor commands, \autoref{fig:motorcompy} show the same trend as the last test, the LQT algorithm
control outputs is overall bigger than the control outputs of the PID system, while this last one displays discontinuities in the commands sent as a consequence of the step demanded.

The data presented in \autoref{tab:tabmotory} indicate  that the control effort increment of the LQT algorithm with respect to the PID controller can get up to more than a 50\%.
\begin{table}[H]
\begin{centering}
\begin{tabular}{|c|c|c|c|c|}
\cline{2-5} 
\multicolumn{1}{c|}{} & $U_{1}\textrm{[\ensuremath{\times10^{12}}]}$ & $U_{2}[\ensuremath{\times10^{12}}]$ & $U_{3}[\ensuremath{\times10^{12}}]$ & $U_{4}[\ensuremath{\times10^{12}}]$\tabularnewline
\hline 
LQT & $5.66$ & $5.13$ & $6.35$ & $4.88$\tabularnewline
\hline 
PID & $4.89$ & $3.71$ & $5.49$ & $3.23$\tabularnewline
\hline 
$\%\left(\frac{U_{LQT}-U_{PID}}{U_{PID}}\right)$ & $15.75\%$ & $38.28\%$ & $15.66\%$ & $51.08\%$\tabularnewline
\hline 
\end{tabular}
\par\end{centering}
\caption{\label{tab:tabmotory}Motor effort comparison when following a unit step in the
Y position.}
\end{table}

\begin{figure}[H]
\begin{centering}
\includegraphics[scale=0.75]{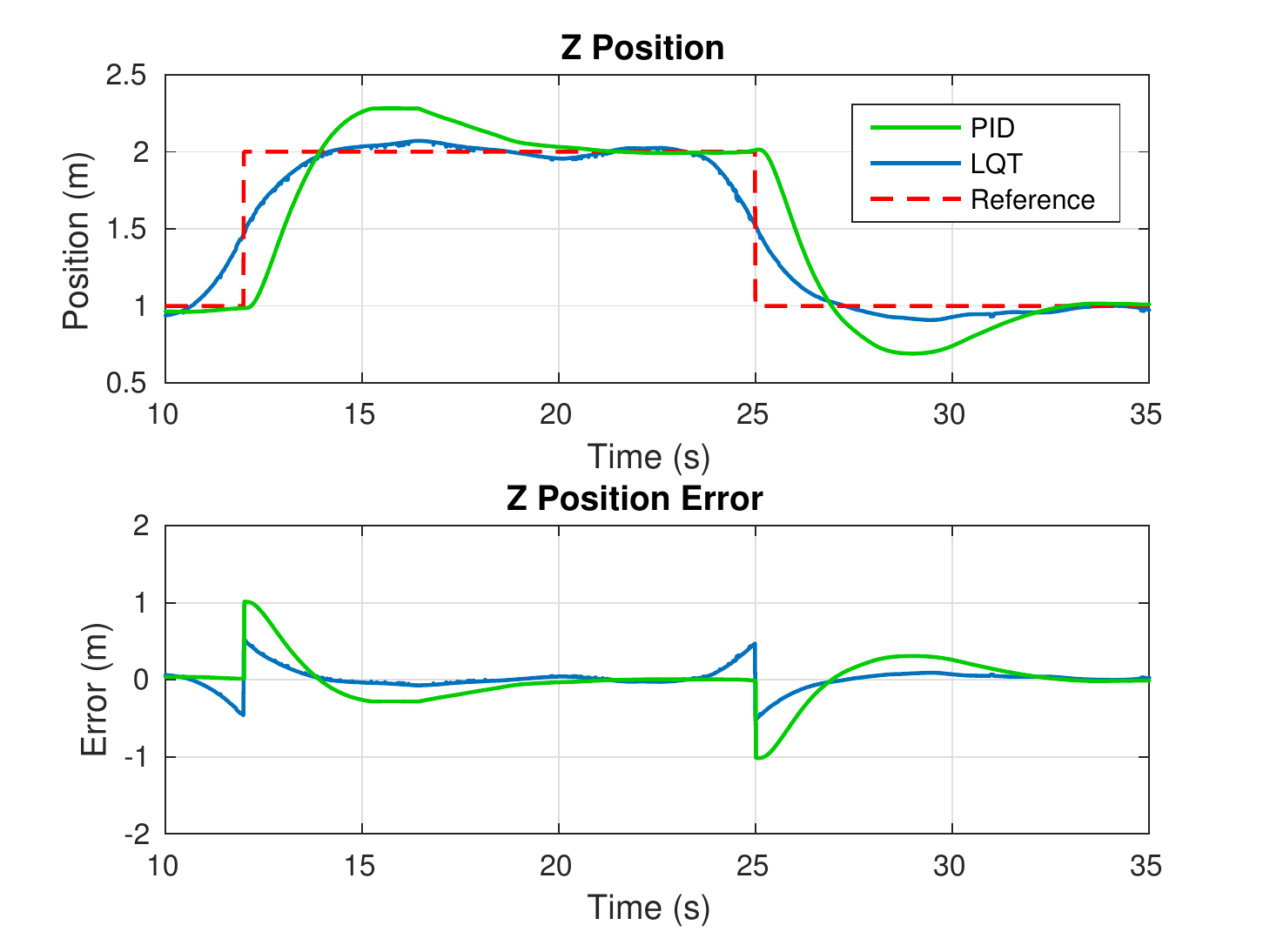}
\par\end{centering}
\caption{\label{fig:compz}Z Position and error comparison when following a unit step.}
\end{figure}

\begin{figure}[H]
\centering
\makebox[0pt]{
\includegraphics[width=1.2\textwidth]{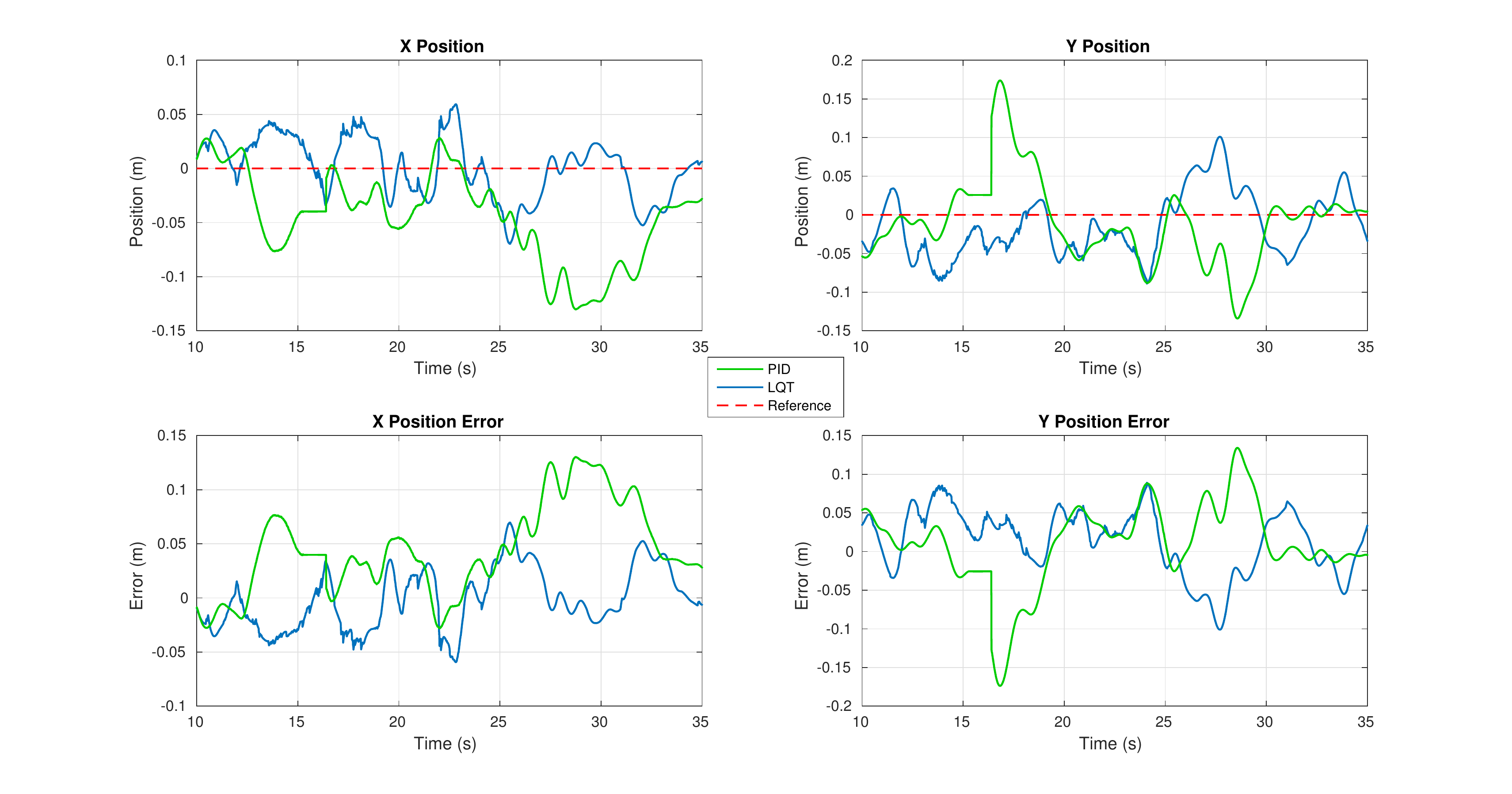}}
\caption{\label{fig:compxz}X-Y Position and error comparison when following a unit step in the
Z position.}
\end{figure}

\begin{itemize}
\item \textbf{Step command in the Z direction}
\end{itemize}
The last comparison for the linear trajectories was done sending a 1 meter step command in the Z coordinate. The test result flight data is exposed in \Cref{fig:compz,fig:compxz}, where all three coordinate comparisons between the two controllers are illustrated.
~\\

The LQT algorithm outperforms by a good margin the PID, by greatly
reducing the overshoot and the response time in the Z position. The LQT also kept a more precise X-Y position, both at a constant altitude and when applying the command to ascend 1 meter.

The analysis of the results is summarized in \autoref{tab:tabcompz}. The error comparison confirms the superiority of the LQT with respect to the PID, with reduced RMS values of error and better performance
indices.
\begin{table}[H]
\begin{centering}
\begin{tabular}{|c|c|c|c|c|c|c|}
\cline{2-7} 
\multicolumn{1}{c|}{} & $\textrm{RMS}\left(e_{x}\right)\,\textrm{[cm]}$ & $\textrm{RMS}\left(e_{y}\right)\,\textrm{[cm]}$ & $\textrm{RMS}\left(e_{z}\right)\,\textrm{[cm]}$ & $\%\xi_{x}$ & $\%\xi_{y}$ & $\%\xi_{z}$\tabularnewline
\hline 
LQT & $2.82$ & $4.35$ & $13.38$ & $100$ & $99.48$ & $81.48$\tabularnewline
\hline 
PID & $6.13$ & $5.46$ & $28.59$ & $86.43$ & $91.83$ & $52.68$\tabularnewline
\hline 
\end{tabular}
\par\end{centering}
\caption{\label{tab:tabcompz}Error comparison when following a unit step in Z position.}
\end{table}

\begin{figure}[H]
\centering
\makebox[0pt]{
\includegraphics[width=1.2\textwidth]{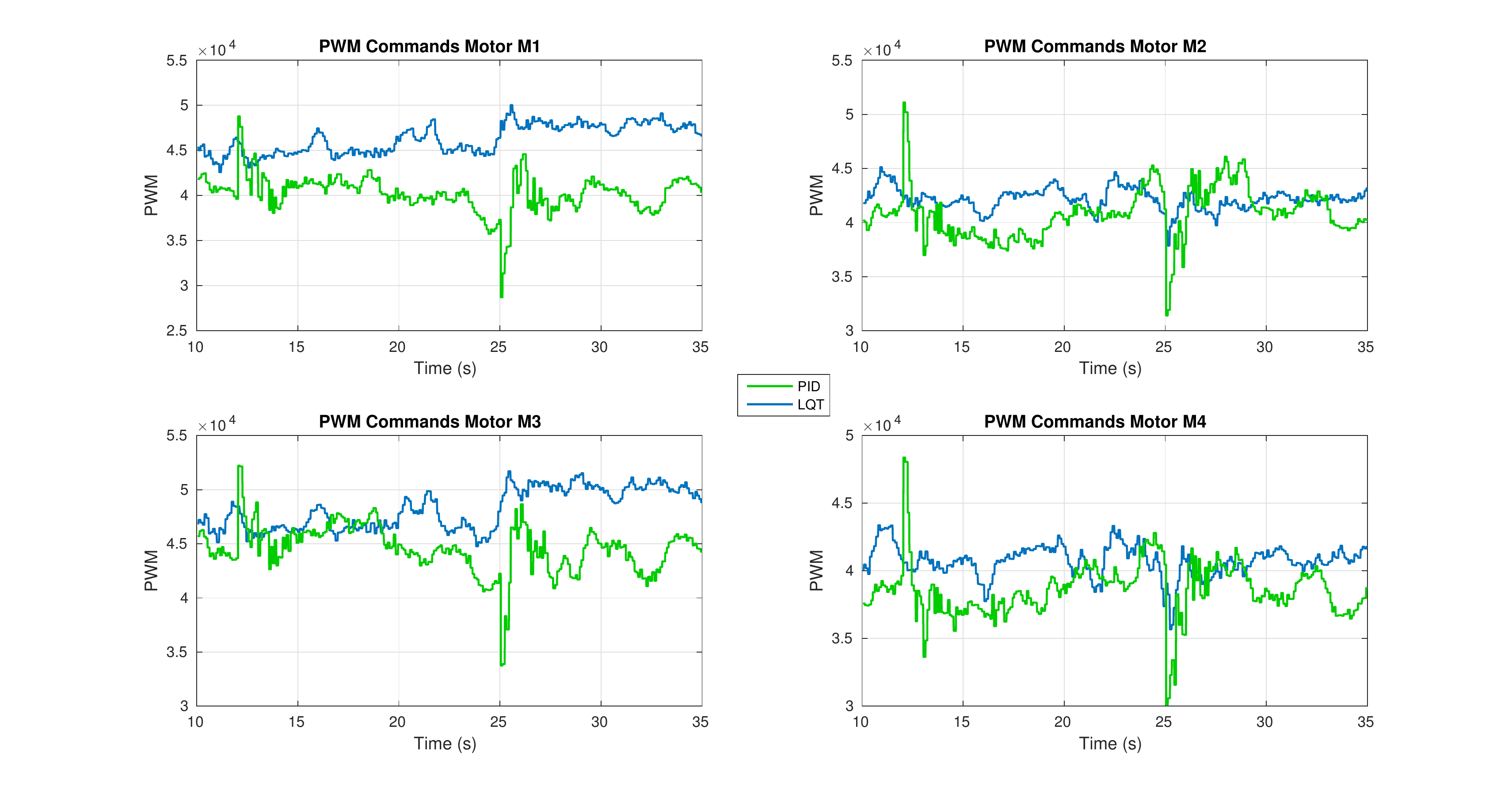}}
\caption{\label{fig:compmotorz}Motor commands comparison when following a unit step in the Z position.}
\end{figure}

The control commands of the motors are visualized in \autoref{fig:compmotorz}. As in all the other tests, the LQT control effort remained greater than that of the PID. However the LQT plots do not present evident
command peaks as the PID when sending the altitude command.

The increase in control effort is exposed by the results in \autoref{tab:tabmotorz}. The numbers reveal a maximum increase of around 31\% in control effort in one of the motors with the LQT controller with respect to the PID.

\begin{figure}[H]
\centering
\makebox[0pt]{
\includegraphics[width=1.2\textwidth]{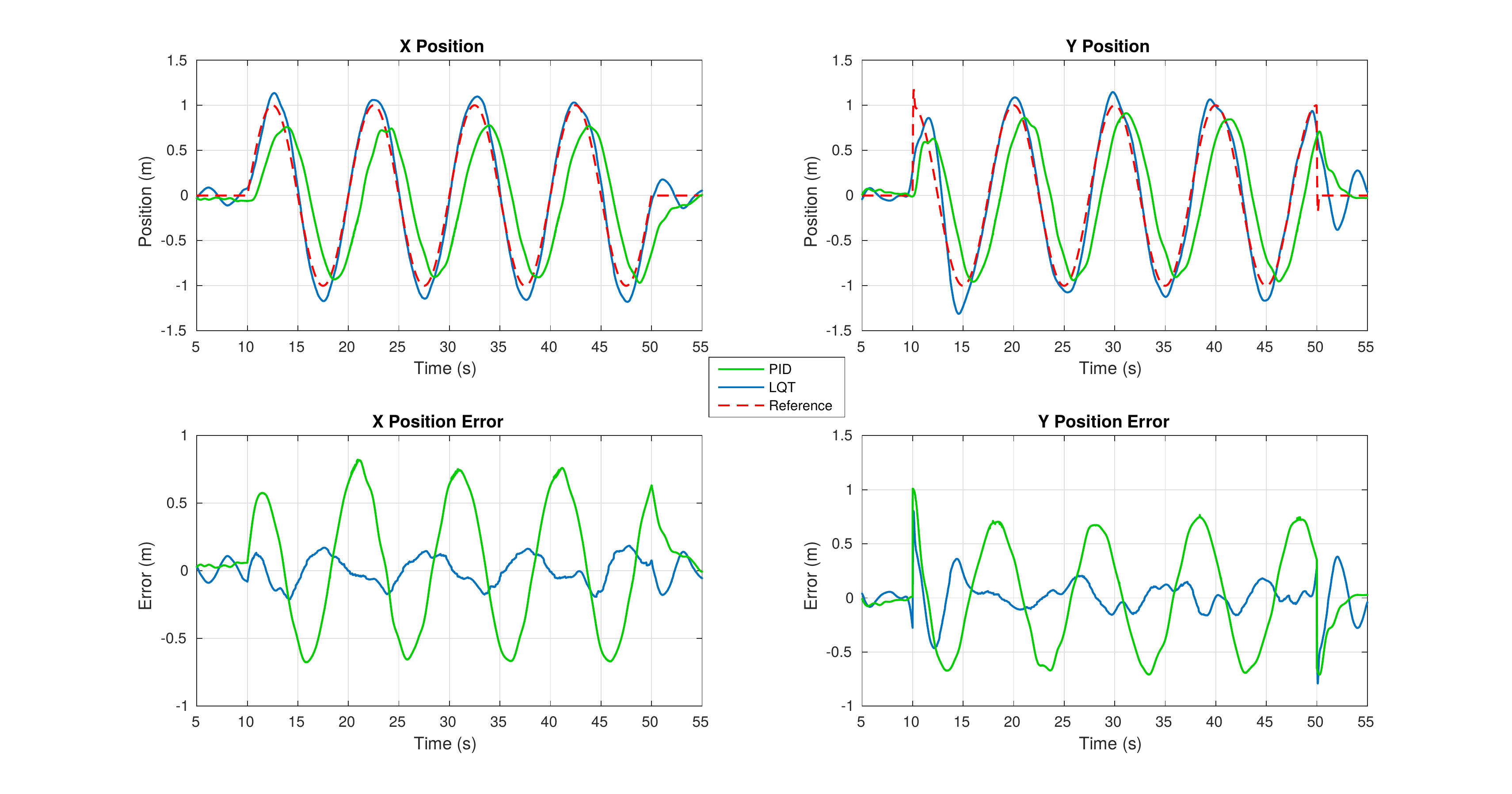}}
\caption{\label{fig:hola}X-Y Position and error comparison when following a circular trajectory.}
\end{figure}

\begin{table}[H]
\begin{centering}
\begin{tabular}{|c|c|c|c|c|}
\cline{2-5} 
\multicolumn{1}{c|}{} & $U_{1}\textrm{[\ensuremath{\times10^{12}}]}$ & $U_{2}[\ensuremath{\times10^{12}}]$ & $U_{3}[\ensuremath{\times10^{12}}]$ & $U_{4}[\ensuremath{\times10^{12}}]$\tabularnewline
\hline 
LQT & $5.32$ & $4.42$ & $5.78$ & $4.13$\tabularnewline
\hline 
PID & $4.05$ & $4.18$ & $4.98$ & $3.73$\tabularnewline
\hline 
$\%\left(\frac{U_{LQT}-U_{PID}}{U_{PID}}\right)$ & $31.36\%$ & $5.74\%$ & $16.06\%$ & $10.72\%$\tabularnewline
\hline 
\end{tabular}
\par\end{centering}
\caption{\label{tab:tabmotorz}Motor effort comparison when following a unit step in the
Z position.}
\end{table}
\begin{itemize}
\item \textbf{Circular trajectory}
\end{itemize}
A circular trajectory composing a circle in the X-Y plane was conducted
with both controllers to test the tracking of rapidly varying trajectories.
The time plot comparisons of the X-Y positions are displayed in \autoref{fig:hola}.

The LQT controller manages to track and stay in phase with the sinusoidal
waves, while the PID controller was not capable of such feat nor of
obtaining the desired 1 meter amplitude. The error plots expose the
improvement in trajectory tracking of the LQT controller with respect
to the PID, and are validated by the data shown in \autoref{tab:compcirc}.

\begin{table}[H]
\begin{centering}
\begin{tabular}{|c|c|c|c|c|}
\cline{2-5} 
\multicolumn{1}{c|}{} & $\textrm{RMS}\left(e_{x}\right)\,\textrm{[cm]}$ & $\textrm{RMS}\left(e_{y}\right)\,\textrm{[cm]}$ & $\%\xi_{x}$ & $\%\xi_{y}$\tabularnewline
\hline 
LQT & $10.32$ & $16.69$ & $55.74$ & $55.00$\tabularnewline
\hline 
PID & $46.05$ & $47.28$ & $14.72$ & $17.68$\tabularnewline
\hline 
\end{tabular}
\par\end{centering}
\caption{\label{tab:compcirc}Error comparison when following a circular position.}
\end{table}
The RMS error is about 4 times greater with the PID controller than
with the LQT algorithm, clearly showing the superiority of the latter
in tracking more complex trajectories.
~\\

\begin{figure}[H]
\centering
\makebox[0pt]{
\includegraphics[width=1.2\textwidth]{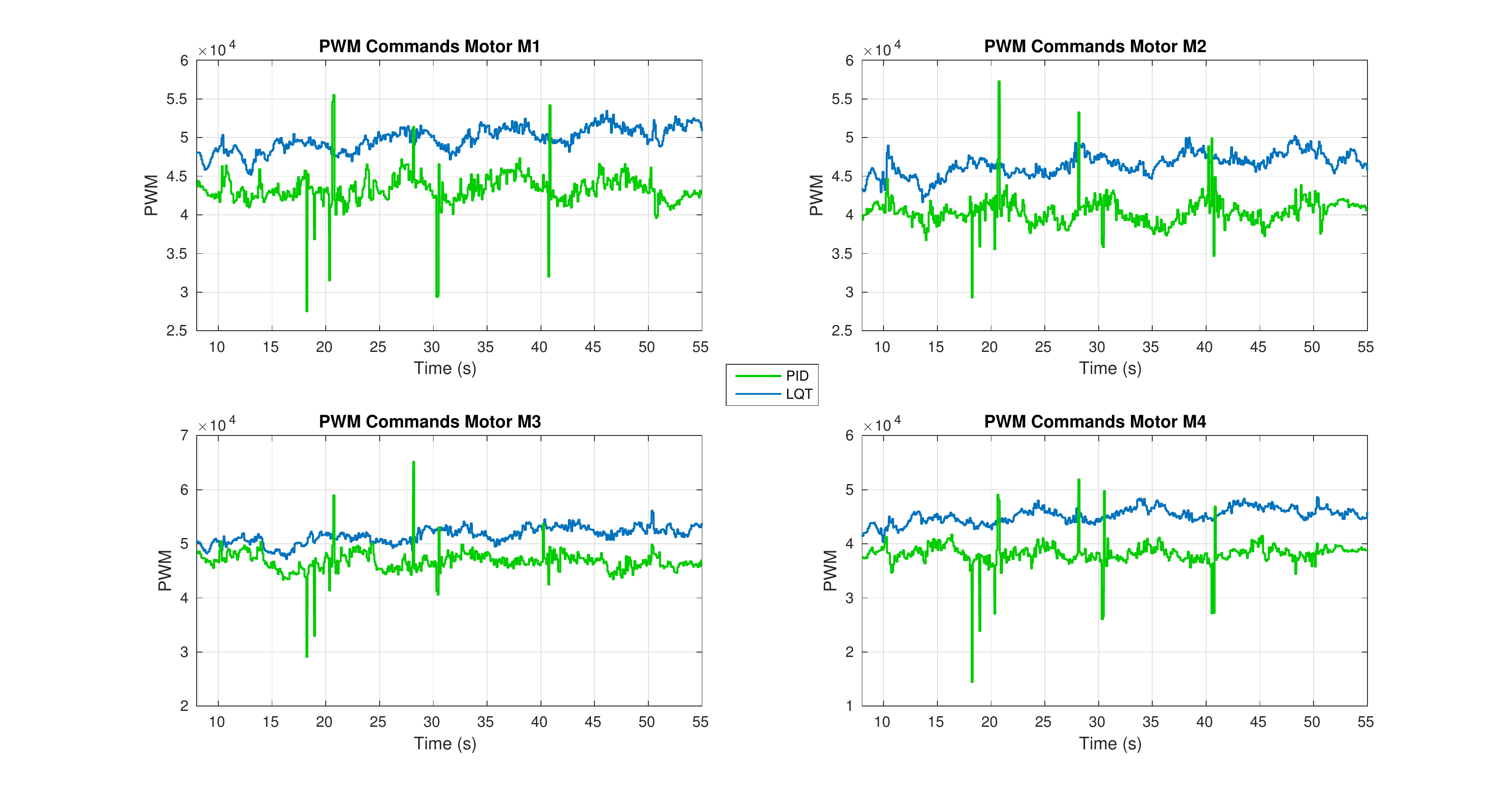}}
\caption{\label{tab:motorcirc}Motor command comparison when following a circular trajectory.}
\end{figure}

The motors' 16-bit PWM signals of both controllers are compared in \autoref{tab:motorcirc}.

The tendency of a greater command effort for the LQT algorithm compared
to the PID remained as before, also confirmed by the values in \autoref{tab:Motor-effort-statistics}. Note that for this trajectory there are important control spikes with the PID controller exactly
in the points where the error reached its maximum points in \autoref{fig:hola}, causing for instance a motor saturation at the 28 second mark in motor M3.

\begin{table}[H]
\begin{centering}
\begin{tabular}{|c|c|c|c|c|}
\cline{2-5} 
\multicolumn{1}{c|}{} & $U_{1}\textrm{[\ensuremath{\times10^{12}}]}$ & $U_{2}[\ensuremath{\times10^{12}}]$ & $U_{3}[\ensuremath{\times10^{12}}]$ & $U_{4}[\ensuremath{\times10^{12}}]$\tabularnewline
\hline 
LQT & $11.68$ & $10.20$ & $12.50$ & $9.65$\tabularnewline
\hline 
PID & $8.82$ & $7.69$ & $10.28$ & $6.88$\tabularnewline
\hline 
$\%\left(\frac{U_{LQT}-U_{PID}}{U_{PID}}\right)$ & $32.43\%$ & $32.64\%$ & $21.60\%$ & $40.26\%$\tabularnewline
\hline 
\end{tabular}
\par\end{centering}
\caption{\label{tab:Motor-effort-statistics}Motor effort comparison
when following a circular trajectory.}
\end{table}

The comparison between the PID and LQT control systems gave insightful data to draw conclusions about their advantages and disadvantages. Starting with the overall performance in position and trajectory tracking, the LQT was superior at keeping low levels of error with respect to the desired position. The goal of improving the tracking performance of the PID controller was achieved with the LQT algorithm. ~\\

As for the command effort of the motors, even though the LQT algorithm does an optimization process to minimize it, the experimental data exposed that the PID used less effort in every trial. The explanation for this phenomenon are the elevated integral gains and weighting factors used with the LQT algorithm to obtain the desired performance. It is intuitive to think that the correct tracking of more demanding trajectories leads to a greater control effort, as for most control systems design this ends up being a compromise issue, in this case between performance and motor power. This increase in control effort translated in a shorter battery life-time while using the LQT controller with respect to the PID.

\section{LQT with UWB Position Estimation}

\begin{figure}[H]
\centering
\includegraphics[scale=0.08]{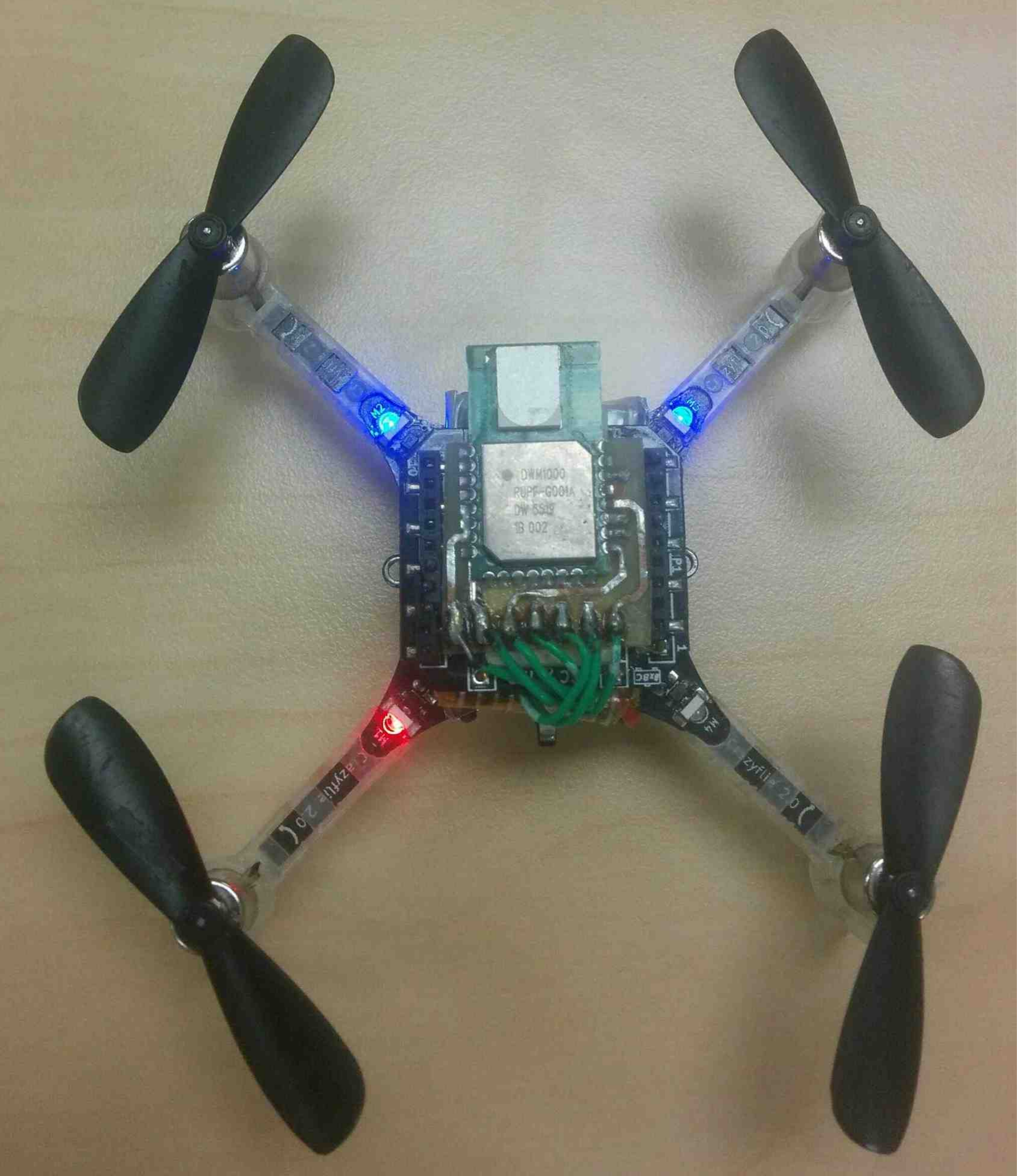}
\caption{\label{fig:uwbmodule}Crazyflie 2.0 with UWB module exposed}
\end{figure}

Using the two-way ranging ultra-wide band system developed in \cite{key-36}, four base stations were placed forming
a 7x4m rectangle at 2.5m of height from the floor. \autoref{fig:uwbmodule} displays how the tag was incorporated to the quadcopter's body.
A series of trajectories were followed using the VICON and then the
UWB system to test the LQT controller. In both cases the flight data was compared using the VICON measurements as the ground truth in order to study the tracking performance. The UWB estimations were used only
for the X and Y position, while the altitude came from the VICON system.
The analysis was focused to the tracking in the X-Y plane, while keeping a 1 meter altitude. With each system a certain amount of flights were executed and the one with best performance is the one showcased in this section, so the comparisons were made in the best-case scenario with each localization system.  Note: while using the UWB system it was decided to lower the position integral gains as it improved the performance.
~\\
~\\

\begin{figure}[H]
\begin{centering}
\includegraphics[scale=0.7]{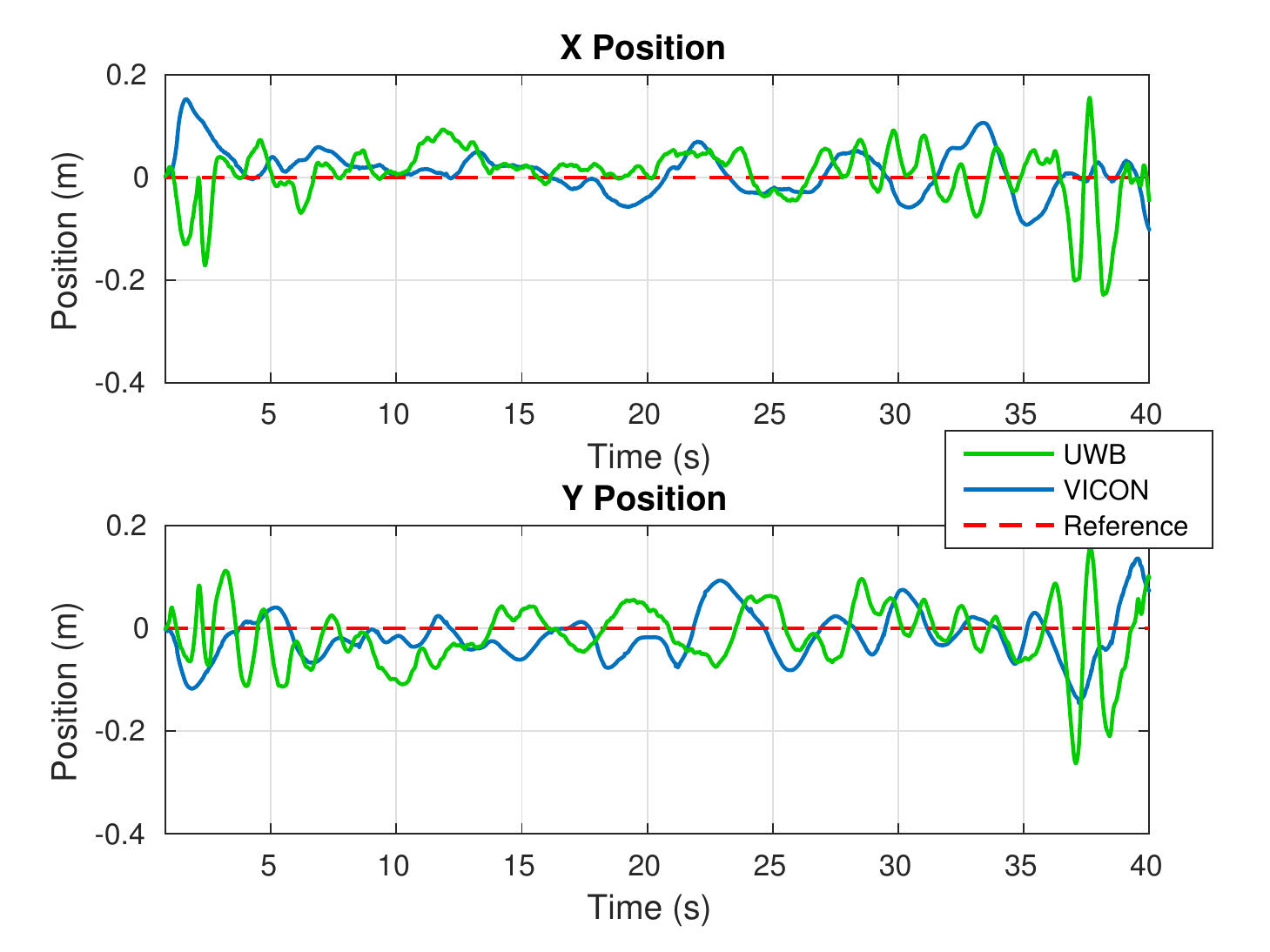}
\par\end{centering}
\caption{\label{fig:hoveruwb}X-Y Position and error comparison while hovering around a point.}
\end{figure}

\begin{itemize}
\item \textbf{Hover}
\end{itemize}
The first test was a simple hover to asses the controller's capability
of keeping a fixed position. \autoref{fig:hoveruwb} presents the experimental results for the hover test.

Although in both cases the controller performed similarly, with the
UWB system there were more oscillations, specially in the take-off
and landing stages at the beginning and end of the time plot. \autoref{tab:tabuwbhover}
quantifies the error values of the hover flight.

\begin{table}[H]
\begin{centering}
\begin{tabular}{|c|c|c|c|c|}
\cline{2-5} 
\multicolumn{1}{c|}{} & $\textrm{RMS}\left(e_{x}\right)\,\textrm{[cm]}$ & $\textrm{RMS}\left(e_{y}\right)\,\textrm{[cm]}$ & $\%\xi_{x}$ & $\%\xi_{y}$\tabularnewline
\hline 
VICON & $4.67$ & $5.03$ & $93.74$ & $94.78$\tabularnewline
\hline 
UWB & $5.90$ & $6.42$ & $92.15$ & $90.27$\tabularnewline
\hline 
\end{tabular}
\par\end{centering}
\caption{\label{tab:tabuwbhover}Error comparison while hovering around a point.}
\end{table}
Both hover flights were solid, although the RMS errors and performance
indices suggests that with the VICON system the drone hold more precisely its position.

\begin{figure}[H]
\centering
\makebox[0pt]{
\includegraphics[width=1.2\textwidth]{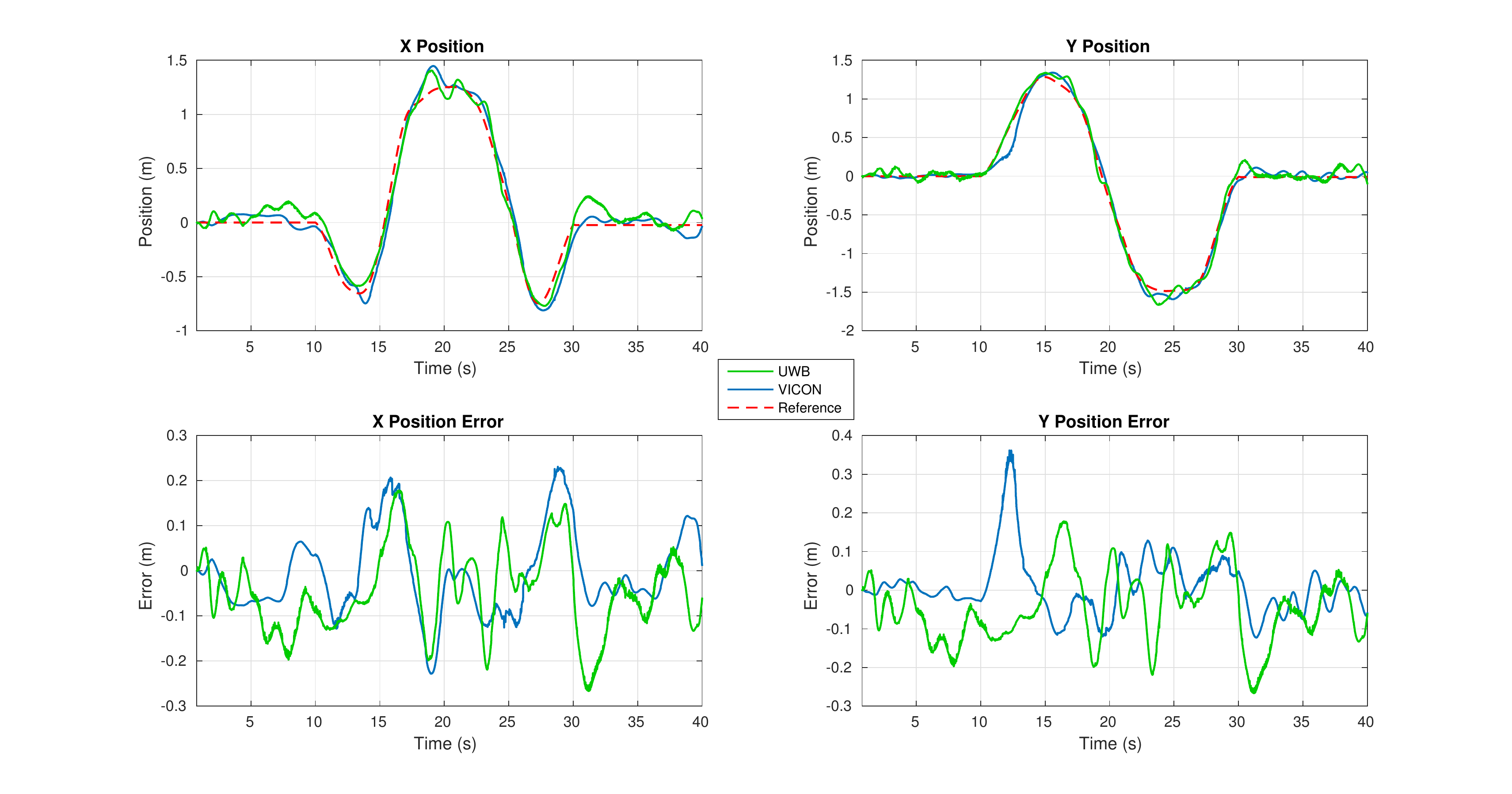}}
\caption{\label{fig:uwb1}X-Y Position and error comparison when following Trajectory \#1.}
\end{figure}

\begin{itemize}
\item \textbf{Trajectory \#1}
\end{itemize}
The results for the first trajectory are displayed in \autoref{fig:uwb1}.

\begin{figure}[H]
\centering
\makebox[0pt]{
\subfloat[Standard view.]{\includegraphics[width=0.55\textwidth]{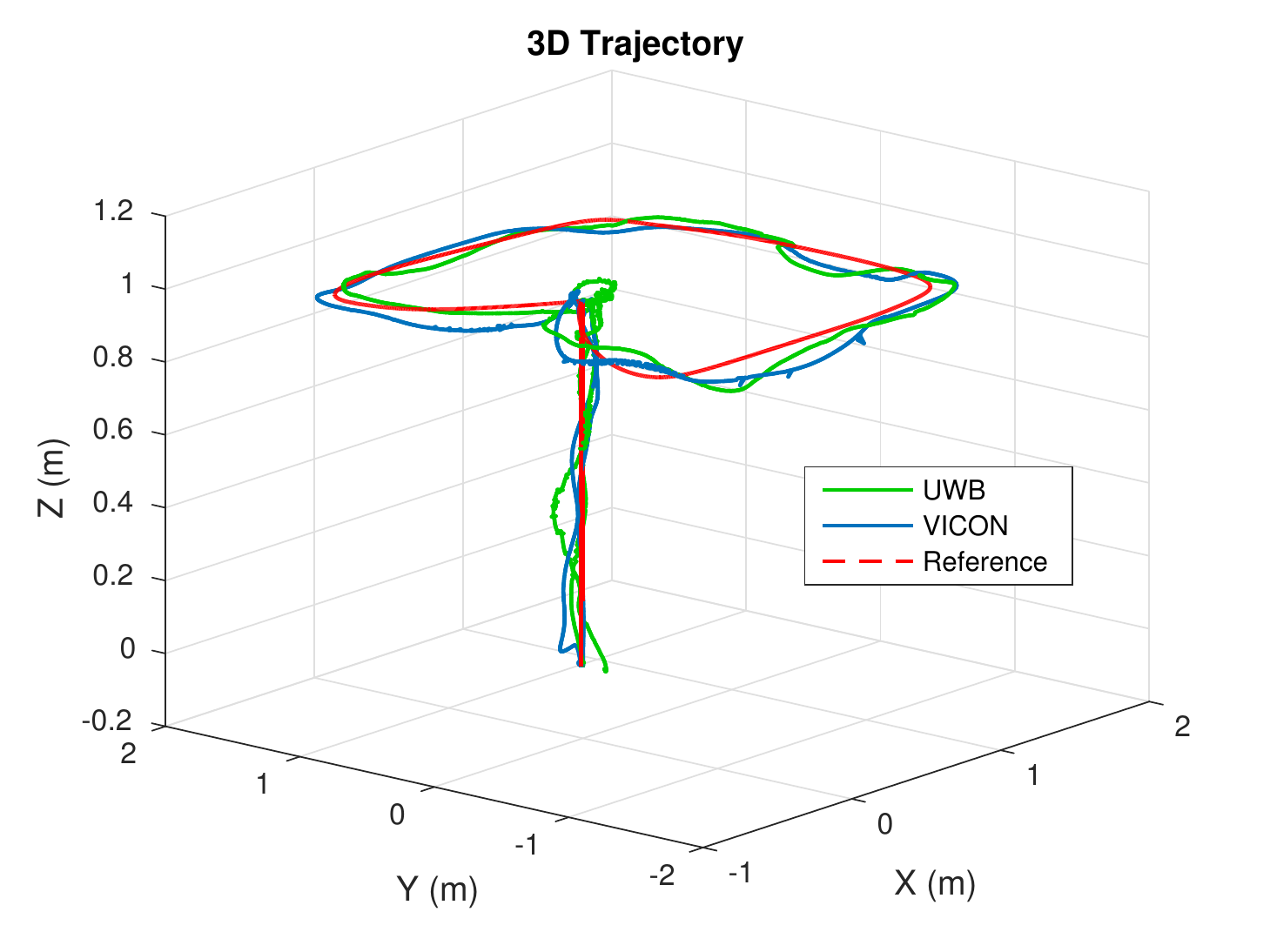}}
\subfloat[XY Plane.]{\includegraphics[width=0.55\textwidth]{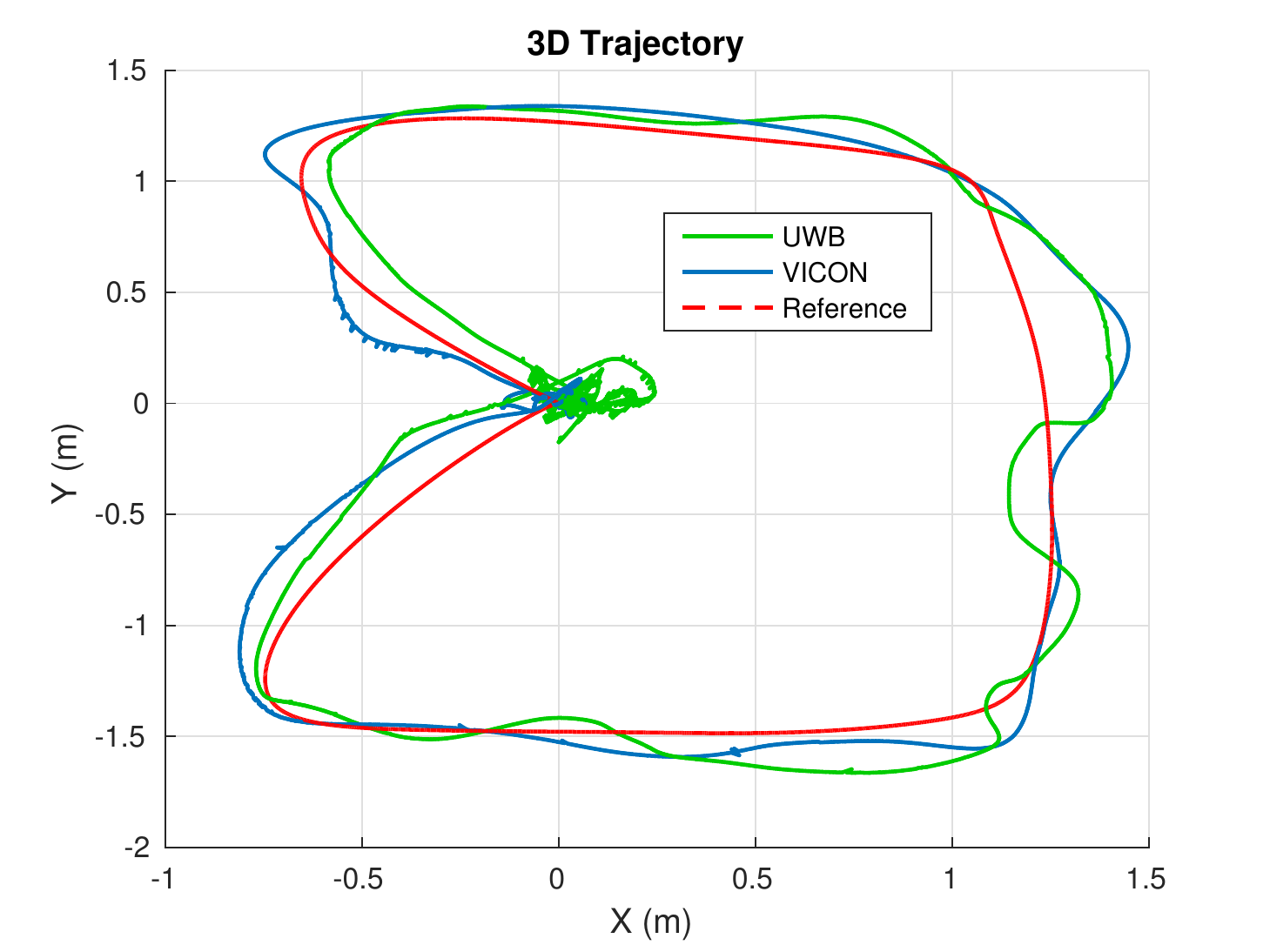}}}
\caption{\label{fig:Comparison-of-3D}Comparison of 3D Trajectory\#1.}
\end{figure}

Once again, the overall performance in both cases in terms of the
errors was similar, but the greater levels of noise of the UWB system
with respect to the VICON is translated into more oscillations in
the position. The 3D perspectives in \autoref{fig:Comparison-of-3D} show
the smoothness of the system with the VICON with respect to the more
oscillating trajectory with the UWB.

As for the performance, the summary in \autoref{tab:Error-statistics-comparison}
indicates that the overall performance with the VICON system is better,
but nonetheless with the UWB system the controller manages to follow
the trajectory with RMS errors around 10 centimeters.

\begin{table}[H]
\begin{centering}
\begin{tabular}{|c|c|c|c|c|}
\cline{2-5} 
\multicolumn{1}{c|}{} & $\textrm{RMS}\left(e_{x}\right)\,\textrm{[cm]}$ & $\textrm{RMS}\left(e_{y}\right)\,\textrm{[cm]}$ & $\%\xi_{x}$ & $\%\xi_{y}$\tabularnewline
\hline 
VICON & $9.16$ & $7.55$ & $73.69$ & $86.39$\tabularnewline
\hline 
UWB & $10.22$ & $7.82$ & $63.44$ & $79.85$\tabularnewline
\hline 
\end{tabular}
\par\end{centering}
\caption{\label{tab:Error-statistics-comparison}Error comparison
while following Trajectory \#1.}
\end{table}

\begin{figure}[H]
\centering
\makebox[0pt]{
\includegraphics[width=1.2\textwidth]{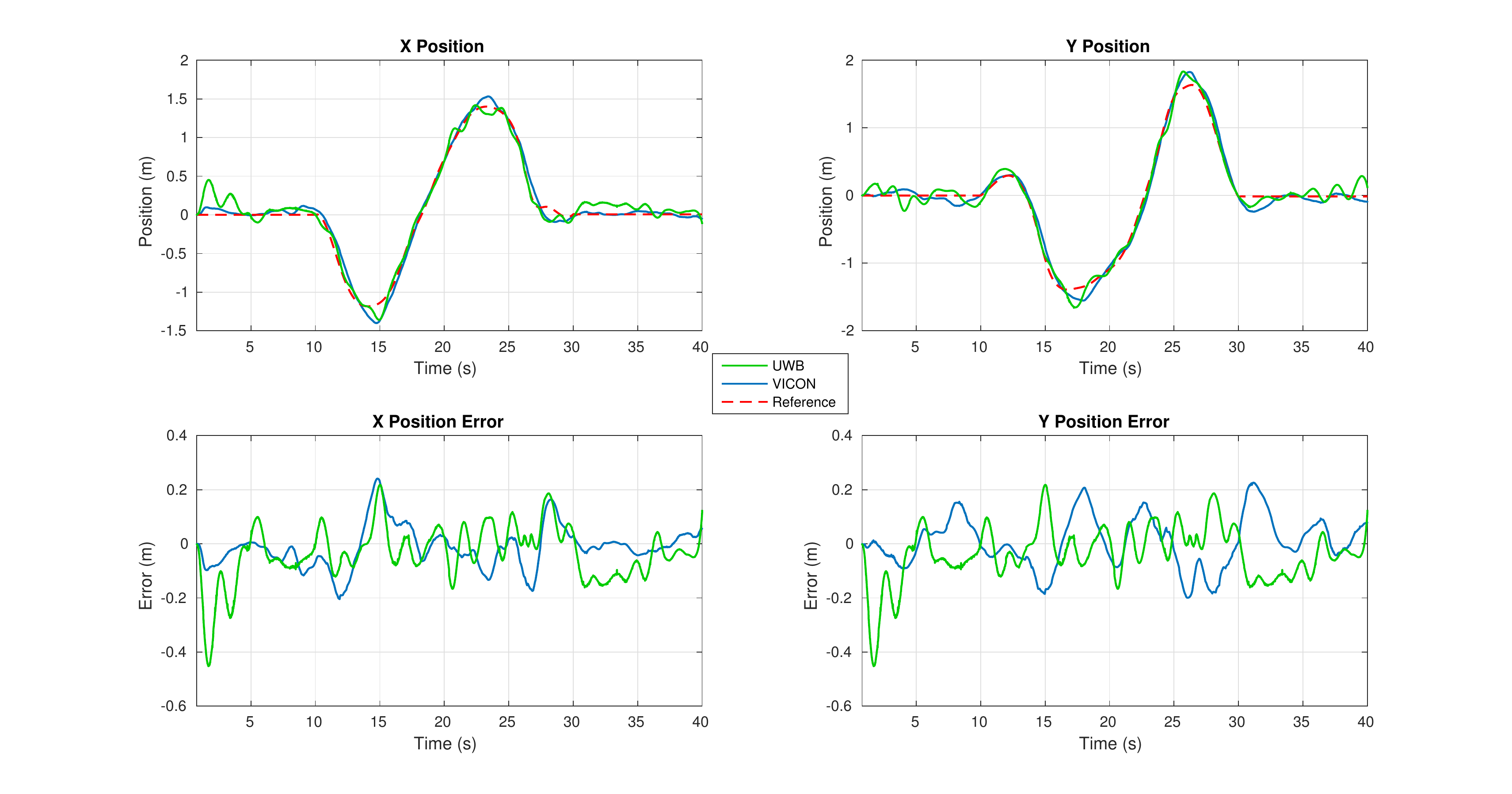}}
\caption{\label{fig:X-Y-Position-and-1}X-Y Position and error comparison when following Trajectory \#2.}
\end{figure}

\begin{itemize}
\item \textbf{Trajectory \#2}
\end{itemize}
While following a second trajectory, similar in complexity as the
first one, the results presented in \autoref{fig:X-Y-Position-and-1}
illustrate the tracking capabilities of the LQT controller in both
cases, despite the higher levels of noise when using the UWB system.
The error plots compare the smooth lines in the VICON flight with
the more oscillating ones of the UWB.

\begin{figure}[H]
\centering
\makebox[0pt]{
\subfloat[Standard view.]{\includegraphics[width=0.55\textwidth]{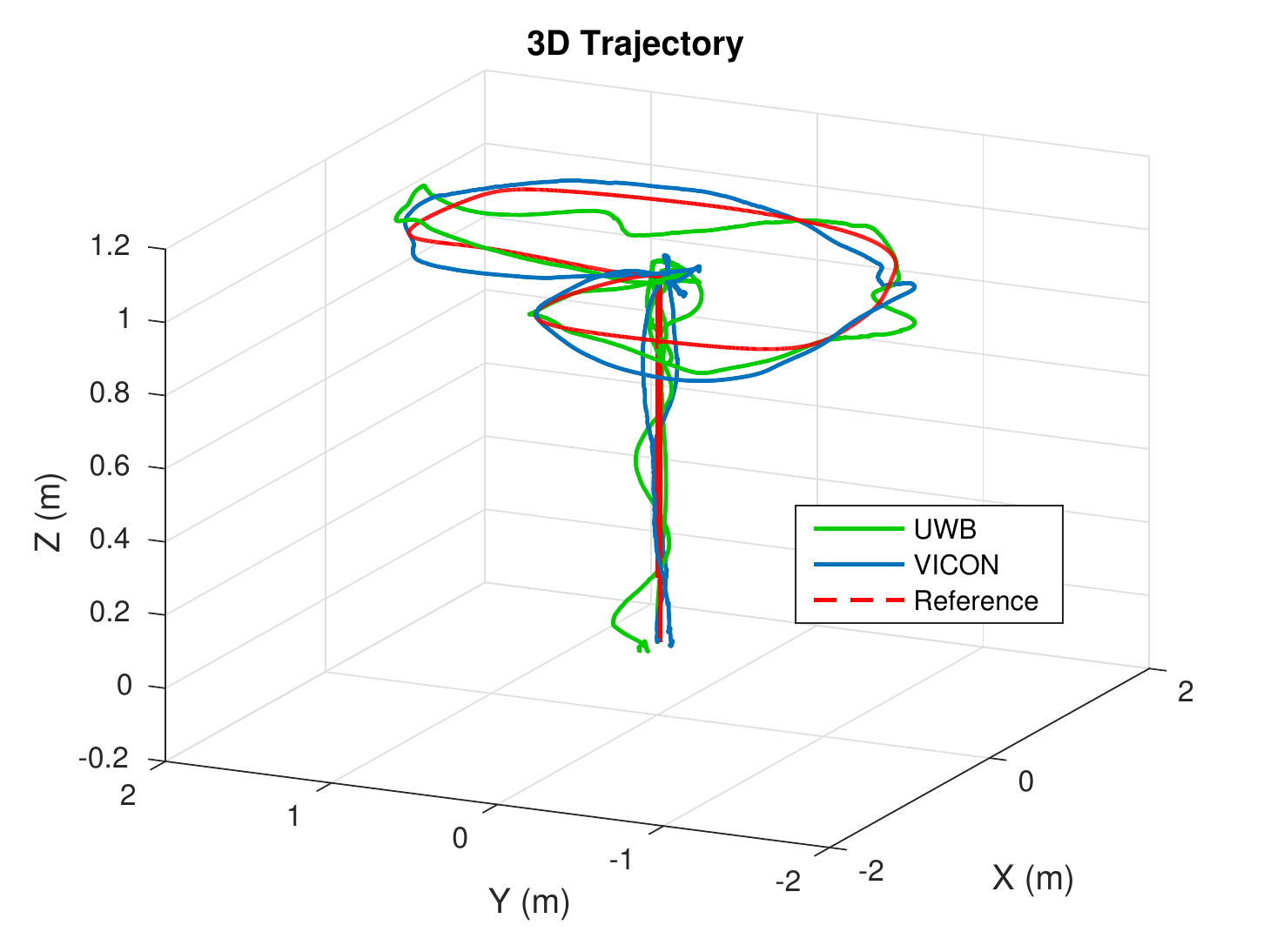}}
\subfloat[XY Plane.]{\includegraphics[width=0.55\textwidth]{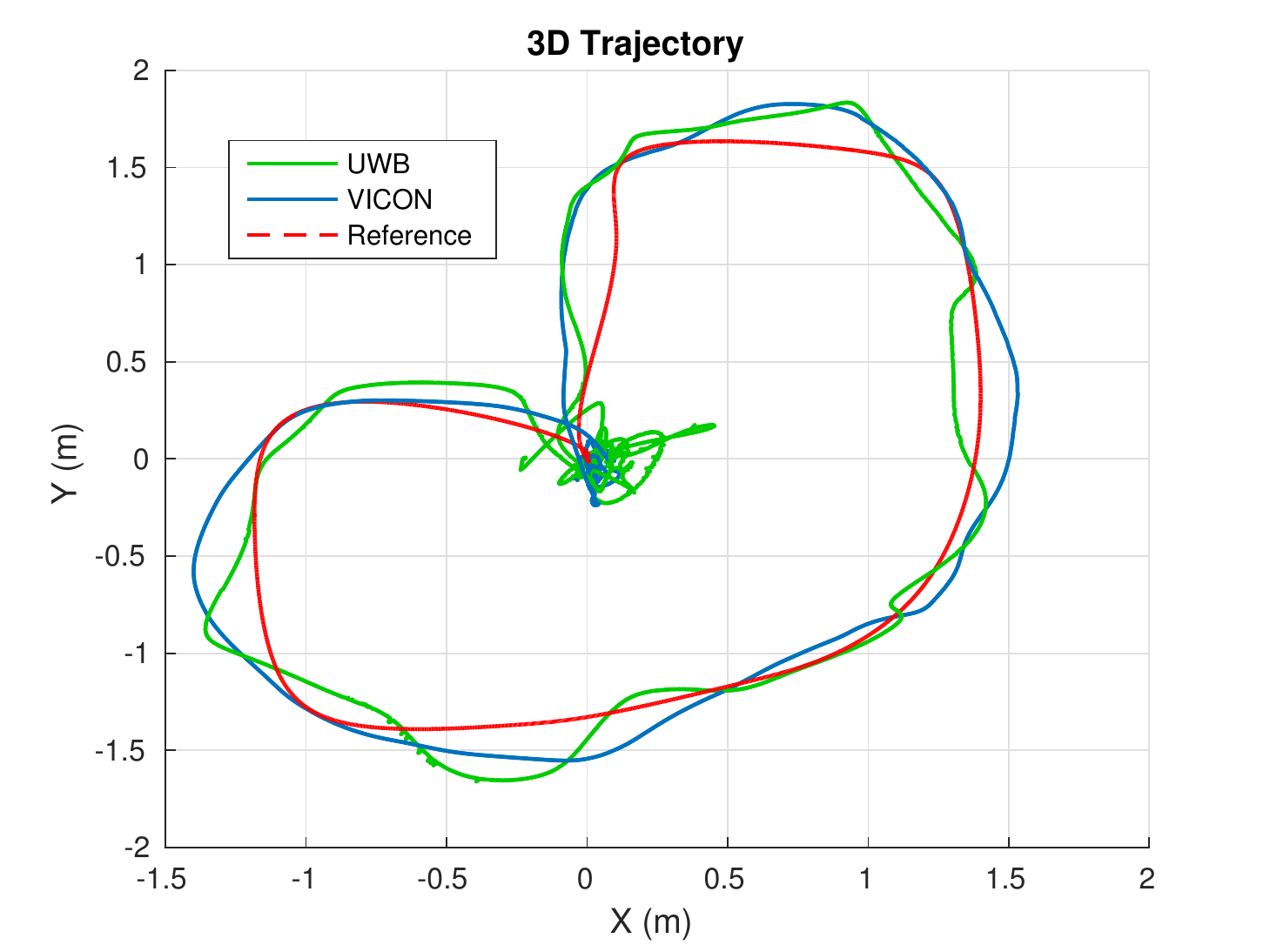}}}
\caption{\label{fig:3duwb2}Comparison of 3D Trajectory\#2.}
\end{figure}

The 3D perspectives in \autoref{fig:3duwb2} illustrate the trajectory followed by the quadcopter in both cases,
and specially the X-Y plane view shows the main difference between
the two flights: the smoothness and stability with which
the quadcopter managed to follow the desired trajectory.

\autoref{tab:Error-statistics-comparison-1} summarizes the performance
in trajectory tracking. The UWB system incurred in a greater RMS error
than the VICON, around 3.4cm more for the X position and 0.6cm for
the Y position. The performance indices $\xi_{x}$ and $\xi_{y}$
were inferior while using the UWB system, but still close enough to be acceptable.

\begin{table}[H]
\begin{centering}
\begin{tabular}{|c|c|c|c|c|}
\cline{2-5} 
\multicolumn{1}{c|}{} & $\textrm{RMS}\left(e_{x}\right)\,\textrm{[cm]}$ & $\textrm{RMS}\left(e_{y}\right)\,\textrm{[cm]}$ & $\%\xi_{x}$ & $\%\xi_{y}$\tabularnewline
\hline 
VICON & $7.71$ & $9.50$ & $82.08$ & $73.17$\tabularnewline
\hline 
UWB & $11.16$ & $10.15$ & $71.04$ & $70.32$\tabularnewline
\hline 
\end{tabular}
\par\end{centering}
\caption{\label{tab:Error-statistics-comparison-1}Error comparison
while following Trajectory \#2.}
\end{table}

\begin{figure}[H]
\centering
\makebox[0pt]{
\includegraphics[width=1.2\textwidth]{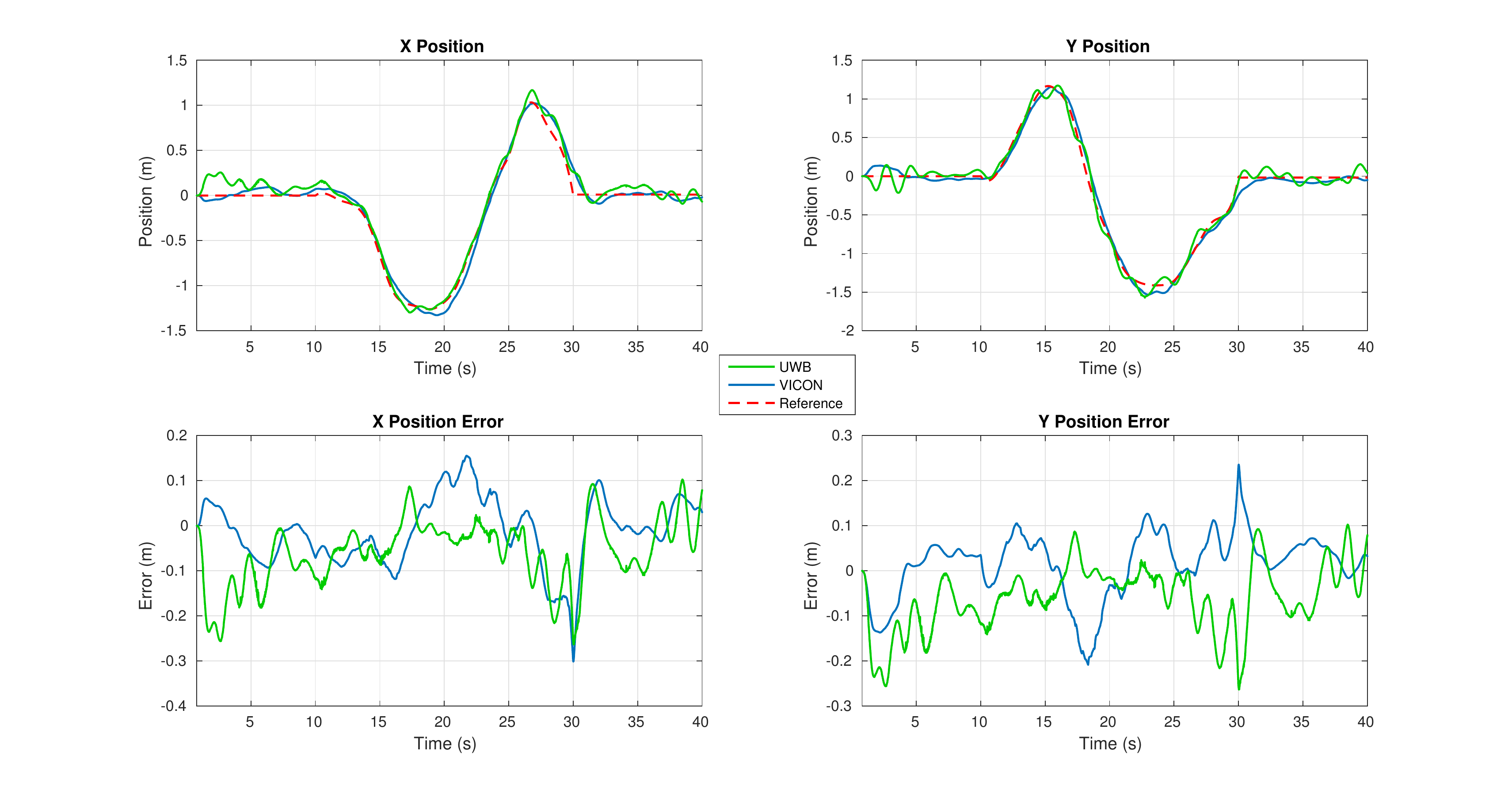}}
\caption{\label{fig:X-Y-Position-and-2}X-Y Position and error comparison when following Trajectory \#3.}
\end{figure}

\begin{figure}[H]
\centering
\makebox[0pt]{
\subfloat[Standard view.]{\includegraphics[width=0.55\textwidth]{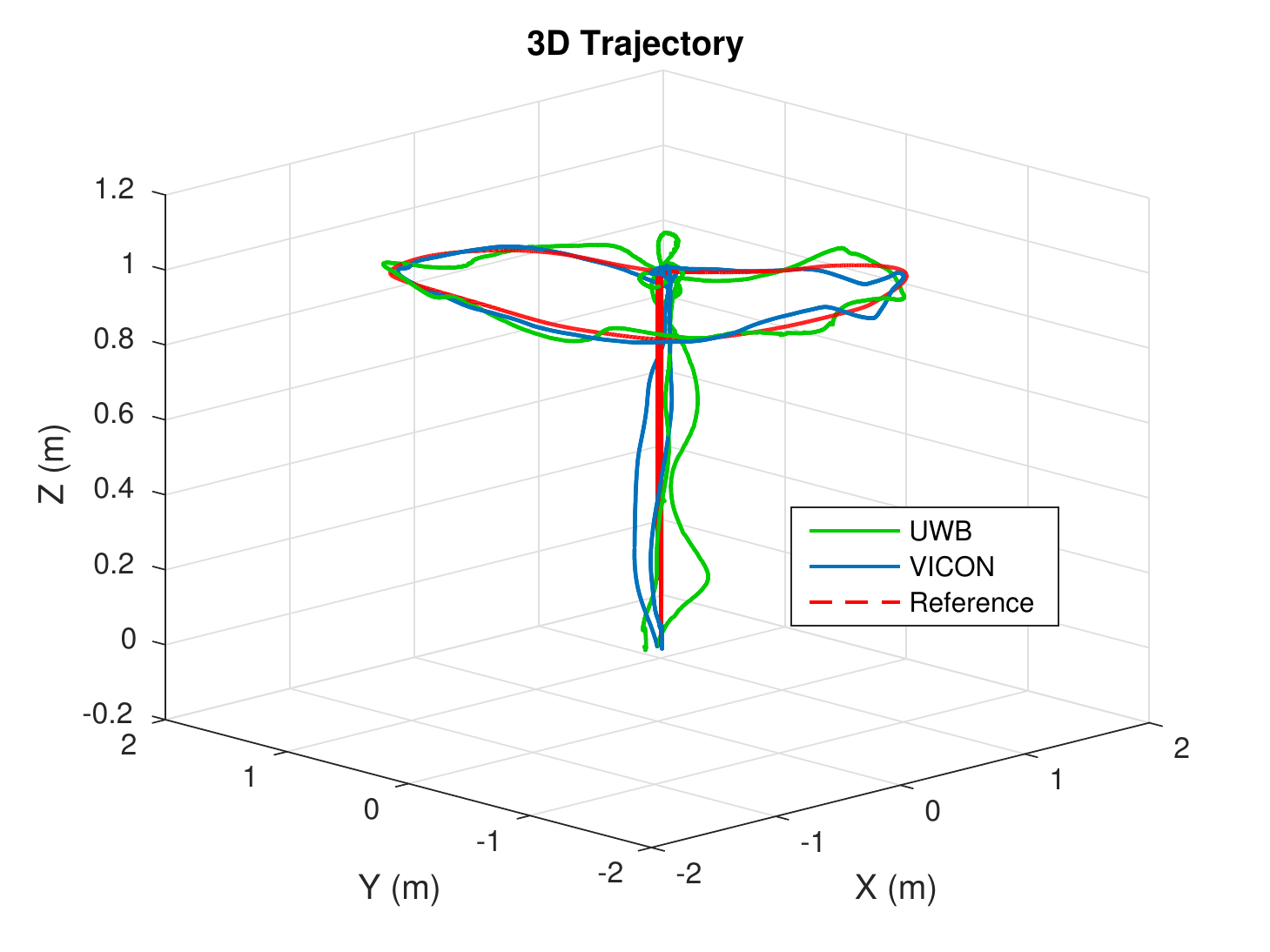}}
\subfloat[\label{fig:XY-Plane1}XY Plane.]{\includegraphics[width=0.55\textwidth]{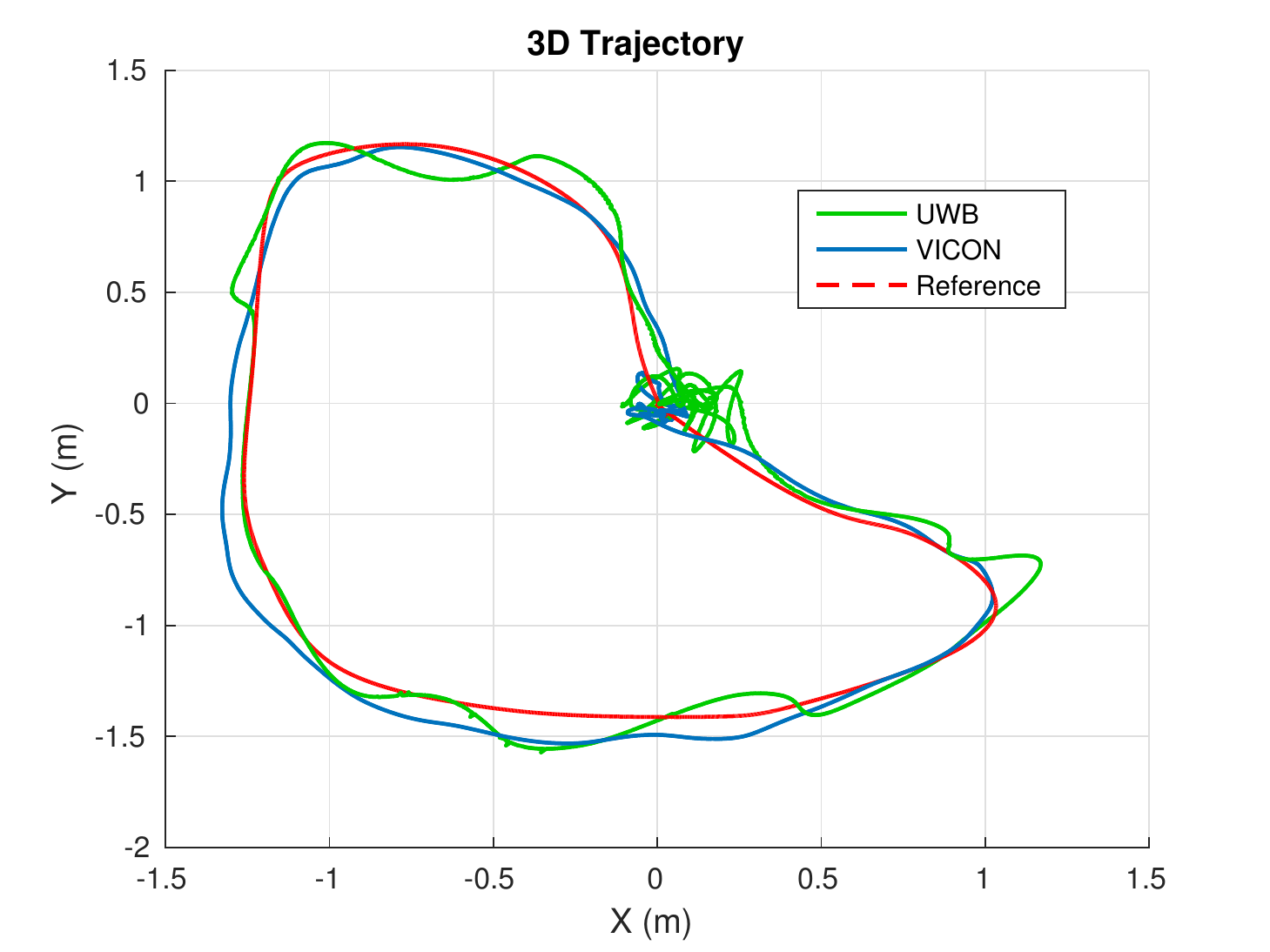}}}
\caption{\label{fig:Comparison-of-3D-1}Comparison of 3D Trajectory\#3.}
\end{figure}

\begin{itemize}
\item \textbf{Trajectory \#3}
\end{itemize}
The third trajectory time plots and 3D path are exposed in figures \Cref{fig:X-Y-Position-and-2,fig:Comparison-of-3D-1}. A similar behavior as before is appreciated, both flights had some level of success in following the commanded trajectory, but the error plots suggests a greater noise in the position while using the UWB system. Nonetheless, the top view in \Cref{fig:XY-Plane1} illustrate the trajectory
tracking capability in both flights.
~\\

The position errors reflected in \autoref{tab:Error-statistics-comparison-2}
suggest a similar performance in the X position, with a more pronounced
discrepancy in the Y position, in both cases being the VICON was the more
precise system.

\begin{table}[H]
\begin{centering}
\begin{tabular}{|c|c|c|c|c|}
\cline{2-5} 
\multicolumn{1}{c|}{} & $\textrm{RMS}\left(e_{x}\right)\,\textrm{[cm]}$ & $\textrm{RMS}\left(e_{y}\right)\,\textrm{[cm]}$ & $\%\xi_{x}$ & $\%\xi_{y}$\tabularnewline
\hline 
VICON & $7.56$ & $7.29$ & $84.69$ & $83.00$\tabularnewline
\hline 
UWB & $9.43$ & $8.11$ & $74.68$ & $78.66$\tabularnewline
\hline 
\end{tabular}
\par\end{centering}
\caption{\label{tab:Error-statistics-comparison-2}Error comparison
while following Trajectory \#3.}
\end{table}

The experimental data confirms the robustness of the control system while using a positioning system with around 100 times greater standard deviation noise than the initial system. The tuning of the Kalman Filter to adapt to this new source of noise was vital to ensure a good compromise between filtering and precision in the estimations. Even though there is a clear advantage on using more precise technology, such as the VICON, the system could still track the desired trajectories with an acceptable degree of precision while using a more cheaper system such as the UWB.~\\

The fact that the position integral gains for the X and Y positions had to be lowered while using the UWB system, was a necessary compromise to ensure less oscillations while following the desired trajectory. Such a compromise did not arise while using the VICON system, with which the quadcopter remained stable and followed smoothly the trajectories for a wide range of gain values.~\\

If the control system could only use the UWB system, then a more detailed study of how to compensate the different sources of noise and biases should be made. For instance, the UWB system looses precision when the tag is close to one of the anchors, or if the tag is facing away from one of the anchors. All this subtleties, if taken in account while designing the control system, could lead to a better performance than the one obtained and presented in this work.
~\\

Finally, the video found in \cite{key-38} shows a summary of the project's simulation and experimental results. 
\chapter{Conclusions and Future Work}

The study was set out to explore the dynamics of an open source nanoquadcopter
named Crazyflie 2.0, as well as creating a simulation environment
for control design and then testing it in the real platform. This
type of unmanned aerial vehicles is becoming the preferred platform
for testing control algorithms of diverse natures, thus the inherent
importance of conceiving a mathematical model of the vehicle that
can predict, up to some extent, how the system will evolve over time. Hence, the project started by a modeling
of the nanoquadcopter and an identification of certain physical parameters,
based in previous work. Working in parallel with the
literature and the quadcopter's embedded firmware was the main key
in describing the system behavior just as it is in the real platform,
an important milestone for future work as the dynamics of a system
is the heart of every simulation environment.

~\\
The second phase of the project was building the simulation that served as the first test-bench of the control architectures proposed. Using both the non-linear dynamics and the linearised
state space realisation of the system, the simulation created is a
solid testing environment to conceive all types of control systems.
It was incredibly useful during the first stages of the project to get a better understanding of how the system worked. In
addition, the simulation was used for designing both the PID position controller and the LQT trajectory tracker.~\\

An important conclusion is that the initial
belief that all dynamics were decoupled as suggested during the linear
modeling was not entirely true in the non-linear system. As observed
in the simulations, there exists some interference between movements
that, for instance, does not allow the quadcopter to describe a perfectly
straight line trajectory when there are more than one movement involved (a yaw rotation for example).

~\\
The position PID tracker was tested for time-varying
trajectories, such as circles and helices. Even though the system
could described these trajectories, there were some drawbacks and performance
issues, for example not getting fast enough to the desired points
which lead to errors in the desired trajectory. The fact
that the task at hand was managed by a position tracker and not a
trajectory tracker was the main reason of these discrepancies. To address the deficiencies of the PID controller, a new control system was conceived using the LQT algorithm, which proved to have interesting characteristics while following step responses, mainly that it started moving before the command was asked in order to reduce the tracking error. The feature was possible thanks to the off-line calculation of the algorithm and the knowledge of the trajectory beforehand.

~\\
The comparisons between the PID and the LQT controller indicate a clear superiority of the LQT in terms of reducing the trajectory tracking error, specially in the more demanding trajectories, in which the LQT algorithm reduced up to 4 times the RMS errors obtained with the PID controller. Directly related to the better tracking, the LQT incurred in higher levels of control effort than the PID, but it also eliminated the great command peaks seen in the motor time plots of the PID, thus getting rid of the undesired motor saturations that could lead to unstable states.
~\\

There are two main drawbacks of the LQT algorithm with respect to the PID: the first one is the inability to specify trajectories for the heading (yaw angle) and the second one is the need to know the trajectory before its execution. Taking in account these shortcomings, it is proposed as future work for this research to incorporate a method to control the yaw angle while keeping the good performance in the LQT algorithm, the author proposes a gain-scheduling method being the yaw angle the scheduling variable as a possible solution for this problem. As for the second drawback of the LQT algorithm, more research should be directed towards an on-line implementation thus making the controller useful in more complex tasks such as planning and execution missions in real-time.
~\\

The GUI created for trajectory generation proved to be a valuable asset to quickly test different types of trajectories, with varying difficulty. But the tool can be improved by adding physical constraints to the trajectory generation, as to assure the trajectory is feasible for the quadcopter to follow. Future work in this area should explore feasible trajectory generation as proposed in works such as \cite{key-25}.

~\\
The simulation versus experimental comparative time plots show that the simulation environment developed in this project was accurate to some extent, serving its purpose as a useful design tool for the controllers synthesized, but it had its limitations mainly due to unmodeled phenomena, which lead to the need of introducing high integral gains in the controllers to compensate the model errors and other perturbations of the system. As future work, it is suggested a more thorough model identification for the quadcopter, for example using numerical methods such as the closed-loop "black box" identification proposed in \cite{key-11}.
 ~\\
 
The Kalman Filter approach for estimating the linear velocities from the position data proved to be successful using both the VICON and the UWB, specially with the latter in which the data had 100 times greater standard deviation noise. The VICON versus UWB experiments suggest that in both cases the LQT tracked the desired position, but with obvious different levels of smoothness and precision. Even though both performances were satisfactory in terms of the scope of this work, future research into improving the control system while using the UWB position system would be ideal. Starting from an identification of different sources of added noise and biases of the UWB system, upto different filtering techniques that are more appropriate than the classic Kalman filter proposed in this project are the author's recommendations to improve the control system performance.
~\\

This work represents a solid base for future research using this platform,
with enough explanation in the calculus for newcomers in the area
to understand the basic functioning of the system. The simulation
environment was developed in a fashion that corresponds exactly with
the equations shown in the mathematical model, which helps in the
quick understanding of how everything works and saves time in comprehending an otherwise complex system, plus it is easily customizable for future users to develop their own controllers. The project successfully fulfilled its ultimate goal of characterizing the provided quadcopter platform and doing all the steps needed to develop an efficient control system for trajectory tracking.

\singlespacing

\chapter*{\hypertarget{appA}{}Appendix A: Firmware Modifications}

\addcontentsline{toc}{chapter}{Appendix A}

The firmware used during this project was ``Release 2016.02'' found
in \url{https://github.com/bitcraze/crazyflie-release/releases},
with the following changes:
\begin{itemize}
\item power\_distribution\_stock.c lines 58-64
\end{itemize}
\lstset{language=C++,
                basicstyle=\ttfamily,
                keywordstyle=\color{blue}\ttfamily,
                stringstyle=\color{red}\ttfamily,
                commentstyle=\color{green}\ttfamily,
                morecomment=[l][\color{magenta}]{\#}
}

\begin{lstlisting}[language=C,basicstyle={\small}]
#ifdef QUAD_FORMATION_X
int16_t r = control->roll / 2.0f;
int16_t p = control->pitch / 2.0f;
motorPower.m1 = limitThrust(control->thrust - r - p - control->yaw);
motorPower.m2 = limitThrust(control->thrust - r + p + control->yaw);
motorPower.m3 = limitThrust(control->thrust + r + p - control->yaw);
motorPower.m4 = limitThrust(control->thrust + r - p + control->yaw);
\end{lstlisting}

\begin{itemize}
\item controller\_pid.c lines 64-84
\end{itemize}
\begin{lstlisting}[language=C++,basicstyle={\tiny}]
attitudeControllerCorrectAttitudePID(state->attitude.roll, -state->attitude.pitch, state->attitude.yaw,
                                setpoint->attitude.roll, setpoint->attitude.pitch, attitudeDesired.yaw,
                                &rateDesired.roll, &rateDesired.pitch, &rateDesired.yaw);

//Bypass Attitude controller if Rate mode active
    if (setpoint->mode.roll == modeVelocity) {
      rateDesired.roll = setpoint->attitudeRate.roll;
    }
    if (setpoint->mode.pitch == modeVelocity) {
      rateDesired.pitch = setpoint->attitudeRate.pitch;
    }
    if (setpoint->mode.yaw == modeVelocity) {
      rateDesired.yaw = setpoint->attitudeRate.yaw;
    }

attitudeControllerCorrectRatePID(sensors->gyro.x, sensors->gyro.y, sensors->gyro.z,
                                 rateDesired.roll, rateDesired.pitch, rateDesired.yaw);

attitudeControllerGetActuatorOutput(&control->roll,
                                    &control->pitch,
                                    &control->yaw);
\end{lstlisting}
\end{document}